\documentclass[onecolumn,fleqn]{revtex4}

\usepackage{ifthen}
\usepackage{ifpdf}

\usepackage{latexsym}
\usepackage{amsmath} 
\usepackage{amssymb} 
\usepackage{bm}
\usepackage{wasysym}

\ifpdf
\usepackage{graphicx}
\usepackage{epstopdf}

\graphicspath{{./Figs/}{./}}
\else
\usepackage{graphicx}
\usepackage{epsfig}

\renewcommand{\includegraphics}[2][0]{FIGURE}
\fi

\usepackage[colorlinks,linkcolor=blue,urlcolor=blue,bookmarks=true]{hyperref}

\def\dbar{{\mathchar'26\mkern-12mu d}}

\newcommand{\const}{\mathrm{const}}
\newcommand{\trc}{\mathsf{trace}}
\newcommand{\ar}{\mathsf{r}}
\newcommand{\im}{\mathrm{Im}}
\newcommand{\re}{\mathrm{Re}}
\newcommand{\eexp}{\mathrm{e}^}

\newcommand{\mass}{\mathsf{m}} 
\newcommand{\gdos}{\mathsf{g}} 
\newcommand{\EPS}{\mathcal{E}}
\newcommand{\Vol}{\mathsf{V}}

\newcommand{\tbox}[1]{\text{#1}}
 
\newcommand{\amatrix}[1]{\begin{matrix} #1 \end{matrix}} 
\newcommand{\pd}[2]{\frac{\partial #1}{\partial #2}}

\newcommand{\bra}[1]{\left\langle #1 \right|}
\newcommand{\ket}[1]{\left| #1 \right\rangle}
\newcommand{\braket}[1]{\left\langle #1 \right\rangle }

\newcommand{\Dn}[1][0.5]{\vspace{#1cm}}
\newcommand{\bitem}{$\bullet$ \ \ \ }

\newcommand{\Cn}[1]{\begin{center} #1 \end{center}}
\newcommand{\mpg}[2][1.0\hsize]{\begin{minipage}[b]{#1}{#2}\end{minipage}}

\newcommand{\putgraph}[2][0.30\hsize]{\includegraphics[width=#1]{#2}}

\newcommand{\beq}{\begin{eqnarray}}
\newcommand{\eeq}{\end{eqnarray}}

\usepackage{tikz}

\newcommand{\ExternalLink}{    \tikz[x=1.2ex, y=1.2ex, baseline=-0.05ex]{        \begin{scope}[x=1ex, y=1ex]
            \clip (-0.1,-0.1) 
                --++ (-0, 1.2) 
                --++ (0.6, 0) 
                --++ (0, -0.6) 
                --++ (0.6, 0) 
                --++ (0, -1);
            \path[draw, 
                line width = 0.5, 
                rounded corners=0.5] 
                (0,0) rectangle (1,1);
        \end{scope}
        \path[draw, line width = 0.5] (0.5, 0.5) 
            -- (1, 1);
        \path[draw, line width = 0.5] (0.6, 1) 
            -- (1, 1) -- (1, 0.6);
        }
    }

\renewcommand{\thesection}{\arabic{section}}
\renewcommand{\thesubsection}{\arabic{subsection}}
\renewcommand{\thesubsubsection}{\arabic{subsubsection}}
\newwrite\tempfile
\immediate\openout\tempfile=sqm.txt
\newcommand{\writeout}[1]{\immediate\write\tempfile{#1}}

\newcommand{\sheadA}[1]
{
\addtocounter{section}{1}
\newpage 
\begin{center} 
\pdfbookmark[1]{#1}{shdA\thesection}
{\bf {\LARGE #1}} 
\end{center} 
\addtocontents{toc}{\protect\contentsline{section}{{\large #1}}{}{}}
\writeout{ }
\writeout{{\noexpand\  \noexpand\newline ======= \noexpand\Large #1. (page \thepage)}}
\phantomsection
}

\newcommand{\sheadB}[1]
{
\addtocounter{subsection}{1}
\setcounter{subsubsection}{0} 
\setcounter{equation}{0}
\pdfbookmark[2]{#1}{shdB\thesubsection}
{\bf\LARGE[\thesubsection] \ #1} 
\nopagebreak
\addtocontents{toc}{\protect\contentsline{subsection}{\thesubsection \ \ #1}{\thepage}{shdB\thesubsection.2}}
\writeout{ }
\writeout{{\noexpand\  \noexpand\newline  \noexpand\bf \noexpand\large #1: }}
\phantomsection
}

\newcommand{\sheadC}[1]
{
\addtocounter{subsubsection}{1}
\vspace{5mm}
{\Large\bf $=\!=\!=\!=\!=\!=\;$ [\thesubsection.\thesubsubsection] \ #1}  
\nopagebreak
\writeout{{#1; }}
\phantomsection
}

\begin{document} 

\title{Lecture Notes in Statistical Mechanics and Mesoscopics}

\author{Doron Cohen} 
 
\affiliation{Department of Physics, Ben-Gurion University, Beer-Sheva 84105, Israel} 

\makeatletter
\def\Dated@name{\,\!}
\makeatother
\date{\href{http://arxiv.org/abs/1107.0568}{\texttt{arXiv:1107.0568}}}

\begin{abstract} 
These are the \href{http://physics.bgu.ac.il/~dcohen/ARCHIVE/sqm.pdf}{lecture notes \ExternalLink} 
for quantum and statistical mechanics \href{https://physics.bgu.ac.il/~dcohen/courses/StatMech}{courses \ExternalLink} 
that are given by \href{https://physics.bgu.ac.il/~dcohen}{DC} at Ben-Gurion University.  
They are complementary to {\em Lecture Notes in Quantum Mechanics} [\href{http://arxiv.org/abs/quant-ph/0605180}{arXiv:quant-ph/0605180}].
Some additional topics are covered, including: introduction to master equations;
non-equilibrium processes; fluctuation theorems; linear response theory; 
adiabatic transport; the Kubo formalism; and the scattering approach to mesoscopics. 
\end{abstract} 

\maketitle

\vspace*{-2cm}

\renewcommand*{\tocname}{}
{\footnotesize \tableofcontents}

\Dn

{Detailed table of contents is available in the last two pages.}

\sheadA{Thermal Equilibrium}

\makeatletter{}\sheadB{The statistical picture of Mechanics}

Before we start discussing the {\em canonical formalism} of statistical mechanics, 
we would like to dedicate the first lecture for some preliminaries regarding:
Random variables and probability functions; 
The statistical picture of classical dynamics in phase space; 
The notion of chaos; Stationary states in general; 
and the canonical state in particular.   

{\em This lecture is quite terse, and possibly will be expanded in the future}.

\sheadC{Random variables}

Here is a list of topics that should be covered by a course in probability theory:
\beq
\text{Random variable/observation} & \ \ \ \ \ \ &  
\hat{x} 
\\
\text{Distribution function} &&  
\rho(x)
\\
\text{for discrete spectrum} &&  
\rho(x) \ \ \equiv \ \ \text{Prob}\left(\hat{x}=x\right)
\\
\text{for continuous spectrum} &&  
\rho(x)dx \ \ \equiv \ \ \text{Prob}\left(x<\hat{x}<x+dx\right)
\\
\text{Changing variables} &&  
\hat{y}=f\left(\hat{x}\right),  \ \ 
\tilde{\rho}\left(y\right)dy=\rho(x)dx
\\
\text{Expectation value of the random variable} &&  
\langle  \hat{x}\rangle \equiv \sum_x\rho(x)x
\\
\text{Expectation value of some other observable} && 
\langle  \hat{A}\rangle \equiv \sum_x\rho(x)A(x)
\\
\text{Variance} &&
\text{Var}(\hat{x}) 
= \langle\left(\hat{x}-\langle \hat{x} \rangle\right)^2\rangle
= \langle \hat{x}^2 \rangle - \langle \hat{x} \rangle^2 
\eeq
\beq
\text{Moment generating function} &\ \ \ \ \ \ & 
Z(\lambda) = \langle \eexp{\lambda \hat{x}} \rangle 
\\
\text{Comulant generating function is defined through}  && 
Z(\lambda) \equiv \exp[g(\lambda)]
\\
\text{Gaussian distribution, definition} &&
\rho(x) \propto \exp\left[-\frac{1}{2}\left(\frac{x-\mu}{\sigma}\right)^2\right]
\\
\text{Gaussian distribution, comulant} &&
g(\lambda) = \mu \lambda + \frac{1}{2}\sigma^2\lambda^2
\eeq

{\bf Legendre transform.-- } 
We can write the probability function as $\rho(x)=\exp(-F(x))$, 
and redefine the comulant generating function as ${G(\lambda)=-g(\lambda)}$. 
We have by definition
\beq
e^{-G(\lambda)} \ \ = \ \ \int_{-\infty}^{\infty}  e^{-F(x) \ + \ \lambda x} \ dx 
\eeq
If we are allowed to use a saddle point approximation, 
it follows that $G(\lambda)$ is related to $F(x)$ by a Legendre transform:
\beq
G(\lambda) \ \ \approx \ \  \min_{x} \Big\{ F(x) - \lambda x \Big\}
\ \ = \ \ F(\bar{x}) - \lambda \bar{x}
\eeq 
where the most probable value~$\bar{x}$ is determined by solving ${ \lambda = F'(x) }$. 
We shall see that this is formally the same mathematics 
as going from the Helmholtz to Gibbs free energy. Below we explain that the 
inverse of this relation is the large deviation theory.

\newpage
\sheadC{Several random variables}

In classical probability theory we can define a joint distribution function for random variables, 
and then characterize this distribution by correlation functions.   
\beq
\text{Joint distribution function of two variables} &&
\rho\left(x,y\right)
\\
\text{Correlation between two variables} && 
C_{xy} =  \langle \hat{x}\hat{y} \rangle -  \langle \hat{x} \rangle \langle \hat{y} \rangle
\eeq
In the quantum framework, known as ``measurement theory", 
it is not possible in general to define joint distribution function.
Instead one defines a probability matrix. See the lecture regarding
the first and the second quantum postulates 
in \href{http://arxiv.org/abs/quant-ph/0605180}{quant-ph/0605180}

If we have a sequence of random variable $\{ \hat{x}_j \}$ it is called a stochastic process, 
and the common notation for the correlation function is $C_{ij}$. 
For time-continuous process the notations is ${C(t',t'')}$ where $t'$ and $t''$ are the 
two ``sampling" times of the ``signal".

{\bf Adding random variables.-- } \\
\begin{minipage}[t]{0.3\hsize}
Adding two independent random variables:
\beq \nonumber
\hat{S} &=& \hat{x}+\hat{y} 
\\ \nonumber
\\ \nonumber
\langle\hat{S}\rangle &=& \langle\hat{x}\rangle+\langle\hat{y}\rangle 
\\ \nonumber
\text{Var}(\hat{S}) &=& \text{Var}(\hat{x}) + \text{Var}(\hat{y})
\\ \nonumber
g_s(\lambda) &=& g_x(\lambda) + g_y(\lambda)
\eeq
\end{minipage}
\hspace*{0.2\hsize} 
\begin{minipage}[t]{0.3\hsize}
Adding $N$ independent and identically distributed random variables:
\beq \nonumber
\hat{S} &=& \sum_{j=1}^{N} \hat{x}_j
\\ \nonumber
\langle\hat{S}\rangle &=& N\mu
\\ \nonumber
\text{Var}(\hat{S}) &=& N\sigma^2
\\ \nonumber
g_s(\lambda) &=& N g(\lambda) 
\eeq
\end{minipage}

The are two useful results for large $N$. 
One is the central limit theorem
and the other is the large deviation theory.

{\bf Central limit theorem.-- } 
We define the scaled variable 
\beq
\hat{y} \ \ \equiv \ \ \frac{\sum_j \hat{x}_j - N\mu}{\sqrt{N} \ \sigma}  
\eeq
The statement is that in the large $N$ limit it has 
a normal distribution with zero average and unit dispersion. 
This follows by taking the limit of 
\beq
g_y(\lambda) \ \ = \ \ N \left[  g\left(\frac{\lambda}{\sqrt{N}\sigma} \right) -  \frac{\lambda \mu}{\sqrt{N}\sigma}  \right]
\eeq

{\bf Large deviation theory.-- } 
Define the scaled variable ${\hat{x}=(1/N)\sum \hat{x}_j}$. Accordingly 
\beq
g_x(\lambda) \ \ = \ \ N g\left(\frac{\lambda}{N} \right) 
\eeq
The sloppy statement regarding its distribution is 
\beq
\rho(x) \ \ \sim \ \ e^{-Nf(x)}, 
\ \ \ \ \ \ \ \ \ \  
f(x) = \max_{\lambdabar} \{ \lambdabar  x - g(\lambdabar) \} 
\eeq
In order to prove this result note that $\Theta(x)<e^{\lambdabar x}$ for any positive $\lambdabar$. Consequently 
\beq
\mbox{Prob}(\hat{x}>x) 
\ \ = \ \ \braket{ \Theta\left[ \left(\sum \hat{x}_j\right) - Nx \right] }   
\ \ < \ \ \braket{ e^{\lambdabar \left[ \left(\sum \hat{x}_j\right) - Nx \right]} }
\ \ = \ \ e^{N \left( g(\lambdabar) - \lambdabar x \right)}
\eeq
A lowest bound is obtained by optimizing the value of ${\lambdabar \in [0,\infty]}$. A complementary inequality is obtained for ${\mbox{Prob}(\hat{x}<x)}$, where the value of ${\lambdabar \in [-\infty,0]}$ is optimized to get the lowest bound. Thus, the unconstrained optimization provides a lowest bound for 
${\tilde{\rho}(x) \equiv \min\{\mbox{Prob}(\hat{x}{<}x), \mbox{Prob}(\hat{x}{>}x)\}}$, 
which is asymptotically similar to $\rho(x)$.  
Note that the optimization parameter $\lambdabar$ is formally like $\lambda/N$, where $\lambda$ is conjugate to the random variable $\hat{x}$.

\sheadC{The statistical description of a classical particle}

The statistical state of a classical particle with one degree of freedom 
is described by a probability function:
\beq
\rho(x,p)  \frac{dxdp}{2\pi\hbar} 
\ \ \equiv \ \  
\text{PROB}
\left(x< \hat{x}< x+dx, p<\hat{p}< p+dp\right) 
\eeq
where the normalization is 
\beq
\iint\frac{dxdp}{2\pi\hbar} \ \rho\left(x,p\right) \ \ = \ \ 1 
\hspace{2cm} \text{[in the next lectures $\hbar=1$]}
\eeq
The generalization of this definition to the case of $d$~freedoms 
is straightforward with Planck cell volume $(2\pi\hbar)^d$.   
The expectation values of observables are defined in the usual way:
\beq
\langle A \rangle 
\ \ = \ \ 
\iint\frac{dxdp}{2\pi\hbar} 
\ \rho\left(x,p\right)
\ A(x,p)
\eeq
We note that in the quantum context one can define 
a quasi distribution that corresponds to $\rho(x,p)$, 
known as the {\em Wigner function}. Furthermore with 
any observable $\hat{A}$ we can associate a phase 
apace function $A(x,p)$ such that the expectation 
value can be calculated using classical look-alike formulas. 
This is known as the {\em Wigner-Weyl formalism}.
This formalism can be regraded as generalization of WKB:
Roughly speaking one may say that each Planck cell in phase space 
can be regarded as representing a quantum state. 
The volume of Planck cell is $(2\pi\hbar)^d$ where $d$ is 
the number of freedoms. Above we have assumed ${d=1}$.
Note that the normalization convention allows a sloppy
interpretation of $\rho\left(x,p\right)$ as the probability 
to occupy a Planck~cell in phase space.
We also remark that the quantum requirement ${\trc(\rho^2) \le 1}$ 
implies that a wavepacket in space space cannot occupy 
a volume that is less than a Planck~cell. 
The probability function of~$x$ is  
\beq
\rho(x) \ \ = \ \ \int\frac{dp}{2\pi} \rho\left(x,p\right) 
\eeq
The "spreading" of a wavepacket is characterize by 
\beq
\sigma_{x}^{2}
\ \ &\equiv& \ \  
\text{Var}(\hat{x}) 
\ \ = \ \ \langle\left(\hat{x}-\langle \hat{x} \rangle\right)^2\rangle
\ \ = \ \ \langle \hat{x}^2 \rangle - \langle \hat{x} \rangle^2 
\\
\sigma_{p}^{2}
\ \ &\equiv& \ \  
\text{Var}(\hat{p}) 
\ \ = \ \ \langle\left(\hat{p}-\langle \hat{p} \rangle\right)^2\rangle
\ \ = \ \ \langle \hat{p}^2 \rangle - \langle \hat{p} \rangle^2 
\eeq
In the quantum context ${\sigma_x\sigma_p > (\hbar/2)}$.
The "energy" of the system is defined as follows: 
\beq
E \ \ = \ \ \langle \mathcal{H}\left(\hat{x},\hat{p}\right) \rangle \ \ = \ \ 
\iint\frac{dxdp}{2\pi\hbar} 
\ \rho\left(x,p\right)
\ \mathcal{H}(x,p) 
\eeq
Later we shall define some other "spectral" functions that are related to $\mathcal{H}$. 
Those can be written as an expectation value of functions of $\mathcal{H}$.

\sheadC{Dynamics in phase space}

The difference between ``classical mechanics" and ``classical statistical mechanics" 
parallels the distinction between ``Heisenberg picture" and ``Schrodinger picture" 
in quantum mechanics. The former describes the evolution of the system 
using a set of dynamical variables that obey some equations of motion, 
while the latter describe the evolution of the associated probability function. 
In order to make the above distinction clear we consider the simplest 
example: a free particle. The Hamiltonian is 
\beq
\mathcal{H} \ \ = \ \ \frac{p^2}{2\mathsf{m}} + V(x),
\hspace{2cm} \text{for free particle} \ V(x)=0 
\eeq
Say that at ${t=0}$ the particle is at ${ (x_0,p_0) }$.
The equations of motion are 
\beq
\dot{x} &=& \frac{\partial{\mathcal{H}}}{\partial{p}} \ \ = \ \ \frac{p}{\mathsf{m}} 
\\
\dot{p} &=& -\frac{\partial{\mathcal{H}}}{\partial{x}} \ \ = \ \ 0 
\eeq
The solution is:
\beq
x(t) &=& x_0 + \frac{t}{\mathsf{m}} p_0  
\\
p(t) &=& p_0 
\eeq
In the Heisenberg picture we regard $\hat{x}_0$ and $\hat{p}_0$ 
as random variables that have some probability function $\rho\left(x,p\right)$.
Then we define new random variables
\beq
\hat{x}_{t} &=& \hat{x}_{0} + \frac{t}{\mathsf{m}} \hat{p_0}  
\\
\hat{p}_{t} &=& \hat{p_0} 
\eeq
It follows from the composition law of random variables 
that there is spreading in space as a function of time:
\beq
\sigma_{x}\left(t\right)
\ \ = \ \ \sqrt{
\sigma_{x}^{2}\left(0\right)
+\left( \frac{\sigma_{p}\left(0\right)}{\mathsf{m}} \right) t^2}
\ \ \sim \ \ 
\frac{\sigma_p(0)}{\mathsf{m}} t
\eeq
It should be clear that ``spreading" is a classical effect
that originates if we assume that there is some dispersion
in the momentum. In quantum mechanics this effect is unavoidable 
because preparations with zero dispersion are non-physical.     

In the optional Schrodinger picture we define $\rho_t\left(x,p\right)$
as the probability distribution of $\hat{x}_t$ and $\hat{p}_t$.
So instead of talking about the time evolution of $\hat{x}$ and $\hat{p}$ 
we talk about the time evolution of $\rho\left(x,p\right)$.  
In statistical mechanics we prefer the latter point of view. 
Evolution takes place in phase space. Liouville theorem applies.
Let us see how we use the ``Schrodinger picture" in the above example. 
Assume that the free particle has been prepared in a ``classical pure state" 
at the point ${(x_0,p_0)}$ in phase space. Accordingly 
\beq
\rho_{t=0}\left(x,p\right) 
\ \ = \ \ 2\pi \delta\left(p-p_0\right) 
\ \delta\left(x-x_0\right) 
\eeq
After time $t$ the state is 
\beq
\rho_{t}\left(x,p\right) 
\ \ = \ \ 2\pi\delta\left(p-p_0\right) 
\ \delta\left(x-\left(x_0+\frac{p_0}{\mathsf{m}}t\right)\right) 
\eeq
If the preparation is not a ``classical pure state", but say a Gaussian wave-packet
that has some finite momentum spread~$\sigma_p$,   
then one observes spreading as explained previously.
More generally we can discuss the spreading of a wavepacket 
in the case of a non-linear oscillator. In such case $V(x)$ 
has either {\em sub-quadratic} or {\em super-quadratic} variation, 
and consequently the oscillation frequency $\omega(E)$ 
depends on the energy: decreases or increases with energy respectively.   
If the initial distribution has some finite spread $\sigma_E$ in energy, 
there will be angular spreading that leads to a quasi-ergodic 
distribution within the energy shell. It is not really ergodic 
because if we started with a mono-energetic distribution (${\sigma_E=0}$) 
it would not fill uniformly the energy surface: here 
the energy surface is merely a one-dimensional ``ellipse". 
For graphical illustrations see figures in the next section.

\newpage
\sheadC{The route to ergodicity}

Let us outline some major observations with regard 
to the dynamics of classical Hamiltonian systems. 

{\bf Simple 1D system:-- } 
The student is expected to be familiar with 
the dynamics of harmonic oscillator; potential well; pendulum.  
In the case of non-linear oscillations we have the 
{\em spreading} effect. In the case of a pendulum 
we have a multi-component phase space with separatrix. 
The dynamics is not chaotic. One can define the 
oscillation frequency $\omega(E)$ as a function of energy.
In the quantum case $\omega(E)$ corresponds to 
the level spacing at the vicinity of the energy~$E$.  
  
{\bf Chaotic system:-- } 
The student is expected to be familiar with 
the dynamics in simple billiards. The visualization 
can be achieved using a Poincare section.
In the case of a Sinai billiard (motivated by the discussion 
of Lorentz gas) the dynamics is fully chaotic, leading to ergodization.
More generally we might have {\em mixed phase space}
that contains "chaotic sea" as well as "islands".  

{\bf Ergodization:-- } 
The evolution of a chaotic system leads to an ergodization 
on the energy shell. This can be mathematically described 
using the Boltzamnn approach: course graining of phase space 
by dividing it into cells; definition of Boltzamnn entropy.
Eventually the system will become stationary-like, as if it 
were prepared in a state that has maximum entropy. 

{\bf Driven system:-- } 
There is a complicated route to chaos in the case 
of driven integrable (1D) systems. In contrast 
to that in the case of driven globally chaotic systems
the picture is qualitatively simple: 
if we prepare the system initially within an energy shell, 
it will "evolve" with this energy shell, along with    
diffusion transverse to the energy shell. This diffusion 
leads in general to increase of the average energy (heating). 

\ \\

\mpg[0.45\hsize]{
\includegraphics[width=\hsize]{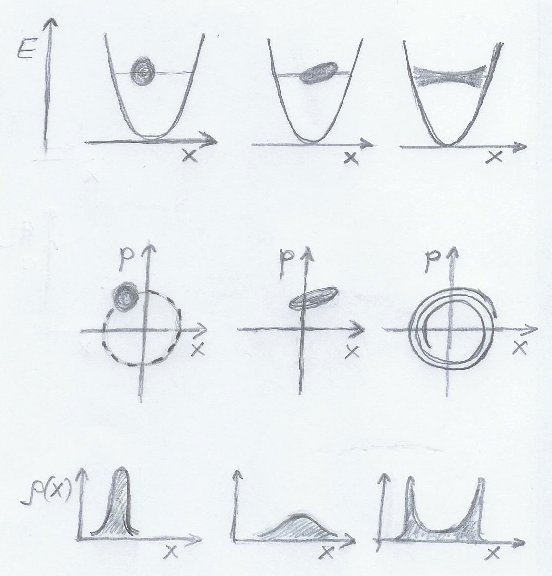} 

{\bf Spreading illustration. -- } 
We consider the evolution of an initial Gaussian 
distribution (left panels) in the case of  
a non-linear oscillator. 
After a short time (middle panels) 
the spreading is like that of a free particle. 
After a long time (right panels) one observes an 
ergodic-like distribution within the energy shell. 
However, this is not really ergodic: if we started with 
a mono-energetic distribution, it would remain 
localized on the energy shell, as in the case of 
an harmonic oscillator. 
}
\hspace{0.1\hsize}
\mpg[0.45\hsize]{
\Cn{\includegraphics[width=0.7\hsize]{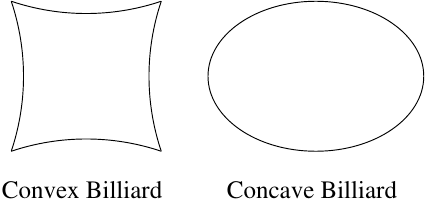}} 
\includegraphics[width=\hsize]{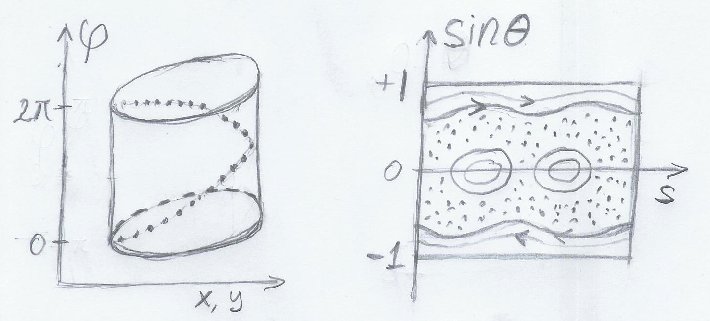} 

{\bf Phase space illustration. -- }
The dynamics of a particle in a convex (Sinai) Billiard 
is completely chaotic. In contrast to that, in the case of 
a concave billiard, we have a mixed phase space that contains 
both quasi-integrable regions and chaotic sea.     
The phase space is 3-dimensional ${(x,y,\varphi)}$ 
where $\varphi$ is the direction of the velocity.
It is illustrated in the left lower panel.  
The dotted line indicates the normal direction on the boundary.
The reflections are specular with regard to this 
direction.  The right lower panel is the two-dimensional ${(s,\theta)}$ 
Poincare section of phase space: each trajectory 
is represented by a sequence of points that 
indicate successive collisions with the boundary, 
where $s$ is the boundary coordinate, 
and $\theta$ is the collision angle (relative to the normal).  
}

\newpage
\sheadC{Stationary states}

The evolution of a statistical state is determined by the Lioville 
equation of classical mechanics, which becomes the von-Neumann Lioville
equation in quantum mechanics.
\beq
\frac{\partial{\rho}}{\partial{t}} \ \ = \ \ [\mathcal{H},\rho]_{\text{PB}} 
\eeq
We consider non-driven bounded systems, and focus 
our attention on {\em stationary} states that do not 
change in time. This means $\partial{\rho}/\partial{t}=0$. 
In the classical language $\rho$ can be regarded 
as a mixture of "energy shells", while in the quantum   
language it means that $\rho \mapsto \text{diag}\{p_r\}$
is a mixture of energy eigenstates labelled by~$r$. 
In particular the classical microcanonical state 
corresponds to an energy eigenstate, and is formally 
written as   
\beq
\rho(x,p) \ \ = \ \ \frac{1}{\gdos(\mathcal{E})} \delta(\mathcal{H}(x,p)-\mathcal{E}) 
\eeq
The canonical state is 
\beq
p_r \ \ = \ \ \frac{1}{Z(\beta)} \ \eexp{-\beta E_r} 
\eeq
and in a classical context it is written as  
\beq
\rho(x,p) \ \ = \ \ \frac{1}{Z(\beta)} \ \eexp{-\beta \mathcal{H}(x,p)} 
\eeq
The density of states and the
partition function are defined as 
\beq
\gdos(E) \ &=& \ \trc(\delta(E-\mathcal{H})) \ \ = \ \ \sum_r \delta(E-E_r) \\
Z(\beta) \ &=& \ \trc(\eexp{-\beta \mathcal{H}}) \ \ = \ \ \sum_r \eexp{-\beta E_r}
\ \ = \ \ \int \gdos(E)dE \ \eexp{-\beta E}
\eeq
We note that the probability distribution of the energy 
can be written as ${\rho(E)=\gdos(E) \ f(E)}$, 
where the occupation probability function is ${f(E)\propto\delta(E-\mathcal{E})}$
and ${f(E)\propto \eexp{-\beta E}}$ in the microcanonical 
and canonical cases respectively.  
If we have a many body system of non-interacting participles 
we can re-interpret $f(E)$ as the occupation function, and 
accordingly $\rho(E)$ becomes the energy distribution of the 
particles (with normalization $N$).

\sheadC{The microcanonical and canonical states}

Let us assume the following total Hamiltonian 
for a universe that consists of system and environment: 
\beq
\mathcal{H}_{\text{total}}
\ \ = \ \
\mathcal{H}\left(Q\right)
+\mathcal{H}_{\text{env}}\left(Q_{\alpha}\right)
+\mathcal{H}_{\text{int}}\left(Q, Q_{\alpha}\right)
\eeq
For sake of presentation we do not write 
the conjugate momenta, so $Q$ stands for $(Q,P)$ 
or it may represent spin freedoms.
If one neglect the interaction the eigenenergies  
are written as ${E_{rR}=E_r+E_R}$, where $r$ labels 
system states and $R$ labels environmental states.

It is argued that the weak interaction 
with the environment leads after relaxation to a canonical state 
which is determined by the parameter 
\beq
\beta \ \ = \ \ \frac{d}{dE} \log(\gdos_{\tbox{env}}(E))
\eeq
where $\gdos_{\tbox{env}}(E)$ is the density of states,
which is assumed to be huge and rapidly growing with energy.  
The argument is based on the assumption 
that the universe (system+environment) 
is (say) a closed system with some total energy $E_{\tbox{total}}$.
After ergodization the system get into 
a stationary-like state that resembles 
a microcanonical states:  
\beq
p_{r,R} \ \  \propto \ \ \delta\Big(E_{\tbox{total}} - (E_r + E_R)\Big)
\eeq
with finite width (broadened) delta function. 
The probability $p_r$ to find the system 
in a state $E_r$ is proportional to 
${\gdos_{\tbox{env}}(E_{\tbox{total}}{-}E_r) 
\approx \gdos_{\tbox{env}}(E_{\tbox{total}})\eexp{-\beta E_r}}$. 
Accordingly 
\beq
p_{r} \ \ = \ \ \frac{1}{Z}\eexp{-\beta E_r}
\eeq
where the so-called partition function 
provides the normalization
\beq
Z(\beta) \ \ = \ \ \sum_{r}\eexp{-\beta E_{r}}
\eeq
The partition function may depend on parameters 
that appear in the system Hamiltonian. Therefore we 
use in general the notation $Z(\beta,X)$.

\sheadC{Mathematical digression}

Sometimes is is more appropriate to expand the log of a function.
Specifically this would be the case if the function is definite 
positive and span many decades. Let us see what is the error 
which is involved in such an expansion: 
\beq
&& f(x) = x^{N} 
\\
&& f\left(x+\delta x\right) = x^{N}+Nx^{N-1}\delta x + \frac{1}{2}N(N-1)x^{N-2} \delta x^{2}
\\
&& \delta x \ll x/N 
\eeq
Optionally we expand the log of the function:
\beq
&& S(x) \equiv \ln f(x)=N\ln(x) 
\\
&& S\left(x+\delta x\right)=N\ln(x) + \frac{N}{x}\delta x - \frac{1}{2} \frac{N}{x^{2}} \delta x^{2} 
\\
&& \delta x \ll x 
\eeq
Thus we have the recipe:
\beq
f\left(x+\delta x\right) \ \ \approx \ \ f(x) \eexp{\beta \delta x} 
\hspace{2cm} \text{where} \ \ 
\beta \ \ \equiv \ \ \frac{\partial \ln f(x)}{\partial x} 
\eeq
In complete analogy we have:
\beq
\mathsf{g}\left(E_{0}+\epsilon\right)
\ \ \approx \ \ 
\mathsf{g}\left(E_{0}\right)\eexp {\beta \epsilon} 
\eeq
where $\beta$ is the log derivative of the density of states.

\newpage
\makeatletter{}\sheadB{Spectral functions}

Various types of spectral functions are defined in mathematical physics. 
In the quantum context they characterize the spectrum $\{E_n\}$ of energies 
of as given Hamiltonian $\mathcal{H}$. In the continuum or classical limit    
it is essential to define a {\em measure}. Below we focus on the most 
popular spectral functions in statistical mechanics: the density of states $\gdos(E)$, 
and the partition function $Z(\beta)$. We shall see later that the state equations 
of a system in equilibrium can be derived from, say, the partition function.
Hence the spectral function serves as a {\em generating~function}.      

In the section below we define $\gdos(E)$ and $Z(\beta)$, 
and show how they are calculated using standard examples: 
Two level system; 
Harmonic oscillator; 
Particle in a box; 
Particle with general dispersion relation; 
The effect of ${A(x), V(x)}$ potential; 
Several particles; 
Identical classical particles, the Gibbs factor;  
Particles with interactions; 
In particular two quantum particles; 
Molecules of type AA and AB (exercise).

\sheadC{The definition of counting and partition functions}

We consider a time independent bounded system which is described 
by a Hamiltonian ${\mathcal{H}}$ whose eigenvalues are ${E_{r}}$.
We can characterize its energy spectrum by the functions  
\beq
\mathcal{N}(E)
\ \ &\equiv& \ \ 
\sum_{r} \Theta\left(E-E_{r}\right)
\ \ = \ \  
\sum_{E_r<E} 1
\\ 
Z(\beta)
\ \ &\equiv& \ \ 
\sum_{r}\eexp{-\beta E_r} 
\eeq
If we have a large system we can smooth $\mathcal{N}(E)$, 
and then we can define the density of states as  
\beq
\gdos(E)
\ \ \equiv \ \ 
\frac{d\mathcal{N}(E)}{dE}
\ \ = \ \ 
\overline{\sum_{r}\delta\left(E-E_{r}\right)}
\eeq
Note that 
\beq
Z(\beta) 
\ \ = \ \ 
\int \gdos(E)dE \ \eexp{-\beta E}
\eeq
For a classical particle in 1D we can use the above definitions with the prescription  
\beq
\sum_{r} \ \ \longmapsto \ \ 
\iint \frac{dxdp}{2\pi\hbar}
\eeq
Each "Planck cell" in phase space represents a state. Accordingly 
\beq
\mathcal{N}(E)
\ \ &=& \ \ 
\iint\frac{dxdp}{2\pi}
\Theta\left(E-\mathcal{H}\left(x,p\right)\right)
\ \ = \ \ 
\iint_{\mathcal{H}(x,p)<E} \frac{dxdp}{2\pi} 
\\ 
Z(\beta)
\ \ &=& \ \ 
\iint\frac{dxdp}{2\pi} \eexp{-\beta \mathcal{H} \left(x,p\right)} 
\ \ = \ \ 
\int \mathsf{g}(E)dE \ \eexp{-\beta E} 
\eeq
In what follows the Gaussian integral is useful:
\beq
\int \eexp{-\frac{1}{2}ax^{2}}dx
\ \ = \ \ 
\left(\frac{2\pi}{a}\right)^{\frac{1}{2}}
\eeq

\sheadC{Two level system or spin}

The Hamiltonian of spin 1/2 in magnetic field is
\beq
\mathcal{H} \ \ = \ \ \frac{1}{2} h \sigma_{z}
\eeq
The eigenstates are ${ |+\rangle }$  and  ${ |-\rangle }$
with eigenvalues ${ E_{\pm}=\pm h/2}$. Accordingly 
\beq
Z(\beta)
\ \ = \ \  
\eexp{-\beta \left(-\frac{h}{2}\right)}
+ \eexp{-\beta\left(\frac{h}{2}\right)}
\ \ = \ \ 
2\cosh\left(\frac{1}{2}\beta h\right)
\eeq
Optionally we can write the energies of 
any two level system as ${E_{r}=\epsilon n}$ with $n=0,1$ then 
\beq
Z(\beta) = \left(1+\eexp{-\beta\epsilon}\right) 
\eeq

\sheadC{Two spins system in interaction}

If we have $N$ is interacting spins the sum over states can be factorized 
and we simply get
\beq
Z_N(\beta) \ \ = \ \ \Big(Z_1(\beta)\Big)^N
\eeq
For two spins in the absence of magnetic field we get ${Z_2=2^2=4}$.
Let us see what happens if there is an interaction:
\beq
\mathcal{H} \ = \ \varepsilon \sigma^{a}\cdot \sigma^{b} \ \ = \ \ \left(2S^{2}-3\right)\varepsilon, 
\hspace*{1cm} S=\frac{1}{2}\sigma^{a}+\frac{1}{2}\sigma^{b}
\eeq
The energy levels are ${E_{singlet}=-3\varepsilon}$
and ${E_{triplet}=\varepsilon}$. With and added magnetic field
the partition function is  
\beq
Z(\beta) \ \ = \ \ \eexp{3\beta\varepsilon}+ \left[\eexp{\beta h} + \eexp{-\beta h}  +1 \right]\eexp{-\beta\varepsilon}
\eeq
which factorized for ${\varepsilon=0}$, but not in general.

\sheadC{Harmonic oscillator}

The Hamiltonian of Harmonic oscillator is 
\beq
\mathcal{H}=\frac{p^{2}}{2\mass}+\frac{1}{2}\mass\omega^2 x^{2}
\eeq
The eigenstates are ${ |n\rangle }$ 
with eigenvalues ${E_{n}=\left(\frac{1}{2}+n\right)\omega}$. 
Accordingly 
\beq
Z(\beta)
\ \ = \ \ 
\sum_{n=0}^{\infty}
\eexp{-\beta\left(\frac{1}{2}+n\right)\omega}
\ \ = \ \ 
\frac{1}{2\sinh \left(\frac{1}{2}\omega\beta\right)}
\eeq
Note that if we write the energies 
as  ${E_{r}=\epsilon n}$ with $n=0,1,2,3,4,...$ then 
\beq
Z(\beta) = \frac{1}{1-\eexp{-\beta\epsilon}} 
\eeq

Now let us see how the classical calculation is done.
\beq
\mathcal{N}(E) 
\ \ &=& \ \ 
\frac{1}{2\pi}
\ \text{ellipse area}
\ \ = \ \ 
\frac{1}{2\pi}
\ \pi\left(\frac{2E}{\mass \omega^2}\right)^{\frac{1}{2}}
\ \left(2\mass E\right)^{\frac{1}{2}}
\ \ = \ \ 
\frac{E}{\omega}
\\
Z
(\beta)
\ \ &=& \ \ \int dx 
\ \eexp{-\beta \frac{1}{2}\mass x^{2}}
\int \frac{dp}{2\pi}
\ \eexp{-\beta\frac{p^{2}}{2\mass}}
\ \ = \ \ 
\left(\frac{2\pi}{\beta \mass \omega^{2}}\right)^{\frac{1}{2}}
\left(\frac{\mass}{2\pi\beta}\right)^{\frac{1}{2}}
\ \ = \ \ 
\frac{T}{\omega}
\eeq
One can verify the validity of WKB quantization.

\sheadC{Particle in a 1D box}

The simplest is to assume periodic boundary conditions
\beq
\mathcal{H} \ \ = \ \ \frac{p^{2}}{2\mass} \ \ \ \ \ \ x\in[0,L] \ \  \ \text{(ring)} 
\eeq
The eigenstates are the momentum states ${ |p\rangle }$ with 
\beq
p \ \ = \ \ \frac{2\pi}{L}n
\ \ \ \ \ \
\text{where} \ n=0,\pm1,\pm2... 
\eeq
Hence the eigenvalues are
\beq
E_n \ \ = \ \ \frac{1}{2\mass}\left( \frac{2\pi}{L} n \right)^{2}
\eeq
The number of states up to energy $E$ is 
\beq
\mathcal{N}(E)
\ \ = \ \ 
2\frac{L}{2\pi}\left(2\mass E\right)^{\frac{1}{2}}
\ \ \equiv \ \  \frac{1}{\pi} k_E L 
\ \ \equiv \ \  2 \frac{L}{\lambda_E}
\eeq
The density of states is  
\beq
\gdos(E)
\ \ = \ \ 
\frac{L}{\pi v_E}
\eeq
The 1D case here is pathological because in general
the density of states grows rapidly with energy.
Nevertheless in the limit of "infinite volume" we may 
treat the spectrum as a continuum: 
\beq
Z(\beta)
\ \ = \ \ 
\sum_{n=-\infty}^{\infty}
\eexp{-\beta E_n} 
\ \ \approx \ \ 
\int_{-\infty}^{\infty}dn
\ \eexp{-\beta \frac{1}{2\mass}\left(\frac{2\pi}{L}\right)^2 n^{2}} 
\ \ = \ \ 
L \left(\frac{\mass}{2\pi\beta}\right)^{\frac{1}{2}}
\ \ \equiv \ \
\frac{L}{\lambda_T}
\eeq

Let us see how the calculation is carried out classically.
We can still consider a ring, or optionally 
we can write the Hamiltonian with a box potential $V_{L}(x)$.
Then we get  
\beq
\mathcal{N}(E)
\ \ &=& \ \ \text{rectangle area}
\ \ = \ \ \frac{1}{2\pi} \times L \times 2\left(2\mass E\right)^{\frac{1}{2}}
\ \ = \ \  2\frac{L}{\lambda_E} 
\\
Z(\beta)
\ \ &=& \ \ 
\int dx \int \frac{dp}{2\pi} \ \eexp{-\beta\frac{p^2}{2\mass}}
\ \ = \ \ L\left(\frac{\mass}{2\pi\beta}\right)^{\frac{1}{2}}
\ \ = \ \ \frac{L}{\lambda_T}
\eeq
One can verify the validity of WKB quantization (but without the $1/2$ shift).

\sheadC{A particle in 3D box, or higher dimensions}

Consider a particle in a $d{=}3$ box of volume $\Vol=L^d$.
\beq
\mathcal{H}
\ \ = \ \ \sum_{i=1}^{3}\frac{p_{i}^{2}}{2\mass} 
\ \ + \ \ [\text{implicit boundary conditions with volume} \ L^d]
\eeq
The eigenstates are 
\beq
\vec{p} \ \ &=& \ \ \frac{2\pi}{L}\left(n_{1},n_{2},n_{3}\right)
\\ 
E_{n_{1}n_{2}n_{3}}
\ \ &=& \ \ 
\frac{1}{2\mass}
\left(\frac{2\pi}{L}\right)^{2}
\left(n_{1}^{2} + n_{2}^{2} + n_{3}^{2}\right)
\eeq
The summation over the states factorizes:
\beq
Z(\beta)
\ \ = \ \ \sum_{n_{1}n_{2}n_{3}}
\eexp{-\beta E_{n_{1}n_{2}n_{3}}}
\ \ = \ \ \left(\sum_{n}\eexp{-\beta E_{n}}\right)^{3}
\ \ = \ \ \frac{\Vol}{\lambda_T^3} 
\eeq
The above calculation gives $Z = (L/\lambda_T)^d$ in $d$ dimensions.
For the counting function we get: 
\beq
\mathcal{N}(E) \ \ = \ \ 
\frac{1}{(2\pi)^{d}}
\frac{\Omega_d}{d}
\left(k_E L\right)^{d}
\ \ = \ \ 
\left\{ \amatrix{ 2 \cr \pi \cr 4\pi/3 } \right\} 
\left(\frac{L}{\lambda_{E}}\right)^{d}
\eeq
and accordingly 
\beq
\gdos(E)
\ \ = \ \ 
\frac{\Omega_d}{(2\pi)^{d}}
\left(k_E L\right)^{d-1}
\frac{L}{v_E}
\ \ \propto \ \ 
E^{(d/2)-1}
\eeq
The factor $\left(k_E L\right)^{d-1}$ can be interpreted 
as the number of open modes. For $d=2$ the DOS is 
independent of energy and reflects the mass of the particle. 

As far as the {\em classical} calculation is concerned, 
$N$~particle systems is formally like one~particle system 
with ${d \mapsto Nd}$. In the quantum treatment 
the Fermonic or Bosonic nature of identical particles 
should be taken into account: see later how the calculation 
is done e.g. for two particles).

\sheadC{Classical particle in magnetic field}

For a particle in an arbitrary scalar potential $V(r)$  in 3D we get 
\beq
Z(\beta)
\ \ = \ \ \int\frac{dr \ dp}{(2\pi)^3} \ \eexp{-\beta\mathcal{H}}
\ \ =\ \ \left(\frac{1}{\lambda_T}\right)^3
\int dr \ \eexp{-\beta V(r)}
\eeq
Let us include also a vector potential:
\beq
\mathcal{H}=\frac{1}{2\mass}\left(p-A(r)\right)^{2} +V(r)
\eeq
\beq
Z \ \ = \ \ 
\int \frac{dr \ dp}{\left(2\pi\right)^{3}}
\ \eexp{-\beta\left[\frac{1}{2\mass}\left(p-A(r)\right)^{2}
+V(r)\right]}
\ \ = \ \ 
\int \frac{dr \ dp'}{\left(2\pi\right)^{2}}
\ \eexp{-\beta\left[\frac{1}{2\mass}\left(p'\right)^{2}
+V(r)\right]}
\eeq
The result does not depend on ${A(r)}$. 
The energy spectrum is not affected from 
the existence of ${A(r)}$.  
The energies are ${E=(1/2)\mass v^{2} + V(r)}$ 
irrespective of ${A(r)}$.
This is no longer the case upon quantization.
Note the implicit assumption of having background 
relaxation processes that make the dynamics irrelevant.

\sheadC{Gas of classical particles in a box}

Let us consider $N$ particles:
\beq
\mathcal{H}
\ \ = \ \ \sum_{\alpha=1}^{N}
\left[
\frac{\vec{p}_{\alpha}^{2}}{2\mass}
+V(r_{\alpha}) 
\right]
\ + \  U(r_1,...,r_N)
\eeq
In the absence of interaction the partition function is 
\beq
Z_{N}(\beta)
\ \ = \ \ 
\Big(Z_{1}(\beta)\Big)^{N}
\ \ = \ \ 
\left[ 
\frac{1}{\lambda_T^3}
\int d^{3}r \ \eexp{-\beta V(r)}
\right]^{N}
\eeq
From now on we assume gas of identical particles 
and therefore include the Gibbs factor:
\beq
Z_{N}(\beta) \ \ \mapsto \ \ \frac{1}{N!}Z_{N}(\beta)
\eeq
For ${N}$ interacting particles we get
\beq
Z_{N}(\beta)
\ \ = \ \ 
\frac{1}{N!}
\left(\frac{1}{\lambda_T^{3}}\right)^N
\int dr_{1}...dr_{N} \ \eexp{-\beta U(r_{1},...,r_{N})}
\eeq

\sheadC{Two quantum identical particles}
   
Let us see what is the partition function for 
a system that consists of two identical particles, 
say in a box. The total energy is written as $E_{ab}=E_{a}+E_{b}$.
The partition function is  
\beq
Z(\beta)
\ \ &=& \ \ \frac{1}{2}\sum_{a \neq b}
\eexp{-\beta \left(E_{a}+E_{b}\right)}
+\left\{\amatrix{1 \cr 0} \right\}
\sum_{a}\eexp{-\beta\left(2E_{a}\right)}
\\
\ \ &=& \ \ 
\frac{1}{2}\left(\sum_{a,b}
\eexp{-\beta \left(E_{a} + E_{b}\right)}
\pm \sum_{a}\eexp{-2\beta E_{a}}\right)
\ \ = \ \ \
\frac{1}{2}\Big[
Z_{1}(\beta)^{2}\pm Z_{1}\left(2\beta\right)
\Big]
\eeq
For a particle in a $d$ dimensional box 
\beq
Z_{1} \ &=& \ \left(\frac{L}{\lambda_{T}}\right)^{d}
\\
Z_{2} \ &=& \ \frac{1}{2}\left(Z_{1}^{2} \pm 2^{-d/2} Z_{1}\right)
\eeq
Note that for $d=3$ we get 
\beq
Z_{2}(\beta) \ \ = \ \ \frac{1}{2}Z_{1}^{2} \times 
\left[1\pm \frac{1}{2^{{3}/{2}}} \left(\frac{\lambda_{T}^{3}}{\text{volume}}\right)\right]
\eeq
The Fermi case is similar to hard sphere:
\beq
Z_{2}(\beta) \ \ = \ \ \frac{1}{2}Z_{1}^{2} \times 
\left[1-\left(\frac{\text{sphere volume}}{\text{box volume}}\right)\right]
\eeq

\sheadC{Two quantum particles in a box with interaction}

The calculation of the partition function $Z_2$ for 
two identical quantum particle in a box, 
is both interesting and later on useful 
for the purpose of calculating the 
second virial coefficient of an $N$~particle gas.
The Hamiltonian is:
\beq
\mathcal{H}=\frac{P^2}{4\mass}+\frac{p^2}{\mass} + V(r)
\eeq
In order to be able to do the calculation using separation of variables 
we cheat with the boundary conditions as follows: 
The center of mass motion is confined to a box of volume $\Vol=(4\pi/3)R^3$, 
and the relative motion is confined by $|r|<R$ independently. 
Accordingly the partition function is factorizes as follows:
\beq
Z_{2} \ \ = \ \ 
\left(2^{3/2} \frac{\Vol}{\lambda_T^{3}}\right)
\left[
\sum_{n\ell m}^{'} \eexp{-\beta E_{n\ell m}}
\right]
\ \ = \ \ 
\left(2^{3/2} \frac{\Vol}{\lambda_T^{3}}\right)
\left[
\sum_b \eexp{-\beta E_{b}} 
+ \int_0^{\infty} \gdos(k)dk \ \eexp{-(\beta/\mass)k^2}
\right]
\eeq
where $(n, \ell, m)$ are the good quantum numbers for the 
relative motion. Ignoring the possibility of spin, 
the sum is over {\em even} or {\em odd} values of $\ell$,  
for Bosons or Fermions respectively. 
In the second equality we separate the bond states 
from the scattering (continuum) states. In order to determine 
the DOS of the latter we recall that the radial wave functions 
are phase shifted spherical Bessel functions. Accordingly the 
box quantization condition for the allowed $k_n$ values is  
\beq
kR - \frac{\pi}{2}\ell + \delta_{\ell} \ \ = \ \ n\pi
\eeq
From here one deduce a formula for the effect 
of the phase shifts on the DOS: 
\beq
\gdos(k) - \gdos^{(0)}(k) \ \ = \ \ 
\frac{1}{\pi}\sum_{\varrho}^{1}\left
(2\ell+1\right)\frac{\partial\delta_{\ell}}{\partial k}
\eeq
Using this result we get after integration by parts 
the following expression for the 
interaction effect on the partition function:
\beq
Z_{2}  - Z_{2}^{(0)}  \ \ = \ \ 
\left(2^{3/2} \frac{\Vol}{\lambda_T^{3}}\right)
\left[ 
\sum_b \eexp{-\beta E_{b}} 
+ \frac{\lambda_T^{2}}{\pi^2} \sum_{\ell}^{'} \int_0^{\infty} kdk  \ \delta_{\ell}(k) \ \eexp{-(\beta/\mass)k^2}
\right]
\eeq

\putgraph[0.2\hsize]{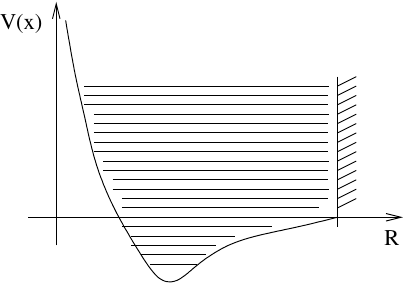}

\newpage
\makeatletter{}\sheadB{The canonical formalism}

\sheadC{The energy equation of state}

Consider some system, for example 
particles that are confined in a box. 
The Hamiltonian is 
\beq
\mathcal{H} \ \ = \ \ \mathcal{H}(\bm{r},\bm{p} ; X)
\eeq
where $X$ is some control parameter, 
for example the length of the box.
Assuming that we are dealing with a {\em stationary} state, 
the energy of the system is  
\beq
E \ \ \equiv \ \ \langle \mathcal{H} \rangle 
\ \ = \ \ \trc(\mathcal{H}\rho) 
\ \ = \ \ \sum_r p_r E_r
\eeq
If the system is prepared in a canonical states, 
then it is a mixture of energy eigenstates with probabilities 
\beq
p_{r}=\frac{1}{Z}\eexp{-\beta E_r}
\eeq
where the partition function is 
\beq
Z(\beta, X) \ \ = \ \ \sum_{r}\eexp{-\beta E_{r}}
\eeq
One observes that the energy of a system 
that is prepared in a canonical state can be derived 
from the partition function as follows:
\beq
E \ \ = \ \ 
\langle \mathcal{H} \rangle 
\ \ = \ \
-\frac{1}{Z}\frac{\partial}{\partial\beta}Z  
\ \ = \ \ -\frac{\partial}{\partial \beta}\ln Z
\eeq
Also one can find expressions for the higher moments, for example
\beq
\langle \mathcal{H}^2 \rangle \ \ = \ \ 
\frac{1}{Z} \frac{\partial}{\partial\beta}\frac{\partial}{\partial\beta} Z  
\eeq
In particular one deduces the relation
\beq
\text{Var}(E) 
\ \ = \ \ 
\langle \mathcal{H}^2 \rangle - \langle \mathcal{H} \rangle^2  
\ \ = \ \
\frac{\partial^2}{\partial\beta^2} \ln Z 
\ \ = \ \  
T^2 \frac{\partial E}{\partial T} \ \ = \ \ T^2 C 
\eeq
where in the latter equality we have defined 
the temperature as $T=1/\beta$ and the heat capacity as $C=dE/dT$. 
The notion of temperature will be discussed further below.

\sheadC{The Equipartition theorem}

In the classical context the Hamiltonian might be a sum of quadratic terms
\beq
\mathcal{H} \ \ = \ \ \sum_j c_{j} q_j^2
\eeq
where $q_j$ are either coordinates of conjugate momenta.
The partition function factorizes, 
where each quadratic term contributes a $\propto T^{1/2}$ term.  
It follows that each quadratic term contributes $T/2$ to 
the energy, and hence $1/2$ to the heat capacity. 

This observation can be applied to the analysis of "balls connected by springs". 
We can always go to normal coordinates. The center of mass degree of freedom 
contributes~$T/2$ to the energy, while each vibrational mode contributes~$T$.

A formal extension of this so-called "Equipartition Theorem" is as follows:
\beq
\left\langle q_i \frac{\partial \mathcal{H}}{\partial q_j} \right\rangle \ \ = \ \ T \delta_{ij}
\eeq
The proof is as follows: The measure of integration over phase space can be written as $dq_idq_jdq'$,
where $q'$ represents all the other coordinates. Applying integration by parts we have 
\beq
\int dq_i dq_j dq' \ q_i \ \frac{\partial \mathcal{H}(q)}{\partial q_j} \eexp{-\beta \mathcal{H}(q)}
= -\frac{1}{\beta} \int dq_i dq_j dq' \ q_i \  \frac{\partial}{\partial q_j} \left[  \eexp{-\beta \mathcal{H}(q)} \right]
= \delta_{ij}\frac{1}{\beta} \int dq_i dq_j dq' \eexp{-\beta \mathcal{H}(q)}
\eeq
and form here follows the Equipartition Theorem. This generalized version is useful 
in discussing particles that have interaction ${u(x_i-x_j)\propto |x_i-x_j|^{\alpha}}$, 
which constitutes a generalization of the harmonic (${\alpha=2}$) case.

\sheadC{Heat capacity}

From the Equipartition Theorem one deduce that the heat capacity 
of an "ideal" system equals to the effective number of freedoms: 
Each independent quadratic term in the Hamiltonian contributes~$1/2$  
to the heat capacity. This simple prescription should be  
refined for two reasons: 
(i)~Degrees of freedom  can "freeze" in the quantum treatment; 
(ii)~In general a many body system is not ideal due to interactions. 
We first discuss the quantum issue referring to spins and oscillators.

{\bf Spin and oscillator.-- } 
For spin (+) or oscillator (-) with 
level spacing $\omega$ we have
\beq
\ln(Z(\beta)) \ \ &=& \ \ \pm \ln(1\pm\eexp{-\beta\omega}) \\
E \ \ &=& \ \ -\frac{\partial \ln Z}{\partial \beta}  \ \ = \ \ \frac{\omega}{\eexp{\beta\omega}\pm1} \\
C(T) \ \ &=& \ \ \frac{dE}{dT} \ \ = \ \ 
\frac{1}{\left[ 2\mbox{csnh}\left(\frac{\omega}{2T} \right) \right]^2} 
\left(\frac{\omega}{T}\right)^2, 
\ \ \ \ \ \ \ \ \mbox{"csnh" is cosh or sinh} 
\eeq
In both case the low temperature behavior of $C$ is identical, 
namely, for $T\ll\omega$ it is dominated by the Boltzmann factor $\eexp{-\beta\omega}$. 
At high temperature $C$ of the spin drop down because energy reaches saturation, 
while $C$ of the oscillator approaches unity reflecting the classical 
prediction ${E \approx T}$. Since ${E=\omega n}$ it is more illuminating 
to re-write the above results as follows:
\beq
\langle n \rangle \ \ &=&  \ \ \frac{1}{\eexp{\beta\omega}\pm1} \ \ \equiv \ \ f(\omega)  \\
\text{Var}(n) \ \ &=& \ \ [1\mp f(\omega)] f(\omega)
\eeq
where $f(\omega)$ is known as the occupation function. In the case 
of an oscillator the result for the number variance can be regarded  
as a sum of a shot-noise particle-like term ${\text{Var}(n) = \langle n \rangle}$, 
and a classical term  ${\text{Var}(n) = \langle n \rangle^2}$. 
In the case of a spin the fluctuations go to zero 
in both the "empty" and "full" occupation limits.
It is customary in quantum-optics to characterize the fluctuations 
by ${g^{(2)} = (\langle n^2 \rangle-\langle n \rangle)/\langle n \rangle^2}$ 
and to say that the bosonic (oscillator) result ${g^{(2)}=2}$ corresponds to bunching,
while the fermionic (spin) result ${g^{(2)}=0}$  corresponds to anti-bunching.
The value ${g^{(2)}=1}$ reflects Poisson statistics and would 
apply in the case of coherent state preparation.

{\bf Debye model.-- } 
Let us refer further to a system that can be described as consisting of 
many harmonic freedoms, e.g. modes of vibrations. The spectral density of the modes 
might be $\propto \omega^{\alpha-1}$. For example in Debay model ${\alpha=d=3}$, 
with some cutoff frequency $\omega_c$. Then we get for the heat capacity 
\beq
C(T) \ \ = \ \ 
\const\int_{0}^{\omega_c}
\frac{1}{\left[ 2\sinh\left(\frac{\omega}{2T} \right) \right]^2} 
\left(\frac{\omega}{T}\right)^2 
\omega^{\alpha-1}d\omega
\ \ = \ \ 
\const \ T^{\alpha} F\left(\frac{\omega_c}{T}\right)
\eeq
where
\beq
F(\nu) \ \ \equiv \ \ 
\int_{0}^{\nu}
\frac{\eexp{x}}{(\eexp{x}-1)^2} x^{\alpha+1}dx
\eeq
The quantum result can be described 
as emerging from "freezing" of freedoms 
due to the quantization of energy.
This phenomena has lead to the birth of quantum mechanics 
in the context of blackbody radiation (Planck law).

{\bf Glasses.-- }
The standard model for glasses regard them 
as a large collection of "two level" entities 
with splitting $\omega$ that has roughly uniform 
distribution. Hence the calculation of the heat 
capacity is formally as in the Debye model  
model with sinh replaces by cosh, and ${\alpha=1}$, 
leading to a linear dependence ${C(T)\propto T}$.

{\bf Quantum gases.-- }
In the classical treatment, disregarding prefactors of order unity, a gas of $N$~particles have total energy ${E \sim NT}$, 
hence the heat capacity is ${C\sim N}$. If we have a gas of Fermions in low temperatures, 
then the number of excited particles is $N_{\text{eff}} \propto T$, 
hence the energy is ${E \propto T^2}$, and the heat capacity is ${C(T)\propto T}$. 
In contrast to that Bosons in 3D condense into the ground states. 
Hence the occupation of an excited state of energy $\epsilon_{r}$   
is formally the same as the occupation of an oscillator with the 
same frequency. Consequently one observes ${C(T)\propto T^{\alpha}}$    
as in Debye model.

{\bf Phase transitions.-- } 
We shall discuss phase transitions in later lectures.
As the temperature is lowered towards a critical temperature~$T_c$  
the system becomes "correlated", which means that the effective 
number of freedoms is reduced. We assume ${T>T_c}$ and note 
that similar picture applies if one approaches $T_c$ from below.   
We can easily explain why the heat capacity diverges  
as $T_c$ is approached.  
For an ideal gas, or better to think about 
a collection of non-interacting oscillators,  
the partition function is ${Z=g^N}$, 
where $N$ is the number of freedoms, 
and ${g \propto T}$ is the number of 
accessible states for a single freedom 
at temperature~$T$.   
For a correlated system ${Z=g^{N_{\tbox{eff}}}}$, 
where  $N_{\tbox{eff}}=N/\xi^d$ is the effective number of 
independent regions, and $\xi$ is called  the correlation length.
The prototype Ising model consist of spins (${g=2}$) rather than oscillators  
and ${\xi \propto |T-T_c|^{-\nu}}$ where ${\nu \approx 1/2}$.  
Either way we can write the expression for the heat capacity as follows:
\beq
C(T) \ \ = \ \ \beta^2 \frac{d^2\ln Z}{d\beta^2} \ \ \equiv \ \ C_g(T) + C_{\xi}(T)
\eeq
where the non-singular $C_g(T)$ originates from the temperature dependence of $g$, 
and equals $N$ for non-interacting oscillators, reflecting the effective 
number of freedoms. The singular term  $C_{\xi}(T)$ originates from the 
temperature dependence of $\xi$. For an Ising system its divergence 
near the critical temperature is described by ${|T-T_c|^{\nu d-2}}$.   
Note the significance of the space dimension~$d$.

\putgraph[0.8\hsize]{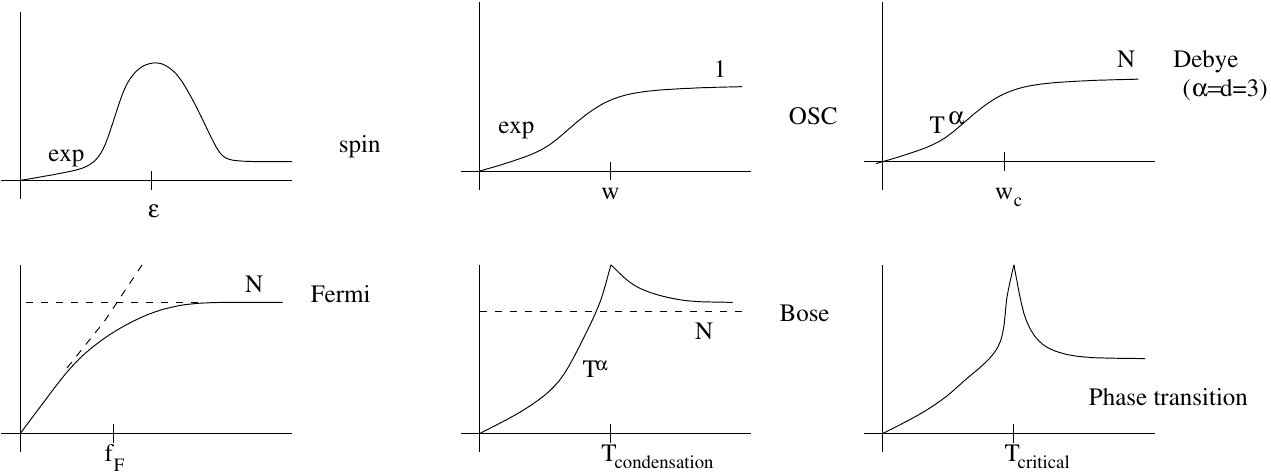}

\newpage
\sheadC{Generalized forces}

Assume that $X$ is a parameter that appears in the Hamiltonian.  
We define the generalized force $\mathcal{F}$ 
which is associated with the parameter $X$ as   
\beq
\mathcal{F} \ = \ -\frac{\partial \mathcal{H}}{\partial X} 
\eeq
This definition will be motivated later on when we 
discuss the notion of {\em work}.
We shall explain that for an isolated system that undergoes a quasi-static adiabatic process 
the change in energy is ${dE=-\braket{\mathcal{F}}dX}$, 
meaning that the work that has been dome by the system is ${dW=\braket{\mathcal{F}}dX}$.
Here are some examples for generalized forces: 

\begin{tabular}{|l|l|}
\hline 
\mbox{\bf parameter} & \mbox{\bf generalized force} \\
\hline
\mbox{piston displacement in cylinder with gas} \ \ \ \ \ & \mbox{Newtonian force}\\
\mbox{volume of a box with gas} & \mbox{Newtonian pressure} \\
\mbox{length of a polymer} & \mbox{Newtonian force (tension)} \ \ \ \ \  \\
\mbox{homogeneous electric field} & \mbox{total polarization} \\
\mbox{homogeneous magnetic field} & \mbox{total magnetization} \\
\mbox{magnetic flux through a ring} & \mbox{electric current} \\
\hline
\end{tabular} 

\ \\

{\bf Flux and Current.-- }
We would like to better clarify why magnetic flux and 
electrical current are conjugate variables. 
Note that for an homogeneous magnetic field 
the flux through a ring is $\Phi=\mathsf{A}\mathcal{B}$, 
and the magnetization is ad-hock defined as $\tilde{M}=\mathsf{A}I$, 
where $\mathsf{A}$ is the area of the ring.
Using the notation $X=\Phi$, the direct identification 
of the conjugate operator $\mathcal{F}$ as the current,  
is rationalized in a simple-minded manner as follows: 
If we make a change $dX$ of the flux during a time $dt$, 
then the electro-motive force (EMF) is $-dX/dt$, 
leading to a current ${\cal I}$ in the ring. The energy 
increase of the ring is the EMF times the charge, 
namely $dE=(-dX/dt)\times({\cal I}dt)=-{\cal I}dX$.    

\ \\

{\bf Magnetic field.-- }
Usually we shall denote the {\em applied} magnetic field by the 
letter $h$, possibly absorbing into it definition the  coupling constant.
For example we write the interaction of a spin with a vertical magnetic 
field as $-h\sigma_z$. But there are circumstance in which the sample 
affect the the magnetic field in a way that cannot be ignored.
For example: if we place a typeI superconductor inside a solenoid, 
it expels sideways the magnetic field, such that the {\em total} 
magnetic field is $\mathcal{B}=0$ inside the sample.  
We therefore have to be careful in how we write the Hamiltonian.
Schematically we write 
\beq
\mathcal{H}_{\text{total}} 
\ \ =&& \ \ 
\sum_{j\in \text{system}} \frac{1}{2\mass_j}(p_j-e_jA)^2 + U(r_1,r_2,...) \\
\ &+& \ \Big[\text{similar expression for the solenoid} \Big] 
\ + \ \frac{1}{8\pi}\int \mathcal{B}(x)^2 \, d^3x 
\eeq 
Our focus is on the system, so we keep only the interaction 
of the system with the solenoid:
\beq
\mathcal{H}_{\text{total}} 
\ \ = \ \ 
\mathcal{H}_{\text{system}}(r_j,p_j; A)  
\ + \frac{1}{8\pi} \int \mathcal{B}(x)^2 \, d^3x 
\ - \int A\cdot J_{\text{solenoid}} \, d^3x 
\eeq 
The current density of the solenoid defines 
the {\em applied} magnetic field through the relation 
\beq
\nabla \times h \ \ = \ \ 4\pi \ J_{\text{solenoid}}   
\eeq
Substitution of this definition into the last term, 
and doing integration by parts, the Hamiltonian that 
described the interaction of the system with the {\em applied} 
magnetic field takes the following form:
\beq
\mathcal{H}_{\text{total}} 
\ \ = \ \ 
\mathcal{H}_{\text{system}}(r_j,p_j; A) 
\ + \frac{1}{8\pi} \int \mathcal{B}(x)^2 \, d^3x  
\ - \frac{1}{4\pi} \int h(x) \mathcal{B}(x) d^3x  
\eeq
Schematically the interaction is described 
by a term that looks like $-h \mathcal{B}$, 
where $h$ is a control parameter, 
and $\mathcal{B}$ is the conjugate dynamical variable. 
This sounds less strange if we think of $h$ 
as an external current. Note that the role 
of the "current" and the "flux" have been switched.
Here the current is the source, and the magnetic "flux"
is a dynamics variable. The expression for the work 
for a bulk sample will take the form 
\beq
dW \ \ = \ \ \frac{\Vol}{4\pi} \, \mathcal{B}(h) \, dh  
\ \ \equiv \ \ \frac{\Vol}{4\pi} \, h \, dh \ + \ \tilde{M}(h) \, dh  
\eeq
If the state equation $\tilde{M}(h)$ is known, 
we can deduce from the above relation how the free 
energy of the sample changes as the magnetic field
is turned on. In the next lecture regarding 
thermodynamics $\mathcal{H}_{\text{total}}$ 
as defined above will be identified as 
a "grand Hamiltonian" with which a "Gibbs" 
free energy can be associated.

\sheadC{Susceptibility and fluctuations}

Given $X$ and assuming that the system is prepared 
in a canonical state characterized by some $\beta$, 
we can derive the average 
value~$y$ of the generalized force $\mathcal{F}$ 
from the partition function as follows:  
\beq
y(X) \ \ \equiv \ \ 
\left\langle \mathcal{F} \right\rangle_X 
\ \ = \ \ 
\sum_r p_r \left(-\frac{dE_r}{dX}\right)
\ \ = \ \ 
\frac{1}{\beta} \frac{\partial \ln Z}{\partial X}
\eeq
The generalized susceptibility describes the 
dependence of $y(X)$ on the the parameter $X$, namely,   
\beq
\chi(X) \ \ \equiv \ \ \frac{\partial y}{\partial X} 
\ \ = \ \ \frac{1}{T} \text{Var}(\mathcal{F})
\eeq
The second equality requires few lines of algebra.  
Let us illuminate this relation, and re-derive it, 
by considering a prototype example: the dependence of 
of the length of a polymer, or the volume of a gas, 
on the applied tension or pressure. In this example  
the total Hamiltonian can be written as 
\beq
\mathcal{H}(\lambda) \ \ = \ \ \mathcal{H} - \lambda V
\eeq  
where the parameter $\lambda$ is the applied field, and $V$ 
is the conjugate dynamical variable (length or volume in the above 
mentioned examples).  Consequently we get in the presence 
of the applied field 
\beq
\langle V \rangle_{\lambda}  
\ \ = \ \ \frac{\trc\left[ V\exp\left(-\beta \mathcal{H}(\lambda)\right)\right] }{\trc\left[ \exp\left(-\beta \mathcal{H}(\lambda)\right)\right]} 
\ \ = \ \  \langle V \rangle + \beta \lambda \left[ \langle V^2 \rangle -  \langle V\rangle^2  \right]
\ + \ \text{higher orders} 
\eeq
where both numerator and denominators have been expanded, 
without much caring about commutation relations.  
From the above we deduce the following {\em classical} relation 
between the compressiblility and the fluctuations:
\beq
\kappa \ \ \equiv \ \ \left[\frac{\partial \langle V \rangle_{\lambda}}{\partial \lambda}\right]_{\lambda=0} 
\ \ = \ \ \frac{1}{T} \text{Var}(V)
\eeq
In a later lecture we shall introduce generalizations of 
this relation that are known as the "Onsager regression theorem" 
and as the "Fluctuation dissipation relation".

The relation ${\kappa=(1/T) \text{Var}(V)}$ parallels 
the relation ${C=(1/T^2) \text{Var}(E)}$ between 
the heat capacity and the fluctuations in energy. 
It automatically implies that these constants 
have to be positive. Another way of looking on it 
is to say that ${C>0}$ and ${\kappa>0}$ are stability conditions.
Negative value  means that that the system 
will undergo a "phase separation" process. 
See discussion of the "Maxwell construction" 
is the "Interactions and phase transitions" lecture.

\newpage

\sheadC{Empirical temperature}

In practice we would like to be able to probe the $\beta$ of the environment. 
For this purpose we use a thermometer. 
The simplest thermometer would be an ideal gas in a box, 
for which the partition function is 
\beq
Z\left(\beta,\Vol\right) \ \ &=& \ \ \Vol^{N}\left(\frac{\mass}{2\pi\beta}\right)^{\frac{3N}{2}}
\\
P \ \ &=& \ \ \frac{1}{\beta}\frac{\partial\ln Z}{\partial \Vol}=\frac{N}{\Vol} \beta^{-1}
\eeq
The empirical temperature is defined as follows:
\beq
\theta \ \ = \ \ \frac{P\Vol}{N} \ \ = \ \ \frac{1}{\beta}
\eeq
We can of course define different thermometers. 
The idea is simply to identify a measurable 
quantity that reflects the parameter~$\beta$.

\sheadC{The Virial theorem}

Somewhat related to the equipartition theorem, is the Virial theorem. 
It is used to relate the expectation value of the ``kinetic" and ``potential" 
terms in Hamiltonian of the type ${\mathcal{H}=K(p)+U(r)}$. 

Consider any observable $G$. It is clear that if the system is prepared 
in a {\em stationary} (not necessarily canonical) state, then the expectation 
value $\langle G \rangle$ is constant in time. By the rate of change 
equation of motion it follows that 
\beq 
\Big\langle [\mathcal{H},G] \Big\rangle \ \ = \ \ 0
\eeq
In particular let us consider the generator of dilations 
\beq 
G \ \ = \ \ \frac{1}{2}\sum_j (r_j\cdot p_j + p_j\cdot r_j)
\ \ \ \ \ \ \ \ \ \ \mbox{[the symetrization is required in the quantum case]} 
\eeq
For the Hamiltonian ${\mathcal{H}=K(p)+U(r)}$ we get
\beq 
\left\langle p \cdot \frac{\partial K}{\partial p} \right\rangle 
- \left\langle r \cdot \frac{\partial U}{\partial r} \right\rangle
\ \ = \ \ 0
\eeq
with implicit summation over~$j$.
If the classical equipartition theorem applies, each term equals $T$ multiplied by the number of freedoms. 
For quadratic $K(p)$ and $U(r)$ the first term equals $2\langle K \rangle$, 
and the second term equals~$-2\langle U \rangle$.
More generally, for two-body interaction of the type 
\beq
U(r) \ \ = \ \ \sum_{\langle ij \rangle} u(r_i-r_j)  
\ \ = \ \   \sum_{\langle ij \rangle} |r_i-r_j|^{\alpha}
\eeq
the second term in the Virial theorem equals $-\alpha\langle U \rangle$.
This is a meaningful statement for ${\alpha>0}$, otherwise 
there should be a ``box" that confines the particles. 
Writing the full Hamiltonian as ${\mathcal{H}=K(p)+U(r)+V_L(r)}$ we deduce that 
\beq
\left\langle p \cdot \frac{\partial K}{\partial p} \right\rangle   
-\left\langle r \cdot \frac{\partial U}{\partial r} \right\rangle
-\left\langle r \cdot \frac{\partial V_L}{\partial r} \right\rangle 
\ \ = \ \ 0
\eeq
In the next section we shall see how this relation helps us 
to derive an expression for the ``pressure" on the walls of the box.

\sheadC{Pressure on walls}

Possibly the simplest point of view about pressure is to regard it 
as arising from collisions of particles with the walls. This is the so 
called the {\em kinetic picture} point of view. However, within 
the framework of the {\em canonical formalism} the pressure is defined 
as the generalized force that is associated with the volume, 
such that $\dbar W = Pd\Vol$. It is quite puzzling that in the formal 
classical calculation the kinetic part factors out and the mass 
of the particles does not appear in the result:
\beq
\ln(Z(\beta,\Vol)) \ \ &=& \ \ -\frac{3N}{2} \ln \beta \ + \ N\ln\Vol \ + \ \const  \\
E \ \ &=& \ \  -\frac{\partial \ln Z}{\partial \beta} 
\ \ = \ \ \frac{3}{2} NT    \\
P \ \ &=& \ \ \frac{1}{\beta}\frac{\partial \ln Z}{\partial \Vol}  
\ \ = \ \ \frac{NT}{\Vol} 
\eeq
With interactions we have to calculated a complicated 
configuration (${dr_1 dr_2...dr_N}$) integral. This calculation will 
be discussed in later sections. In the absence of interactions 
we see that the pressure is related to the kinetic energy, 
namely ${P=(2/3)E/\Vol}$. Below we generalize this relation 
using the Virial theorem: we shall see that quite generally, 
both classically and quantum mechanically, the pressure is related 
to the kinetic and potential energy of the gas. 
  
The volume deformation of a box is represented by a deformation field $D(r)$. 
To be specific let us write the Hamiltonian of $N$ gas particles in a box as follows: 
\beq
\mathcal{H} \ \ = \ \ K(p) + U(r) + V_L(r- \lambda D(r)) 
\eeq 
Here $K(p)$ is the kinetic term, and $U(r)$ are the interactions, 
and $V_L(r)$ is box defining potential, and ${\lambda}$    
is the deformation parameter. We want $\lambda$ to equal the 
extra volume due to the deformation, such that ${\Vol=\Vol_0+\lambda}$.
We therefore normalize the displacement field such that 
\beq 
\oiint D \cdot ds \ \ = \ \ 1, 
\hspace{2cm}
\text{standard choice: } \ D(r)=\frac{1}{3\Vol_0} \ r
\eeq
Accordingly the definition and the expression for the pressure are  
\beq
P \ \ = \ \ \left\langle  -\frac{\partial \mathcal{H}}{\partial \Vol}  \right\rangle  
\ \ = \ \ \left\langle  -\frac{\partial \mathcal{H}}{\partial \lambda}  \right\rangle_{\lambda{=}0} 
\ \ = \ \ \frac{1}{3\Vol} \left\langle r \cdot \frac{\partial V_L}{\partial r} \right\rangle 
\ \ = \ \ \frac{1}{3\Vol} \left[ 
\left\langle p \cdot \frac{\partial K}{\partial p} \right\rangle   
-\left\langle r \cdot \frac{\partial U}{\partial r} \right\rangle
\right]
\eeq
where in the last equality we have used the Virial theorem. 
Note that this extends the discussion of the Virial theorem 
in previous section. The case of inter-atomic interactions 
with ${\alpha>0}$ (bounded system with no walls) 
can be regarded formally as a special case of the above relation with ${P=0}$.
If  ${\alpha<0}$ there is non-zero pressure.  
We can use the equipartition theorem to obtain in the classical case
\beq
P \ \ = \ \ \frac{1}{\Vol} \left[ 
NT - \frac{1}{3}\left\langle r \cdot \frac{\partial U}{\partial r} \right\rangle
\right]
\eeq
where the first term is the same as in the law of ideal gases,
while the second is due to the interactions, 
and can be expressed using moments of the inter-particle separation.

\newpage
\sheadC{Tension of a polymer}

The calculation of a tension of a polymer is very similar 
to the calculation of pressure. The parameter 
in the Hamiltonian is the length $X$ of the polymer, 
which is analogous to the length or the volume of the box 
that contains the gas particles. In both cases the 
formal result depends only on the configuration integral, 
while the kinetic term factors out. Thus in both cases 
the result does not depend on the mass of the gas particles 
or on the mass of the monomers from which the polymer is composed.
The partition function in the case of a polymer is  
\beq
Z(\beta, X) \ \ = \ \ \text{[kinetic term]} \times 
\sum_{\text{conf.}} \ \delta(X-(r_1+r_2+...+r_N)) \ \eexp{-\beta U(\text{configuration})} 
\eeq   
For simplicity we assume a one-dimensional configuration,
such that each monomer is like a link of a chain or small spring 
with potential energy $u(r)$.
Accordingly the total potential energy can be written as ${U=u(r_1)+u(r_2)+...+u(r_N)}$.  
For hard-links, in analogy with the case of hard-spheres, the potential energy 
merely restricts the space of allowed configurations. 
Without the extra $X$~restriction the summation would give a value $Z(\beta)$.
One observes that the ratio ${Z(\beta, X)/Z(\beta)}$ 
would be the probability of observing length~$X$ 
if the polymer were unconstrained at its endpoints. 
According to the central limit theorem, for a long polymer 
\beq
\text{P}(X) \ \ =  \frac{Z(\beta, X)}{Z(\beta)} 
\ \ \propto \ \ \exp\left[-\frac{1}{2}\left(\frac{X}{L_0}\right)^2\right] 
\eeq   
Above we assumed that the polymer can stretch either sides, 
hence its average "algebraic" length is ${\langle X \rangle=0}$, 
while the RMS average is denoted~$L_0$.  
The force that is exerted on the endpoint obeys Hooke's law:  
\beq
\braket{\mathcal{F}}_X  \ \ = \ \  \frac{1}{\beta}\frac{\partial \ln Z}{\partial X}   \ \ = \ \  -(T/L_0^2) \ X 
\eeq
Optionally, if we insist on calculating 
directly the partition sum, we can write
\beq
Z(\beta, X) \ \ = \ \ 
\int \frac{dk}{2\pi}
\int dr_1 dr_2...dr_N \ \eexp{i(X-(r_1+r_2+...+r_N))k} \ \eexp{-\beta U(r_1,r_2,...,r_N)} 
\ \ \equiv \ \  \int \frac{dk}{2\pi} \ \eexp{i\, k \, X}  \ \tilde{Z}(\beta,k)
\eeq
Above the we have omitted the irrelevant kinetic term.
The integral over all possible configurations factorizes 
Notably for hard-links it is the ``volume" of the possible 
configurations that have zero potential energy. 

Assume that a tension~$f$ is applied on the polymer:
this can be regarded as an ``electric" field that is 
applied on the endpoint of the polymer, or as an applied ``weight".  
The Hamiltonian becomes ${\mathcal{H}_G(\hat{r};f)=\mathcal{H}+f\hat{X}}$, 
where ${\hat{X}=\sum r_n}$. In the new configuration $\hat{X}$ is an un-constrained dynamical variable, and the equilibrium point $X_{\text{eq}}$ is determined by the 
condition ${\mathcal{F}(X)=f}$. Note that in order to keep sign consistency we have defined the applied field in the negative direction. 
If fluctuations are neglected we expect ${\braket{X}_f = X_{\text{eq}}}$ 
that is derived from $\mathcal{H}_G$,
to be consistent with $\braket{\mathcal{F}}_X=f$  
that is derived from $\mathcal{H}$.
If we blur the distinction between the tension $\braket{\mathcal{F}}$ in the sense 
of expectation value, and the tension $f$ in the sense of an external parameter (applied force), then, under the same assumption, the relation ${dE_G = Xdf}$ is consistent with ${dE = -fdX}$. In other words: if the conjugate of~$X$ is~$f$, then in the Gibbs-Hamiltonian framework
the conjugate of~$f$ is~$-X$.  The Gibbs partition function is
\beq
Z_G(\beta, f) \ \ = \ \ 
\int dr_1 dr_2...dr_N \ \eexp{-\beta [U(r_1,r_2,...,r_N)+(r_1+r_2+...+r_N)f]} 
\ \ = \ \ \int Z(\beta,X) \ \eexp{-\beta f \, X} \ dX  
\eeq
The factorization of this partition function
implies that the total length ${\langle X \rangle}$ 
of the polymer, for a given applied field~$f$, 
is the sum of lengths of the monomers for the same field 
(the field determines the tension of the polymer).  
We realize that~$Z(\beta,X)$ and~$Z_G(\beta,f)$ 
are related by a Laplace transform.
From strict mathematical point the former  
is like the probability function, 
and the latter is like the associated moment generating function. 
What we were doing is in fact a generalization 
of the "convolution theorem", as used in the 
derivation of the central limit theorem.   

Finally, in the large $N$ limit the relation between~$Z_G(\beta,f)$ 
and~$Z(\beta,X)$ can be formulated as a Legendre transformation.
We shall encounter the Legendre transformation
in the next section, in a formally identical context,   
as the relation between the Gibbs free energy $G(T,P)$ 
and the Helmholtz free energy $F(T,\Vol)$. 
Later we use the same trick in the analysis
of quantum gases, when we go from the canonical 
to the so called ``grand-canonical" framework.

\sheadC{Polarization}

The polarization is the generalized force that is associated
with electric field. Let us assume that we have a bounded system 
of particles with an added uniform electric field:
\beq
\mathcal{H} \ \ &=& \ \ \sum_{\alpha} \frac{p_{\alpha}^2}{2\mass_{\alpha}} 
\ + \ \text{interactions} \ + \ \text{potential} \ - \sum_{\alpha} q_{\alpha}\mathcal{E}x_{\alpha}
\\
\hat{P} \ \ &=& \ \  -\frac{\partial \mathcal{H}}{\partial \mathcal{E}}  
\ \ = \ \ \sum_{\alpha}q_{\alpha}\hat{x}_{\alpha}
\eeq
The polarization $\tilde{P}$ is the expectation value of $\hat{P}$.
One simple example is the calculation of the polarization of an "atom", 
where we have (say) a negative particle that is bounded by a "spring" 
to a positive charge. Another simple example concerns a diatomic 
molecule that has a permanent dipole moment~$\mu$. Here the Hamiltonian is 
\beq
\mathcal{H}(\theta, \phi, p_{\theta}, p_{\phi}) 
\ \ = \ \ \frac{p_{\theta}^2}{2I} + \frac{p_{\phi}^2}{2I\sin^2(\theta)} \ - \mu\mathcal{E}\cos(\theta)
\eeq
where $I$ is the moment of inertia. For the polarization we get 
\beq
\tilde{P} \ \ = \ \  \frac{1}{\beta}\frac{\partial \ln Z}{\partial\mathcal{E}}
\ \ = \ \ \mu \left[\coth\left(\frac{\mu\mathcal{E}}{T}\right)  \ - \ \left(\frac{\mu\mathcal{E}}{T}\right)^{-1} \right]
\eeq    
Note that expansion for weak field implies the electric susceptibility $\chi=(1/3)\mu^2/T$.

\sheadC{Magnetization}

The magnetization is the generalized force that is associated
with magnetic field. It is either due to having spin degree 
of freedom (Pauli) or due to the orbital motion. Here we 
clarify the definition using the three simplest examples.

{\bf Pauli magnetism.-- }
Consider a collection of $N$ spins. We denote the 
magnetic filed by~$h$. The Hamiltonian is  
\beq
\mathcal{H} \ \ &=& \ \ -\sum_{\alpha=1}^N g_{\alpha}h S_z^{\alpha} 
\\
\hat{M} \ \ &=& \ \  -\frac{\partial \mathcal{H}}{\partial h}  \ \ = \ \ \sum_{\alpha=1}^N g_{\alpha}S_z^{\alpha} 
\eeq
The magnetization $\tilde{M}$ is the expectation value of $\hat{M}$.
For a single spin~1/2  entity we get the following result:
\beq
\tilde{M}  \ \ = \ \  \frac{1}{\beta}\frac{\partial \ln Z}{\partial h}
\ \ = \ \ \frac{g}{2}\tanh\left(\frac{gh}{2T}\right)
\eeq
Note that expansion for weak field implies the magnetic susceptibility $\chi=(1/4)g^2/T$.
Note also that in the classical limit ("large spin") the problem becomes 
formally identical to that of calculating polarization of electric dipoles.

{\bf Orbital magnetism (classical).-- }
In the following we shall identify what is the magnetization $\hat{M}$ for charged spinless particles, 
using the formal definition ${-\partial \mathcal{H}}/{\partial h}$. In the 1D case (ring) it is identified 
as arising from a circulating current. In the 2D case it is more convenient to bypass the question 
what is $\hat{M}$ and to go directly to the $\tilde{M}$ calculation via the partition function. In the 
classical case one obtains $\tilde{M=0}$. But in the quantum calculation one obtains finite result. 
The classical result is puzzling because we would like to interpret $\hat{M}$ as arising from circulating 
currents as in the 1D case. Indeed such interoperation is possible. The point to realize that within 
the bulk we indeed have circulating electrons that give rise to a diamagnetic response. But this is compensated 
by "Hall currents" that flow along the boundary. The exact cancellation of these two contributions 
is spoiled upon quantization, instead we get the de Haas van Alphen (dHvA) oscillations. Details below.

{\bf Orbital magnetism (1D).-- }
Consider a spinless particle in a ring of length~$L$, 
and area $\mathsf{A}$. The magnetic flux is $\Phi=h\mathsf{A}$.
The Hamiltonian, the velocity-operator, the current-operator and the magnetization-operator are   
\beq
&& \mathcal{H} \ = \ \frac{1}{2\mass}\left(p-e\frac{\Phi}{L}\right)^2 + V(x) \\
&& \hat{v} = i[\mathcal{H}, x] \ \ = \ \ \frac{1}{\mass}\left(p-e\frac{\Phi}{L}\right) \\
&& \hat{I} \ = \  -\frac{\partial \mathcal{H}}{\partial \Phi}  \ \ = \ \ \frac{e}{L}\hat{v} \\
&& \hat{M} = -\frac{\partial \mathcal{H}}{\partial h}  \ \ = \ \ \mathsf{A} \hat{I} 
\eeq
The magnetization $\tilde{M}$ is the expectation value of~$\hat{M}$, 
or optionally we can refer to the circulating current~$I$,
which is the  expectation value of~$\hat{I}$.

{\bf Orbital magnetism (2D).-- }
The more interesting case is the magnetization of electrons 
in a 2D box (3rd dimension does not play a role) due to the formation 
of Landau levels. We recall again that classically the energy spectrum 
of the system is not affected by magnetic field. 
But quantum mechanically Landau levels are formed 
(see "Lecture notes in Quantum mechanics").
Let us consider a box of area~$\mathsf{A}$ that contains $N$ spinless electrons.   
In the {\em bulk}, the energy of a Landau state 
that belongs to the ${\nu}$ level is ${\varepsilon_{\nu} = (\nu+(1/2))\omega_B}$       
where $\omega_B = e\mathcal{B}/\mass$ is the cyclotron frequency.
The degeneracy of each Landau level is ${g_B=e\mathcal{B}\mathsf{A}/2\pi}$.
The calculation of the single particle partition function 
is the same as that of harmonic oscillator (multiplied by the degeneracy). 
Assuming $N$ electrons that can be treated as an ideal Boltzmann gas we get 
\beq
\tilde{M} \ \ = \ \ -\frac{N}{12}\left(\frac{e}{\mass}\right)^2 \frac{\mathcal{B}}{T} 
\ + \ \mathcal{O}(\mathcal{B}^3)
\eeq
This result does not hold for a low temperature electron gas, 
because the Fermi statistics of the occupation becomes important. 
Assuming zero temperature we define ${\mathcal{B}_n}$ with ${n=1,2,3,...}$  
as the threshold value for which $n$ Landau levels are fully filled.   
This values are determined by the equation ${n g_B = N}$.  
Considering first strong field ${\mathcal{B} > \mathcal{B}_1}$, 
the energy of the system is $E_0^{(N)}=N\omega_B/2$ and hence 
\beq
\tilde{M} \ \ = \ \ -\frac{\partial E_0^{(N)}}{\partial \mathcal{B}} \ \ = \ \ -N\frac{e}{2\mass},
\ \ \ \ \ \ \ \ \ \ \ \ \ \ \mbox{for} \ \mathcal{B} > \mathcal{B}_1
\eeq
This result has a simple interpretation using "Bohr picture" of an orbiting electron: 
each electron performs a minimum energy cyclotron motion with unit angular momentum~$L$, 
and associated magnetic moment $-(e/2\mass)L$. 
If the magnetic field is ${\mathcal{B}_{n+1} < \mathcal{B} < \mathcal{B}_{n} }$, 
one has to sum the energy of the electrons in $n$ filled Landau levels, 
where the upper one is only partially filled. One obtain a quadratic 
expression from which it follows that the magnetization 
grows linearly from $-N(e/2\mass)$ to $+N(e/2\mass)$. 
Hence there is saw-tooth dependence of $\tilde{M}$ on the field,    
which is known as the de Haas van Alphen (dHvA) oscillations.

{\bf Semiclassical interpretation.-- } 
There is a simple way to understand the dHvA result. 
For this purpose assume that~$\mathsf{A}$ looks like a circle.
Each "Landau state" occupies a thin {\em strip} that has a finite width. 
Within each strip there is a diamagnetic cyclotron motion 
whose net effect is like having an inner anticlockwise current (${I^{\circlearrowleft}>0}$), 
and an outer clockwise current (${I^{\circlearrowright}<0}$).
In the bulk the net current of a strip is zero, but nevertheless  
it has a diamagnetic contributions to the magnetization,  
because $I^{\circlearrowright}$ encloses a larger area compared with $I^{\circlearrowleft}$.
As we come close to the boundary, near the potential wall, 
the net current of the strip becomes positive, 
and its value is determined by the potential gradient.
This is known as  Hall effect. 
In the case of hard wall there is a nice semi-classical 
illustration of the trajectories that bounce along the boundary. 
Upon quantization the "strips" support so-called "edge states". 
When $\mathcal{B}$ crosses $\mathcal{B}_{n}$ we get a jump in the 
magnetization that corresponds to the occupation of an additional edge states:
The total Hall conductance of $n$ Landau levels is $G_H=(e/2\pi)n$, 
residing in a region that experiences a potential difference $\omega_B$.
Hence the drop in the magnetization is ${(G_H \omega_B)\times \mathsf{A} = N(e/\mass)}$. 
It is now easy to understand why in the classical limit we do not 
have magnetization: the Hall current along the edges compensates 
the diamagnetic currents of the bulk. It is only upon quantization 
that the balance is violates, and instead we have the dHvA oscillations
as a function of $\mathcal{B}$.

\newpage
\makeatletter{}\sheadB{Thermodynamics}

\sheadC{Absolute temperature and entropy}

Let us {\em formally} vary the parameters $X$ and $\beta$. 
The implied change in the energy is
\beq
dE \ \ = \ \ 
\sum_{r}dp_{r}E_{r} + \sum_{r} p_{r}dE_{r} 
\ \ = \ \ 
\left[
\left(\sum_r \frac{dp_r}{d\beta} E_r\right)d\beta 
+ \left(\sum_r  \frac{dp_r}{dX} E_r \right)dX  
\right] \ + \ \left[ \left(\sum_r p_r \frac{dE_r}{dX}\right) dX\right]
\eeq
The second term in the formal $dE$ expression  
is identified as the work $\dbar W$
that would be done on the system during a reversible 
quasi-static process: 
\beq
\sum_{r} p_{r} dE_{r} \ \ = \ \  \left( \sum_r p_r \frac{dE_r}{dX} \right) dX \ \ = \ \  - y(X) \ dX
\eeq
In the next section we shall identify 
the first term in the formal $dE$ expression 
as the heat $\dbar Q$ that would be absorbed during 
a reversible quasi-static process.
This expression is not an ``exact differential", 
but it has an {\em integration factor} that depends 
only on the empirical temperature. In fact this 
integration factor turns out to be $\beta$, 
hence we define the the {\em absolute temperature}:  
\beq
T \ \ = \ \ \text{inverse integration factor} \ \ = \ \ \frac{1}{\beta} 
\eeq
such that $(1/T)\dbar Q$ is the differential of a so-called entropy function: 
\beq
\sum_{r}dp_{r}E_{r} \ \ = \ \ 
\left(\sum_r \frac{dp_r}{d\beta} E_r\right)d\beta 
+ \left(\sum_r  \frac{dp_r}{dX} E_r \right)dX  
\ \ = \ \ TdS
\eeq
The implied {\em definition} of the thermodynamic entropy is  
\beq
S \ \ = \ \ -\sum p_{r}\ln p_{r} 
\eeq
Note that the thermodynamic entropy is an extensive 
quantity in the thermodynamic limit. It should not be 
confused with other types of ``entropy". In particular 
we shall discuss the "Boltzmann entropy" in a later 
section with regard to the 2nd law of thermodynamics.

We see that the formal expression for $dE$ 
can be written as follows:
\beq
dE \ \ = \ \  TdS - ydX
\eeq
It is important to emphasize that the above formal expression 
is a valid mathematical identity that holds irrespective 
of whether it reflects an actual physical process.  
However, it is only for a reversible quasi-static process 
that $ydX$ is identified as the work, and $TdS$ as the heat. 
For a non-reversible process these identifications are false.

\sheadC{The Thermodynamic potentials}

From the basic relation ${dE = TdS - ydX}$ one concludes 
that if~$E$ is formally expressed 
as a function of~$S$ and~$X$, then we can derive from it 
the state equations $T(S,X)$ and $y(S,X)$.  
Accordingly ee say that ${E(S,X)}$ is a {\em thermodynamic potential}. 
At this stage it is convenient to define also 
the Helmholtz thermodynamic potential:
\beq
F(T,X) 
\ \ \equiv \ \ 
-\frac{1}{\beta}\ln Z(\beta;X)
\eeq
Within the framework of the canonical formalism the energy 
is obtained taking to the derivative of~$Z$ with respect to~$\beta$.   
This translates to the relation ${E=F+TS}$. 
The relation between ${F(T,X)}$ and ${E(S,X)}$ 
is formally a Legendre transform.
Consequently ${dF = -SdT - ydX}$ and the associated state equations are  
\beq
S = -\frac{\partial F}{\partial T},
\hspace{2cm} 
y = -\frac{\partial F}{\partial X}, 
\eeq

Within the framework of the thermodynamic formalism {\em state equations} 
that describe physical systems 
are derived from {\em thermodynamic potentials}. 
The latter should be expressed using  
their canonical variables. 
The common thermodynamic potentials are:
\beq
E\left(S,X\right) 
&\hspace{2cm}& dE=TdS-ydX
\\
F\left(T,X\right) = E-TS,     
&& dF=-SdT-ydX
\\
G\left(T,y\right) \equiv F+yX,  
&& dG=-SdT+Xdy
\\
S\left(E,X\right), 
&& dS=\frac{1}{T}dE+\frac{y}{T}dX
\eeq
The derivatives of the state equations are know as 
the "thermodynamic constants" though they are not really constant...  
\beq
C \equiv T\frac{\partial S}{\partial T}
\hspace*{3cm}
\chi \equiv \frac{\partial y}{\partial X}
\eeq
In the context of gases 
\beq
\text{Compressibility} &\equiv& -\frac{1}{\Vol} \frac{\partial \Vol}{\partial P} \ \ \ \ \ \ \text{[common notation - "beta" or "kappa"]}
\\
\text{ExpansionCoeff} &\equiv& \frac{1}{\Vol} \frac{\partial \Vol}{\partial T} \ \ \ \ \ \ \ \ \ \text{[common notation - "alpha"]}
\eeq

\sheadC{The Gibbs Hamiltonian approach}

It is customary in thermodynamics to define ``thermodynamic potentials" 
that are obtained from the Helmholtz free energy by means of Legendre transform.
This can be regarded as a formal mathematical trick for switching the role 
of conjugate variables, but it also can be motivated physically. 
It is the same procedure that we had discussed regarding the calculation 
of the tension of a polymer. Here we repeat it  with regard 
to a gas in a box with piston.   

Let us regard the position of the piston (the parameter~$X$) 
as a dynamical variable (let us call it~$x$). We can apply force, 
say ``electric" field $f$ on the piston. 
Accordingly the ``Gibbs Hamiltonian" of the system is 
\beq    
\mathcal{H}_G \ \ = \ \ \mathcal{H}(\cdots,x) +  f x \ + \ \text{[optional kinetic term]}
\eeq
The optional kinetic term is required if the piston has finite mass, 
but its inclusion will not affect the analysis because it factors out of the calculation. 
Given ${x=X}$ the force that the system exerts on the piston is ${y(X) =\braket{ -\partial \mathcal{H}/\partial x}_X}$. Once $x$ becomes a dynamical variable, and $f$ is introduced, the equilibrium point of the piston is determined by the equation $f = y(x)$, hence the sign convention for the second term in $\mathcal{H}_G$.

The partition function of $\mathcal{H}_G$ is related to that of $\mathcal{H}$ by Laplace transform:
\beq    
Z_G(\beta,f) 
\ \ = \ \ \sum_{x,r} \eexp{-\beta E_{x,r}} 
\ \ = \ \ \sum_x  Z(\beta, x) \ \eexp{ -(\beta f) x }  
\eeq
This can be written as 
\beq    
\eexp{-G(T,f)/T} 
\ \ = \ \ \sum_x  \exp\left[-\frac{F(T,x)+f x}{T}\right] 
\eeq
In the thermodynamic limit fluctuations can be neglected, 
and a saddle point approximation implies
\beq
G(T,f) \ \ \approx \ \  \min_{x} \Big\{ F(T,x) + f x \Big\}
\ \ = \ \ F(T,\bar{x})+f \bar{x}
\eeq 
where the most probable value~$\bar{x}$ is determined 
by solving the state equation ${ f = -F'(x) }$. 
Accordingly we realize that $G(T,f)$ is the Legendre transform
of $F(T,X)$. The roles of the conjugate variable $X$ and $f$ have been 
switched. If $X$ and $f$ are the volume $\Vol$ and the pressure $P$, 
then $G(T,P)$ is known as the Gibbs function.

\sheadC{The chemical potential}

Consider a gas that consists of $N$ {\em identical} particles.
This can be either {\em classical} or {\em quantum} gas
(contrary to prevailing misconception, quantum mechanics is 
irrelevant to this issue - this will be explained 
in the "chemical equilibrium" lecture).  
Within the framework of the canonical formalism 
we define the chemical potential as follows:
\beq
\mu\left(T,\Vol,N\right)
\ \ \equiv \ \ \frac{\partial F}{\partial N}
\eeq
Accordingly we have 
\beq
dF \ \ &=& \ \ -SdT - Pd\Vol + \mu dN
\\
dG \ \ &=& \ \ -SdT  + \Vol dP +  \mu dN 
\eeq
The above definition of the chemical potential can be motivated 
by adopting a "grand Hamiltonian" perspective.
Let us define  a "grand system" that consists of the system 
and of a reservoir of particles. This reservoir consists of a huge 
collection of sites that hold a huge number of particles with binding energy $\mu$. 
If we transfer $N$ particle from the reservoir to the system the energy of 
the "grand system" becomes 
\beq
\mathcal{H}_G \ \ = \ \ \mathcal{H} \ - \ \mu N
\eeq
The so called grand partition function  $Z_G(\beta,\mu)$ of the Grand system 
will be discussed in future lecture.

\sheadC{The extensive property}

At this stage it is appropriate to remark on a relation between 
the Gibbs function and the chemical potential that holds 
is the so-called thermodynamic limit. In this limit the system 
acquires an {\em extensive} property that can be formulated mathematically.  
Relating to the Gibbs function $G(T,P;N)$, one observes that if $N$ is multiplied 
by some factor, then the volume $\Vol$ and the entropy $S$ for the same $(T,P)$ 
are expected to be multiplied by the same factor, 
and hence also $G$ should be multiplied by the same factor. 
We therefore write 
\beq
G(T,P,N) \ = \ N \  G(T,P,1) 
\eeq
From $\mu=-dG/dN$ we deduce that the chemical potential 
is merely the Gibbs energy per particle. 
Consequently from the expression for $dG$ it follows that 
\beq
d\mu \ \ = \ \ -\frac{S}{N}dT \ + \ \frac{\Vol}{N}dP
\eeq

\newpage
\sheadC{Work}

In the definition of work the system and the environment 
are regarded as one driven {\em closed} unit. 
If we change $X$ in time then from the ``rate of change formula" 
we have the following exact expression: 
\beq
\frac{dE}{dt} \ = \  
\left\langle\frac{\partial \mathcal{H}}{\partial t} \right\rangle \ = \ 
-\langle \mathcal{F} \rangle_t \ \dot{X}
\eeq
it follows that 
\beq
\mathcal{W} \ \ \equiv \ \ 
\text{work done on the system} \ \ = \ \ 
E_{\tbox{final}}-E_{\tbox{initial}} 
\ \ = \ \  
- \int \langle \mathcal{F} \ \rangle_t \ dX
\eeq
This is an exact expression. 
Note that $\langle \mathcal{F} \rangle_t$ 
is calculated for the time dependent 
(evolving) state of the system.
In a quasi-static adiabatic process one 
replaces  $\langle \mathcal{F} \rangle_t$ 
by  $\langle \mathcal{F} \rangle_{X(t)}$, 
where the notation  $\langle \mathcal{F} \rangle_{X}$
implies that the system is assumed to be 
in a canonical state at any moment. 
More generally, within the framework of linear response theory 
\beq
\langle \mathcal{F} \rangle_t \ \ \approx \ \ \langle \mathcal{F} \rangle_X \ - \eta \dot{X} 
\ \ = \ \  y(X) - \eta \dot{X} 
\eeq
The first terms is the conservative force, 
which is a function of $X$ alone. 
The subscript implies that the expectation 
value is taken with respect to the instantaneous 
adiabatic state. The second term is the leading 
correction to the adiabatic approximation. 
It is the ``friction" force which is proportional 
to the rate of the driving. The net conservative work 
is zero for a closed cycle while the ``friction" leads 
to irreversible dissipation of energy with a rate  
\beq 
\dot{{\cal W}}_{\tbox{irreversible}} \ = \ \eta \dot{X}^2
\eeq
More generally it is customary to write 
\beq 
{\cal W}  \ \ = \ \ -W  + {\cal W}_{\tbox{irreversible}} 
\eeq
where the first term is the conservative work, 
or so to say ``the work which is done by the system"
\beq
W \ \  = \ \ \int \langle \mathcal{F} \rangle_X \ dX 
\ \ = \ \ \int_{X_A}^{X_B} y(X) \ dX
\eeq
The two main examples that illustrate the above discussion are:  \\

\begin{minipage}{0.35\hsize}

{\bf Example 1:} box with piston \\

\includegraphics[height=3.8cm]{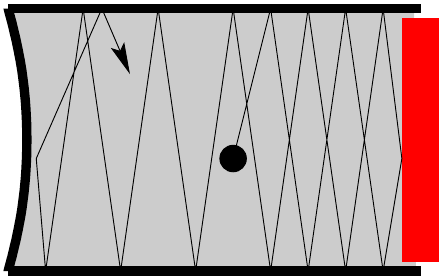} 

\beq
& X &= \text{position of a wall element (or scatterer)} 
\nonumber \\
& \dot{X} &= \text{wall (or scatterer) velocity}
\nonumber \\
& \langle \mathcal{F} \rangle &= \text{Newtonian force} 
\nonumber \\
& -\eta \dot{X}  &= \text{friction law}
\nonumber \\
& \eta \dot{X}^2 &= \text{rate of heating}
\nonumber
\eeq

\end{minipage}
\hspace*{0.1\hsize}
\begin{minipage}{0.4\hsize}

{\bf Example 2:} ring with flux \\

\includegraphics[height=4cm]{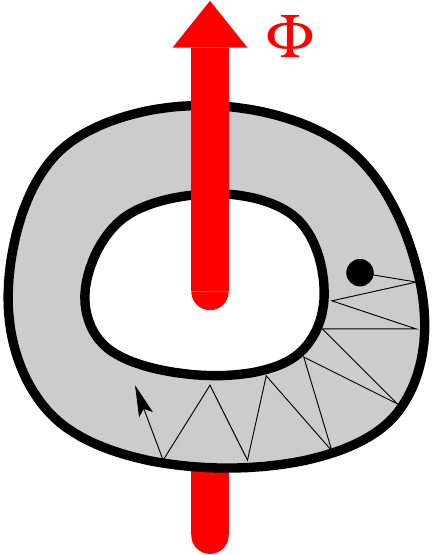}

\beq
& X &= \text{magnetic flux through the ring} 
\nonumber \\
& -\dot{X} &= \text{electro motive force}
\nonumber \\
& \langle \mathcal{F} \rangle &= \text{electrical current} 
\nonumber \\
& -\eta \dot{X}  &= \text{Ohm law}
\nonumber \\
& \eta \dot{X}^2 &= \text{Joule law}
\nonumber
\eeq

\end{minipage}
\hspace*{0.1\hsize}

In the first example instead of having a displaceable wall ("piston") 
we can have a moveable object {\em inside} the box ("scatterer").
In the latter case there is friction while the conservative force is 
zero (because the volume of the box is not changing).

\sheadC{Heat}

In order to understand which type of statements 
can be extracted form the canonical formalism 
we have to discuss carefully the physics 
of work and heat. We distinguish between the system 
and the environment and write the Hamiltonian 
in the form
\beq
\mathcal{H}_{\tbox{total}}
=\mathcal{H}(\bm{r},\bm{p};X(t))+\mathcal{H}_{\tbox{int}}+\mathcal{H}_{\tbox{env}}
\eeq
It is implicit that the interaction term is extremely 
small so it can be ignored in the calculation of the total energy. 
The environment is characterized by its temperature.
More generally we assume that the environment consists 
of several ``baths" that each has different temperature,  
and that the couplings to the baths can be switched on and off.  
Below we consider a {\em process} in which both the initial 
and the final states are equilibrated with a single bath. 
This requires that at the end of the driving process there 
is an extra waiting period that allows this equilibration. 
It is implied that both the initial and the final states 
of the system are canonical. Now we define 
\beq
\mathcal{W} = \text{work}  
&\equiv& 
\Big(
\langle \mathcal{H}_{\tbox{total}} \rangle_{B}
-\langle \mathcal{H}_{\tbox{total}}\rangle_{A}
\Big)
\\
\mathcal{Q} = \text{heat}
&\equiv &
-\Big(\langle  
\mathcal{H}_{\tbox{env}}\rangle_{B}
-\langle  \mathcal{H}_{\tbox{env}}\rangle_{A}
\Big)
\\
E_{\tbox{final}}-E_{\tbox{initial}}  
& \equiv & 
\langle \mathcal{H}\rangle_{B}
-\langle \mathcal{H}\rangle_{A}
\ \ = \ \ \mathcal{W} \ + \ \mathcal{Q}
\eeq
It is important to emphasize that the definition of 
work is the same as in the previous section, 
because we regard  $\mathcal{H}_{\tbox{total}}$ 
as describing an isolated driven system. 
However, $E$ is redefined as the energy of the system only, 
and therefore we have the additional term $\mathcal{Q}$ 
in the last equation.

{\bf Note.-- }
It is possible to treat work and heat on equal footing. For this purpose one should define ``work agents" in analogy to ``heat baths". The work agent is described by an Hamiltonian  $\mathcal{H}_{\tbox{agent}}(X,P)$, and $\mathcal{W}$ is defined as the change of its energy. For example, a piston is described by ${[1/(2M)] P^2 + fX}$, where $f$ is a weight. Assuming a large mass~$M$, the work is stored in the form of potential energy of the weight.

\sheadC{Quasi static process}

In general we have the formal identity:
\beq
dE \ \ = \ \ 
\sum_{r} dp_{r} E_{r}
+\sum p_{r} dE_{r}
\eeq
We would like to argue that for an ideal (reversible) 
quasi-static process we can identify the 
first term as the heat $\dbar Q$ and the second 
term is the work $-\dbar W$.  
One possible scenario is having no driving.
Still we have control over the temperature 
of the environment. 
Assuming a {\em volume preserving} quasi-static process we have
\beq
dX &=& 0
\\
dE &=& \sum dp_{r}E_{r}=TdS
\\
\dbar Q &=& TdS
\\
\dbar W &=&0
\\
\mathcal{Q} &=& [E(B)-E(A)] 
\\
W &=& 0
\eeq
A second possible scenario is having 
an isolated system going through 
an adiabatic process: 
\beq
dp_{r}&=&0
\\ 
dE&=& \sum_r p_{r}dE_{r}=-ydX
\\
\dbar Q&=&0
\\
\dbar W &=& ydX
\\
\mathcal{Q} &=& 0
\\
W &=& - [E(B)-E(A)] 
\eeq
Any general quasi-static process can be 
constructed from small steps as above, 
leading to   
\beq
\mathcal{Q} &=& \int_{A}^{B} T dS
\\
W &=& \int_{A}^{B} y(X) dX 
\eeq
In particular for {\em isothermal} process we get
\beq
\mathcal{Q} &=& T \times [S(B)-S(A)]
\\
W &=& - [F(B)-F(A)] 
\eeq
If a process is both isothermal (constant $T$) 
and isobaric (constant $P$) we can still get 
work being done by changing some other parameter $X$. 
For example $X$ might be a "reaction coordinate". Then we get 
\beq
\mathcal{Q} &=& T \times [S(B)-S(A)]
\\
W &=& - [G(B)-G(A)] 
\eeq

\sheadC{Cycles}

It is possible to design cycles in $(X,T)$ space, 
such that the net effect is to convert heat into work (engine) 
or in reverse (heat pump). Consider for example 
a gas in a cylinder with a piston. If there is no restriction 
on the availability of baths the simplest engine could 
work as follows: Allow the gas to expand at high temperature;   
Lower the temperature; Compress the gas back by moving the piston 
back to its initial position; Raise back the temperature. 
The net effect here is to convert heat into work. This is known 
as the {\em Stirling cycle}. 
A traditional version of a {\em Stirling engine} 
is displayed in the following figure [left panel taken from Wikipedia]:

\reflectbox{\includegraphics[height=0.3\hsize]{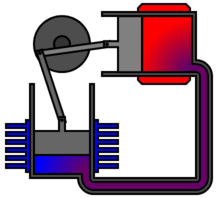}}
\hspace{2cm}
\includegraphics[height=0.3\hsize]{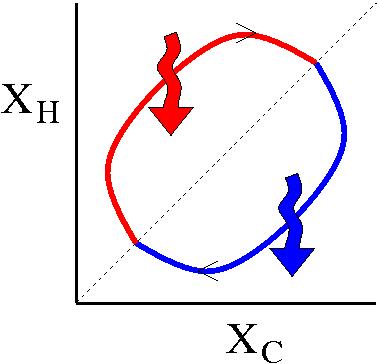}

In order to see the relation between the engine and the cycle 
it is proposed to analyze the operation as follows. 
Denote by $X_\text{H}$ and  $X_\text{C}$ the volumes 
of the hot and cold cylinders. As the wheel is rotated 
it defines a cycle in $(X_\text{C},X_\text{H})$ space.
The $X_\text{H}>X_\text{C}$ segment of the cycle represents 
expansion of gas during the stage when most of it is 
held in high temperature.
The  $X_\text{H}<X_\text{C}$ segment represents the 
compression of the gas during the stage when most of it is 
held in low temperature.

The disadvantage of the Stirling cycle is that in order 
to realize it in a reversible manner we need infinitely 
many intermediate baths in the heating and cooling stages.
The way to do it in practice is to use a ``heat exchange" device.
This device can be regarded as layered structure that is attached  
in one end to the hot bath and in the other end to the cold bath.
As a result each layer is held in a different temperature.
We assume that the layers are quesi-isolated from each other.
The trick is to couple the pipes that lead the gas between the 
hot and the cold cylinders to this layered structure, 
such that they can exchange heat with the layers without 
net effect on the temperature of the layer.        

\includegraphics[height=0.2\hsize]{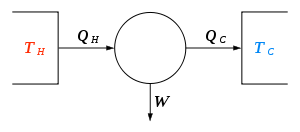}
    
If we want to use a reversible cycle that uses two baths only,    
we can consider the Carnot cycle. See block diagram above [taken form Wikipedia]. 
Note that if we operate this cycle in reverse we get a heat pump instead of an engine.
Let us analyze what happens during a Carnot cycle.  
Assuming that the levels become more dense as $X$ is increased, 
it follows that the result of an adiabatic process would be 
a lower temperature (adiabatic cooling). 
To make this point clear consider just two levels $E_1$ and $E_2$  
with occupation probabilities $p_1$ and $p_2$ respectively. 
The implied temperature is ${T=(E_2-E_1)/[-\ln(p_2/p_1)] }$.
In an adiabatic process the probabilities do not change, 
hence as the level get closer the implied temperature become lower.  
If the process is isothermal rather than adiabatic there will be 
heat absorption (isothermal absorption) and re-distribution 
of the probabilities such that ${p_2/p_1 = \exp[-(E_2-E_1)/T_0]}$ . 
These ``rules of thumb" allow to gain intuition with 
regard to the operation of engines and heat-pumps.

\includegraphics[width=0.9\hsize]{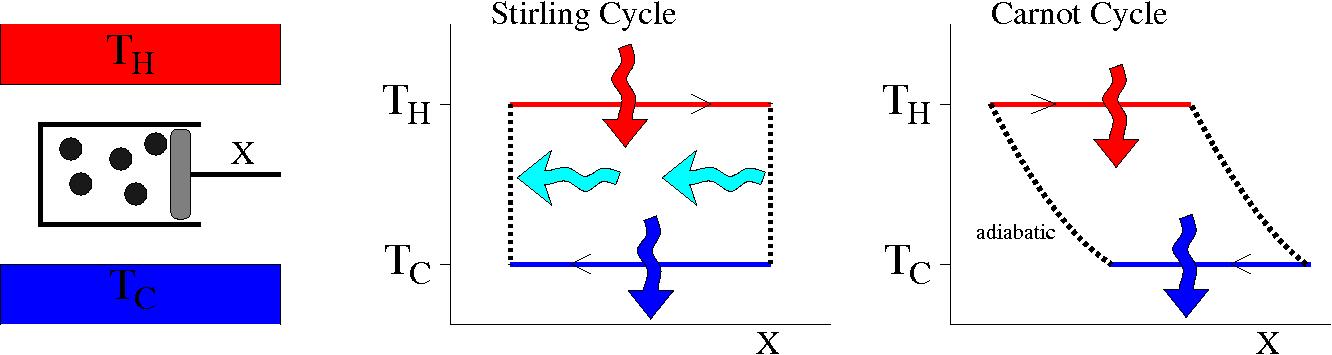}

Besides the piston example, the other simplest example 
for a thermodynamic cycle concerns spin $S\gg1$ in magnetic field. 
In order to be consistent with the piston 
example we define $X=-|h|$, so larger $X$ is like larger volume, 
i.e. higher density of states.
We consider a cycle that consists of 4 stages: 
adiabatic cooling to lower temperature; 
isothermal absorption stage (${Q_\text{C}>0}$);
adiabtic heating to higher temperature;
and isothermal emission stage (${Q_\text{H}<0}$).   
The net effect is is to transfer heat 
from the cold bath to the hot bath, 
which requires to invest work.
  
At each stage the work $W$ is positive or negative depending 
on whether the occupied levels go down or up respectively.
The inclination is to say that during the adiabatic cooling
stage the work is positive. This is true in the piston
example, but not in general, as can be verified with the spin example.
It should be clear that doing work on the system does not 
imply that its temperature becomes higher: the simplest 
demonstration would be to take an isolated container with gas to the top 
of Eifel Tower: it requires work but the temperature is not affected.   
What is essential for the operation of the cycle is the variation 
in the {\em density} of the levels, irrespective of whether they 
go up or down during the cycle.

\newpage
\makeatletter{}\sheadB{Chemical equilibrium and the Grand Canonical state}

\sheadC{The Gibbs prescription}

In this lecture we are going to discuss chemical equilibrium.
We shall see that the condition of chemical equilibrium
involves the chemical potentials of the participating gases.
For the purpose of calculating $\mu$ it is essential 
to find how the partition function depends on the number of particles.   
Classically the calculation of $Z_N$ for a gas of {\em identical} 
particles is done using the Gibbs prescription:
\beq
Z_{N}\text{[Gibbs]} \ \ = \ \ \frac{1}{N!}\mathbb{Z}_{N}\text{[distinguishable particles]}
\eeq
We shall motivate this prescription in the following 
discussion of chemical equilibrium. For an ideal gas we get 
\beq
Z_{N}  =  \frac{1}{N!}Z_{1}^{N},
\ \ \ \ \ \ \ \ \text{where} \ \
Z_{1} =  \frac{\Vol}{\lambda_{T}^{3}}
\sum \eexp{-\beta\varepsilon_{\tbox{bound}}}
 \equiv  g_0  \frac{\Vol}{\lambda_{T}^{3}}
\eeq
The summation is over the non-translational freedoms of the particle. 
Hence we get 
\beq
\mu \ \ = \ \  \frac{\partial F}{\partial N} 
\ \ = \ \ T\ln\left(\frac{N}{Z_{1}}\right)
\ \ = \ \ \varepsilon_0 + T\ln\left(\frac{N}{\Vol} \lambda_{T}^{3}\right)
\eeq
where in the last equality we have assumed that the particle 
has a single well defined binding energy. 
The inverse relation is 
\beq
N \ \ = \ \ Z_1 \ \eexp{\beta\mu}  \ \ = \ \ \frac{\Vol}{\lambda_{T}^3} \ \eexp{-(\varepsilon_0-\mu)/T}
\eeq

{\bf The notion of identical particles:-- }
The notion of {\em identical} particles does not require extra explanations  
if they are {\em indistinguishable} as in the quantum mechanical theory.  
Still we can ask what would happen if our world were classical. 
The answer is that in a classical reality one still has to maintain 
the Gibbs prescription if one wants to formulate a {\em convenient} 
theory for Chemical equilibrium. 
Namely, the condition for "chemical equilibrium" that we derive below 
has a simple form if we adopt the Gibbs prescription.
Without the Gibbs prescription one would be forced to formulate  
an equivalent but non-friendly version for this condition.

\sheadC{Chemical equilibrium}

Consider the following prototype problem of chemical equilibrium:
\beq
A[a] \rightleftharpoons A[b]
\eeq
where "a" and "b" are two phases, say to be in one of two regions in space, 
or to be either in the bulk or on the boundary of some bounded region. 
Given $N$ identical particles we characterize the macroscopic 
occupation of the two phases by a reaction coordinate $n$, 
such that $N-n$ particles are in phase [a] 
and $n$ particles are in phase [b]. 
The partition function is 
\beq
\mathbb{Z}_N^{ab} 
\ \ = \ \ 
\sum_n \left\{\frac{N!}{(N-n)! \ n!}\right\} \ \mathbb{Z}_{N-n}^{a} \mathbb{Z}_{n}^{b}
\eeq
The combinatorial "mixing" factor 
in the curly brackets counts the number of possibilities 
to divide $N$ particles into two groups. 
It should be excluded if the particles are {\em indistinguishable}, 
as in the quantum theory. In the classical theory, 
where the particles are {\em distinguishable} it 
should be included, but it can be absorbed 
into the definition of the partition function.
This is what we call the ``Gibbs prescription".  
Using the Gibbs prescription the above sum can be re-written as follows:
\beq
Z_N^{ab} 
\ \ = \ \ 
\sum_n Z_{N-n}^{a} Z_{n}^{b}
\eeq
The probability to have $(N{-}n,n)$ occupation is proportional 
to the $n$th term in the partition sum:  
\beq
p(n) \ \ = \ \ \left\{\frac{N!}{(N-n)!n!}\right\}  
\times \frac{\mathbb{Z}_{N-n}^{a} \mathbb{Z}_{n}^{b}}{\mathbb{Z}_N^{ab}} 
\ \ = \ \ \frac{Z_{N-n}^{a}Z_{n}^{b}}{Z_N^{ab}} 
\ \ = \ \ 
C \exp\left[-\beta \left(F^{a}\left(N-n\right)+F^{b}\left (n\right)\right)\right]
\eeq
One should appreciate the usefulness 
of the Gibbs prescription. It is thanks to this prescription 
that the Free Energy is {\em additive}. If we did not use 
the Gibbs prescription we would be compelled to add in~$F$ 
a term that reflects {\em "mixing entropy"}. 
The most probable value $\bar{n}$
is determined by looking for the largest term. 
This leads to the Chemical equilibrium condition:
\beq
&& F^{a}\left(N-n\right)+F^{b}\left (n\right) \ \ = \ \ \text{minimum}
\\
\leadsto \ \ \ \ \ && -\mu^{a}\left(N-n\right)+\mu^{b}\left(n\right) \ = \ 0
\eeq
Let us consider the case of ideal gases.
Using the expression for $\mu$ we get 
\beq
\frac{n}{N-n} = \frac{Z_{1}^{b}}{Z_{1}^{a}}
\ \ \ \ \ \ \ \ \ \leadsto \ \ \ \ \ \ \ \ \ 
\bar{n} = N \frac{Z_{1}^{b}}{Z_{1}^{a}+Z_{1}^{b}}
\eeq
This example is simple enough to allow a determination 
of the average value $\langle  n \rangle$ too.
The probability distribution of the reaction coordinate is 
\beq
p(n) \ \ = \ \ \frac{N!}{(N-n)!n!}
\frac{
\left(Z_{1}^{a}\right)^{N-n}
\left(Z_{1}^{b}\right)^{n}}
{(Z_1^{a}+Z_1^{b})^N}
\eeq
leading to 
\beq
\langle  n \rangle 
\ \ = \ \ \sum_n p(n) \ n 
\ \ = \ \  \bar{n}
\eeq
We see that the expectation value of $n$
coincides with its typical (most probable) value. 
In the more general case of chemical equilibrium, 
as discussed below, this is an approximation that 
becomes valid for $N\gg1$ in accordance with 
the central limit theorem.

\sheadC{The law of mass action}

This procedure is easily generalized. 
Consider for example
\beq
2C \rightleftharpoons 5A+3B
\eeq
Given that initially there are 
$N_{\tbox{A}}$ particles of type A,
$N_{\tbox{B}}$ particles of type B,
and $N_C$ particles of type C
we define a macroscopic reaction coordinate $n$ 
such that $N_C{-}2n$ is the number of particles of type C, 
and $N_{\tbox{A}}{+}5n$ is the number of particles of type A, 
and $N_{\tbox{B}}{+}3n$ is the number of particles of type B.
Accordingly 
\beq
Z^{abc} \ \ = \ \ \sum_{n} Z_{N_C{-}2n}^{c} Z_{N_{\tbox{A}}{+}5n}^{a} Z_{N_{\tbox{B}}{+}3n}^{b}
\eeq
and
\beq
p(n) \ \ = \ \ \const \ \eexp{-\beta \left(F^{c}(N_C{-}2n)+F^{a}(N_{\tbox{A}}{+}5n)+F^{b}(N_{\tbox{B}}{+}3n)\right)}
\eeq
leading to the equation 
\beq
-2\mu^{c}(N_C{-}2n)+5\mu^{a}(N_{\tbox{A}}{+}5n)+3\mu^{b}(N_{\tbox{B}}{+}3n) \ \ = \ \ 0
\eeq
which with Boltzmann/Gibbs approximation becomes  
\beq
\frac{(N_{\tbox{A}}{+}5n)^{5} (N_{\tbox{B}}{+}3n)^{3}}{(N_C{-}2n)^2}
\ \ = \ \ \frac{
\left(Z_{1}^{a}\right)^{5}
\left(Z_{1}^{b}\right)^{3}
}
{(Z_{1}^{c})^{2}}
\eeq
or, assuming that [a],[b],[c] are all {\em volume} phases, 
\beq
\frac{\left(\frac{N_{\tbox{A}}{+}5n}{\Vol}\right)^{5}
\left(\frac{N_{\tbox{B}}{+}3n}{\Vol}\right)^{3}}
{\left(\frac{N_C{-}2n}{\Vol}\right)^{2}}
\ \ = \ \ \kappa(T)
\eeq
where the reaction rate constant $\kappa(T)\propto \eexp{-\varepsilon/T}$ 
depends on the reaction energy ${\varepsilon = 5\varepsilon_a +3\varepsilon_b - 2\varepsilon_c}$.
In this sign convention $\varepsilon<0$ means exotermic reaction.

\sheadC{Equilibrium in pair creation reaction}

Consider the reaction
\beq
\gamma+\gamma \ \ \rightleftharpoons \ \ \eexp{+} \ + \ \eexp{-}
\eeq
This can be analyzed like a chemical reaction ${C \rightleftharpoons A+B}$, 
which is of the same type as considered in the previous version.
The important point to notice is that $Z^c$ is independent of~$n$, 
and therefore the chemical potential of the electromagnetic field 
is formally ${\mu^c=0}$. The electromagnetic field is like a "bath", 
and we can regard it as part of the environment, hence we could have 
written ${\text{vacuum} \rightleftharpoons \eexp{+} + \eexp{-}}$.
In any case we get at equilibrium 
\beq
\mu^{\eexp{+}}(n_{1}) \ + \ \mu^{\eexp{-}}(n_{2}) \ \ = \ \ 0
\eeq
where in the Boltzmann/Gibbs approximation 
\beq
\mu(n) \ \ \approx \ \ 
\mass c^{2} + T\ln\left(\frac{n\lambda_{T}^{3}}{\Vol}\right)
\eeq
leading to
\beq
n_{1}n_{2}
=\left(\frac{\Vol}{\lambda_{T}^{3}}\right)^{2}
\eexp{-2\mass c^{2}/T}
\eeq
This problem is formally the same as that of a semiconductor 
where $\eexp{+}$ and $\eexp{-}$ are the holes and the electrons, 
and $2\mass c^{2}$ corresponds to the energy gap between the 
valance and the conduction bands. 
Accordingly, an optional derivation of the latter equilibrium condition 
can be based on a grand-canonical perspective (see next lecture) 
with regard to the occupation of the electrons.

\newpage
\sheadC{Equilibrium in liquid-gas system}

The equilibrium between a liquid phase and a gaseous phase 
is just another example for a chemical equilibrium. 
We can write the equation that determines the coexistence 
curve in $(T,P)$ diagram as ${[\mu_a(T,P)-\mu_b(T,P)]=0}$.
By implicit differentiation of this equation with respect to~$T$ 
we get the Clausius-Clapeyron relation     
\beq
\left. \frac{dP}{dT} \right|_{\text{coexistence}} 
\ \ = \ \ -\frac{\partial_T [\mu_a-\mu_b]}{\partial_P [\mu_a-\mu_b]} 
\ \ = \ \ \frac{\Delta S }{\Delta V} \ \ = \ \ \frac{1}{T} \ \frac{\text{[Latent heat]}}{\text{[Volume change]}}
\eeq
Outside of the coexistence curve either $\mu_a$ or $\mu_b$ are 
smaller, and accordingly all the particles occupy one phase only.

\sheadC{Site system}

The chemical potential can be calculate easily 
for a system of $N$ identical particles that occupy a set 
of $M$ sites (or modes) that have the same binding energy $\varepsilon$.
Since we assume that the biding energy is the same for all sites, 
it follows that estimating $Z_1$ is essentially 
a combinatorial problem. We assume $n\gg1$ so we can 
approximate the derivative of $\ln(n!)$ as $\ln(n)$. 
We also write the result for the most 
probable $n$ which is obtained given $\mu$. 
Note that the result for ${\bar{n}}$ is meaningful 
only for large ${M}$.

\putgraph{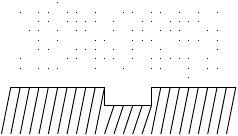}

{\bf Fermionic site:-- } 
Each site can have at most one particle
\beq
Z_{n} &=& \frac{M!}{n! (M-n)!} \eexp{-\beta\varepsilon n}
\\
\mu &=& \varepsilon+T\ln\left(\frac{n}{M-n}\right)
\\
\bar{n} &=& M (\eexp{\beta(\varepsilon-\mu)}+1)^{-1}
\eeq

{\bf Bosonic site:--} 
Each site can have any number of particles.
The combinatorial problem is solved by asking how many
ways to divide $n$ particles in a row with ${M-1}$ partitions. 
If the particles were distinct the result would be ${(n+(M-1))!}$.
Taking into account that the particles are indistinguishable we get 
\beq
Z_{n}&=&\frac{(n+M-1)!}{n! (M-1)!} \eexp{-\beta\varepsilon n}
\\
\mu &=& \varepsilon+T\ln\left(\frac{n}{\left(M-1\right)+n}\right)
\\
\bar{n} &=& (M-1) (\eexp{\beta(\varepsilon-\mu)}-1)^{-1}
\eeq

{\bf Electromagnetic mode:-- } 
Each mode of the electromagnetic field 
can be regarded as a Bosonic site 
that can occupy photons with binding energy $\omega$. 
Since $n$ is not constrained 
it follows formally that 
\beq
\mu &=&  0 
\\
\bar{n} &=& (\eexp{\beta \omega}-1)^{-1}
\eeq

{\bf Boltzmann approximation:-- } 
Assuming dilute occupation ($1\ll n \ll M$) 
we get a common approximation 
for both Fermi and Bose case:
\beq
Z_{n} &=& \frac{M^n}{n!} \eexp{-\beta\varepsilon n} 
\\
\mu&=&\varepsilon+T\ln\left(\frac{n}{M}\right)
\\
\bar{n} &=& M \eexp{-\beta(\varepsilon-\mu)}
\eeq

{\bf General system of sites:-- } 
If we want to consider the partition function of $N$ particles 
in $M$ sites that have different binding energies we have to calculate 
\beq
Z_{N}(\beta) \ \ =  \ \ \sum_{n_1+...+n_M=N} \eexp{-\beta(\varepsilon_1 n_1+...+\varepsilon_M n_M)} 
\eeq
Because of the constraint the sum cannot be factorized.
We therefore adopt the "Grand Hamiltonian" strategy and calculate  
the Grand partition function $\mathcal{Z}(\beta,\mu)$ that 
corresponds to ${\mathcal{H}_G=\mathcal{H}-\mu N}$. 
In principle we can get $Z_{N}(\beta)$ from $\mathcal{Z}(\beta,\mu)$ 
via an inverse transform, but in practice it is more convenient 
to stay with the Grand Hamiltonian framework.

\sheadC{The grand canonical formalism}

We can regard the grand canonical formalism 
as a special case of the canonical formalism, 
where the Grand Hamiltonian  ${\mathcal{H}_G=\mathcal{H}-\mu N}$
describes a Grand system that consists of the gas particles  
and a hypothetical reservoir. Optionally we can 
motivate the introduction of a the grand canonical formalism 
following the same justification strategy as in the 
case of the canonical formalism. First we have to specify 
the many body eigenstates $R$ of the system: 
\beq
\hat{N}|R\rangle &=& N_{R}|R\rangle 
\\
\hat{H}|R\rangle &=& E_{R}|R\rangle 
\eeq
Then we assume that the system can exchange particles 
as well as energy with the environment. 
The probability of a many body eigenstate $R$ is 
\beq
p_{R} \ = \ \frac {\eexp{-\beta E_{R}} Z_{\bar{N}-N}^{\tbox{env}}} {Z^{\tbox{sys+env}}},
\ \ \ \ \ \ \ \ \ \
\text{with} \ \ \
Z_{\bar{N}-N}^{\tbox{env}} \ \ \propto \ \ \eexp{\beta \mu N}
\eeq
We deduce that
\beq
p_{R} \ \ = \ \ \frac{1}{\mathcal{Z}} \eexp{-\beta \left(E_{R}-\mu N_{R}\right)}
\eeq
where the normalization constant is 
\beq
\mathcal{Z}(\beta,\mu) \ \ \equiv \ \ \sum_R  \eexp{-\beta \left(E_{R}-\mu N_{R}\right)}
\eeq

The Grand Canonical $\mathcal{Z}(\beta,\mu)$ is defined in complete 
analogy with the canonical case as sum over the many body states "R". 
For some purposes it is convent to write is as a function $\mathcal{Z}(z; \beta)$ 
of the fugacity:  
\beq
z \equiv \eexp{\beta\mu}, 
\hspace*{3cm}
\frac{1}{\beta}\frac{\partial}{\partial \mu}=z \frac{\partial}{\partial z}
\eeq
The Grand Canonical $\mathcal{Z}(\beta,\mu)$ can serve 
as a generating function as follows:
\beq
&& N \equiv \langle  \hat{N}\rangle 
= \sum_R p_{R}N_{R}
=\frac{1}{\beta}\frac{\partial \ln \mathcal{Z}}{\partial \mu}
\\
&& E-\mu N
=-\frac{\partial \ln \mathcal{Z}}{\partial \beta}
\\
&& P \equiv 
\left\langle  -\frac{\partial H}{\partial \Vol}\right\rangle 
=\frac{1}{\beta}\frac{\partial \ln \mathcal{Z}}{\partial \Vol}
\eeq
Equivalently 
\beq
&& F_{G}(T,\Vol,\mu)  \equiv -\frac{1}{\beta}\ln \mathcal{Z}
\\
&& N = -\frac{\partial F_{G}}{\partial\mu}
\\
&& P = -\frac{\partial F_{G}}{\partial \Vol}
\\
&& S = - \frac{\partial F_{G}}{\partial T}
\\
&& E=F_{G}+TS+\mu N
\eeq
In the thermodynamic limit ${F_{G}}$ is extensive, 
also in the case of non ideal gas. Consequently 
\beq
&& F_{G}(T,\Vol,\mu) \ = \ -\Vol P(T,\mu)
\\
&& dP \ = \ \frac{S}{\Vol}dT + \frac{N}{\Vol}d\mu
\eeq
In other words rather then using the notation ${F_{G}}$, 
we can regard $P(T,\mu)$ as the generating function.  
Note that this is the "Grand canonical" version 
of the "canonical" Gibbs function relation 
\beq
d\mu \ = \ -\frac{S}{N}dT + \frac{\Vol}{N}dP
\eeq
For constant $T$, a variation in the chemical potential 
is related to a variation $dP=nd\mu$ in the pressure, 
where ${n=N/\Vol}$ is the density. In the canonical 
setup $N$ is fixed, while in the grand-canonical setup $\Vol$ is fixed.  
The compressibility of the gas can be expressed as follows: 
\beq
\kappa_T 
\ \ = \ \ -\left.\frac{1}{\Vol}\frac{d\Vol}{dP}\right|_N 
\ \ = \ \ \left.\frac{1}{N}\frac{dN}{dP}\right|_{\Vol} 
\ \ = \ \ \frac{1}{n}\frac{dn}{dP} 
\ \ = \ \ \frac{1}{n^2}\frac{dn}{d\mu}
\eeq

\sheadC{Fermi occupation}

A site or mode can occupy ${n=0,1}$ particles. 
The binding energy is ${\epsilon}$.
the site is in thermochemical equilibrium 
with a gas in temperature ${\beta}$ and chemical 
potential ${\mu}$. 
\beq
N_n &=& n 
\\
E_n &=& n\epsilon
\\
p_{n}&=&\frac{1}{\mathcal{Z}}\eexp{-\beta(\epsilon-\mu)n}
\eeq
and accordingly, 
\beq
\mathcal{Z}(\beta,\mu) &=& \left(1+\eexp{-\beta(\epsilon-\mu)}\right)
\\
N(\beta,\mu) 
&=& \langle  \hat{n} \rangle 
\ = \ \sum_{n}p_{n}n
\ = \ \frac{1}{\eexp{\beta(\epsilon-\mu)}+1}
\equiv f(\epsilon-\mu)
\\
E(\beta,\mu) 
&=& \langle  \hat{n}\epsilon \rangle 
\ = \ \epsilon f(\epsilon-\mu)
\eeq
We have defined the Fermi occupation function ${0 \leq f(\epsilon-\mu) \leq 1}$

\sheadC{Bose occupation}

A site or mode can occupy ${n=0,1,2,3...}$ particles. 
The binding energy is ${\epsilon}$.
the site is in thermochemical equilibrium 
with a gas in temperature ${\beta}$ and chemical 
potential ${\mu}$. 
\beq
N_n &=& n 
\\
E_n &=& n\epsilon
\\
p_{n}&=&\frac{1}{\mathcal{Z}}\eexp{-\beta(\epsilon-\mu)n}
\eeq
and accordingly, 
\beq
\mathcal{Z}(\beta,\mu) &=& \left(1-\eexp{-\beta(\epsilon-\mu)}\right)^{-1}
\\
N(\beta,\mu) 
&=& \langle  \hat{n} \rangle 
\ = \ \sum_{n}p_{n}n
\ = \ \frac{1}{\eexp{\beta(\epsilon-\mu)}-1}
\equiv f(\epsilon-\mu)
\\
E(\beta,\mu) 
&=& \langle  \hat{n}\epsilon \rangle 
\ = \ \epsilon f(\epsilon-\mu)
\eeq
We have defined the Bose occupation function ${0 \leq f(\epsilon-\mu) \leq \infty}$. 
If ${\epsilon < \mu }$ then ${\langle n\rangle \rightarrow \infty}$. 
If ${\epsilon = \mu }$ then the site may have any occupation. 
If ${\epsilon < \mu }$ then ${\langle n\rangle}$ is finite.

\sheadC{Bosonic mode occupation}

The occupation of a mode of vibration, say the number photons 
in an electromagnetic mode, or the number of phonons in a vibration mode,   
are described by the canonical ensemble, by can be optionally 
regarded as described by the grand-canonical ensemble with ${\mu=0}$.
With slight change in notations we have:
\beq
N_n &=& n 
\\
E_n &=& n\omega
\\
p_{n}&=&\frac{1}{Z}\eexp{ - \beta \omega n}
\eeq
and accordingly, 
\beq
Z(\beta) &=& \left(1-\eexp{-\beta \omega}\right)^{-1}
\\
N(\beta) 
&=& \langle  \hat{n} \rangle 
\ = \ \sum_{n}p_{n}n
\ = \ \frac{1}{\eexp{\beta \omega}-1}
\equiv f(\omega)
\\
E(\beta) 
&=& \langle  \hat{n}\omega \rangle 
\ = \ \omega f(\omega)
\eeq

\newpage
\makeatletter{}\sheadB{Quantum ideal gases}

\sheadC{Equations of state}

In what follows, unless written otherwise $\epsilon=0$ 
is the ground state and 
\beq
\sum_{r} \rightarrow
\int_{0}^{\infty} \gdos(\epsilon)d\epsilon
\eeq
The stationary states of the multi particle system 
are occupation states 
\beq
|\bm{n}\rangle  \ \ = \ \ |n_{1},n_{2},n_{3},...,n_{r},...\rangle
\eeq
where ${n_{r}=0,1}$ for Fermi occupation 
and ${n_{r}=0,1,2,3,4,...}$ for Bose occupation. 
For these states we have 
\beq
N_{\bm{n}} &=& \sum_{r}n_{r}
\\
E_{\bm{n}} &=& \sum_{r}n_{r}\epsilon_{r}
\\
p_{\bm{n}}
&\propto&  
\eexp{-\beta \sum_{r} (\epsilon_{r}-\mu)n_r }
\eeq
which can be factorized. This means 
that each site or mode can be treated 
as an independent system. 
We use ${E}$ and ${N}$ without index for the 
expectation values in an equilibrium state. 
For the Fermionic and Bosonic case we 
have respectively ($\pm$)
\beq
\ln \mathcal{Z} 
&=& \pm \sum_r \ln (1\pm \eexp{-\beta(\epsilon-\mu)}) 
= \beta \int_0^{\infty} \mathcal{N}(\epsilon) d\epsilon \ f(\epsilon-\mu)
\\
N
&=&\sum_{r} \langle \hat{n}_{r} \rangle 
= \sum_{r} f(\epsilon_r-\mu)
= \int_0^{\infty} \gdos(\epsilon) d\epsilon \ f(\epsilon-\mu)
\\
E
&=&\sum_{r} \epsilon_{r} \langle \hat{n}_{r} \rangle 
= \sum_{r}  f(\epsilon_r-\mu) \epsilon_r
= \int_0^{\infty} \gdos(\epsilon) \epsilon d\epsilon \ f(\epsilon-\mu)
\\
P 
&=& \frac{1}{\beta} \frac{\ln \mathcal{Z}}{\Vol}
= \frac{1}{\Vol}\int_0^{\infty} \mathcal{N}(\epsilon) d\epsilon \ f(\epsilon-\mu)
\eeq
It is good to remember that $P(T,\mu)$ can serve as a generating function 
for all other state equations.  This would be true also if the gas were not ideal.
In particular $N/\Vol = dP/d\mu$ relates the density to the chemical potential, 
which implies a relation between the the pressure $P$ and the density~$N/\Vol$.

\sheadC{Explicit expressions for the state equations}

We assume one particle states ${ |r\rangle}$ that have the density 
\beq
\gdos(\epsilon) \ = \ \Vol c \ \epsilon^{\alpha-1}, 
\hspace*{2cm}
\mathcal{N}(E) \ = \ \frac{1}{\alpha} \ \epsilon \ \gdos(\epsilon)
\eeq
For a particle in $d$ dimensional box $\alpha=d/\nu$ 
where $\nu$ is the exponent of the dispersion 
relation $\epsilon \propto |p|^{\nu}$, 
and $c$ is a constant which is related to the mass~$\mass$.
For example, in the case of spin ${1/2}$ particle 
in ${3D}$ space we have 
\beq
\gdos(\epsilon) = 2 \times \Vol
\frac{(2\mass)^{3/2}}{(2\pi)^{2}}
\epsilon^{\frac{1}{2}}
\eeq
The following integral is useful (upper sign for Bose, lower sign for Fermi):
\beq
F_{\alpha}(u) \ \ \equiv \ \ 
\int_{0}^{\infty} 
\frac{x^{\alpha-1} \ dx}{\eexp{x-u}\mp1}
\ \ \equiv \ \
\pm \Gamma(\alpha) \mathrm{Li}_{\alpha}(\pm z), 
\ \ \ \ \ \ \ \ \ \ z \equiv \eexp{u}
\eeq
where the upper/lower sign refers to the Bose and the Fermi case respectively.
Details of the {\em Polylogarithm function} $\mathrm{Li}_{\alpha}(z)$ 
can be found in Wikipedia. In the physics community   
it is commonly denoted as $g_{\alpha}(z)$.  Note that  
\beq
\mathrm{Li}_{\alpha}(z)
\  \equiv \  \sum_{\ell=1}^{\infty}
\frac{1}{\ell^{\alpha}}z^{\ell} \ = \ z + ..., 
\ \ \ \ \ \ \ \ \ \ \ \ 
\mathrm{Li}_{\alpha}(1)\equiv \zeta(\alpha), 
\ \ \ \ \ \ \ \ \ \ \ \ 
\frac{d}{dz}\mathrm{Li}_{\alpha}(z)=\frac{1}{z}\mathrm{Li}_{\alpha{-}1}(z)
\eeq
As $u$ becomes larger the function $F_{\alpha}(u)$
grows faster in the case of a Bose occupation, 
and it either diverges or attains a finite value as $u \rightarrow 0$. 
The finite value ${F_{\alpha}(0) = \Gamma(\alpha) \zeta(\alpha)}$ is attained for ${\alpha>1}$.  
In particular we have ${\Gamma(3/2)=\sqrt{\pi}/2}$ and ${\zeta(3/2) \approx 2.612}$.
For ${\alpha=1}$ one obtains ${\mathrm{Li}_{1}(z)=-\ln(1-z)}$, 
which has logarithmic divergence as $z \rightarrow 1$.
For ${\alpha<1}$ it is easily shown that ${F_{\alpha}(u) \sim [1/(1{-}\alpha)](-u)^{-(1{-}\alpha)}}$
as $u$ approach zero from below.
In the Fermi case the integral is 
always finite. Using the step-like 
behavior of the Fermi occupation function 
we obtains for~$z\gg1$ the so-called Sommerfeld expansion: 
\beq
F_{\alpha}(u) 
\ \ = \ \  
\frac{1}{\alpha} u^{\alpha} \left[1+\alpha(\alpha{-}1)\frac{\pi^2}{6}\left(\frac{1}{u}\right)^2+ ... \right] 
\eeq

\putgraph{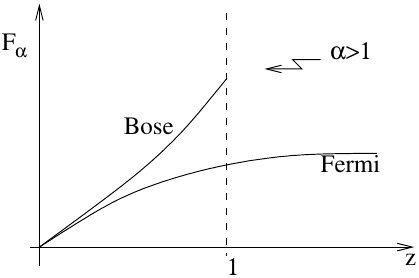}

We can express the state equations using this integral, 
where $z$ is identified as the fugacity. We get
\beq
\frac{N}{\Vol} &=& c T^{\alpha} \ F_{\alpha}\left(\frac{\mu}{T}\right)
\\
\frac{E}{\Vol} &=& c T^{\alpha{+}1} \ F_{\alpha{+}1}\left(\frac{\mu}{T}\right)
\eeq
while $P$ is related trivially to the energy: 
\beq
P \ \ = \ \ 
\frac{1}{\beta} \left(\frac{\ln \mathcal{Z}}{\Vol}\right)
\ \ = \ \ 
\frac{1}{\alpha} \left(\frac{E}{\Vol} \right)
\eeq
The grand-canonical free-energy is ${F_G=-VP}$, from which one can derive the entropy ${S=-(dF_G/dT)_{\mu}}$.  
Optionally the canonical free energy for $N$ particles can be calculated via Legendre transform ${F=F_G+\mu N}$. 
The specific results in the case of a spinless non-relativistic Bose particles are [see also Huang p.231-232;242]: 
\beq
\frac{N}{\Vol} &=& \frac{1}{\lambda_T^3} \ \mathrm{Li}_{3/2}(z)
\\
\frac{E}{\Vol}  &=& \frac{3}{2} \ \frac{T}{\lambda_T^3} \ \mathrm{Li}_{5/2}(z),
\ \ \ \ \ \ \ \ \ \ \ 
P = \frac{2}{3} \left(\frac{E}{\Vol} \right) \ =  \ \frac{T}{\lambda_T^3} \ \mathrm{Li}_{5/2}(z)
\eeq

\sheadC{Ideal gases in the Boltzmann approximation}

We take $\epsilon=0$ as the ground state energy of the one-particle states. 
The Boltzmann approximation is 
\beq
f(\epsilon-\mu) \ \ \approx \ \ \eexp{-\beta(\epsilon-\mu)}
\eeq 
It holds whenever the occupation is $f()\ll 1$.
If it is valid for the ground state ${\epsilon=0}$,  
then it is valid globally for all the higher levels. 
Accordingly the validity condition is $z\ll1$, 
meaning ${\eexp{\beta\mu} \ll 1}$. Under such condition 
one can make the approximation $\mathrm{Li}(z) \approx z$.  
In the case of standard 3D gas the Boltzmann approximation 
condition can be rewritten as 
\beq
N \lambda_T^3  \ \ \ll \ \ \Vol
\hspace{1cm} \leadsto \ T\gg \frac{1}{\mass\ell^2}
\hspace{1cm} \leadsto \ T\gg T_c \ \text{(Bosons)}, 
\hspace{1cm} \leadsto \ T\gg T_{\tbox{F}} \ \text{(Fermions)}
\eeq 
where $\ell=(V/N)^{1/3}$ is the typical distance between particles.  
Is later sections we shall defined the condensation temperature ($T_c$) 
and the Fermi energy ($T_{\tbox{F}}$).
Within the framework of the Boltzmann approximation  
we can re-derive the classical equation of an ideal gas:
\beq
\frac{N}{\Vol} &=& \frac{1}{\lambda_T^3} \ z \ = \ \frac{1}{\lambda_T^3} \ \eexp{\mu/T}
\\
\frac{E}{\Vol} &=& \frac{3}{2} \frac{T}{\lambda_T^3} \ z \  =  \ \frac{3}{2}\frac{N}{\Vol}T
\\
P &=&  \frac{T}{\lambda_T^3} z  \ = \ \frac{N}{\Vol} T  
\eeq 
Note that within this approximation 
$E$ and $P$ do not depend 
on the mass of the particles.

\putgraph[0.5\hsize]{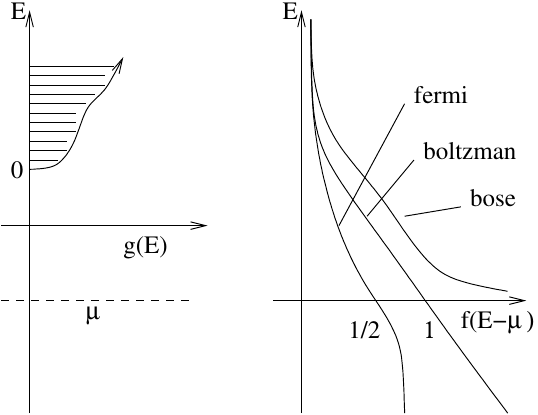}

\sheadC{Bose Einstein condensation}

Let us write again the general expression for the 
occupation of the orbitals:  
\beq
N(\beta,\mu)
\ \ = \ \ \langle n_0 \rangle  + \sum_{r>0} \langle n_r \rangle 
\ \ = \ \ \frac{1}{\eexp{-\beta \mu}-1} \ + \   
c\Vol \int \epsilon^{\alpha{-}1}d\epsilon 
\left(\frac{1}{\eexp{\beta(\epsilon-\mu)}-1}\right)
\eeq
In the limit $\mu\rightarrow 0^{-}$ this expression 
always diverges, so we can invert it and find $\mu$ 
as a function of~$N$. But the physics is more illuminating 
if the ground-orbital occupation ($n_0$) is {\em dropped} from the above expression. 
Then we realize that for ${\alpha > 1}$, notably for ${\alpha=3/2}$,  
the total occupation remains finite, namely ${N(\mu\rightarrow 0^{-})=c\Vol\Gamma(\alpha)\zeta(\alpha) T^{\alpha}}$.
It is implied that the excited states can accommodate 
only a {\em finite} fraction $N/\Vol$ of particles 
in the thermodynamic limit ($\Vol\to\infty$). 
Any additional amount of particles forces ${\mu=0}$,   
and has to condense into the ground state orbital.
The conclusion if different for ${\alpha < 1}$. 
For clarity we change notation to ${\Vol=L^d}$ and ${\alpha=d/2}$.
The integral is dominated by the implicit lower cutoff ${k \sim 1/L}$.
Hence we get ${ N(\mu\rightarrow 0^{-}) \propto L^d (L^2)^{1-(d/2)}T}$.
It is implies that the excited states can accommodate {\em any} fraction $N/L^d$ of particles 
in the thermodynamic limit ($L\to\infty$). So in the latter case condensation is not forced.
The figure below illustrates the reasoning of extracting $\mu$ versus $T$ for a given $N$ in both cases .

\putgraph[0.7\hsize]{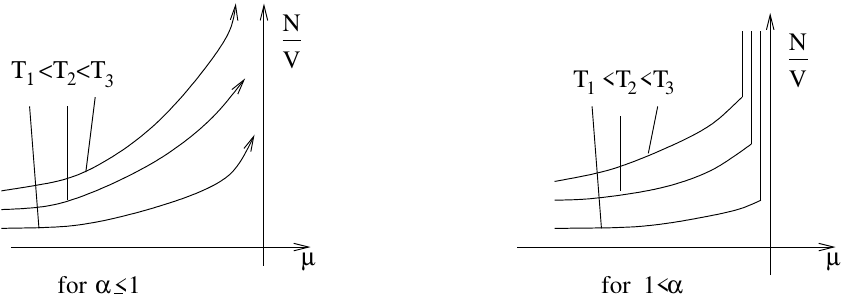}

\putgraph[0.7\hsize]{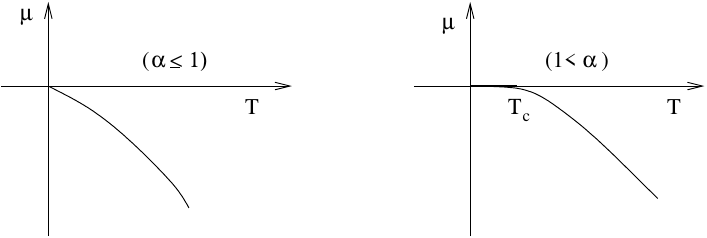}

Considering ${\alpha > 1}$, having ``${\mu=0}$" implies that 
\beq
N \ &=& \ n_0 \ + \  c\Vol \Gamma(\alpha)\zeta(\alpha) T^{\alpha}
\\
E  \ &=& \ c\Vol \Gamma\left(\alpha{+}1\right)\zeta\left(\alpha{+}1\right)T^{\alpha{+}1}
\\
P  \ &=& \ \frac{1}{\alpha} \left(\frac{E}{\Vol}\right)
\eeq
In particular the standard results for condensation in 3D are
\beq
N &=& n_{0}+\Vol\zeta\left(\frac{3}{2}\right)\left(\frac{m}{2\pi}\right)^{\frac{3}{2}}T^{\frac{3}{2}}
\\
P &=& \zeta\left(\frac{5}{2}\right)\left(\frac{\mass}{2\pi}\right)^{\frac{3}{2}}T^{\frac{5}{2}}
\eeq
The pressure $P$ is independent of the total number of particles, 
because the condensate does not have any contribution. 
Hence the compressibility ${\kappa \propto ( {\partial P}/{\partial \Vol} )^{-1} = \infty}$. 
If we change the volume the extra/missing
particles just come from the ground state, 
which is like a reservoir of ${\mu=0}$ particles.

Given $T$, if we push $N$ particles into a box, the condition 
to have condensation is ${N > N(\beta,\mu\rightarrow 0^{-})}$.
The condensation temperature, below which ${\mu=0}$, is
\beq
T_{c} \ \ = \ \ \left(\frac{1}{c\Gamma \left(\alpha\right)\zeta
\left(\alpha\right)}\frac{N}{\Vol}\right)^{{1}/{\alpha}}
\ \ \sim \ \ \frac{1}{\mass \ell^2}
\eeq
where $\ell$ is the average distance between the particles.  
Given $N$, if one tries to eliminate $\mu$, 
and writes it as a function of~$T$, 
then one observes that below the condensation 
temperature ${\mu}$ is forced to become zero. 
Under such circumstances all the particles 
that cannot be occupied in the excited states 
have to condense in the ground state:
\beq
\langle n_0 \rangle 
\ \ = \ \ 
N \ - \ N\left(\beta,\mu\rightarrow 0^{-}\right)
\ \ = \ \ \left(1-\left(\frac{T}{T_{c}}\right)^{\alpha}\right)N
\eeq
The common phrasing is that a macroscopic fraction of the particles 
occupies the ground state. This fraction is determined by $(T/T_c)^{\alpha}$
or equivalently by $[\Vol/\lambda_T^3]/N$. Note that $T\gg T_c$ is an 
optional way to write the Boltzmann condition.

\sheadC{Fermi gas at low temperatures}

At zero temperatures the Fermi function is a step function.
At finite temperatures the step is smeared over a range~$T$.
In order to find explicit expressions for the state functions 
we have to perform an integral that involves the product 
of $f(\epsilon)$ with a smooth function $g(\epsilon)$.
The latter is the density of states  $\gdos(\epsilon)$ 
if we are interested in~$N$, or $\epsilon\gdos(\epsilon)$ 
if we are interested in~$E$. The Sommerfeld expansion is 
a procedure to get an approximation, say, to second-order in~$T$. 
For this purpose we first define the zero temperature result
\beq
G(\mu) \ \ \equiv \ \ \int_{-\infty}^{\mu} g(\epsilon) d\epsilon 
\eeq
And then proceed with the finite temperature calculation 
using integration by parts: 
\beq
\int_{-\infty}^{\infty}  d\epsilon \ g(\epsilon) \ f(\epsilon-\mu)
\ \ &=& \ \ 
\int_{-\infty}^{\infty} d\epsilon \ G(\epsilon) \ [-f'(\epsilon-\mu)]
\ \ \equiv \ \ 
\int_{-\infty}^{\infty} d\epsilon \ G(\epsilon) \ \delta_T(\epsilon-\mu)
\\
\ \ &=& \ \ 
\int_{-\infty}^{\infty}  d\epsilon  \ \left[G(\mu)+ G'(\mu)(\epsilon-\mu) + \frac{1}{2}G''(\mu)(\epsilon-\mu)^2 + ... \right] \ \delta_T(\epsilon-\mu)
\\
\ \ &=& \ \ 
G(\mu) \ + \ \frac{\pi^{2}}{6}T^{2}G''(\mu) \ + \ \mathcal{O}(T^4)
\eeq
We can apply this formula to the ${N=\mathcal{N}(\mu)}$ calculation.
First we do the zero temperature integral, 
and from it eliminate $\mu$ as a function of $N$. 
This zero temperature result is known as the Fermi energy $\mu=\epsilon_{F}$. 
Then we substitute ${\mu=\epsilon_{F}+\delta \mu}$ 
in the above second order expression, 
expand ${G(\mu)\approx G(\epsilon_{F})+\gdos(\epsilon_{F})\delta\mu}$ , and find 
\beq
\mu(T) \ \ \approx \ \ \epsilon_{F} \ - \ \frac{\pi^{2}}{6}
\frac{\gdos'\left(\epsilon_{F}\right)}{\gdos\left(\epsilon_{F}\right)}T^{2} \ + \ \mathcal{O}(T^4)
\eeq

\putgraph[0.6\hsize]{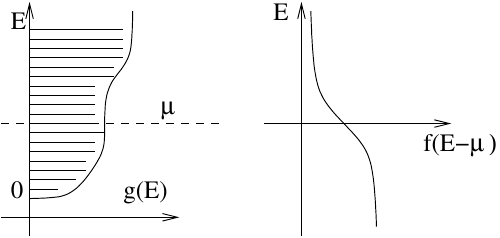}

\putgraph[0.6\hsize]{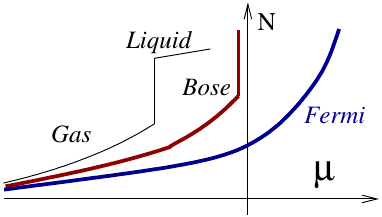}

The specific result for $N$ fermions in system with $\epsilon^{\alpha-1}$ density of orbitals, 
and in particular for spinless non-relativistic fermions in 3D box is:
\beq
N \ \ = \ \ 
\frac{1}{\alpha}c\Vol \mu^{\alpha}\left[1+\alpha\left(\alpha-1\right)\frac{\pi^{2}}{6}\left(\frac{T}{\mu}\right)^{2}+...\right]
\ \ = \ \ 
\Vol \frac{1}{6\pi^{2}}\left(2m\right)^{\frac{3}{2}}\mu^{\frac{3}{2}}\left[1+\frac{\pi^{2}}{8}\left(\frac{T}{\mu}\right)^{2}+...\right]
\eeq
leading after elimination to
\beq
\epsilon_F \ &=& \ \left(\frac{\alpha}{c}\frac{N}{\Vol}\right)^{\frac{1}{\alpha}} 
\ \ = \ \ 
\frac{1}{2\mass}\left(6\pi^{2}\frac{N}{\Vol}\right)^{\frac{2}{3}}
\\
\mu \ &=& \
\left[1-\left(\alpha{-}1\right)\frac{\pi^{2}}{6}\left(\frac{T}{\epsilon_{F}}\right)^{2}+...\right] \epsilon_{F}
\ \ = \ \ 
\left[1-\frac{\pi^{2}}{12}\left(\frac{T}{\epsilon_{F}}\right)^{2}+...\right] \epsilon_{F}
\eeq
For the energy we get
\beq
E \ \ = \ \ \Vol \frac{3}{5}\frac{1}{6\pi^{2}}\left(2m\right)^{\frac{3}{2}}\mu^{\frac{5}{2}} \left[1+\frac{5\pi^{2}}{8}\left(\frac{T}{\mu}\right)^{2}+...\right]
\ \ = \ \  \left[1+\frac{5\pi^{2}}{12}\left(\frac{T}{\epsilon_{F}}\right)^{2}+...\right] \frac{3}{5}N\epsilon_{F}
\eeq
The pressure is given by the equation
\beq
P \ \ = \ \ \frac{2}{3}\left(\frac{E}{V}\right) 
\ \ = \ \ \frac{1}{5}\left(6\pi^{2}\right)^{\frac{2}{3}}\frac{1}{\mass} 
\left(\frac{N}{\Vol}\right)^{\frac{5}{3}} \ + \ \mathcal{O}(T^2)
\eeq
The grand-canonical free-energy is ${F_G=-VP}$ from which one can derive the entropy ${S=-(dF_G/dT)_{\mu} \propto T}$.  
The canonical free energy for $N$ particles can be calculated via Legendre transform ${F=F_G+\mu N}$ leading to 
\beq
F(T,V,N) \ \ = \ \ \left[1-\frac{5\pi^{2}}{12}\left(\frac{T}{\epsilon_{F}}\right)^{2}+...\right] \frac{3}{5}N\epsilon_{F}
\eeq
From here one can recover the expression for the entropy ${S=-(dF/dT)_{N}}$, 
and additionally calculate the heat capacity ${C_V = T(dS/dT)_{V,N} \propto T}$ 
for a closed system of fermions.

\sheadA{Systems with interactions}

\makeatletter{}
\sheadB{Interactions and phase transitions}

{\bf Energy scales:-- }
With regard to the prototype models of systems with interactions 
there are generically two energy scales. 
One is the interaction strength~${\varepsilon}$, 
and the other is the temperature~$T$.
For ${T \gg \varepsilon}$ a perturbative treatment 
is appropriate. See below the cluster expansion. 
For ${T \ll \varepsilon}$ it is advised to 
re-model the system with Hamiltonian that describes 
its collective excitations. The interesting
regime is ${T \sim \varepsilon}$ where the 
phase transition takes place.

{\bf Models of interest:-- }
It is natural to start with the discussion with the phenomenology 
of the gas-liquid phase transition, as implied by the Van-der-Waals 
equation of state. Later one realizes that the essential physics  
is captured by the "lattice gas" version, which is formally equivalent 
to the "Ising model" that describes a ferromagnetic phase transition. 
Its generalization is known as the "Potts model".
The system consists of $\mathcal{N}$ sites. 
At each site there is a "spin" that can be in one of 
$s$~possible states. The Ising model is a special case 
with $s=2$, and the interaction is $\sigma_i \sigma_j$, 
where $\sigma=\pm1$. 
The "Ising model" has a discrete up/down symmetry.
Its Field theory version is known as the Landau model.
The Heisenberg model is a different lattice model  
that has $S_i\cdot S_j$ interaction. This interaction 
has a continuous symmetry with respect to rotations. 
Its 2D version is known as the XY model.       
There are also corresponding Field theory models 
that are known as non-linear sigma models.

{\bf First order phase transition.-- }
There are systems where there are (say) two families of states, 
such that each family has different DOS with different minimum. 
In such case a control parameter (call it $h$) might induce a crossover 
from the dominance of one family to the dominance of the second family.
This crossover is reflected in the partition function and hence 
in the heat capacity and in the state equations. 
In the thermodynamic limit the crossover might be abrupt. 
In such case it is a "first order phase transition". 
If a change in a parameter leads to a {\em bifurcation} in the calculation 
of the partition function, it is called a "second order phase transition".  
The prototype example for phase transition is ferromagnetism 
where the magnetization might be "up" or "down".

{\bf Order parameter.-- }
In order to analyze a second order phase transition 
it is useful to identify the "order parameter", 
which is a field $\varphi(x)$ that describes the coarse grained 
state of the system. 
In the prototype example of ferromagnetism 
it is the magnetization density in the sample. 
Defining an entropy functional $S[\varphi]$ that 
reflects the number of microscopic states that have the same 
field configuration, we can express the partition function as 
\beq
Z \  =  \ \sum_{\varphi} \eexp{-A[\varphi]}, 
\ \ \ \ \ \ \ \ \ \ 
A[\varphi] \ = \ \frac{1}{T} \Big[ E[\varphi] - TS[\varphi] \Big] 
\eeq

{\bf Symmetry breaking.-- }
Second order phase transition is due spontaneous symmetry breaking
leading to long range order. At ${T=0}$ the definition of 
symmetry breaking is very simple. It means that $E[\varphi]$ 
attains (say) two distinct minimum that are described 
by different field configurations (different "order"). 
However, at finite temperature the canonical state is not 
the minimum of the energy functional $E[\rho]$ 
but of the free energy functional $F[\rho]=E[\rho]-TS[\rho]$.
Accordingly entropic contribution may wash away the broken symmetry. 

There is an implicit assumption with regard 
to the possibility to observe "symmetry breaking". 
It is assumed that $\varphi(x)$ has slow dynamics. 
If a magnet is prepared (say) with "up" magnetization 
then it takes a huge time until it flips 
to the quasi degenerate "down" magnetization.

{\bf Long range order.-- }
In the prototype examples at high temperatures  
there is no "order" meaning that the correlation 
function ${g(r)=\langle \varphi(r) \varphi(0) \rangle}$
decays exponentially. As the critical temperature   
is approached from above the correlation length 
diverges. Below the critical temperature 
there is "long range order" and the 
correlation function should be re-defined 
with respect to the new order.
There is a possibility to witness 
"infinite order phase transition" 
where below the critical temperature 
there is no long range order, but instead   
the correlation function become powerlaw.
See discussion of the XY model.

{\bf Formal analysis.-- }
Disregarding a few models that possess exact solutions, 
the analysis of the partition function can be done 
by adopting the following stages:
{\bf \ (1)} Primitive mean field theory evaluates 
the partition function by calculating $A[\varphi]$
for the field configuration that minimizes it. 
This corresponds mathematically to an inferior saddle point approximation.
{\bf \ (2)} Proper mean field theory is based 
on proper saddle point approximation, which means 
that the calculation takes into account the Gaussian 
fluctuations around the minimum.  
{\bf \ (3)} Renormalization Group (RG) treatment
is required in the critical regime, 
whenever the Gaussian approximation in not valid. 
It explains the scaling anomalies that are witnessed 
in the vicinity of the critical temperature.

\sheadC{Gas of weakly interacting particles}

Consider a classical gas of interacting particles:
\beq
\mathcal{H} \ \ = \ \ \sum_{i=1}^{N} \frac{\vec{p_{i}}^{2}}{2\mass}
\ + \ \sum_{\langle ij \rangle} u\left(\vec{x}_{i}-\vec{x}_{j}\right)
\eeq
The partition function without the Gibbs factor is  
\beq
\mathsf{Z}_{N}\left(\beta,\Vol\right) 
\ \ &=& \ \ 
\left(\frac{1}{\lambda_{T}^3}\right)^{N}
\int dx_1...dx_N  \ \exp\left[-\beta \sum_{\langle ij \rangle} u\left(x_{ij}\right)\right]
\\
\ \ &\equiv& \ \ 
\left(\frac{1}{\lambda_{T}^3}\right)^{N} 
\int dx_1...dx_N \ \prod_{\langle ij \rangle} \left(1+f\left(x_{ij}\right)\right), 
\hspace*{2cm} f(r)\equiv \eexp{-\beta u(r)}-1
\eeq
Note that the configuration space integral has the dimensions of $\Vol^N$. 
It equals $\Vol^N$ if there are no interaction. 
If there are interactions we can regard the $f(r)$ as a perturbation.
Then we can expand the product and perform integration term by term. 
The result can be organized as an expansion:
\beq
\mathsf{Z}_{N}\left(\beta,\Vol\right) \ \ = \ \ 
\left(\frac{\Vol}{\lambda_{T}^3}\right)^{N} 
\ \left[1 
+ \text{coef}_2\left(\frac{N}{\Vol}\right) 
+ \text{coef}_3\left(\frac{N}{\Vol}\right)^2 
+ ... 
\right]^N
\eeq
Note that we have raised an $N$ using ${(1+NS)\approx(1+S)^N}$, 
such that $S$ is an expansion in powers of the density $(N/V)$. 
From here we can derive the so called Virial expansion for the pressure:
\beq
P 
\ \ = \ \ 
\frac{NT}{\Vol} 
\left[
1 + a_2 \left(\frac{N}{\Vol}\right) 
+ a_3 \left(\frac{N}{\Vol}\right)^2 
+ ... 
\right]
\ \ = \ \ 
T \
\sum_{\ell=1}^{\infty} a_{\ell}(T) \left(\frac{N}{\Vol} \right)^{\ell}
\eeq
The $a_{\ell}$ are known as the virial coefficients.
From the above it is implied that 
\beq
a_2 \ \ = \ \ 
-\frac{1}{2} \int f(r) d^3r \ \ = \ \  
\frac{1}{2}\int  \left[1 - \eexp{-\beta u(r)}\right] d^3r  
\hspace*{3cm} \mbox{[classical]}
\eeq
More generally it is implied from the discussion in the next sections 
that in order to get $a_2$ we just have to calculate 
the two-body partition function $\mathsf{Z}_2$. Namely:
\beq
a_2 \ \ = \ \ -\frac{(\lambda_{T}^3)^2}{\Vol} \ \frac{1}{2!}\left[\mathsf{Z}_2-\mathsf{Z}_1^2\right]
\hspace*{3cm} \mbox{[general, no Gibbs prescription here!]}
\eeq
The calculation of $\mathsf{Z}_2$ for two interacting 
quantum particles, given the scattering phase-shifts, 
has been outlined in a past lecture.
In the classical case it is standard to assume that the gas particles 
are like hard spheres, each having radius $R$, 
with some extra attractive part that has depth $\sim\epsilon_0$, 
similar to Lenard-Jones potential. Using high temperature 
expansion in $\beta$ we get in leading order 
\beq
a_2 
\ \ \approx \ \ \frac{1}{2}\Big[1 - \frac{\epsilon_0}{T}\Big] \frac{4\pi}{3}(2R)^3
\ \ \equiv \ \  \bar{b} -\frac{\bar{a}}{T} 
\hspace*{3cm} \mbox{[Van-der-Waals]}
\eeq
The coefficients $\bar{a}$ and $\bar{b}$ appear in the phenomenological Van-der-Waals equation of state
that we shall discuss in a later stage. They are related to the attraction 
between the particles, and to their hard-core radius.  
Note that~$\bar{b}$ is the excluded volume per particle multiplied by $2^{d-1}$, where $d=3$. 
Contrary to a common misconception it is only in 1D that~$\bar{b}$ equals the excluded volume.

\sheadC{The grand canonical perspective}

It is simplest to deduce the Virial expansion from the grand canonical formalism.
From now on the dependence on the temperature is implicit, 
and we emphasize the dependence on the fugacity~$z$. 
The grand canonical partition function using the Gibbs prescription is 
\beq
\mathcal{Z}(z) \ \ = \ \ \sum_{N=0}^{\infty} \frac{1}{N!} \mathsf{Z}_N z^{N}, 
\hspace{3cm} \mbox{[Here $\mathsf{Z}_N$ is defined without Gibbs factor]}
\eeq
For an ideal classical gas all the $\mathsf{Z}_N$ are determined 
by the one-particle partition function, namely $\mathsf{Z}_N=\mathsf{Z}_1^N$.
Accordingly $\ln(\mathcal{Z})$ includes a single term, 
namely  ${\ln(\mathcal{Z})=\mathsf{Z}_1 z}$. It makes sense to assume 
that interactions and quantum effects will add higher order terms. 
Hence we postulate an expansion
\beq
\ln \mathcal{Z}(z)
\ \ = \ \ 
\sum_{n=1}^{\infty} \frac{1}{n!} \mathsf{B}_{n} z^{n} 
\eeq
The relation between the $\mathsf{B}_{n}$ and the $\mathsf{Z}_{n}$ 
is formally the same as the relation between commulants and moments 
in probability theory:
\beq
\mathsf{Z}_1 &=& \mathsf{B}_{1}
\\
\mathsf{Z}_2 &=& \mathsf{B}_{1}^{2} + \mathsf{B}_{2}
\\
\mathsf{Z}_3 &=& \mathsf{B}_{1}^{3} + 3\mathsf{B}_{1}\mathsf{B}_{2} + \mathsf{B}_{3}
\eeq
Or backwards:
\beq
\mathsf{B}_{1} &=& \mathsf{Z}_1
\\
\mathsf{B}_{2} &=& \mathsf{Z}_2-\mathsf{Z}_1^{2}
\\
\mathsf{B}_{3} &=& \mathsf{Z}_3-3\mathsf{Z}_2\mathsf{Z}_1+2\mathsf{Z}_1^{3}
\eeq
We can use these relations both directions: 
First we can evaluate a few $\mathsf{Z}_N$, typically $\mathsf{Z}_1$ and $\mathsf{Z}_2$,  
in order to get the leading order $\mathsf{B}_{n}$ coefficients, 
say $\mathsf{B}_{1}$ and $\mathsf{B}_{2}$.
Once the leading order $\mathsf{B}_{n}$ coefficients are known, 
we can generate from them a generalized 
Gibbs approximation for all(!) the $\mathsf{Z}_N$.

\sheadC{The cluster expansion}

Our objective is to calculate the $\mathsf{B}_n$ coefficients
in the expansion of $\ln(\mathcal{Z})$. For convenience 
we define their scaled versions $b_n$ through the following substitution:
\beq
\frac{1}{n!}\mathsf{B}_n \ \ \equiv \ \ \Vol  \left(\frac{1}{\lambda_{T}^3}\right)^n \ b_{n}(T) 
\eeq
We turn to outline a general diagrammatic procedure to evaluate 
the $b_{n}$ for a classical gas of interacting particles. 
A graph (network, diagram) is a set of vertices (nodes) 
that are connected by edges (connectors, bonds).  
In the present context each diagram represents an integral. 
The sum over all the connected diagrams that have $n$ nodes gives the 
expansion coefficient $\mathsf{B}_n$ of the "comulant" generating 
function~$\ln(\mathcal{Z})$, while the sum over all 
diagrams (including reducible diagrams) gives the 
expansion coefficient $\mathsf{Z}_N$ of the moments generating function~$\mathcal{Z}$.
Formally we write 
\beq
\mathsf{Z}_N \ \ = \ \ 
\left(\frac{1}{\lambda_{T}^3}\right)^{N} 
\int dx_1...dx_N \ \prod_{\langle ij \rangle} \left(1+f\left(x_{ij}\right)\right)
\ \ = \ \ 
\left(\frac{1}{\lambda_{T}^3}\right)^{N} 
\sum [\text{diagrams with} \ N \ \text{nodes}]
\eeq
In this expression a diagram represents an integral of the type
\beq
C[3',1,2,3] \times \int [f(x_{12}) f(x_{23})] \ [f(x_{56})] \ [f(x_{78}) f(x_{89}) f(x_{97})] \ dx_{1}...dx_{9}
\eeq
where $C$ is a combinatorial factor that arise because we identify 
diagrams that differ only in the labelling of the vertices. 
One should realize that if a diagram is reducible, say ${N=n_1+n_2+n_3}$, 
then $C[n_1,n_2,n_3]=[N!/(n_1!n_2!n_3!)]C[n_1]C[n_2]C[n_3]$.
In the above example ${C[3']=3}$ is the number of ways to have 
a triangle with 2~bonds, while $C[3]=1$.     
Using this observation it is not difficult to prove that 
\beq
\mathsf{B}_n \ \ = \ \ 
\left(\frac{1}{\lambda_{T}^3}\right)^{n} 
\sum [\text{connected diagrams with} \ n \ \text{nodes}]
\eeq
The implied expression for the $b_n$ is the same diagrammatic sum, 
but the prefactor is replaced by $1/(n!\Vol)$.
The expressions for the leading coefficients are:
\beq
b_{1} \ \ &=& \ \ 
\frac{1}{\Vol} \int dx \ \ = \ \ 1
\\
b_{2}  
\ \ &=& \ \   
\frac{1}{2!\Vol} \int f(x_{12})\ dx_1dx_2 
\ \ = \ \ 
\frac{1}{2!}   \int f(r)\ dr 
\\
b_{3}  \ \ &=& \ \  
\frac{1}{3!\Vol} 
\int \left[ 3f(x_{12}) f(x_{23}) + f(x_{12})f(x_{23})f(x_{31}) \right] \ dx_1dx_2dx_3 
\eeq

\sheadC{The Virial coefficients}

Having found the $b_n$ the grand canonical partition function is  
\beq
\ln \mathcal{Z}(z)
\ \ = \ \ 
\Vol \sum_{n=1}^{\infty} b_{n}(T) \left(\frac{z}{\lambda_{T}^3}\right)^n
\eeq
where $b_1=1$, and $b_n$ has the dimension of length$^{n-1}$. 
Note that for an ideal Bose or Fermi gas 
one obtains ${b_{n} = (\pm1)^{n{+}1} n^{-5/2} (\lambda_{T}^3)^{n-1}}$.
We would like to find a procedure 
to determine these coefficients if there 
are weak interactions between the particles. 
Once they are known we get the state equations from  
\beq
N &=& z\frac{\partial}{\partial z} \ln \mathcal{Z}
\\
P &=& \frac{T}{\Vol} \ln \mathcal{Z} 
\eeq
leading to 
\beq
\frac{N}{\Vol} 
&=& 
\sum_{n=1}^{\infty} nb_{n}(T) \ \left(\frac{z}{\lambda_{T}^3}\right)^n
\\
\frac{P}{T} 
&=&  
\sum_{n=1}^{\infty} b_{n}(T) \ \left(\frac{z}{\lambda_{T}^3}\right)^n
\eeq
It is customary to eliminate $z/\lambda_{T}^3$ from the first 
equation and to substitute into the second equation, 
thus getting  the virial expansion with the coefficients   
\beq
a_{1} &=& b_{1}=1
\\
a_{2} &=& -b_{2}
\\
a_{3} &=& 4b_{2}^{2}-2b_{3}
\eeq

\sheadC{The Van-der-Waals equation of state}

Consider a classical gas that is composed of $N$ particles in volume $\Vol$. 
The particles have hard core of radius $R$, and the two-body interaction 
is assumed to be attractive, with depth $\sim \epsilon_0$. 
We have formally obtained from the virial expansion the following equation of state:
\beq
P \ \ \approx \ \ \frac{NT}{\Vol} \left[1  +  \left(\bar{b}-\frac{\bar{a}}{T}\right) \frac{N}{\Vol} \right] 
\eeq
where ${\bar{b} \sim R^3}$ and ${\bar{a} \sim \epsilon_0 R^3}$. 
The effect of hard-core repulsion is under-estimated
in this leading order perturbative expansion.
The add-hock correction is to re-write the equation of state as follows:
\beq
P \ \ = \ \ \frac{NT}{\Vol - N\bar{b}} - \left(\frac{N}{\Vol}\right)^2 \bar{a} 
\eeq  
Roughly this equation can be derived by assuming that the partition function 
is like that of an ideal gas, where each particle experiences volume ${\Vol_{\text{eff}} = (\Vol - N\bar{b})}$, 
and mean potential ${\langle U \rangle = -N\bar{a}/\Vol}$.   
Optionally the $a$ term could have been deduced from the virial theorem, 
using the estimate ${\langle r \cdot ({\partial U}/{\partial r}) \rangle \sim N^2 \, ({\epsilon_0 R^3}/{\Vol}) }$. 
If we plot $P$ versus $\Vol$ we find that it becomes non-monotonic 
if the temperature is lower than a critical value. For a detailed analysis 
see {\bf [Huang, section~2.3]}. The critical value of the temperature is 
\beq
T_c \ \ = \ \ \frac{8\bar{a}}{27\bar{b}} \ \ \sim \ \ \epsilon_0
\eeq 
The $P$ dependence for $T<T_c$ is illustrated in the figure below {\bf [taken from Wikipedia]}.
From the relation $P=-dF/d\Vol$ one can deduce the free energy $F(\Vol)$. 
One can argue that there is a $\Vol$ range of instability where the free energy can 
be lowered via phase separation. This is known as Maxwell construction (details below). 
A similar reasoning can be applied to the ferromagntic phase transition 
where the role of $\Vol$ is played by the magnetization.   

\includegraphics[height=5.5cm]{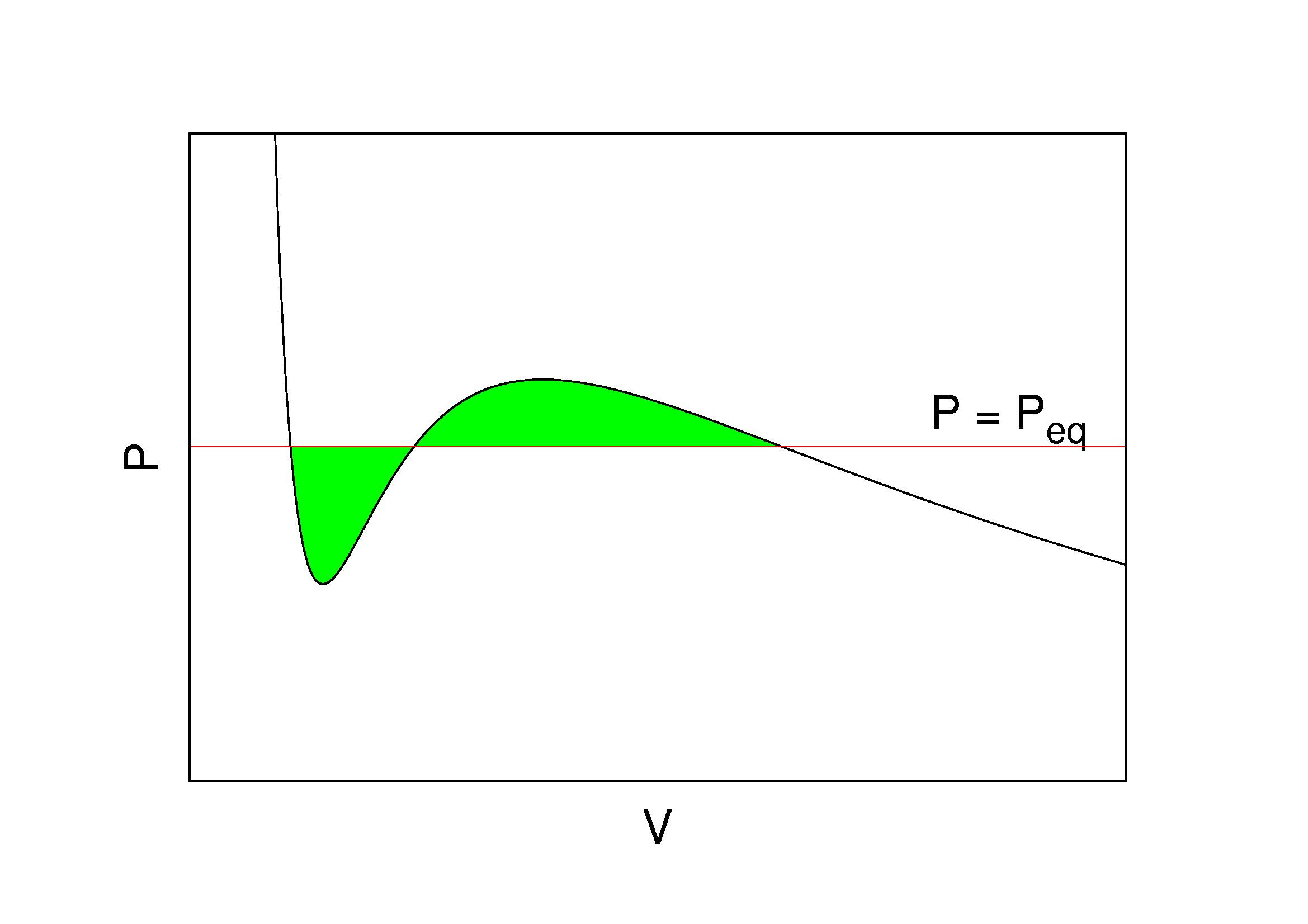}
\ \ \ \ \ \ \ \ \ 
\includegraphics[height=5.5cm]{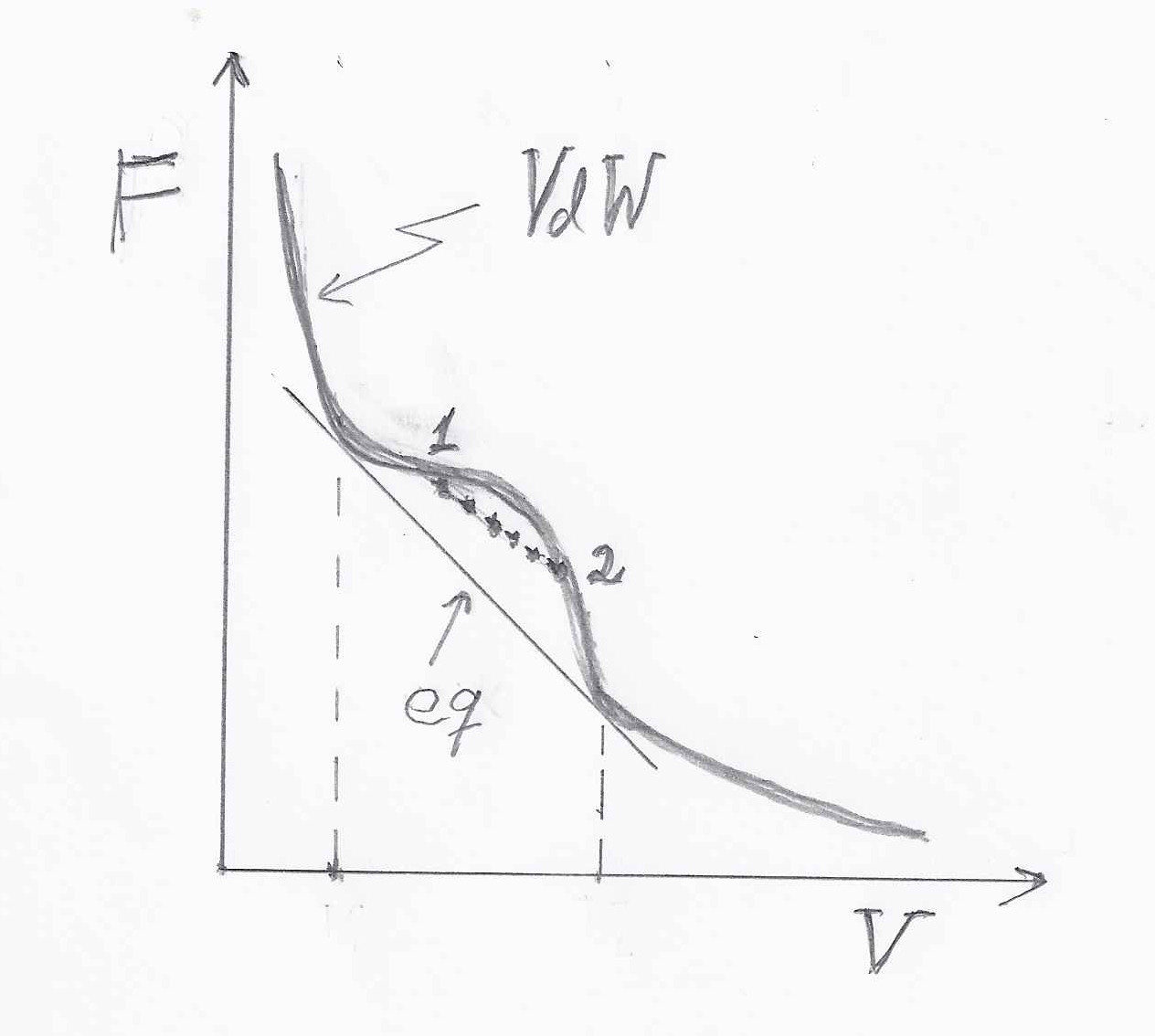}

{\bf Maxwell construction.-- }
Assume that we divide the volume such that ${V=\tilde{V}_1+\tilde{V}_2}$, and the particles 
are partitioned such that ${N=\tilde{N}_1+\tilde{N}_2}$. Using the extensive 
property of $F(\Vol;N)$ we deduce that the free energy of the mixed phase is  
\beq
F_{\text{mix}} \ \ = \ \ 
\frac{\tilde{N}_1}{N}F\left(\frac{N}{\tilde{N}_1}\tilde{V}_1\right)
+ \frac{\tilde{N}_2}{N}F\left(\frac{N}{\tilde{N}_2}\tilde{V}_2\right)
\ \ \equiv \ \ x F(V_1) + (1-x) F(V_2)
\eeq 
where $x\in[0,1]$ is the fraction of particles in phase1, while $V_1$ and $V_2$ are the volumes 
that would be occupied if all the particles were in phase1 or in phase2 respectively.
Observing that the mixture occupies a volume ${V=xV_1+(1{-}x)V_2}$ we deuce that the mixture      
is represented by a point that is located on a chord that connects point1 and point2 
of the $F(V)$ plot above (right panel). It follows that any concave segment of $F(V)$ is unstable:
the free energy can be lowered via phase separation. Observing that $P(V)$ is the derivative 
of $F(V)$, the concave segment can be optionally determined by the "equal area" law.
In reality we expect, as the volume is increased, to go along the constant 
pressure equilibrium line, until all the particles evaporate from the "liquid" phase 
to the "gas" phase. 
        
A possibly simpler perspective on Maxwell construction is to regard the applied $P$ as the free variable, 
and see how $V$ depends on it. The grand Hamiltonian is ${\mathcal{H}_G = \mathcal{H}+PV}$, 
and the grand partition function is related by Laplace transform:
\beq
Z_G(P) \ = \ \int dV \ \eexp{-A(V;P)} 
\ \ \ \ \ \ \ \ \ \ \ \ \  
A(V;P) \equiv \frac{1}{T}\left( F(V)+PV \right)  
\eeq 
Note that a plot of $A(V)$ versus $V$ is related trivially to the plot of $F(V)$.
The integral is dominated by the minimum of $A(V)$, which provide 
the most probable value of~$V$. This leads to the standard Legendre prescription
for the determination of the Gibbs function. 
But here the situation is somewhat subtle. As $P$ is increased we get 
at some point two minima that represent stable and meta-stable solutions. 
As $P$ is further increased, at some stage the two minima will swap, 
implying a jump at~$V$. This swap corresponds to the Maxwell construction.    
We note that the volume~($V$) as a function of the pressure~($P$) 
is analogous to the magnetization~($M$) as a function of the field~($h$), 
which we discuss in more detail later on.

\sheadC{From gas with interaction to Ising problem}

Consider classical gas with interactions ${U\left(\vec{r_{1}},...,\vec{r_{N}}\right)}$.
The $N$ particle partition function is 
\beq
Z_N \ \ = \ \ \frac{1}{N!}\left(\frac{1}{\lambda_T}\right)^{3N}
\int d^{3N}r \ \eexp{-\beta U\left(r_{1},...,r_{N}\right)}
\eeq
We see that the kinetic part factors out, 
hence the whole physics of the interactions  
is in the configuration integral.
Therefore, without loss of generality we can 
consider "static gas". To further simplify the 
treatment we consider a "lattice gas" version:
\beq
\mathcal{H}
\ \ = \ \ U\left(\vec{r_{1}},...,\vec{r_{N}}\right)
\ \ = \ \ \sum_{\braket{x,x'}} 
u\left(x,x'\right)
n(x) n\left(x'\right)
\eeq
We can represent graphically the interaction between 
two sites $x$ and $x'$ by ``bonds".  
The notation  $\langle x,x' \rangle$ means summation 
over all the bonds without double counting. 
In the simplest case there are interactions only 
between near-neighbor sites. 
The grand partition function is 
\beq
\mathcal{Z} \ \ = \ \  
\sum_{n(\cdot)}
\exp\left[ 
-\beta
\left(
\sum_{\braket{x,x'}}
u\left(x,x'\right) 
n(x) n\left(x'\right)
-\mu \sum_x n(x)
\right)
\right]
\eeq
where $n(x)=0,1$. We define 
\beq
n(x)=\frac{1+\sigma(x)}{2}, \ \ \ \ \ \ \ \ \ \ \ \ \ \ \ \sigma(x)=\pm 1
\eeq
Then we get 
\beq
\mathcal{Z} =
\sum_{\sigma(x)}
\exp\left[  
-\beta
\left(
-\sum_{\langle x,x' \rangle} 
\varepsilon\left(x,x'\right)
\sigma(x) \sigma\left(x'\right)
-h \sum_x \sigma(x) + \const
\right)
\right]
\eeq
where ${h=[\mu-\bar{u}]/2}$. Here $\bar{u}$ is the interaction energy per site for full occupation.
Note that ${h=0}$ implies that a fully occupied lattice has the same energy as an empty lattice.
We also have changed notation ${u(x,x')=-4\varepsilon(x.x')}$, with $\varepsilon>0$ corresponding to attractive interaction.

We see that the calculation of ${\mathcal{Z}}$ for static lattice gas 
is formally the same as calculation of ${Z}$ for an Ising model. 
The following analogies should be kept in mind
\beq
\mbox{occupation $N$} & \longleftrightarrow & \mbox{magnetization $M=2N-\mathcal{N}$}
\\
\mbox{chemical potential $\mu$} & \longleftrightarrow & \mbox{magnetic field $2h$}
\\
\mbox{fugacity $z=\eexp{\beta\mu}$} & \longleftrightarrow & \mbox{define $z=\eexp{2\beta h}$}
\\
\mbox{grand canonical $\mathcal{Z}(\beta,\mu)$} & \longleftrightarrow & \mbox{canonical $Z(\beta,h)$}
\eeq
From now on we refer to Ising model, but for the formulation of 
some theorems in the next section it is more convenient to use 
the lattice gas language for heuristic reasons. Note also that $N$ 
is more convenient than $M$ because it does not skip in steps of~$2$.  

\ \\

\hspace*{2cm}
$M$ vs $h$ \ \ \
\hspace*{5cm}
$N$ vs $\mu$ \\ 
\includegraphics[height=3cm]{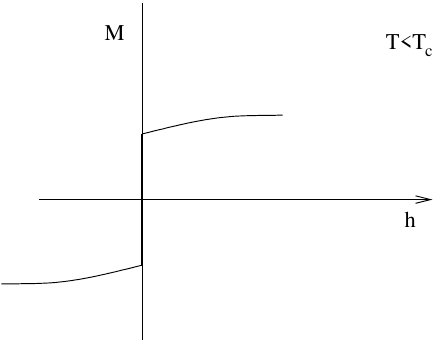}
\hspace{3cm}
\includegraphics[height=3cm]{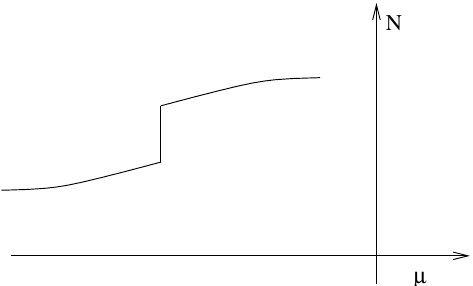}

\ \\

\includegraphics[height=4cm]{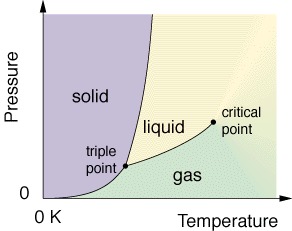}
\hspace{3cm}
\includegraphics[height=4cm]{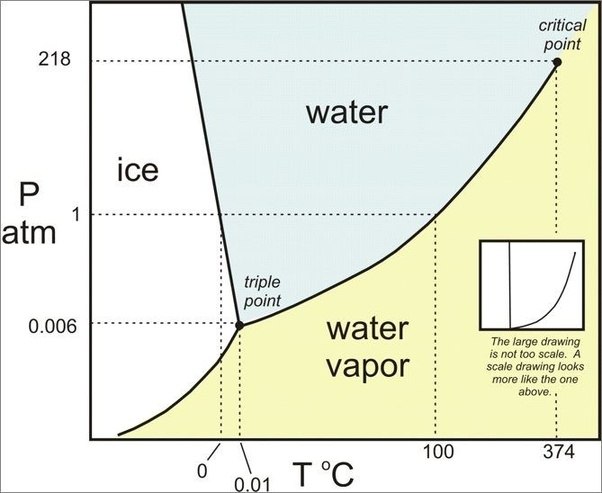}

\includegraphics[height=4cm]{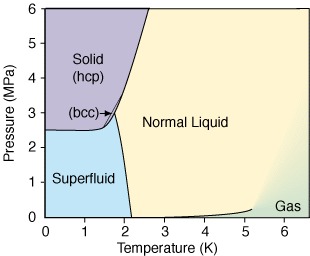}
\hspace{3cm}
\includegraphics[height=4cm]{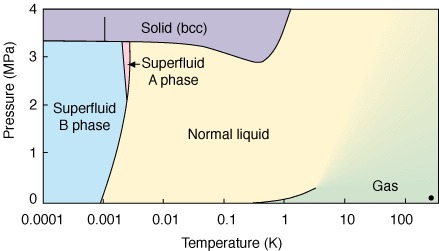}

We shall clarify that in the thermodynamic limit (large $\mathcal{N}$) 
the magnetization density $M/\mathcal{N}$, when plotted as a function of $h$,  
might exhibits a jump at $h=0$. This is called 1st order phase transition. 
Similarly, in the lattice gas model, the density $N/\mathcal{N}$, 
when plotted as a function of $\mu$, might exhibits a jump at $\mu=\bar{u}$.
This can be regarded a gas-to-liquid phase transition.  
In the canonical perspective we fix~$N$ and solve for~$\mu$.
It is best to illustrate the procedure graphically. 
Having a plot of $N$ vs $\mu$, we get a "gas" or a "liquid" state 
provided the horizontal $N$ line does not intersect the discontinuity. 
Otherwise there is a phase separation, where the gas and the liquid phases 
coexist with the same chemical potential~${\mu=\bar{u}}$.

In the phase diagram ${(T,h)}$ of the Ising model 
it is customary to indicate the discontinuity of the 
first order transition by a solid line that stretches 
from ${(0,0)}$ to  ${(T_c,0)}$.
Similarly, in the phase diagram ${(T,\mu)}$ of the lattice gas 
the discontinuity is indicated by a solid line that stretches 
from ${(0,\bar{u})}$ to  ${(T_c,\bar{u})}$.
However in practice it is customary to use a ${(T,P)}$ phase 
diagram. Here we bring the phase diagrams for  
conventional gas-liquid-solid transition, 
for water, for Helium-II and for Helium-III [taken from the web]:

\sheadC{Yang and Lee theorems}

Consider the lattice gas or the equivalent Ising model.\
We can use $n(x)$ or $\sigma(x)$ in order to specify 
whether a cell is filled (spin "up") or empty (spin "down").
The probability of a given configuration is determined  
by the grand-canonical energy ${E_{\sigma}-\mu N_{\sigma}}$, namely 
\beq
p_{\sigma} \ \ \propto \ \ \eexp{-\beta (E_{\sigma} -\mu N_{\sigma}) }
\eeq
Here $E_{\sigma}$ is the Ising energy due to the interactions, 
and ${N_{\sigma}=\sum_x n(x)}$ is the number of "up" spins, 
which we call below "total magnetization".   
In fact the total magnetization is ${2N_{\sigma}-\mathcal{N}}$, 
hence $\mu$ in the lattice-gas model is like $2h$ in the Ising model.  
Either way we use the fugacity as the free variable and write 
the probability of a given configuration as follows: 
\beq
p_{\sigma} \ \ = \ \  \frac{1}{\mathcal{Z}}\left[\eexp{-\beta E_{\sigma}}\right] \  z^{N_{\sigma}}, 
\ \ \ \ \ \ \ \ \ \ \  z \equiv \eexp{\beta \mu} \equiv \eexp{2\beta h}
\eeq
Note that $z=1$ means zero field. The partition function is 
\beq
\mathcal{Z}\left( z;\beta \right) \ \ = \ \ \sum_{N=0}^{\mathcal{N}} Z_N(\beta)  z^{N}
\eeq
where $Z_N$ sums over all the configurations that 
have $N$ spins up, and $\mathcal{Z}$ sums over all 
the possible values of the "total magnetization".  
The Helmholtz function is 
\beq
F\left( z;\beta \right) \ \ = \ \ -\frac{1}{\beta} \ln \mathcal{Z}(z;\beta)
\eeq
The expectation value of the "total magnetization" is 
\beq
\langle N \rangle \ \ = \ \ -\beta z \frac{\partial}{\partial z} F\left(z;\beta\right)
\eeq
As we increase $z$ we expect the magnetization $\langle N \rangle$ 
to grow, and $\langle N \rangle/\mathcal{N}$ to reach a well 
defined value in the limit $\mathcal{N}\rightarrow\infty$.  
Moreover, below some critical temperature we expect to find  
a phase transition. In the latter case we expect $\langle N \rangle$ 
to have a jump at zero field ($z=1$). The Yang and Lee theorems 
formulate these expectations in a mathematically strict way.  
Given $\mathcal{N}$ it is clear that we can write 
the polynomial $\mathcal{Z}$ as a product over its roots:  
\beq
\mathcal{Z}\left(z\right) \ \ = \ \ \const \times \prod_{r=1}^{\mathcal{N}} (z-z_{r})
\eeq  
Consequently 
\beq
F\left(z\right) \ \ &=&  \ \ -\frac{1}{\beta} \sum_{r=1}^{\mathcal{N}} \ln (z-z_{r}) + \const
\\
\langle N \rangle \ \ &=& \ \  z \sum_{r=1}^{\mathcal{N}} \frac{1}{z-z_{r}} 
\eeq
There is a strict analogy here with the 
calculation of an electrostatic field in a 2D geometry.  
In the absence of interactions (infinite temperature) 
we get that all the roots are at $z=-1$. Namely,  
\beq
\mathcal{Z}\left(z;\beta\right) 
\ \ = \ \ \sum_{N=0}^{\mathcal{N}} C^N_{\mathcal{N}} \ z^N 
\ \ = \ \ (1+z)^{\mathcal{N}} 
\ \ \ \ \ \ \ \ \ \ \mbox{[non-interacting sites]}
\eeq  
So we do not have phase transition since the physical axis is $0<z<1$, 
where this function is analytic. The questions are what happens 
to the distribution of the roots as we increase the interaction 
(lower the temperature), and what is the limiting distribution in 
the thermodynamics limit ($\mathcal{N}\rightarrow\infty$).  
There are three statements that give answers to these questions 
due to Yang and Lee. The first statement is regarding the existence 
of the thermodynamics limit:
\beq
\lim_{\mathcal{N} \rightarrow \infty}  
\frac{F\left(z\right)}{\mathcal{N}} \ \ = \ \ \text{exists}
\eeq
The second statement relates specifically to the standard Ising model, 
saying that all the roots are lying on the circle $|z_r|=1$. 
In general other distributions are possible. 
The third statement is that below the critical temperature 
the density of roots at $z=1$ becomes non-zero, 
and hence by Gauss law $\langle N \rangle/\mathcal{N}$
has a jump at zero field. 
This jump is discontinuous in the thermodynamic limit.

\putgraph[0.4\hsize]{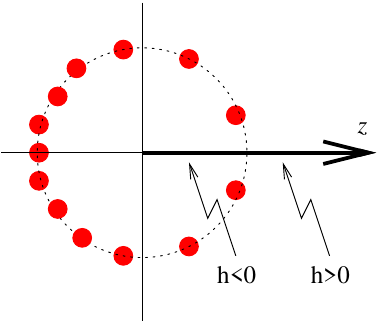}

\newpage
\makeatletter{}
\sheadB{The Ising model}

\sheadC{Model definition}

The energy of a given Ising model configuration state is  
\beq
E[\sigma\left(\cdot\right)]
\ \ = \ \ -\sum_{\langle x,x' \rangle} \varepsilon\left(x,x'\right)
\sigma(x) \sigma\left(x'\right)
- \sum_{x} h(x) \sigma(x)
\eeq
The canonical state is 
\beq 
p[\sigma\left(\cdot\right)] \ \ = \ \ 
\frac{1}{Z}
\eexp{-\beta E\left[\sigma\left(\cdot\right)\right]}
\eeq
where the partition function is 
\beq
Z[h\left(\cdot\right),\beta]
\ \ = \ \ \sum_{\sigma\left(\cdot\right)}
\exp\left[
\beta
\left(\sum_{\langle x,x' \rangle}
\varepsilon\left(x,x'\right)
\sigma(x) \sigma\left(x'\right)
+\sum_{x} h(x) \sigma(x)
\right)
\right]
\eeq
We expand the Helmholtz function as 
\beq
F[h\left(\cdot\right),T]
\ \ = \ \ 
F_0\left(T\right)
-\frac{1}{2T}\sum_{x,x'}
G(x,x') h(x) h\left(x'\right)
+\mathcal{O}\left(h^{4}\right)
\eeq
In the absence of external field $F(T)=F_0(T)$ 
and we have the usual relations 
\beq
E\left(T\right) \ &=& \ F\left(T\right) + T S\left(T\right)
\\
S\left(T\right) \ &=& \ -\frac{\partial F\left(T\right)}{\partial T}
\\
C\left(T\right) \ &=& \ T\frac{\partial S}{\partial T} = \frac{dE}{dT}
\eeq
Next we assume weak field, leading to 
a linear response relation between ${\langle \sigma(x)\rangle}$ and~$h(x)$. 
Namely, 
\beq
\langle \sigma(x)\rangle 
&=&-\frac{\partial F}{\partial h(x)}
\ \ = \ \ \frac{1}{T} \sum_{x'} G(x,x') h\left(x'\right)
\\
\langle\sigma(x)\sigma\left(x'\right)\rangle_0
&=&-T 
\left.
\frac{\partial F}{\partial h(x)\partial h\left(x'\right)}
\right|_{0} \ \ = \ \ G(x,x')
\eeq
For an homogeneous field we get 
${\langle \sigma(x)\rangle=\chi h}$, 
and ${F(T,h) = F_0(T) + (1/2) \mathcal{N} \chi h^2 }$,  
where 
\beq
\chi \ = \ \frac{1}{T}\sum_{r} G(r), 
\hspace{2cm} \mbox{[fluctuation-response relation]}
\eeq
For the total magnetization we get 
\beq
\langle \tilde{M} \rangle
\ \ = \ \  \sum_{x} \langle\sigma(x) \rangle  
\ \ = \ \ \mathcal{N} \chi h
\eeq
Optionally we could get the same result by replacing ${h(x)\mapsto h}$, 
and using ${\tilde{M}=-{\partial F}/{\partial h}}$.
For the fluctuations of the total magnetization we get
\beq
\langle \tilde{M}^2 \rangle 
\ \ = \ \ \sum_{x,x'} \langle  \sigma(x) \sigma(x') \rangle 
\ \ = \ \ {\mathcal{N}} \sum_{r} G(r)
\eeq
Form here we deduce that 
\beq
\chi \ = \ \frac{1}{T} \ \frac{\langle \tilde{M}^2 \rangle}{\mathcal{N}}
\hspace{2cm} \mbox{[fluctuation-response relation]}
\eeq
This is merely another version of the same "fluctuation-response relation".

\sheadC{The spatial correlation function}

It is possible to measure ${G(r)}$ via a scattering experiment.  
Given a configuration $\sigma(x)$ the intensity of the scattering 
in the Born approximation is
\beq
I\left(q\right) \ \ \propto \ \ 
\left|
\int \sigma(x) 
\eexp{-i\vec{q} \cdot \vec{x}} d\vec{x}
\right|^{2}
\eeq
If we average over configurations we get  
\beq
I\left(q\right) 
\ \ \propto \ \  
\int dxdx' 
\ \langle \sigma(x) \sigma\left(x'\right) \rangle 
\ \eexp{-iq\cdot\left(\vec{x}-\vec{x}'\right)}
\ \ \propto \ \  \tilde{G}(q)
\eeq
Here $\tilde{G}(q)$ is the FT of the correlation function 
${G(r) = \langle \sigma(x) \sigma\left(x'\right) \rangle}$,
where $r=|x-x'|$.

We would like to discuss what happens to ${G(r)}$ as 
the temperature is lowered. Specifically we would like 
to illuminate what is the fingerprints of approaching 
a critical temperature of a phase transition, 
below which the system is ``ordered". We note that 
all the discussion below can be repeated if we 
apply an infinitesimal field ${h=+0}$ and approach 
the critical temperature from below. In the latter 
scenario the correlation function should be redefined 
by subtracting the constant $\langle  \sigma \rangle^{2}$.

We shall see in the next section that Landau's approach 
in the Gaussian approximation leads to the Ornstein-Zernike expression 
for the FT of the correlation function:
\beq
\tilde{G}(q) \ \ = \ \ \left(\frac{(1/\xi)}{q^{2}+(1/\xi)^{2}}\right)
\eeq
This leads to 
\beq
G(r) \sim & \exp({-r}/{\xi}) \ \ \ & \text{if} \ \xi<\infty 
\\
G(r) \sim & {1}/{r^{d-2}} \ \ \ & \text{for} \ d{>}2 \ \text{if} \ \xi=\infty 
\eeq
Using the scaled variable $\mathsf{r} = r/\xi$ the exact FT can be expressed in terms of the modified Bessel function of the second kind: 
\beq
G(\mathsf{r}) \ \ = \ \  \frac{1}{(2\pi)^{d/2}} \left(\frac{1}{\mathsf{r}}\right)^{(d/2){-}1} \bm{K}_{(d/2){-}1}(\mathsf{r})  
\eeq
In 1D it is $G(\mathsf{r})=[1/2]e^{-\mathsf{r}}$, and in 3D it is $G(\mathsf{r})=[1/(4\pi\mathsf{r})]e^{-\mathsf{r}}$.

The information about order-disorder transition is in ${G(r)}$.  
If ${\xi< \infty }$ there is no long range order, and we get ${\chi<\infty}$. 
As ${\xi \rightarrow \infty }$ the susceptibility diverges,  
which implies a phase transition. 
Note that for finite $\xi$ the total magnetization $\tilde{M}$ 
can be regarded as a sum of random variables, 
its variance scales like $\mathcal{N}$,   
and consequently $\chi$ comes out finite, 
as implied by the "fluctuation-response relation". 
At the critical temperature the fluctuations are strongly correlated 
over arbitrarily large distances, and $\chi$ diverges.

\sheadC{Critical behavior and the scaling hypothesis}

Below we display the phases diagram in $(T,h)$ space, 
and qualitative plots of the state equations.
For the ${2D}$ Ising model with near neighbor interactions ${T_{c} \approx 2.27 \epsilon}$.

\includegraphics[width=0.4\hsize]{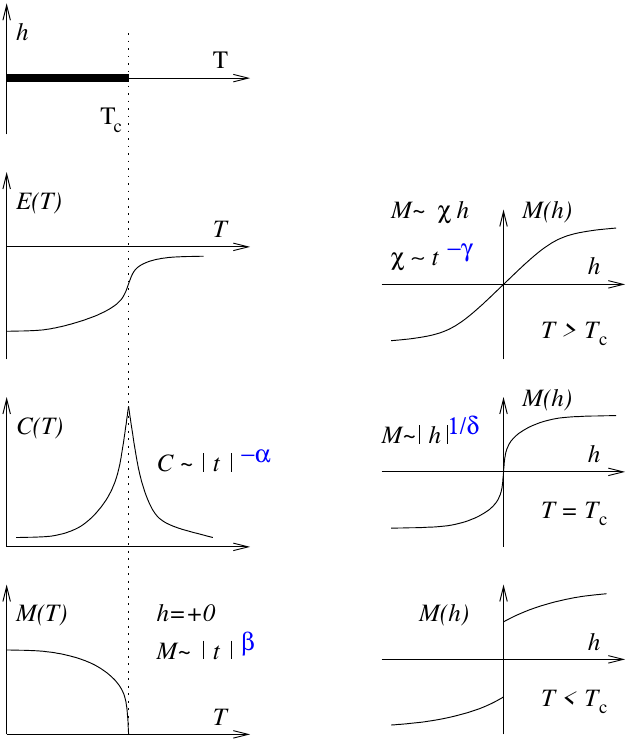}

The state equations in the critical region are characterized by 
the exponents $\alpha, \beta. \gamma, \delta$ (see below). 
Two other exponents $\nu$ and $\eta$ are defined via 
the critical behavior of the correlation function, 
which is assumed to be a variation on the Ornstein-Zernike expression. 
Namely, one conjectures that the divergence of the 
correlation length as ${T\rightarrow T_c}$ is described by 
\beq
\xi \ \ \sim \ \ |T-T_c|^{-\nu}
\eeq
and that the correlation function is 
\beq
G(r) \ \ \sim \ \ \frac{1}{r^{d-2+\eta}}  \ \exp({-r}/{\xi}) 
\eeq
Here we combined the ${T=T_c}$ and the ${T>T_c}$ 
into one expression. This expression describes 
the long range behavior. Note that the "microscopic" short 
range behavior is not too interesting because it is 
bounded by ${G(0)=1}$. The divergence of $\chi$ is due 
to the slow power-law tails. 
Below  ${T_{c}}$ the behavior is similar to ${T > T_{c}}$ 
provided the correlation function is properly defined.
Going on with the same phenomenology the conjecture 
is that away from $T_c$ the correlation distance $\xi$ 
is the only relevant length scale in the thermodynamic limit.
This means that each "correlated block" of the system 
has the same contribution to the Free energy irrespective 
of the temperature, hence  
\beq
F_0(T) \ \ \sim \ \ \frac{\Vol}{\xi^d} \ \ \propto \ \ |T-T_c|^{\nu d}
\eeq
where $d$ is the dimensionality. 
It is customary to define the scaled temperature as ${t=|T-T_c|/T_c}$, 
and to summarize the above conjectures as a {\em scaling hypothesis} 
that involves the critical exponents:  
\beq
G(sr,s^{-1/\nu}t) &=& s^{-(d-2+\eta)} G(r,t)
\\
F_0(s^{-1/\nu}t) &=& s^{-d} F_0(t)
\eeq
From here it follows that 
\beq
\chi(s^{-1/\nu}t) &=& s^{2-\eta} \chi(t)
\\
C(s^{-1/\nu}t) &=& s^{(2/\nu)-d} C(t)
\eeq
From the combined scaling relation 
\beq
F\left(s^{-1/\nu}t ,s^{-(d+2-\eta)/2} h \right)  \ \ =  \ \ s^{-d} F(t,h)
\eeq
we can deduce similar relations for the magnetization.
These scaling relations allow to deduce the critical 
exponents ${\alpha, \beta. \gamma, \delta}$ from ${d,\nu,\eta}$, 
leading to
\beq
C \sim & |t|^{-\alpha}, \,\,\,  & \alpha=2-\nu d
\\
M \sim & |t|^{\beta}, \,\,\,    & \beta=(d-2+\eta)\nu/2
\\
\chi \sim & t^{-\gamma}, \,\,\, & \gamma=(2-\eta)\nu
\\
M \sim & |h|^{1/\delta}, \,\,\, & \delta=(d+2+\eta)/(d-2+\eta)
\eeq
The so called ``classical" mean-field exponents 
that we derive later are 
\beq
\nu = 1/2, \hspace{1cm}
\eta = 0, \hspace{1cm} 
\alpha = 0, \hspace{1cm}
\beta = 1/2, \hspace{1cm} 
\gamma = 1, \hspace{1cm} 
\delta = 3
\eeq
In order to get a non-trivial result for $\alpha$
we have to take into account Gaussian fluctuations 
around the mean field leading to ${\alpha = [2-(d/2)]}$, 
in consistency with the scaling relations.
However, one observes that the classical mean-field exponents satisfy  
the other scaling relations with ${d=4}$, and not with ${d=3}$. 
This implies that we have to go beyond mean field theory in order to establish the experimentally 
observed scaling behavior.

\sheadC{Digression regarding scaling}

A function of one variable has a scaling property if 
\beq
F(sx) = s^{D_F} F(x)
\eeq
where $D_F$ is the scaling exponent. 
It follows that ${F(x)=\const \ x^{D_F}}$. 
For example ${F(x)=x^{2}}$ 
has the scaling exponent ${D_F=2}$.
If we have say two variables then the 
more general definition is 
\beq
F(s^{D_x}x,s^{D_y}y) = s^{D_F} F(x,y)
\eeq
Note that the scaling exponents 
can be multiplied by the same number, 
and still we have the same scaling relation.    
It follows that there is a scaling function 
such that 
\beq
F(x,y) = y^{D_F/D_y} f\left(\frac{x}{y^{D_x/D_y}}\right)
\eeq
For example ${F\left(x,y\right)=x^{2}+y^{3}}$
has the scaling exponents ${D_x=1/2, D_y=1/3,D_F=1}$.
More generally any ``physical" function 
has an ``engineering" scaling property that 
follows trivially from dimensional analysis.

\newpage
\sheadC{Solution of the 1D Ising Model}

Assuming only near neighbor interactions
\beq
E\left[\sigma\right]
\ \ = \ \ -\varepsilon \sum_{\langle ij\rangle }
\sigma_i \sigma_j
- \sum_{i} h_{i}\sigma_{i}
\eeq
The partition function is  
\beq
Z[h,\beta] \ \ = \ \ \sum_{\sigma(\cdot)} \eexp{-\beta E[\sigma]}
\eeq
For $\varepsilon=0$ we get
\beq
Z[h,\beta] \ \ = \ \ \prod_{i=1}^{N} 2\cosh\left(\beta h_i\right)
\eeq
and hence
\beq
F[h,T] \ \ = \ \ -T\sum_{i=1}^{N}\ln \left(2 \cosh \left(\frac{h_i}{T}\right)\right)
\ \ \approx \ \ 
-NT \ln \left(2\right) -\frac{1}{2T} \sum_{i=1}^{N} h_{i}^{2}
\eeq
The correlation function is 
\beq
G(r) \ \ = \ \  
-T\frac{\partial F}{\partial h_{i} \partial h_{j}}
\ \ = \ \ \delta_{ij} \ \ = \ \ \delta_{r,0}
\eeq
and hence the susceptibility is 
\beq
\chi \ \ = \ \ \frac{1}{T}\sum_{r} G(r) \ \ = \ \ \frac{1}{T}
\eeq
The magnetization is 
\beq
\tilde{M}
\ \ = \ \ -\frac{\partial F}{\partial h}
\ \ = \ \ N \tanh \left(\frac{h}{T}\right)
\ \ \approx \ \  
N \chi h +\mathcal{O}\left(h^{3}\right)
\eeq
We turn now to the case $\varepsilon \ne 0$. Without an external field 
the calculation is very easy. We can define 
$s_{\langle ij \rangle} = \sigma_i\sigma_j$.
Then the interaction can be written as $-\varepsilon\sum_b s_b$.
Instead of summing over spins, we can sum over the bonds $s_{b}$.
Assuming a chain of $N$ spins the sum factorizes and we    
get ${Z=2[2\cosh(\beta\varepsilon)]^{\mathcal{N}-1}}$.
Next we would like to assume that there is non zero homogeneous field~$h$. 
The calculation becomes somewhat more complicated,  
and requires the so called ``transfer matrix" method.  
Let us define the matrix 
\beq
T_{\sigma' \sigma''} \ \ \equiv \ \  
\exp\left[
\tilde{\varepsilon} \sigma' \sigma'' 
+ \frac{1}{2} \tilde{h} \left(\sigma'+\sigma''\right)
\right]
\ \ = \ \ 
\left(
\begin{array}{cc}
\eexp{\tilde{\varepsilon}+\tilde{h}} & \eexp{-\tilde{\varepsilon}} \\
\eexp{-\tilde{\varepsilon}} & \eexp{\tilde{\varepsilon}-\tilde{h}} \\
\end{array}
\right),
\hspace{1cm}
\ \tilde{\varepsilon} \equiv \beta \varepsilon, 
\ \tilde{h} \equiv \beta h
\eeq
The eigenvalues of this matrix are  
\beq
\lambda_{\pm}
\ \ = \ \ \eexp{\tilde{\varepsilon}} \cosh \left(\tilde{h}\right)
\pm \eexp{-\tilde{\varepsilon}} 
\sqrt{1+\eexp{4\tilde{\varepsilon}} \sinh^{2}\left(\tilde{h}\right)}
\eeq
The partition function of $\mathcal{N}$ site Ising model on a ring can be calculated as 
\beq
Z\left(\beta,h\right) 
\ \ = \ \ \sum_{\sigma(\cdot)} T_{\sigma_0,\sigma_1} T_{\sigma_1,\sigma_2} ... T_{\sigma_{\mathcal{N}{-}1},\sigma_0}
\ \ = \ \ \trc\left(T^\mathcal{N}\right)
\ \ = \ \ \lambda_{+}^{\mathcal{N}}+\lambda_{-}^{\mathcal{N}}
\eeq
and hence for very large $\mathcal{N}$ we get 
\beq
F\left(T,h\right) \ \ = \ \ -\mathcal{N}T \ln \left(\lambda_{+}\right)
\eeq
Expanding we get 
\beq
F\left(T,h\right) \ \ \approx \ \  
-\mathcal{N}T \ln \left(2\cosh \left(\frac{\varepsilon}{T}\right)\right)
-\frac{1}{2} \mathcal{N} \frac{\exp\left(2\frac{\varepsilon}{T}\right)}{T} h^{2}
\eeq
Hence 
\beq
\chi = \frac{1}{T} \exp\left(2\frac{\varepsilon}{T}\right)
\eeq
Now we would like to calculate the correlation function at zero field.
\beq
G(r) \ \ \equiv \ \ \langle  \sigma_{0}\sigma_{r}\rangle 
\ \ = \ \ 
\frac{1}{Z}
\sum_{\sigma_{0}\sigma_{r}}
\sigma_{0}
T_{\sigma_{0}\sigma_{r}}^{r}
\sigma_r
T_{\sigma_{r}\sigma_{0}}^{\mathcal{N}-r}
\eeq
We have
\beq
T_{\sigma_{'}\sigma_{''}}
=
\left(\begin{array}{cc}
\frac{1}{\sqrt{2}} & \frac{1}{\sqrt{2}} \\
\frac{1}{\sqrt{2}} & -\frac{1}{\sqrt{2}} 
\end{array}
\right)
\left(
\begin{array}{cc}
\lambda_{+} & 0 \\
0 & \lambda_{-} 
\end{array}
\right)
\left(
\begin{array}{cc}
\frac{1}{\sqrt{2}} & \frac{1}{\sqrt{2}} \\
\frac{1}{\sqrt{2}} & -\frac{1}{\sqrt{2}} 
\end{array}
\right)
\eeq
with 
\beq
\lambda_{+} &=& 2 \cosh \left(\tilde{\varepsilon}\right) \\ 
\lambda_{-} &=& 2 \sinh \left(\tilde{\varepsilon}\right)
\eeq
Using standard Pauli matrix notations and denoting the digonalized 
matrix $T$ as $\Lambda$ we get 
\beq
G(r) 
= 
\frac{1}{Z} 
\trc\left[
\sigma_z
T^{r}
\sigma_z
T^{\mathcal{N}-r}
\right]
=
\frac{1}{Z} 
\trc\left[
\sigma_x
\Lambda^{r}
\sigma_x
\Lambda^{\mathcal{N}-r}
\right]
=
\frac{\lambda_{+}^{r}\lambda_{-}^{\mathcal{N}-r}+\lambda_{-}^{r}\lambda_{+}^{\mathcal{N}-r}}
{\lambda_{+}^{\mathcal{N}}+\lambda_{-}^{\mathcal{N}}}
\eeq
For very large $\mathcal{N}$ we get 
\beq
G(r)
\ \ = \ \ \left(\frac{\lambda_{-}}{\lambda_{+}}\right)^{r}
\ \ = \ \ \eexp{-{r}/{\xi}}
\eeq
where 
\beq
\xi \ \ = \ \ \left[\ln \left(\coth\left(\frac{\varepsilon}{T}\right)\right)\right]^{-1} 
\ \ \approx \ \ \frac{1}{2}\eexp{2\varepsilon/T} 
\eeq
The calculation of $\sum G(r)$ involves a geometric summation, 
and it can be verified that it agree with the the result for~$\chi$.
The same result as the exact one is obtained 
from the approximated exponential expression 
if the summation is replaced by an integral.

\sheadC{Solution of the 2D Ising model}

The full details of the Onsager solution for this problem 
is in Huang. Also here the transfer matrix approach is used.
Recall that the zero field solution of the 1D model is 
\beq
\frac{1}{\mathcal{N}}\ln Z
\ \ = \ \ \ln\left(2\right)
\ + \ \ln\left(\cosh\left(\tilde{\varepsilon}\right)\right)
\eeq
The 2D solution is 
\beq
\frac{1}{\mathcal{N}}\ln Z
\ \ &=& \ \ \ln \left(2\right)
\ + \ \frac{1}{2}
\int\int \frac{d\theta d\theta'}{\left(2\pi\right)^{2}}
\ln\left[
\left(\cosh\left(2\tilde{\varepsilon}\right)\right)^{2}
+\sinh\left(2\tilde{\varepsilon}\right) \left(\cos\theta+\cos\theta'\right)
\right]
\\
\ \ &=& \ \ \ln\left(2\right) 
\ + \ \ln\left(\cosh\left(\tilde{\varepsilon}\right)\right)
\ + \ \frac{1}{2}
\int\int \frac{d\theta d\theta'}{\left(2\pi\right)^{2}}
\ln\left[
1+\frac{\kappa}{2}\left(\cos\theta+\cos\theta'\right)
\right]
\eeq
The integral is determined by the dimensionless parameter 
\beq
\kappa \ \ \equiv \ \  
\frac{2\sinh\left(2\tilde{\varepsilon}\right)}
{\left(\cosh\left(2\tilde{\varepsilon}\right)\right)^{2}}
\ \ \le \ \ 1 
\eeq
The value $\kappa=1$, for which $\ln Z$ exhibits discontinuity 
in its derivative, is attained for $\sinh(\tilde{2\varepsilon})=1$, 
from which it follows that the critical value of the interaction 
is ${\tilde{\varepsilon}=0.44}$, leading to ${T_c=2.27\varepsilon}$.
This is almost half compared with the heuristic ``mean field" 
value ${T_c \approx  4\varepsilon}$ that will be derived in the next lecture.

\newpage
\makeatletter{}
\sheadB{Phase transitions - heuristic approach}

\sheadC{The ferromagnetic phase transition}

The standard Ising Hamiltonian is
\beq
\mathcal{H} \ \ = \ \ -\varepsilon \sum_{\langle  ij\rangle }\sigma_{i}\sigma_{j} \ -h\sum_{i}\sigma _{i}
\eeq
Let us assume that in equilibrium  we can regard 
the spins as quasi-independent, each experiencing 
an effective field~$\bar{h}$, such that the 
effective Hamiltonian for the spin at site~$i$ 
is $\mathcal{H}^{(i)} = -\bar{h}\sigma_{i}$.
This means that the equilibrium state is 
\beq
p_{\sigma_{1}...\sigma_{N}} \ \ \propto \ \  
\exp\left[ \beta \bar{h} \sum_{i}\sigma_{i} \right]
\eeq
We have to find what is $\bar{h}$. The formal way is to use 
a variational scheme. We shall introduce this procedure later.
In this section we guess the result using a self-consistent 
picture. By inspection of the Hamiltonian if the 
mean magnetization of each spin is $\langle \sigma \rangle$, 
then it is reasonable to postulate that     
\beq
\bar{h} \ \ = \ \ h + \varepsilon  \sum_{\tbox{neighbors}} \langle \sigma_{j} \rangle 
\ \ = \ \   h + c\varepsilon \langle \sigma \rangle   
\eeq
where $c$ is the coordination number.
Form  $\mathcal{H}^{(i)}$ we get the 
self-consistent requirement 
\beq
\langle  \sigma \rangle 
\ \ = \ \ \tanh \left(\frac{1}{T} 
\left(h+c\varepsilon  \langle\sigma\rangle \right) 
\right)
\eeq
This equation should be solved for ${\langle\sigma\rangle}$, 
and then we get $\bar{h}$ as well.

\putgraph[0.5\hsize]{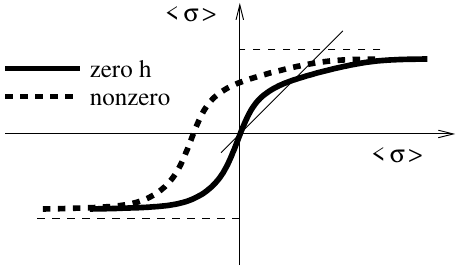}

By inspection of the plot we observe that for ${h=0}$  
the condition for getting a non trivial 
solution is ${ c\varepsilon/T > 1 }$.  
Therefore ${T_{c}=c\varepsilon}$.
If we want to explore the behavior in the critical 
region it is  convenient to re-write the equation 
in the following way:
\beq
h \ \ = \ \ T \tanh^{-1} \langle\sigma\rangle \ - \ T_c \langle\sigma\rangle 
\eeq
and to approximate it as  
\beq
h \ \ = \ \ \left(T-T_{c}\right)\langle  \sigma\rangle 
\ + \ \frac{1}{3}T_c \langle\sigma\rangle ^{3}
\eeq
For ${T > T_{c}}$ we get the Curie-Weiss law, which implies a critical exponent ${\gamma=1}$. Namely,  
\beq
\langle\sigma\rangle \ \ = \ \ \frac{1}{T-T_{c}}h
\eeq
For ${T=T_{c}}$ the dependence of $h$ is characterized by the critical exponent ${\delta=3}$. Namely, 
\beq
\langle\sigma\rangle \ \ = \ \ \left(\frac{3}{T_{c}}h\right)^{\frac{1}{3}}
\eeq
For zero field (${h=+0}$), below ${T_{c}}$,    
the temperature dependence is characterized by ${\beta=1/2}$. Namely,  
\beq
\langle\sigma\rangle \ \ = \ \ \left(3\frac{T_{c}-T}{T}\right)^{\frac{1}{2}}
\eeq

In the mean field approximation the spins are 
independent of each other, 
and therefore ${\langle \sigma_i \sigma_j \rangle = \langle \sigma_i \rangle \langle \sigma_j \rangle}$.
It follow that the energy is 
\beq
E \ \ = \ \ \langle H\rangle \ \ = \ \ -\frac{1}{2} c\mathcal{N} \ \varepsilon \ \langle \sigma \rangle^{2}
\eeq
For the heat capacity we get 
\beq
C(T) \ \ = \ \ \left. \frac{d E}{d T}  \right|_{h=0}
\ \ = \ \  -c\epsilon\mathcal{N} 
\ \left.\langle \sigma \rangle\frac{\partial \langle \sigma \rangle}{\partial T}\right|_{h=0}
\eeq
For ${T>T_c}$ we get ${C(T)=0}$, and from below we approach a constant value. 
The implied critical exponent is ${\alpha=0}$. 
To get the non-trivial mean-field result ${\alpha=[2-(d/2)]}$ 
we have to take into account Gaussian fluctuations.

\putgraph{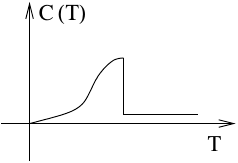}

\sheadC{The anti-ferromagnetic phase transition}

Let us consider a less trivial example for the use of the heuristic approach.
An anti-ferromagnet is described by the Ising Hamiltonian with $\varepsilon\mapsto -\varepsilon$. 
Specifically we consider a 2D square lattice that consists 
of to sub-lattices (for other lattices we might have frustrations).
We mark the magnetization of the two sub lattices by ${M_{a}}$ and ${M_{b}}$. We define 
\beq
M &=& \frac{1}{2}\left(M_{a}+M_{b}\right) \\
M_s &=& \frac{1}{2}\left(M_{a}-M_{b}\right)
\eeq
Without the magnetic field, the problem is the same 
as the ferromagnetic one with ${M_s}$ as the order parameter. 
With magnetic field~$h$ the heuristic mean field equations become 
\beq
M_{a} = \tanh\left(\frac{1}{T}\left(h-T_c M_{b}\right)\right)
\hspace{3cm}
M_{b} = \tanh\left(\frac{1}{T}\left(h-T_c M_{a}\right)\right)
\eeq
Following the same algebraic simplification procedure 
as in the ferromagnetic case, we get after 
addition and subtraction of the two resulting equations,  
\beq
\left(T-T_{c}\right)M_{s}+\frac{1}{3}T_c\left(3M^2M_{s} + M_{s}^{3}\right) &=& 0
\\
\left(T+T_{c}\right)M+\frac{1}{3}T_c\left(3M_{s}^{2}M + M^{3}\right) &=& h
\eeq
From here it follows that (see graphical illustration):
\beq
M_{s}=0 \ \ \ \ \text{or} \ \ \ \  3M^{2}+M_{s}^{2}&=&3\left(\frac{T_{c}-T}{T}\right)
\\
(2+M_{s}^{2}) M + \frac{1}{3} M^{3} &=& \frac{h}{T_c}
\eeq
As expected from the second equation we get ${M=0}$ in the absence 
of an external field, and from the first equation we get the order 
parameter $M_s(T)$, which satisfies the same equation as in the 
ferromagnetic problem.  If we switch on the magnetic field $T_c$ 
is shifted to a lower temperature.

\putgraph[10cm]{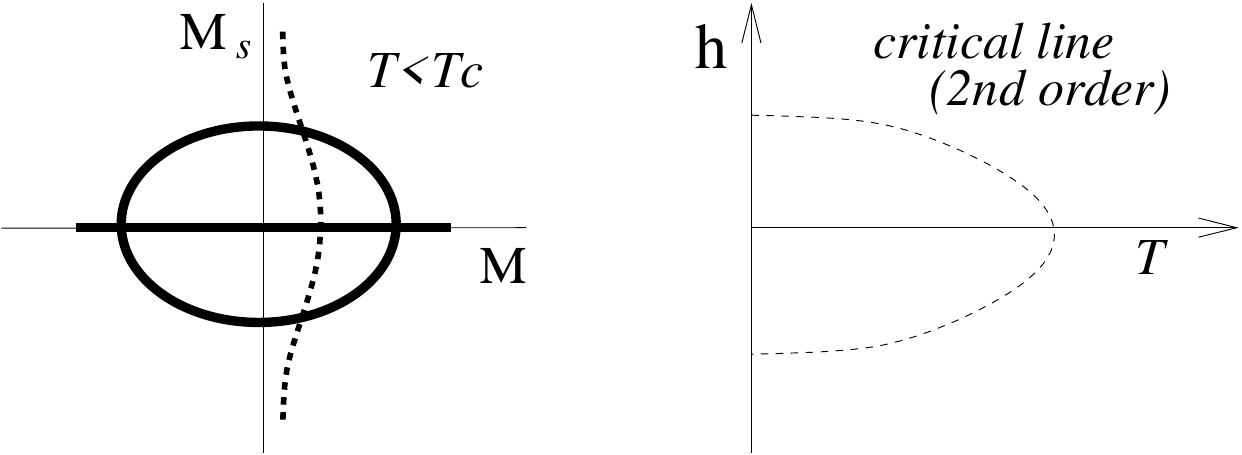}

If the magnetic field ${h}$ is strong enough, it destroys 
the anti-ferromagnetic order and causes ${M_{s}=0}$. 
This is implied by the identification of the ground state:
\beq
E\left(\uparrow \downarrow \uparrow \downarrow\right) 
\ \ &=& \ \ 
\mathcal N\cdot\left(-\frac{1}{2}c\epsilon\right),
\vspace*{2cm} \text{[for weak field]}
\\
E\left(\uparrow \uparrow \uparrow \uparrow\right)
\ \ &=& \ \ 
\mathcal N\cdot \left(\frac{1}{2}c\epsilon-h\right),
\vspace*{2cm} \text{[for strong field]}
\eeq
In the region where ${T\sim T_{c}}$ and ${h\sim 0}$ 
we get for the magnetization
\beq
M \ \ = \ \ 
\left(\frac{1}{T_{c}+T\left(1+\frac{1}{6}M_{s}\left(T\right)^{2}\right)} \right)h 
\ \ \equiv \ \ \chi h
\eeq
We can get a better general expression for all of the temperature
range by differentiation of the heuristic equations
\beq
\chi=\frac{1}{T_{c}+T \cosh ^{2}\left(\frac{T_{c}}{T}M_{s}\left(T\right)
\right)}
\eeq
In the region ${T\sim T_{c}}$ substitution of ${M_{s}\left(T\right)}$ gives
\beq
\chi=\left\{ \begin{array}{ll}
\frac{1}{T_{c}+T} & T_{c}<T \\
\frac{1}{4T_{c}-2T} & T<T_{c}
\end{array}\right.
\eeq

\sheadC{Beyond the Ising model} 

We now make a slight generalization of the Ising model. 
We consider coupled non-inertial oscillators, 
meaning that the kinetic term in the Hamiltonian is neglected: 
\beq
\mathcal{H} \ \ = \ \  \sum_j \left[ U(s_j) - h s_j \right]  \ -\varepsilon \sum_{\langle  ij\rangle }s_{i}s_{j} 
\eeq
For the Ising model ${U(s)=0}$ for ${s=\pm1}$ and ${U(s)=\infty}$ otherwise. 
But more generally we assume, say, ${U(s)=(\alpha/2)s^2 + (u/4)s^4}$. 
In the absence of interaction 
\beq
M \ \equiv \ \braket{s}  \ \ = \ \ \text{TANH}( \chi h)
\eeq
where TANH is a function that has by definition slope unity at the origin ($\text{TANH}'(0)=1$), 
and $\chi$ is a constant that has the meaning of zero-field susceptibility.
For zero non-linearity (${u=0}$) we get $\text{TANH}(x)=x$ with slope ${\chi=1/\alpha}$. 
Otherwise the TANH becomes a concave function with temperature dependent $\chi(T)$. 
For the Ising model $\chi(T)=1/T$. The heuristic mean field equation is 
\beq 
M \ \ = \ \ \text{TANH}\left[\chi\left(h+ c \epsilon M \right) \right] 
\eeq
In the absence of an external field it possesses a non-trivial solution 
provided ${c \epsilon\chi(T)  > 1}$, leading to a finite $T_c$.  
But if we consider dynamical degrees of freedom (see discussion of coupled rotors below), 
the susceptibility might be finite also at zero temperature due to quantum fluctuations.    
Thus, if ${c \epsilon\chi(0)  < 1}$, phase transition does not take place. 
Then, by tuning the model parameters at ${T=0}$, we can witness a {\em quantum phase transition} 
once we cross to a regime where ${c \epsilon\chi(0)  > 1}$. 

{\bf Note:} Considering again zero non-linearity (${u=0}$), the heuristic approach implies that the system becomes unstable for ${\alpha < c\epsilon}$. This condition becomes more illuminating if we wrote the interaction between to oscillators as $(\epsilon/2)[s_i-s_j]^2$. The price for that is to write   ${U(s)=(a/2)s^2}$ instead of ${U(s)=(\alpha/2)s^2}$ with ${a=\alpha-c\epsilon}$. The condition for instability becomes simply ${a<0}$.

\sheadC{The mean-field Hamiltonian} 

The heuristic approach for phase-transition can be regarded as the formal outcome of a mean-field approximation for the system Hamiltonian. We set in the Hamiltonian $s_j=M+\delta s_j$, expand the interaction term, and get
\beq
\mathcal{H} \ \ \approx \ \   \sum_j H^{(j)}  \ -\varepsilon \sum_{\langle  ij\rangle }\delta s_{i} \delta s_{j} 
\eeq
were the first sum is the mean field Hamiltonian with  
\beq
H^{(j)} \ \ = \ \  U(s_j) \ - (h+c\varepsilon M) s_j  \ + \frac{1}{2} c\varepsilon M^2
\eeq
Assuming that the fluctuation are uncorrelated we deduce 
\beq
E \ \ = \ \ \braket{\mathcal{H}} \ \  \approx \ \ \mathcal{N} \ \left[ \braket{U(s)}-\frac{1}{2}c\varepsilon M^2 -hM \right]
\eeq  
where $\braket{U(s)}$ should be calculated from the mean field Hamiltonian. 
Note that it is zero for the Ising model, while here we consider a more general class of systems. 
If the non-linear term and the fluctuation are neglected $\braket{U(s)}=(\alpha/2)M^2$, 
hence the symmetry breaking is implied if ${a=(\alpha-c\epsilon)<0}$. 
Taking non-linearity into account, symmetry breaking is implied if the temperatures is low enough.

\newpage
\sheadC{Coupled rotors} 

We now consider coupled rotors. The rotors are dynamical entities, they have finite mass.
We define $\gamma$ as the inverse moment of inertia. 
Note that if we started with ${[\varphi,p]=i\hbar}$, 
then with ${n=p/\hbar}$ we get that ${\gamma\propto \hbar^2}$. 
Accordingly infinite mass is like taking the classical limit. 
The Hamiltonian is 
\beq
\mathcal{H} \ \ = \ \  \sum_j \left[ \frac{\gamma}{2}n_j^2 - h \cos(\varphi_j) \right]  \ -\varepsilon \sum_{\langle ij\rangle } \cos(\varphi_j-\varphi_i)  
\eeq
If we ignore the kinetic term it is formally like coupled non-inertial oscillators with ${s_j=\cos(\varphi_j)}$.  
In a classical context if we take the kinetic term into account it has no effect
because it factorizes out of the partition function. The mean field Hamiltonian is 
\beq
H^{(j)} \ \ = \ \ \frac{\gamma}{2}n_j^2  \ - (h+c\varepsilon M) \cos(\varphi_j)  \ + \frac{1}{2} c\varepsilon M^2
\eeq
In the quantum treatment the energy shift of the ground-sate is 
not ${-(1/2)c\varepsilon M^2}$ because of {\em quantum fluctuations}: 
the price of small~$\varphi$ is large uncertainty in the conjugate momentum~$n$.
The implications is that quantum fluctuations are able to diminish~$M$ 
at zero-temperature.

{\bf Quantum phase transition.-- } 
Let us find the condition for diminished ``order" at zero-temperature.
The simplest perspective is the heuristic approach. 
At zero temperature standard quantum-mechanical calculation using second order perturbation theory 
shows that the zero temperature susceptibility of a rotor is ${\chi = 2/\gamma }$.
It follows that symmetry-breaking is avoided if
\beq
2c\varepsilon \ \ <  \ \ \gamma 
\hspace{3cm} \mbox{[Mott phase]}
\eeq
We see that zero-temperature ``order" is diminished 
either by having $T$ or $\gamma$ that are larger than $\sim\varepsilon$, 
reflecting strong quantum or thermal fluctuations respectively.
An equivalent way to deduce the above condition is to consider 
the ground state energy of $E_0(h)$ of ${H_0=(\gamma/2)n^2-h\cos(\varphi)}$.
For large $h$ using harmonic-oscillator approximation ${E_0=-h+(1/2)\sqrt{\gamma h}}$, 
but for small $h$ using 2nd order perturbation ${E_0=-h^2/\gamma}$. 
Using the latter result we get at the vicinity of ${M=0}$
that the mean-field energy per rotor is
\beq
E=\braket{H^{(j)}}= \frac{a}{2}M^2, 
\ \ \ \ \ \ \ \ \ \ \mbox{with} \ \ 
a=-2\frac{(c\varepsilon)^2}{\gamma} +c\varepsilon
\eeq
Symmetry-breaking is avoided if ${a>0}$.

\sheadC{The variational approach}

A different way to derive the heuristic mean-field equations 
is to use the variational approach. 
The canonical state minimizes the free energy functional. 
Accordingly we look for a solution to the variation problem
\beq
F\left[\rho\right] \ \ \equiv \ \  
\langle H \rangle - T S\left[\rho\right] \ \ = \ \ \text{minimum}
\eeq
with implicit constraint on the normalization. 
In the mean-field approach the canonical state 
is assumed to be well approximated by $\rho=\{p_{\sigma}\}$, 
where 
\beq
p_{\sigma_{1}...\sigma_{N}} \ \ = \ \ 
\frac{1}{\left(2 \cosh \left(\beta \bar{h}\right)\right)^{\mathcal N}} 
\exp\left[ -\beta \bar{h} \sum_{k}\sigma_{k} \right]
\eeq
Here the variational parameter ${\bar{h}}$ is the effective mean field.
We would like to determine the optimal value of ${\bar{h}}$
for which $F\left[\rho\right]$ is minimal. 
For the calculation we use the identity 
${F\left[\rho\right]=F_{0}\left[\rho\right]+\langle H-H_{0}\rangle}$, 
where ${H_{0}=-\bar{h} \sum_i\sigma_i}$, leading to  
\beq
F\left[\rho\right]
=\mathcal{N}\left[
f(\bar{h})-\frac{1}{2}c \epsilon \, m(\bar{h})^2 -\left(h-\bar{h}\right) m(\bar{h})
\right]
\eeq
where $f(\bar{h})=-T\ln \left(2\cosh \left({\bar{h}}/{T}\right)\right)$  
and $m(\bar{h})=-f'(\bar{h})$ is the mean-field magnetization.   
The variational equation for ${\bar{h}}$ is as expected 
\beq
\bar{h} \ = \ h + c\epsilon \tanh\left(\frac{\bar{h}}{T}\right)
\eeq
Hence, we get the variational free energy
\beq
F\left(T,h\right)
\ \ = \ \ \mathcal{N}
\left[
-T\ln \left(2\cosh \left(\frac{\bar{h}}{T}\right)\right)
+\frac{1}{2} c \epsilon \left(\tanh\left(\frac{\bar{h}}{T}\right)\right)^{2}
\right]
\eeq
This is not a pleasant expression because the dependence on~$h$ is implicit in $\bar{h}$.
We can differentiate this equation to find $\tilde{M}$, which involves $\partial \bar{h}/\partial h$. 
The calculation is lengthy, but we can skip it because the result is obvious 
\beq
\tilde{M} \ \ = \ \ -\frac{\partial F\left(T,h\right)}{\partial h}
\ \ = \ \ \mathcal{N} \tanh\left(\frac{\bar{h}}{T}\right)
\eeq
To make calculations of the state equations more convenient 
we notice that $F(T,h)$ depends in a very simple way on~$\tilde{M}$, 
hence it is useful to make the Legendre transformation  
\beq
A(T,\tilde{M}) \ \ \equiv \ \ F(T,h) + h\tilde{M}
\eeq
such that $dA=-SdT+hd\tilde{M}$. Note that the mean field equation for $\bar{h}$ implies that   
\beq
h \ \ = \ \ T \tanh^{-1}(M) - c \epsilon M
\hspace{3cm}
\text{where} 
\ \ M \equiv \frac{\tilde{M}}{\mathcal{N}} = \langle \sigma \rangle
\eeq
Using the identity $\tanh^{-1}(x)=(1/2)\ln((1+x)/(1-x))$ one obtains 
\beq
A\left(T,M\right)
&=&
\mathcal{N} \left[-T\ln 2+\frac{1}{2}T\ln \left(1-M^{2}\right)
+\frac{1}{2}TM \ln \left(\frac{1+M}{1-M}\right)-\frac{1}{2}c\epsilon M^{2}\right]
\\
&=&
\mathcal{N}T \left[\frac{1+M}{2}\ln \frac{1+M}{2}+\frac{1-M}{2}\ln\frac{1-M}{2}\right]
-\mathcal{N} \frac{1}{2}c\epsilon M^{2}
\eeq
From this expression it is convenient to derive explicit results
for the state equations. In particular ${S=-\partial A/\partial T}$
and one can recover the result for the heat capacity.

\newpage
\sheadC{The Bragg Williams formulation}

Consider an Ising model with $\mathcal{N}$ sites,
at any dimension, and with any coordination number. 
Given a spin configuration define
\beq
\mathcal{N} &=& \text{total number of spins} \\
m &=& \text{total magnetization} \\
M &=& m/\mathcal{N} \\
N_{+} &=&  \text{number of up spins}   = \frac{1}{2}(\mathcal{N}+m) = \frac{1}{2}\mathcal{N}(1+M) \\
N_{-} &=&  \text{number of down spins} = \frac{1}{2}(\mathcal{N}-m) = \frac{1}{2}\mathcal{N}(1-M) \\
N_{+-} &=& \text{number of bonds connecting spins with opposite direction}
\eeq
The total number of bonds is $(1/2)c\mathcal{N}$, where $c$ is the coordination number.
It follows that 
\beq
\sum \sigma_{i} &=& m \\
\sum_{\langle ij \rangle} \sigma_{i} \sigma_{j} &=&  \frac{1}{2}c\mathcal{N}-2N_{+-}
\eeq
If we look on two connected spins, there is a probability $(N_+/\mathcal{N})$ 
to have the first up, and  a probability $(N_-/\mathcal{N})$ to have the 
the second down. Or we can have the first down and the second up.  
This motivates the Bragg Williams approximation: 
\beq
N_{+-} \ \ \approx \ \ 
2 \left(\frac{N_{+}}{\mathcal{N}}\right)\left(\frac{N_{-}}{\mathcal{N}}\right) \frac{\mathcal{N}c}{2} 
\ \ = \ \  \frac{\mathcal{N}c}{4}(1-M^2)
\eeq
Assuming that it holds for typical configurations we 
approximate the energy functional as    
\beq
E[\sigma] \ \ \approx \ \ -\mathcal{N} \times \left(\frac{1}{2} c\varepsilon M^{2} + hM\right)
\eeq
We note that this expression with $c=\mathcal{N}$ 
if formally exact for a fully connected cluster of spins. 
The number of configuration with total magnetization $m$ is 
\beq
g_{m} \ \ = \ \ \frac{\mathcal{N}!}{\left(N_{+}\right)!\left(N_{-}\right)!}
\ \ \approx \ \ 
\const \ \exp\left[-\mathcal{N}\left(\frac{1}{2}M^{2}+\frac{1}{12}M^{4}+...\right)\right]
\eeq
In order to derive the latter approximation note that 
\beq
-\frac{\partial}{\partial m}\ln g_m \ \ = \ \ 
\frac{1}{2}\left(\ln(1+M)-\ln(1-M)\right) \ \ \approx \ \  M+\frac{1}{3}M^3+\frac{1}{5}M^5+...
\eeq
With this approximation we get
\beq
Z \ \ = \ \ \sum_{m} \sum_{\sigma \in m} 
\eexp{-\beta E\left[\sigma\right]}
\ \ \approx \ \ 
\sum_{m} g_{m} \eexp{-\beta E\left(m\right)}
\ \ = \ \ 
\sum_{M} \eexp{-A\left(M\right)}
\\ 
A(M)
\ \ = \ \ \mathcal{N} \times \left[
\frac{1}{2}
\left(1-\beta c\varepsilon\right)M^{2}+\frac{1}{12}M^{4}-\beta hM 
\right]
\eeq
In the next section we are going to clarify the following points:
{\bf (1)} The sum can be evaluated via Gaussian integration.
{\bf (2)} This Gaussian approximation can be justified if $\mathcal{N}$ is large.
{\bf (3)} Phase transition is implied. 
{\bf (4)} But nevertheless the result is false for ${d=1}$. 
In the next lecture we shall further explain that the Bragg-Williams 
formulation fails in providing the correct description of 
the symmetry-breaking if the critical temperature is approached.

\sheadC{The Gaussian approximation}

The expression that we have obtained for $Z$ using the Bragg Williams formulation 
is a typical approximation that can obtain for various models. 
We rewrite  it as follows:
\beq
Z \ = \ \int d\varphi \ \eexp{-A(\varphi)} 
\hspace{2cm} 
A(\varphi) \ = \ \mathcal{N} \times 
\left[
\frac{a}{2} \varphi^{2} + \frac{u}{4} \varphi^{4} - h\varphi 
\right]
\eeq
This sum can be evaluated via Gaussian integration.
The dominant contribution comes from the~$\varphi$ 
for which $A(\varphi)$ is minimal.
One can easily verify that ${A'(\varphi)=0}$ coincides 
with the heuristic mean field equation that 
has been discussed in a previous lecture.
Non trivial solutions appear for ${a<0}$
which implies ${T_c=c\varepsilon}$. 
Note that ${a \approx (T-T_c)/T_c}$.

Above the critical temperature there is 
a single minimum at ${\bar{\varphi}=(1/a)h}$  
and one obtains 
\beq
Z \ \ \approx \ \ \left(\frac{2\pi}{\mathcal{N}a}\right)^{1/2} \ \exp\left[\frac{\mathcal{N}}{2a}h^2 \right]
\eeq
In the absence of an external field, as $a$ becomes negative,  
the trivial minimum  ${\bar{\varphi}=0}$
bifurcates into two minima, namely  ${\bar{\varphi}=\pm(|a|/u)^{1/2}}$.
For these values 
\beq  
A(\bar{\varphi}) \ \ = \ \ -\frac{a^2}{4u} \ \mp \ \left(\frac{|a|}{u}\right)^{1/2}h
\eeq
Approximating $Z$ as the sum of two Gaussian integrals, 
one realizes, after expanding ${ A(\bar{\varphi}+\tilde{\varphi})}$, 
that the coefficient of the $\tilde{\varphi}^2$ term is the same 
as above $T_c$, with ${a}$ replaced by $2|a|$ (positive).  
We get that the  partition function is like that of a spin:
\beq
Z \ \ \approx \ \ \left(\frac{\pi}{\mathcal{N}|a|}\right)^{1/2} 
\ \exp\left[ \mathcal{N}\frac{a^2}{4u} + \frac{\mathcal{N}}{4|a|}h^2 \right]
\ 2\cosh\left[ \mathcal{N} \left(\frac{|a|}{u}\right)^{1/2} h \right]
\eeq
From here one deduces that for ${T<T_c}$ 
the susceptibility becomes ${\chi= [1/(2|a|)] + \mathcal{N}|a|/u}$ 
instead of ${\chi=1/a}$.

At this point one can ask whether it was allowed to ignore 
the quartic term in $A(\varphi)$. 
This should be checked self consistently. 
For ${a>0}$ the dispersion of $\varphi$
in the Gaussian approximation is $(\mathcal{N} a)^{-1/2}$.
The quartic term can be neglected if ${u|\varphi|^4 \ll a|\varphi|^2}$ 
leading to the condition $a \gg (u/\mathcal{N})^{1/2}$. 
This condition is always satisfied if $\mathcal{N}$ is large enough.
The same condition also guarantees that for ${a<0}$  
the dispersion is much smaller compared with 
the non-zero mean field ${\bar{\varphi}}$. 
Generalization of this condition in the field-theory treatment 
will be discussed later and lead to the Ginzburg criterion.

It is now appropriate to point out that the above treatment 
implies a phase-transition in the thermodynamic limit. 
We first note that the existence of the thermodynamic limit 
for $A(\varphi)/\mathcal{N}$ could have been anticipated 
from general considerations. 
From $Z$ we can get the free energy $F(h)/\mathcal{N}$ that 
will have a thermodynamic limit too. The question is 
whether the subsequent limits ${h\rightarrow+0}$ and ${h\rightarrow-0}$
lead to the same magnetization, or optionally whether the susceptibility $\chi$ 
beomes infinite below~$T_c$. Indeed this is what we found.

\newpage
\sheadC{The importance of fluctuations}
 
The above analysis is misleading. The Bragg Williams approximation 
underestimates the effect of fluctuations. We already know from the exact
solution of the Ising model that in 1D there is no phase transition 
at finite temperature. We would like to explain in detail why the 
fluctuations in 1D smear away the phase transition. Later we shall see 
that also for ${d=2,3}$  the fluctuations are important: 
they do not smear away the phase transition, but they modify the
state equations in the critical region, 
which explains the failure of mean field theory there, 
and the observed {\em anomalous} values of the scaling exponents. 
\\

\putgraph[0.5\hsize]{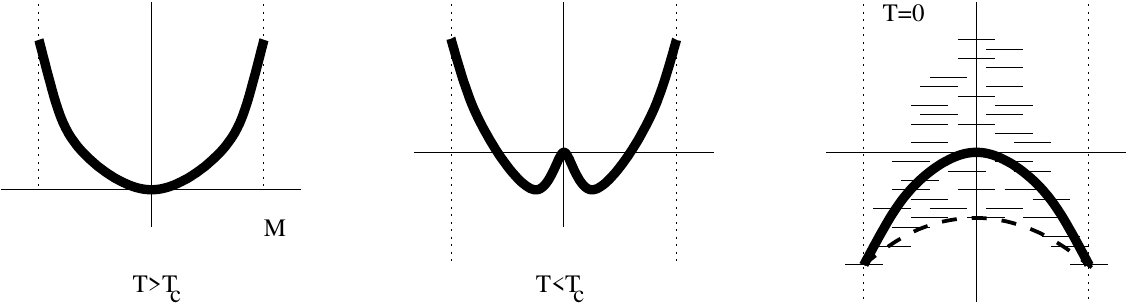}

The problem with the Bragg Williams approximation 
is implied by the figure above. The action $A(\varphi)$ is plotted. 
It is determined by the energy term $E[\varphi]$, 
and by the entropy term $S[\varphi]$. Recall that  
\beq
p(\varphi) \ \ = \ \ \frac{1}{Z}\eexp{-A(\varphi)} 
\ \ \propto  \ \ \exp\left[ -\frac{1}{T}E[\varphi] + S[\varphi] \right]  
\eeq 
In the ${T=0}$ panel 
the energy of the states is indicted by bars.  
By taking the "typical" value of the energy (thick solid line) 
we ignore a dominant fraction of less typical states that 
have a very low energy. These states corresponds 
to configurations where the spins a bunched in "zones". 
The simplest arrangement in 1D has two zones and its energy 
is ${E_b=E_0+2\epsilon}$, where ${E_0=-\mathcal{N}\epsilon}$ 
is the ground state energy.  Consequently the effective barrier 
between the "all up" and "all down" states is very low
(thick dashed line), and the symmetry breaking is avoided 
at any finite temperature. The formal argument is outlined below. 
In contrast to that, in 2D the effective barrier 
is ${E_b=E_0+ \mathcal{N}^{1/2}\epsilon}$ and therefore 
symmetry breaking is realized in the thermodynamic limit.

{\bf Domain walls.-- }
It is possible to argue that in 1D there is no phase 
transition at finite temperature. The argument goes 
as follows: consider a mixture of states with one 
domain wall. For such state $E[\rho]=E_0+2\epsilon$, 
where $\epsilon$ is the cost of the domain wall. 
But the entropic contribution is $S[\rho]=\ln[\mathcal{N}]$
where $\mathcal{N}$ is the length of the chain. 
It follows that for any finite $T$ the ground state,  
or any (exclusive) mixture of ground-like states, 
do not minimize $F[\rho]$ at the thermodynamic limit.
We can lower $E[\rho]$ by adding states that have 
with equal probability any magnetization. 
Consequently we get huge fluctuations whose relative 
amplitude does not diminish, 
in contrast with the ${(\mathcal{N})^{-1/2}}$ prediction 
of the mean-field Gaussian estimate. 
Therefore spontaneous magnetization at finite temperature is impossible.

The above argument fails in 2D because the energy cost  
of a domain $E[\rho]=E_0+\mathcal{N}^{1/2}\epsilon$ 
domeiniate over the entropic contribution.  
In fact it is possible to refine the domain wall approach 
and deduce that for the ${d=2}$ Ising model there exists 
spontaneous magnetization at finite temperature [see Huang p.349], 
in consistency with the exact solution. 
{\em However}, the possibility to witness phase transition at $d=2$
is limited to systems with discrete symmetries.
The Mermin-Wagner theorem states that a continuous symmetry  
cannot be spontaneously broken at finite temperature 
in $d\le2$ systems with short-range interactions.
See discussion of the Heisenberg model.

\newpage
\makeatletter{}
\sheadB{Phase transitions - field theory}

\sheadC{The Landau model}

We would like to take into account the spatial fluctuations of the 
magnetization $\varphi(x)$ in the calculation of the partition function. 
We therefore use a refined version of the Bragg-Williams approach. 
Namely, first we sum over all microscopic configuration that corresponds 
to a magnetization $\varphi(x)$, and then we are left with 
a so-called functional integral:
\beq
Z[h, \, \text{parameters}; \, \Lambda, L] 
\ \ =  \ \ \sum_{\varphi(\cdot)} \eexp{-A[\varphi(\cdot)]}
\ \ = \ \ \int D\varphi \ \eexp{-A[\varphi(\cdot)]}
\eeq
where the sum over configurations becomes an integral with the measure 
\beq
D\varphi \ = \ \prod_{x \in L^d} d\varphi_x \ = \  \prod_{|k|<\Lambda} d\tilde{\varphi}_k
\ ,
\hspace*{2cm} \tilde{\varphi}_k \equiv \frac{1}{L^d}\int \varphi(x) \ \eexp{-ikx} dx 
\eeq
In this definition $\tilde{\varphi}_k$ are the Fourier components of $\varphi(x)$, 
and we are sloppy about an uninteresting overall prefactor.
It is implicit that one assumes a finite volume $L^d$, 
and a finite momentum cutoff $\Lambda$, 
otherwise the functional-integral that gives~$Z$ is ill defined. 
To have a momentum cutoff is like to assume 
that space is discretized with lattice spacing $2\pi/\Lambda$. 
Accordingly the number of freedoms of the model 
is $\mathcal{N}=(L \Lambda/2\pi)^d$. 
Technical remark: for presentation purpose it is more convenient 
to work with the complex "exp" Fourier basis, but from mathematical 
point of view the meaning of the $D\varphi$ integration is 
more obvious if we work with real "sin" and "cos" Fourier basis, 
corresponding to the real and imaginary parts of $\tilde{\varphi}_k$.
Either way the $\varphi$ field is represented 
by $\mathcal{N}$ independent real amplitudes.

In the Landau model the assumed action is 
\beq
A[\varphi(\cdot)]
\ \ = \ \ 
\int d\bm{x} \left(
\frac{c}{2} \left(\nabla\varphi\right)^{2}
+\frac{a}{2} \varphi^{2}
+\frac{u}{4} \varphi^{4}
-h\varphi \right)
\ \ = \ \ 
L^d \sum_k \left(
\frac{1}{2} (ck^2 + a) |\tilde{\varphi}(k)|^2
+... \right)
\eeq
The summation over the $k$ components 
of the field is conventionally written as 
an integral with the measure $[L/(2\pi)]^ddk$. 
The convention ${c=1}$  with regard to the prefactor 
of the first term  fixes the dimensions of $\varphi$,  
and hence of all the other model parameters. 
We write these dimensions as $L^d$, accordingly 
\beq
d_{\varphi} = -\frac{d-2}{2}, 
\ \ \ \ \ \
d_{h}  = -\frac{d+2}{2},
\ \ \ \ \ \
d_{a}  = -2,
\ \ \ \ \ \
d_{u}  = -(4-d),
\eeq
The model has a thermodynamic limit, hence $L$ is not significant, 
and we can calculate the Helmholtz free energy~$F$ per unit volume.
In contrast to that $\Lambda$ is significant. In particular we note 
that the model contains {\em two} significant dimensionless parameters 
that are related to the underlying microscopic Hamiltonian:
\beq
\tilde{a} \ &=& \ a/\Lambda^2  
\hspace*{2cm} 
\mbox{Note: later we see that $\xi=a^{-1/2}$ is the correlation length} 
\\
\tilde{u} \ &=& \ u/\Lambda^{4-d}
\hspace*{2cm} 
\mbox{Note: later we see that $a/u^{2/(4{-}d)}$ is the Ginzburg parameter} 
\eeq
Relating to the Bragg-Williams approximation we 
identify $a \propto (T-T_c^{(0)})$, where $T_c^{(0)}$ 
is the mean field critical temperature. In fact we shall 
see that the field theory analysis implies that 
for ${2 \le d <4}$ the actual critical temperature 
is pushed down due to the fluctuations ($a_c<0$). 
In the ${d=1}$ case there is no phase transition. 

{\bf Coarse graining.-- }
The Landau model can be regarded as the outcome of coarse graining on scale $\Lambda$. 
Therefore its parameters $a(\Lambda)$ and $u(\Lambda)$ are ``running coupling constants".
The cutoff $\Lambda$ is in the range ${[\Lambda_0,\Lambda_{\infty}]}$, 
where $\Lambda_{\infty}$ reflects a limiting microscopic scale, 
while $\Lambda_0$ reflects the maximal spatial range over which coarse-graining 
is meaningful. Clearly it is the correlation distance, hence ${\Lambda_0 \sim 1/\xi}$.  
We shall see that as $\Lambda$ is decreased, as the result of successive 
course-graining operations, we get $\tilde{u}(\Lambda) \rightarrow \tilde{u}_c$, 
where $\tilde{u}_c=1/9$ for 3D. So in some sense there is only one relevant 
parameter ($a$) in this model, and results for different values of $a$ are related by scaling.

\sheadC{Related models}

The Landau model stands by itself as a prototype model.
It might have different physical interpretations.
Possibly the simplest is to regard it as the continuum 
model of "coupled oscillators". 
If ${a=u=0}$ it is formally like the Debye model. 
Having ${a>0}$ means that the oscillators have 
a positive spring constant that stabilizes them at ${\varphi=0}$.   
For ${a<0}$ each oscillator is pushed 
away from ${\varphi=0}$ and its new equilibrium position 
is determined by the nonliterary~$u$ of the spring.  
Thanks to the non-linearity of the springs the ${\varphi}$ 
cannot diverge to infinity.  

Above we have regarded the Landau model as a coarse grained 
version of the Ising model, using the Bragg-Williams 
approximation for each coarse-grained cell. There is an optional 
possibility to motivate the Landau model as an approximation 
for the Ising model using a somewhat more direct procedure. 
For this purpose one replaces the discrete summation over ${\sigma=\pm1}$ 
by an integration over $\varphi$ with a weight function:  
\beq
\sum_{\sigma\left(\cdot\right)}
\ \ \rightarrow \ \  
\int \prod_{x} d\varphi_x
\ \eexp{-\frac{1}{4}u\left(\varphi_x^{2}-1\right)^{2}}
\eeq
One should realize that the ferromagnetic interaction $-\sigma(x)\sigma(x')$ corresponds 
to differences ${ (\varphi(x)-\varphi(x'))^2 }$, and hence translates
to the gradient term in the Landau model.

The field theory that corresponds to the Ising model contains 
a real field. It reflects the discrete mirror (up/down) symmetry  
of the system. More generally we can consider a system that 
has a continuous rotational symmetry. 
In such case the Action is $A[S]$ with vector field ${S =(S_1,S_2,S_3)}$. 
Of particular interest is to have a gauge-invariant Action  $A[\Psi]$
with a complex field ${\Psi =(\psi_1,\psi_2) \equiv \sqrt{n}\exp(i\varphi)}$. 
Note that in two dimensions gauge-invariance can be regarded as a rotational-invariance.

\sheadC{The Gaussian approximation}

Let us start with the simplest possibility of having ${u=0}$.
Regarded as an approximation it is meaningful only if ${a>0}$.
If we had ${c=0}$ the result would be the same as that of 
the Bragg-Williams model. If ${c}$ is non-zero 
(equal to unity by convention) the summation still 
factorizes, but in $k$ space. Assuming for simplicity ${h=0}$, 
and not caring about a global prefactor we get the following:
\beq
Z \ \ = \ \ 
\ \prod_k
\int d\tilde{\varphi}_k \,
\exp\left[
-\frac{L^d}{2} \left(k^{2} + a\right) \tilde{\varphi}_k^2 
\right]
\ \ = \ \ 
\prod_{k} \left(\frac{1}{k^{2}+a}\right)^{1/2}
\eeq
The free energy in the mean-field approximation was ${F = T A(0) =0}$. 
Now we have taken the Gaussian fluctuations into account. 
Consequently we get a non-trivial result for the free energy:  
\beq
F(T) \ \ = \ \ \frac{T}{2} \sum_k \ln\left(k^{2}+a\right) 
\eeq
In particular we can derive from this expression 
the Gaussian prediction for the heat capacity.
Contrary to the mean field approximation, it is no longer zero.  
The singular contribution at the vicinity of $T_c$ 
originates from the second derivative with respect to~$a$. 
Accordingly      
\beq 
C(T) 
\ \ = \ \ -T\frac{d^2F}{dT^2}
\ \ \sim \ \ \int_0^{\Lambda} \frac{k^{d-1}dk}{(k^2+a)^2} 
\ \ \sim \ \ |T-T_c|^{-(4-d)/2}, 
\hspace{3cm} \text{for} \ d<4
\eeq
Hence the modified mean field exponent $\alpha=0$
is replaced by $\alpha=[2-(d/2)]$.  

Below $T_c$ we can perform a Gaussian approximation 
around the mean-field $\bar{\varphi}$ that will be discussed 
in the next section. The calculation is essentially the same, 
with an offset $A[\bar{\varphi}]$ that is added to the action, 
hence $F(T)\mapsto F(T)+TA[\bar{\varphi}]$. 
The non-singular contribution of this additional "mean field" term 
implies a discontinuity of $C(T)$ at ${T=T_c}$  
as discussed in past lecture, 
and has no effect on the Gaussian value of $\alpha$.

\sheadC{Digression - Gaussian integrals}

The partition function can be calculated exactly whenever 
the action is a quadratic form. The so-called Gaussian 
integral reduces to the product of $\mathcal{N}$ one 
dimensional integrals if we transform it to a basis 
in which the quadratic form is diagonal. For a system 
that has translation symmetry it is momentum space.
\beq
\int D\varphi \ \eexp{-\frac{1}{2}\sum_{i,j}A_{ij}\varphi_i\varphi_j + \sum_i h_i\varphi_i} 
\ \ &=& \ \ 
\prod_k \int d\tilde{\varphi}_k \ \eexp{-\frac{1}{2} a_k \tilde{\varphi}_k^2  + \tilde{h}_k \tilde{\varphi}_k} 
\\
\ \ &=& \ \ 
\prod_k \left(\frac{2\pi}{a_k}\right)^{1/2} \eexp{\frac{1}{2}\left(\frac{1}{a_k}\right)\tilde{h}_k^2}
\ \ = \ \ \sqrt{\det(2\pi G)} \ \exp\left[\frac{1}{2}\sum_{i,j}G_{ij}h_ih_j\right]
\eeq  
here $G=A^{-1}$, and note that $\det(G)=1/\det(A)$. 
Note also that going back to the original basis, 
in the case of position-to-momentum transformation 
implies that $G(r)$ is the Fourier transform of $1/a(k)$.

From the above result, it follows that $G(r)$ 
is the correlation function $\langle \varphi(r)\varphi(0)\rangle$
for ${h=0}$. Otherwise $\langle \varphi \rangle$ is non-zero, 
and it equals to field $\bar{\varphi}$ that minimizes 
that action. It satisfies the equation ${A\varphi=h}$, 
whose solution is $\bar{\varphi}=Gh$. 
Hence $G$ can be regarded as the "Green function".

\sheadC{The mean field equation}

We define the mean field ${\bar{\varphi}}$ via
the equation ${A\left(\varphi\right)=\text{minimum} }$. 
This gives the equation
\beq
\left(-\nabla^{2}+a\right)\varphi+u\varphi^{3} \ = \ h(x)
\eeq
The mean field for an homogeneous ${h(x)=h}$ 
is obtained from ${a\varphi+u\varphi^{3}=h}$.
In particular for ${h=\pm0}$ we get
\beq
\bar{\varphi}_0 \ \ = \ \ 
\left\{ \begin{array}{ll}
0, & \text{for} \ a>0
\\
\pm\left(\frac{-a}{u}\right)^{\frac{1}{2}},  & \text{for} \ a<0
\end{array}
\right.
\hspace{3cm}
A[\bar{\varphi}_0] \ \ = \ \ 
\left\{ \begin{array}{ll}
0, & \text{for} \ a>0
\\
-\frac{a^2}{4u},  & \text{for} \ a<0
\end{array}
\right.
\eeq
For ${a>0}$ we neglect the non linear term in the action, 
define $\xi = a^{-1/2}$, and write the mean field equation 
as ${\left(-\nabla^{2}+(1/\xi)^2\right) \varphi(x) = h(x)}$.      
For ${a<0}$ we make the substitution ${\varphi \mapsto \bar{\varphi}_0 + \varphi}$, 
expand the action around the new minimum, 
and then neglect the non-linear term. 
The mean-field equations takes the same form as for ${a>0}$, with ${\xi=(-2a)^{1/2}}$.  
Accordingly the first-order solution in ${h(x)}$ is 
\beq
&& \bar{\varphi}(x) \ \ = \ \ \bar{\varphi}_0
+\int G(x-x') h\left(x'\right)dx' + \mathcal{O}(h^2)
\\
&& 
\text{where} 
\hspace{1cm}
G(x-x')=\int\frac{dq}{\left(2\pi\right)^{d}}\,\,
\frac{\eexp{iq\left(x-x'\right)}}{q^{2}+\left({1}/{\xi}\right)^{2}}
\hspace{2cm}
\xi \ \ = \ \ 
\left\{ \begin{array}{ll}
a^{-\frac{1}{2}}, &  \text{for} \ a>0
\\
\left(-2a\right)^{-\frac{1}{2}}, & \text{for} \ a<0
\end{array}\right.
\eeq
We recall that $G(r)$ is both the Green function and 
the correlation function in the Gaussian approximation. 
Hence the Gaussian critical exponents are ${\nu=1/2}$ and ${\eta=0}$.
This $\nu$ value is consistent with what we have obtained for 
the heat capacity exponent ${\alpha}$. In 3D we get ${\alpha=1/2}$.

\sheadC{Symmetry breaking }

Let us first assume that $a>0$ and ask whether 
we can trust the Gaussian approximation.
Higher non-Gaussian terms in the expansion 
of ${ A[\varphi]}$ around ${\bar{\varphi}}$ 
were neglected in the above treatment.
The condition for this neglect is found in the same 
way as in the Bragg-Williams analysis.
Namely, the neglect of the non-Gaussian term 
is justified if ${u \varphi^4 \ll a \varphi^2}$.
This leads to the condition  $\text{Var}(\varphi) \ll (a/u)$.
As in the Bragg-Williams analysis 
the same condition is deduced if we approach ${a=0}$
from below, from the condition
$\sqrt{\text{Var}(\varphi)} \ll |\bar{\varphi}|$. 
So we would like to estimate the fluctuations  
and see if this condition is satisfied.
Within the framework of the Gaussian approximation 
the variance of each Fourier component of the field is 
\beq
\text{Var}(\tilde{\varphi}_k) \ \ = \ \ \frac{1}{L^d} \left(\frac{1}{k^{2} + (1/\xi)^2}\right)
\eeq
The field amplitude $\varphi(x)$ at a given point in space 
is the sum of $\mathcal{N}$ independent Fourier components, 
and accordingly 
\beq
\text{Var}(\varphi) \ \ = \ \ \sum_k \text{Var}(\tilde{\varphi}_k)  \ \ = \ \ G(0)
\eeq
If we kept only the ${k=0}$ contribution, as in the Bragg-Williams analysis, 
we would get $\text{Var}(\varphi) \sim \xi^2/L^d$, which would imply that the Gaussian 
approximation is always OK in the thermodynamic limit. 
If on the other hand we keep all the terms in the above sum, 
we get for ${d>2}$ a huge result that depends on $\Lambda$. 
This {\em bare} estimate of the variance has no significance because 
it reflects the renormalization of~$a$ by the large $k$ fluctuations 
as discussed by [\href{http://iopscience.iop.org/0022-3719/7/18/020}{Amit 1974}].  
Consequently, as suggested by Ginzburg the effective cutoff for 
the purpose of estimating the Gaussian fluctuations is $\Lambda_0 \sim 1/\xi$, 
hence 
\beq
\text{Var}(\varphi) 
\ \ \sim \ \ \int_{0}^{1/\xi} \frac{k^{d-1}dk}{k^2 + (1/\xi)^2 }
\ \ \sim \ \ \frac{1}{\xi^{d-2}} \ \ \sim \ \ G(\xi)
\eeq
This value is determined by $(L/\xi)^d$  effective modes 
that each contribute to the variance $\xi^2/L^d$, 
hence it is $\Lambda$ independent unlike the bare value $G(0)$.
Substitution into the condition $\text{Var}(\varphi) \ll (a/u)$ 
leads to the Ginzburg Criterion 
\beq
|T-T_{c}^{(0)}| \ \ \gg \ \ C \ u^{2/(4-d)}
\eeq
where $C$ is a constant. This condition defines 
the border of the critical region.
Within the critical region the Gaussian approximation 
breaks down because of non-Gaussian fluctuations. 

\putgraph[0.5\hsize]{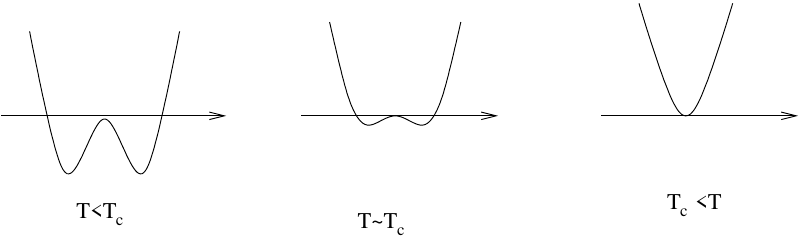}

We now turn to the question whether 
there is a {\em phase transition} 
within the critical region. The other possibility 
is that the fluctuations smear away the 
the phase transition leading to a smooth r
rather than abrupt crossover.  
Namely, for $T\ll T_c^{(0)}$ there is a symmetry breaking, 
such that the mean field $\bar{\varphi}$ jumps 
from positive finite value to negative finite value as $h$ is 
varied across ${h=0}$. Obviously, in order to observe 
a phase transition we have to require that this abrupt jump
is not smeared by the (non-Gaussian) fluctuations within 
the critical region.  
We shall discuss below two cases in which fluctuations 
completely destroy the possibility to observe a phase transition.

{\bf Goldstone excitations. --}
First we refer to the case where the order parameter has a continuous 
rather than a discrete symmetry. To be specific let us assume 
a complex order parameter ${\Psi =(\psi_1,\psi_2)}$.  
In such case the potential ${V(\Psi) = (1/2)a|\Psi|^2+(1/4)|\Psi|^4}$
looks like a Mexican hat for ${r<0}$. It means that instead 
of having two minima we have a continuous set of minima.
This implies that the low frequency excitations of the 
system are phonon-like, called magnons in the ferromagnetic 
context. In general the excitations that appear if a continuous
symmetry is broken are known as Goldstone excitations.  
They have some dispersion $\omega=c|k|$. Coming back to 
the partition sum, we see that we have to include the Goldstone excitations
in the  $\text{Var}(\Psi)$ calculation, leading formally 
to an integral over $1/k^2$ instead of $1/(k^2+(1/\xi)^2)$. 
The integral is "infrared divergent" unless ${d>2}$.
We conclude that the fluctuations destroy the possibility 
to observe a phase transition at ${d=2}$. This is know as the Mermin-Wagner theorem.

{\bf Non-Gaussian fluctuations. --}
Going back to the real field case, there are no Goldstone excitations, 
still we might have non-Gaussian excitations that smear the phase-transition. 
For ${d>1}$ the implication of such fluctuation is not too dramatic: 
the critical point is shifted down (${a_c<0}$) 
but remains finite (see discussion of the RG analysis).
For ${d=1}$ we already have discussed the absence of a finite temperature 
phase-transition using a "domain walls" perspective. 
Let us re-phase the explanation using a field-theory perspective.
It is clear that if we have a phase transition, 
then {\em formally} $\text{Var}(\varphi)\sim \bar{\varphi}^2$ 
on the critical line. The question is whether it 
becomes $\text{Var}(\varphi) \ll \bar{\varphi}^2$ 
for any finite~$h$ away from the critical line. 
This depends on the height of the "barrier" 
between the two minima, and therefore cannot be deduced 
from the Ginzburg criterion: the latter is based on a 
local Gaussian approximation that does not know anything 
about the height of the "barrier".  
We shall see in the next section that in the ${d=1}$ case 
the crossover has a finite width: there is no abrupt 
change in $\langle \varphi \rangle$ as ${h=0}$ is crossed.

{\bf Quantum phase transition. --}
The Landau model is ``classical" in the sense that its Hamiltonian 
commutes with the order parameter. This is not the case e.g. for 
coupled rotors (see previous lecture). In a field theory treatment 
the  partition function $Z=\trc[\eexp{-\beta H}]$ can be written 
as a Feynman path integral over the field $\varphi(x,\tau)$, 
where ${\tau \in [0,\beta]}$ is the so-called imaginary time. 
The integral is over all field configurations in $([0,\beta] \times [0,L]^d)$. 
Accordingly the analysis of the ground state in the thermodynamic 
limit (${\beta,L \rightarrow \infty}$) maps formally to a classical field theory 
with ${d_{cl} = d+1}$ dimensions. This implies that 
is is feasible to observe a zero temperature ``quantum phase transitions", 
as a control parameter is varied, even for ${d=1}$ and notably at ${d=2}$.

\sheadC{The one dimensional model}

The one-dimensional field model can be solved exactly. 
This is merely a variation on the "transfer matrix" method.
The $D\varphi$ integral is sliced and written as a trace 
over the product of $\mathcal{N}$ matrices. Each matrix 
can be written as $\exp(-dx H)$ where $H$ is the "Hamiltonian".
One realize that this is nothing else but the Feynman 
path integral in "imaginary time". Let us define  
\beq
H \ \ = \ \ -\frac{1}{2}\frac{\partial^2}{\partial \varphi^2} + V(\varphi) 
\ \ = \ \ -\frac{1}{2}\frac{\partial^2}{\partial \varphi^2} + \Big[\frac{1}{2}a\varphi^2+\frac{1}{4}\varphi^4\Big]
\eeq 
using the notation $\dot{\varphi}=d\varphi/dx$ and $x=\tau$
with periodic boundary conditions over ${[0,L]}$, 
the calculation of the partition function goes as follow: 
\beq
Z \ \ = \ \ \int D\varphi \ \eexp{-\int_0^L \frac{1}{2}\dot{\varphi}^2 + V(\varphi)d\tau}
\ \ = \ \ \trc(\eexp{-L H}) \ \ = \ \ \sum_n \eexp{-L \mu_n} 
\eeq
where $\mu_n$ are the eigenvalues of $H$. In the thermodynamic 
limit ${F(T,h)= LT \mu_0}$ where $\mu_0$ is the ground state 
energy of $H$. Similarly $\langle \varphi \rangle$ is just the ground 
state expectation value. For the correlation function we get 
\beq
G(r) \ \ \propto \ \ \trc\left[ \eexp{-(L-r)H} \varphi  \eexp{-rH} \varphi  \right]
\ \ \propto \ \ \sum_{n}  |\langle n | \varphi | 0 \rangle|^2 \eexp{-r (\mu_n-\mu_0)} 
\eeq
where in the last equality we already dropped the terms 
that vanish in the thermodynamic limit. We see that the 
long tails are characterized the correlation length ${\xi=1/(\mu_1-\mu_0)}$.
This correlation length does not diverge, reflecting 
that the variation of $\langle \varphi \rangle$ is smooth.
The crossover at $h=0$ has a width that equals 
the tunnel splitting $(\mu_1-\mu_0)$.

\newpage
\sheadC{Coarse graining and scaling}

The free energy $F(a,u,c;\Lambda,L)$ of an homogeneous system 
that is described by the Landau model, 
and the associated correlation function $G(r,a,u,c;\Lambda,L)$  
depend on the following parameters: 
\beq
L &=& \text{linear size of the model} \\
\Lambda &=& \text{largest momentum scale} \\
(c{=}1,a,u) &=& \text{microscopic related parameters} \\
r &=& \text{distance between two test points} 
\eeq  
Schematically we write the free energy as ${F(g;\Lambda,L)}$, 
where $g$ stands for any of the action parameters. 
The ${c=1}$ convention fixes the units of the field $\varphi$,    
as well as the engineering dimension $d_g$ of any 
of the action parameters.  
The microscopic-related parameters have been determined, 
as in the Bragg-Williams approximation, by summation over 
all the microscopic configurations that correspond to 
the same coarse-grained $\varphi(x)$. Accordingly these 
parameters depend on the value of $\Lambda$. To emphasize this 
aspect one may use the notation $g(\Lambda)$.   

Assume that we have used a coarse-graining cutoff $\Lambda$ 
to construct the action. But later we might prefer to work 
with an action that corresponds to a somewhat lower cutoff $\Lambda'$.
Obviously the result of the calculation should be the same.
Accordingly we write 
\beq
F(g; \Lambda,L) 
\ \ = \ \ F(g';\Lambda',L)
\ \ = \ \ F(s^{d_g} g'; \Lambda, sL)
\ \ = \ \ s^{d} F(s^{d_g}g'; \Lambda,L) 
\eeq 
In the second equality we scaled the units by factor $s=\Lambda'/\Lambda$ in order to 
restore the original $\Lambda$ cutoff, and in the last equality we have used the 
thermodynamic limit in order to restore the original $L$ cutoff. 
Using a compact notation we have deduced the scaling relation   
\beq
F(g) \ = \ s^{d} \ F(g_s),
\hspace{2cm}
g_s \equiv s^{d_g} \ g(s\Lambda) 
\eeq 
With regard to the correlation function we note that $G$, unlike $F$, 
does not depend on $L$, but the units of the field have been modified, 
hence $s^{d}$ should be replaced by a different scaling factor that 
we discuss in the next section.

It should be clear that the units of length are arbitrary, hence $F$ 
should be a well defined function of the dimensionless model parameters. 
It follows that we can write the scaling relation for the microscopic related  
parameters without giving explicit reference to $\Lambda$.  Namely,    
\beq
\tilde{g}_s \ \ = \ \ R(s) \ \tilde{g},
\hspace{2cm}
\tilde{g}_s \equiv \Lambda^{d_g} \ g_s
\eeq
where $R(s)$ is a non-linear transformation that depends on~$s$.  
This transformation relates values of $F$ along a trajectory in parameter space,  
and by definition has the semi-group property ${R(s_2)R(s_1)=R(s_2s_1)}$.
Using the parametrization ${s=\eexp{-\tau}}$ we can write the transformation 
as  ${\tilde{g}_{\tau}=R(\tau)\tilde{g}_0}$ and the semi-group property as ${R(\tau_2)R(\tau_1)=R(\tau_2+\tau_1)}$.
Clearly we can generate $R(\tau)$ from infinitesimal steps, 
so we define a $\beta$ function via the expansion 
${R(\tau)\tilde{g} = \tilde{g} + \tau \beta(\tilde{g}) + \mathcal{O}(\tau^2)}$, 
and write the so called renormalization group (RG) equation as       
\beq
\frac{d\tilde{g}}{d\tau} \ \ = \ \ \beta(\tilde{g}),
\ \ \ \ \ \ \ \ \ \ \ \ \
[\text{opposite sign convention if} \ \ d\tau \mapsto d\ln\Lambda]
\eeq
Increasing the course graining parameter $\tau$, we get 
a flow in $g$ space. A fixed point of this flow represents  
a critical point of the model, where the system look-alike 
on any scale. 
If we start the RG trajectory at a point close 
to the fixed point it will flow away, meaning that 
on coarse-grained scale the system looks like having larger ${|T-T_c|}$. 
For the Landau model in 3D we shall see below 
that the RG equation for $a$ becomes    
\beq
\frac{d\tilde{a}}{d\tau} \ \ = \ \ \frac{1}{\nu} (\tilde{a}-\tilde{a}_c), 
\ \ \ \ \ \ \ \ \tilde{a}_c=-\frac{1}{5}, \ \ \nu=\frac{3}{5}  
\eeq
Defining $t=(a-a_c)$ we get the solution $t_s = s^{-1/\nu}t$, 
leading to the scaling relation ${ F(t)= s^d F(s^{-1/\nu}t) }$.
Hence for the heat capacity exponent we get ${\alpha=2-\nu d = 1/5}$, 
and not the mean-filed value ${\alpha=0}$, 
neither the Gaussain value ${\alpha=1/2}$.

\sheadC{Renormalization Group (RG) analysis}

We outline the RG procedure 
that is used in order to find the $\beta(g)$ function, 
where $g$ stands for the parameters $(a,u,c)$ of the Landau model
with the convention ${c=1}$. 
For extra technical details see Section 18.7 of Huang.  

{\bf Step1 of RG.-- } 
Perturbation theory allows to integrate 
the high Fourier components within 
a shell ${\Lambda'<k<\Lambda}$.
where ${\Lambda'=\Lambda{-}\delta\Lambda}$.
Namely any field configuration can be 
written as a sum of smooth and erratic components:
\beq
\varphi(x) \ \ = \ \ 
\sum_{|k|<\Lambda'} \varphi_k \eexp{ikx} 
+ \sum_{\Lambda'<|k|<\Lambda} \varphi_k \eexp{ikx} 
\ \ \equiv \ \ \bar{\varphi}(x) \ + \ \tilde{\varphi}(x)    
\eeq
The action can be expanded with respect to $\tilde{\varphi}$ 
up to quadratic order. This is allowed because $\delta \Lambda$ 
is chosen as arbitrarily small. Using abstract notation
with regard to field indexes we write the expansion as   
\beq
A[\varphi(\cdot)] \ \ = \ \ A[\bar{\varphi}(\cdot)] 
\ + \ h_{\text{induced}}[\bar{\varphi}] \, \tilde{\varphi} + a_{\text{induced}}[\bar{\varphi}] \, \tilde{\varphi}^2 
\eeq
Now it is possible to use Gaussian integration over $\tilde{\varphi}$ 
to get an effective expression for $A[\varphi(\cdot)]$ that involves 
new values for the model parameters. Accordingly 
\beq
F(a,u,c; \Lambda,L) \ \ = \ \ F(a',u',c'; \Lambda',L)
\eeq
Doing the algebra the result is 
\beq
a' \ &=& \ a + \delta \Lambda \Big[ 3\Omega_d \left(\Lambda^{d-3}u-\Lambda^{d-5}au \right)\Big] \\
u' \ &=& \ u - \delta \Lambda \Big[ 9\Omega_d  \Lambda^{d-5} u^2 \Big] \\
c' \ &=& \ c 
\eeq
Note that the $u$ of Huang should be identified with our~$u/4$, 
and $r$ of Huang is identified with our~$a$.  
Though not the case here, one should be aware that in general 
the elimination of the high Fourier components might spoils 
the $c=1$ convention.

{\bf Step2 of RG.-- } 
In "step2" of the RG procedure the original value of $\Lambda$
is restored vis engineering scaling, and then the thermodynamic 
limit is assumed to restore $L$ as well. Accordingly  
\beq
F(a,u,c; \Lambda,L) \ \ = \ \ s^{d} \ F(s^{d_a}a',s^{d_u}u', c'; \Lambda,L), 
\hspace{2cm} 
s \equiv \eexp{-\tau}, \ \ \tau = \delta\Lambda/\Lambda \ll 1 
\eeq
Note that according to the common convention ${d_c=0}$.

{\bf Step3 of RG.-- } 
In "step3" of the RG procedure the field $\varphi$ is re-scaled 
such that the convention ${c=1}$ is restored.
Using the notation $c'=s^{-\eta}$, 
and dropping reference to the restored parameters, we get       
\beq
F(a,u) \ \ &=& \ \ s^{d} \ F(a_s, u_s) \\ 
G(r,a,u) \ \ &=& \ \ s^{-2d_{\varphi}+\eta}  \ G(sr, a_s, u_s)
\eeq
Here we suppressed $c$, because it has been restored to unity. 
In the Landau model ${c'=c}$ and therefore $\eta=0$.
For the two other parameters we get 
\beq
a_s \ \ &=& \ \ s^{-2} a' \ \ = \ \  a \ + \ \tau \Big[2a + 3\Omega_d \left(\Lambda^{d-2}u-\Lambda^{d-4}au \right)\Big]  \\
u_s \ \ &=& \ \ s^{-(4-d)} u' \ \ = \ \ u  \ + \ \tau \Big[(4-d)u - 9\Omega_d  \Lambda^{d-4} u^2 \Big]
\eeq
where ${s=1-\tau}$ applies for an infinitesimal step.     
Without the high frequency contribution the above RG relation 
is just a complicated way to write how the parameters are affected 
by engineering scaling.

{\bf RG equation.-- } 
We can illustrate the RG flow in the $(a,u)$ space. 
Increasing~$\tau$ means lower resolution description of the system, 
with effective parameters $(a_s,u_s)$.   
It is convenient to use dimensionless parameters ${\tilde{g}=\Lambda^{d_g}g}$, 
such that the transformation $R(s)$ becomes free of $\Lambda$. 
Considering an infinitesimal $\tau$ one finds that the RG-equations 
of the Landau model are
\beq
\frac{d\tilde{a}}{d\tau} \ \ = \ \ 2\tilde{a}+3\tilde{u}-3\tilde{a}\tilde{u}
\\
\frac{d\tilde{u}}{d\tau} \ \ = \ \ (4-d)\tilde{u} - 9\tilde{u}^2
\eeq
where $\tilde{a}=a/\Lambda^2$ and $\tilde{u}=\Omega_d u/\Lambda^{4-d}$.

{\bf RG flow.-- }
The RG equation defines flow in ${(a,u)}$ space. 
This flow is illustrated in the figure below. 
In the  Landau model we have two fixed points.
The Gaussian fixed point is for ${\tilde{u}_0=0}$ at $\tilde{a}_0 =0$. 
The nontrivial fixed point is 
\beq
\tilde{u}_c \ \ &=& \ \ \frac{(4{-}d)}{9} \\
\tilde{a}_c \ \ &=& \ \  -\left[ 1-\frac{(4{-}d)}{6} \right]^{-1} \frac{(4{-}d)}{6}
\eeq
For ${d<4}$ the Gaussian fixed point is unstable 
and the flow is dominated by the nontrivial fixed point.
One observes that the critical temperature ($a_c$) 
is shifted below the mean field Gaussian value.  
 
\includegraphics[width=8cm]{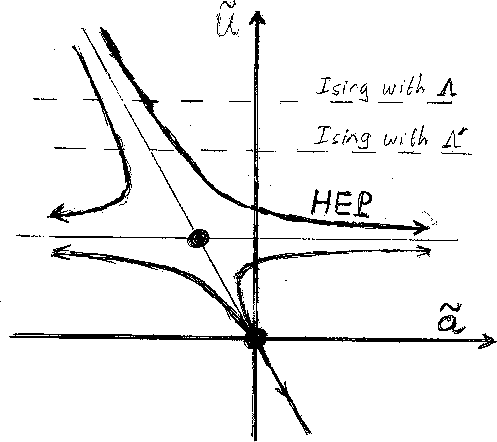} 

\ \\

\sheadC{Implications of the RG results}

The results of the RG analysis are used slightly differently 
in Statistical Mechanics ("Ising") and in high energy physics ("HEP").  
We first would like to explain the subtle difference, 
and then to focus on the the "Ising" context. 
   
{\bf HEP.-- }
In the HEP context $\sqrt{a}=\mass$ is the so-called "bare mass", 
and $u$ is the "bare interaction".
They are associated with the momentum exchange ${q\sim\Lambda}$ 
in, say, electron-electron scattering calculation. 
The low energy scale~$\Lambda_0$ is the Compton wavelength 
that corresponds to the electron mass. 
For larger $\Lambda$ the scattering involves "loop" corrections 
reflecting the virtual appearance of electron-positron pairs. 
Consequently the mass and the interaction become ``running coupling constants".  
With different choices of $\Lambda$ one associates different $a(\Lambda)$ and $u(\Lambda)$.
The physical mass and the physical interaction for $q$-scattering 
are defined through the measured dispersion-relation 
and through the cross-section respectively. 
Both are functions of ${(q; a,u)}$, and are calculated from the bare parameters. 
With different choices of $\Lambda$ we can associate different points 
in $(\tilde{a},\tilde{u})$ space.
It follows that the relevant physics is along 
a {\em specific} HEP line of the RG flow (see figure).
Accordingly what we get from the RG analysis is 
how $a(\Lambda)$ and $u(\Lambda)$ depend on $\Lambda$.
It turns out that the ${d=4}$ Landau model has 
essentially the same beta function as in QED. 
Namely, the beta function for the QED coupling $g=e^2/(4\pi)$ 
is $\beta(g)=Cg^2$ with $C=2/(3\pi)$ instead of $C=1/9$. 
One obtains   
\beq
g(\Lambda) \ \ = \ \  \frac{g_0}{1-C g_0 \ln(\Lambda/\Lambda_0)} \ \ \equiv \ \  \frac{1}{C \ln(\Lambda_{\infty}/\Lambda)} 
\eeq 
where $\Lambda_0$ corresponds to the Compton length of the electron 
and $g_0\approx1/137$ is the asymptotic QED coupling.    
At large distances (${q < \Lambda_0}$) the interaction strength
is renormalized to a universal value that is independent 
of the bare parameters. For short distances the interaction 
becomes stronger, and diverges at the Landau pole $\Lambda_{\infty}$. 
Note that this divergence is possibly not physical because 
the derivation of the RG equation assumes that $g$ is small. 

{\bf Ising.-- }
In the Ising context the reasoning is different. 
After $\Lambda$ coarse-graining we get an action 
with $a$ and $u$, which we represent as a point 
in $(\tilde{a},\tilde{u})$ space. If we vary the temperature 
it is like going along a horizontal line. If we use a lower 
cutoff $\Lambda'$ it would takes us to a lower horizontal line (see figure).      
Assuming that we have done enough coarse-graining, the value 
of $u$ would become $u \approx u_c$. We therefore say that $u$ 
is "irrelevant". Without loss of generality we can set ${u=u_c}$
in our calculations.  We end up with a single RG equation for 
the variable $t=(a-a_c)$, namely    
\beq
\frac{dt}{d\tau} \ = \ (2-3u_c) \, t
\hspace{2cm}
\mbox{generic notation:} \ \ 
\frac{dg}{d\tau} = \lambda \, g
\eeq
The solution of this equation is ${t_{\tau}=\eexp{\lambda\tau}t_0}$ 
hence we deduce the scaling relation ${F(t) = s^{d} F(s^{-\lambda}t)}$. 
We therefore conclude that ${F(t)\propto |t|^{d/\lambda}}$, 
from which the dependence of the heat capacity on the temperature can be derived.   

{\bf The critical exponents.-- }
We want to understand how the RG flow 
explains the scaling hypothesis.
In the vicinity of the fixed point we can 
linearize the RG equation. 
After linear transformation and shift 
we get an equation of the type ${dg/d\tau=\lambda g}$, 
whose solution is ${g_{\tau}=\eexp{\lambda\tau}g}$.
Parameters with negative eigenvalues ${\lambda<0}$ 
vanish from the model due to the coarse graining  
and therefore are called "irrelevant".
We keep only the relevant parameters and deduce that 
\beq
F(g) \ \ = \ \ s^{d} \ F(s^{D_g} g), 
\ \ \ \ \ \ \ \ D_g\equiv-\lambda
\eeq
The anomalous dimension $D_g$ of Huang is defined with opposite sign 
(he is using inverse length units).
In general "$g$" stands for a collection of (relevant) parameters, 
each having its own "dimension". 
In the Landau model the relevant parameter is related 
to the temperature, namely ${t = (a-a_c) \propto (T-T_c)}$. 
From the RG equation we deduce 
\beq
D_{t} \ \ = \ \ -\left[2-3u_c\right] 
\ \ = \ \ -\left[2-\frac{(4{-}d)}{3}\right] \ \ \equiv \ \ -\frac{1}{\nu}  
\eeq 
Hence the scaling relations take the familiar form 
\beq
F(t) \ \ &=& \ \  s^{d} \ F(s^{-1/\nu} t) \\
G(r,t) \ \ &=& \ \  s^{d-2+\eta} \ G(sr, s^{-1/\nu} t)
\eeq 
This means that the coarse grained system looks 
less and less critical as $\tau$ is increased.
Having determined $\nu$ and $\eta=0$ using the RG procedure, 
we can deduce all the other critical exponents from the scaling relations.

\newpage
\sheadC{The Heisenberg model}
 
The Heisenberg model relates to a three component vector field $\bf{S}$.  
It is described by the Hamiltonian 
\beq
\mathcal{H} \ \ =  \ \ -J \sum_{\left\langle {i,j} \right\rangle} \mathbf{S}_i \cdot \mathbf{S}_j 
\eeq
The field theory version of the Heisenberg model 
is the nonlinear sigma model (NLSM), 
where $S(\bm{r})$ is a vector field that obeys the constraint ${|S(\bm{r})|=1}$. 
The NLSM energy functional is commonly written as 
\beq
E[S] \ \ = \ \ \frac{1}{2g} \int |\nabla S|^2 d\bm{r}
\eeq

{\bf The Mermin-Wagner theorem:-- }
In the Heisenberg and Landau-Ginzburg models 
the order parameter has a continuous symmetry 
with respect to spin rotations. 
This is different from the Ising and Landau models 
where the symmetry is discrete (up/down).
The Mermin-Wagner theorem states that continuous symmetries cannot 
be spontaneously broken at finite temperature in $d\le2$ systems 
with short-range interactions.
This is due to long-range fluctuations, 
corresponding to massless "Goldstone modes". 
These fluctuations cost little energy, 
while contribute large entropy. 
Hence at finite temperatures the ${T=0}$ 
broken symmetry is washed away.

Let us outline the argument that explains the dimensionality issue.
Assume that there is a long range order. Use the Gaussian approximation  
to calculate the correlation function around the selected minimum. 
The Goldstone modes are massless (which is like ${r=0}$ in the Landau model).
Hence ${g(r) = \text{FT}[1/k^2]}$. The area under $g(r)$ 
diverges due to the low wavelength fluctuations 
if $d\le2$, indicating that the minimum is washed away.

\sheadC{The XY model}

The XY model relates to a two component vector field $\bf{S}$. 
It can be regarded as describing coupled rotors. 
The orientation of the $i$th spin in the XY plane is $\varphi_i$.
Accordingly we can write
\beq
\mathcal{H} \ \ = \ \ -J \sum_{\left\langle {i,j} \right\rangle} \cos(\varphi_i - \varphi_j)
\eeq
The associated continuous field version is defined by the energy functional  
\beq
E[\varphi] \ \ = \ \ \frac{K}{2\pi} \int (\nabla \varphi)^2 d\bm{r} \ \ = \ \ \frac{K}{2\pi} \int |\Psi'(\bm{r})|^2 d\bm{r}
\eeq
where $d\bm{r}$ integrates over space, and ${\Psi(\bm{r})=\eexp{i\varphi(\bm{r})}}$. 
Note that the Landau-Ginzburg model has an order-parameter ${\Psi(r)=\sqrt{n(\bm{r})}\eexp{i\varphi(\bm{r})}}$,  
and can be regarded as a variation of the XY model, where $n(\bm{r})$ is not constrained to unity.  
The low lying excitations of the XY model in 2D are vortexes. 
For example, a single vortex at the origin is represented by 
\beq
\varphi(\bm{r}) \ \ &=& \ \ q \ \arg(x,y) \ \ = \ \ q \tan^{-1}(y/x)
\\
|\nabla \varphi| \ \ &=& \ \ (q/r) \ \ \text{[tangential]} 
\\
E[\varphi] \ \ &=& \ \  K q^2 \ln(L/a)
\eeq
where $L$ is the radius of the system, and $a$ is the lattice spacing, 
and $q$ is an integer, say $q=\pm 1$.  
One realizes that $\vec{F}=\nabla \varphi$ 
is like a 90deg rotated version of a field 
dues to a charge $q$ in a 2D electrostatic problem, 
while $E[\varphi]$ corresponds to the electroststic energy.
Accordingly if we have several vortexes we get
\beq
E[\varphi] \ \ &=& \ \ \const -  K \sum_{\left\langle {i,j} \right\rangle} q_i q_j \ln|\bm{r}_i - \bm{r}_j| 
\eeq
which is formally the expression for the energy of a Coulomb gas in 2D.

{\bf Note about complex representation.-- }
Consider a field ${\vec{F}=(F_x,F_y)}$ in 2D that has no sources (${\nabla \cdot F = 0}$) 
and no circulations (${\nabla \times F = 0}$).
These two conditions are the Cauchy-Riemann equations  
that allow to represent the field by 
a complex differentiable function ${f(z)=F_x-iF_y}$, where ${z=x+iy}$.
Note that in sloppy notations $\vec{F}=f(z)^*$.  
Note that such functions are called holomorphic and 
hence they are analytic, i.e. have a convergent Taylor 
expansion at any point. 
Note also that a holomorphic function can be regarded 
as a conformal map $w=f(z)$ that induces a 
local transformation $df=Cdz$, where the 
local derivative $C$ can be regarded as a 
rotation (due to its phase) combined with dilation (due to its modulo). 
From the Cauchy-Riemann no-circulation condition it follows that 
the field can be derived from a scalar potential (${\vec{F}=-\nabla V}$).
Optionally from the Cauchy-Riemann no source condition it follows that 
the field can be derived from a vector potential (${\vec{F}=\nabla \times (0,0,A)}$).
We can summarize these two options by 
defining a complex potential $\Psi(z)=V(z)+iA(z)$
such that $f(z)=-\Psi'(z)$.
The lines of constant $A(z)$ are called "stream lines", 
while the lines of constant $V(z)$ are called "propagation fronts". 
Differences of $A$ have the meaning of "flux", 
while differences of $V$ have the meaning of "work".  
For a point charge in a 2D electrostatic problem 
the Coulomb field is $f(z)=q/z$ and $\Psi(z)=-q\ln(z)$ 
corresponding to $V(x,y)=-q\ln(r)$.  
The vortex in the XY model is just a rotated version of 
a Coulomb field with $f(z)=-iq/z$ and $\Psi(z)=-iq\ln(z)$.

{\bf The Kosterlitz-Thouless transition.-- }
Considering the XY model in 2D space, 
let us see what happens if we have a  collection of vortexes.
The entropy which is associated with the number of possibility 
to position the vortex in the lattice is $S = \ln[(L/a)^2]$. Accordingly 
the introduction of a vortex into the system implies 
\beq
F[\rho] \ \ = \ \ E[\rho] - TS[\rho]  \ \ = \ \ (K-2T) \ln[(L/a)]
\eeq
Hence for $T>(K/2)$ the creation of a vortex is favoured, 
so we expect to have a gas of quasi-independent vortexes. 
For $T<(K/2)$ these vortexes have to pair, which is like not 
having vortexes. 
The above argumentation implies an "infinite order phase transition", 
known as the  Kosterlitz-Thouless transition.
In consistency with the Mermin-Wagner theorem
it does not lead to "long range order".
Rather it describes a transition from 
"conventional disordered phase" at high temperature, 
to "quasi-long-range ordered phase" at low-temperature.
The correlation function goes from exponential above the critical temperature, 
to powerlaw at the critical temperature and below.
At the critical temperature $g(r)=1/r^{\eta}$, with ${\eta=1/4}$. 
Below the critical temperature a Gaussian estimate gives ${\eta=T/(2K)}$.

\sheadA{Fluctuations and Response}

\makeatletter{}
\sheadB{Fluctuations}

In order to know the expectation value of an operator we need 
only spectral information which is present in $\gdos(E,X)$ 
or in  $Z(\beta,X)$. Note that these functions contains 
only spectral information about the system 
(no information on the dynamics). 
Still it is enough for the calculation of the 
conservative force. For example, in case of a canonical preparation 
\beq
\langle \mathcal{F} \rangle_0 
\ = \ 
\left\langle -\frac{\partial \mathcal{H}}{\partial X} \right\rangle
\ = \ 
\sum_n p_n \left(-\frac{\partial E_n}{\partial X} \right) 
\ = \ 
\frac{1}{\beta}\frac{\partial \ln(Z)}{\partial X}
\eeq
In contrast to that, the fluctuations 
of $\mathcal{F}(t)-\langle \mathcal{F} \rangle_0$ 
require knowledge of the dynamics, 
and cannot be calculated from the partition function.  
For simplicity of presentation we assume below  
that the fluctuating quantity of interest is re-defined 
such that ${\langle \mathcal{F} \rangle_0=0}$.

\sheadC{The classical power spectrum}

Consider a stationary stochastic classical variable $F(t)$, 
and define its correlation function as 
\beq
C(t_2-t_1) \ \ = \ \ \langle F(t_2) F(t_1) \rangle
\eeq
The power spectrum $\tilde{C}(\omega)$ is the Fourier transform of $C(\tau)$, 
where ${\tau=t_2-t_1}$ is the time difference. 
In practice a realization of $F$ within time interval ${0<t'<t}$ 
can be Fourier analyzed as follows:
\beq
F_{\omega} \ \ = \ \ \int_{0}^{t} F(t') \eexp{i\omega t'} dt'
\eeq
and we get the "Wiener-Khinchin theorem"
\beq 
\langle |F_{\omega}|^2 \rangle \ \ = \ \ \tilde{C}(\omega) \times t
\eeq
where we assume that $t$ is much larger 
compared with the correlation time.

A related perspective concerns the intensity of the power spectrum:
\beq 
\nu \ \ = \ \ \int_{-\infty}^{\infty} C(\tau) d\tau \ \ = \ \ \tilde{C}(\omega{=}0)
\eeq
If we regard $F(t)$ as the velocity ("steps") in a random-walk process, 
then the zeroth Fourier component is 
\beq
r(t) \ \ = \ \ \int_0^t F(t') \ dt' \ \ = \ \ \mbox{total displacement}
\eeq  
On the average $\braket{r}=0$ but the variance is linear in time:
\beq
\mbox{Var}(r) \ \ = \ \  \int_0^t \int_0^t  \braket{F(t')F(t'')} \, dt'' dt' 
\ \ = \ \ \nu \, t  \ \ \equiv \ \ 2D \, t
\eeq
Accordingly the intensity $\nu$ is trivially related to the coefficient~$D$
of a diffusion process that is generated by the noisy signal~$F(t)$.

\sheadC{The quantum power spectrum}

We consider a system whose dynamics is generated 
by Hamiltonian $\mathcal{H}$. We assume that it is 
prepared in a stationary state, possibly but not necessarily 
a thermal states. Our interest is in the fluctuations 
of an observable $\mathcal{F}$. We use the common 
interaction picture notation ${\mathcal{F}(t)= \eexp{i\mathcal{H}t} \mathcal{F} \eexp{-i\mathcal{H}t}}$.
The non-symmetrized correlation function of $\mathcal{F}$ is defined as 
\beq
S(t) \ = \ 
\langle \mathcal{F}(t)\mathcal{F}(0) \rangle 
\eeq
This function is complex, but its Fourier transform 
is real and can be calculated as follows:  
\beq
\tilde{S}(\omega) \ = \ \int_{-\infty}^{\infty} S(t) \eexp{i\omega t} dt
\ = \ \sum_n p_n \sum_m |\mathcal{F}_{mn}|^2 
\ 2\pi \delta\left(\omega-\frac{E_m-E_n}{\hbar}\right)
\eeq
In the case of a {\em microcanonical} preparation 
at some energy~$E$, this is the same object that appears 
in the Fermi-Golden-rule (FGR) for rate of decay 
due to transitions to other levels, namely
\beq
\Gamma_{\tbox{FGR}} \ = \  \tilde{S}_E(\Omega) \times f_0^2, 
\ \ \ \ \ \ 
\text{for} \ \mathcal{H}-f(t) \mathcal{F}, 
\ \ \text{with} \ f(t)=f_0\eexp{-i\Omega t} 
\eeq
See further discussion of the FGR in the Kubo formula section.  
In the above formula $\omega>0$ corresponds 
to absorption of energy (upward transitions), 
while $\omega<0$ corresponds to emission (downward transitions).

\sheadC{The detailed balance relation}

It is a straightforward algebra to show that 
for a {\em canonical} preparations at temperature~$T$, 
where ${p_n\propto \exp(-E_n/T)}$,  
there is a detailed balance relation:
\beq
\tilde{S}_T(-\omega) \ \ = \ \ \exp\left(-\frac{\hbar\omega}{T}\right) \ \tilde{S}_T(\omega)
\eeq
This implies that if we couple to the system 
another test system (e.g. a two level ``thermometer")  
it would be driven by the fluctuations 
into a canonical state with the same temperature.

The disadvantage of $\tilde{S}_T(\omega)$ is that it has 
meaning only in the quantum mechanical context. 
We want a formulation that treat the quantum and the classical 
on equal footing. We therefore define spectral functions  
that have well defined classical limit:
\beq
\tilde{C}(\omega) 
\ \ &\equiv& \ \ \mbox{FT} \ \frac{1}{2} \Big\langle \mathcal{F}(t)\mathcal{F}(0) + \mathcal{F}(0)\mathcal{F}(t) \Big\rangle
\ \ = \ \ \frac{1}{2}\left(\tilde{S}(\omega)+\tilde{S}(-\omega)\right)
\\
\tilde{K}(\omega) 
\ \ &\equiv& \ \ \mbox{FT} \ \frac{i}{\hbar} \Big\langle [\mathcal{F}(t),\mathcal{F}(0)] \Big\rangle 
\ \ = \ \ \frac{i}{\hbar}\left(\tilde{S}(\omega)-\tilde{S}(-\omega)\right)
\eeq
and deduce that at thermal equilibrium they are related as follows: 
\beq
\tilde{K}(\omega)  \ = \  i \frac{2}{\hbar}\tanh\left(\frac{\hbar\omega}{2T}\right)  \  \tilde{C}(\omega)
\eeq
We shall see later that $\tilde{K}(\omega)$ determines 
the absorption coefficient of the system, hence the above 
relation is going to be the basis for a ``fluctuation-dissipation relation".

It is interesting to look on the classical limit 
of the detailed balance relation.
The classical canonical version
can be regarded as the low frequency limit 
of the quantum relation:
\beq
\tilde{K}_T(\omega)  \ \ = \ \  i\omega \times
\frac{1}{T} \  \tilde{C}_T(\omega)
\hspace{4cm}
\text{[classical canonical version]}
\eeq
It looks very nice in time domain:
\beq
K_T(t)  \ = \ -\frac{1}{T} \  \dot{C}_T(t)
\eeq
We shall use the latter version in order to derive 
what we call later the ``DC version" of the 
a generalized fluctuation-dissipation relation.

\sheadC{The classical version of "detailed balance"}

The purpose of the present section is to show how 
a general relation between $K(t)$ and $C(t)$ can 
be derived within the framework of classical mechanics, 
assuming that the system is prepared in some 
arbitrary stationary state ${\rho(x,p)=f(H(x,p))}$. 
The classical canonical version that we have deduced
in the previous version by taking the ${\hbar\omega \rightarrow 0}$ limit 
can be regarded as a special case with ${f(H) \propto \exp(-\beta H) }$.    
We first define $K(t)$ and $C(t)$ in a more general way:
\beq
C(t) \ \ &=& \ \ \trc\left\{ f(H) \ A_t B \right\} \\   
K(t) \ \ &=& \ \ \trc\left\{ f(H) \ [A_t, B] \right\}  
\eeq
The trace means $dxdp$ phase-space integration, 
with obvious generalization to more than one freedom.
We define ${z=(x,p)}$ and $\partial=(\partial_x,\partial_p)$.
The Poisson brackets of two functions are:
\beq
[A, B]_{\text{PB}} \ \ = \ \ (\partial_x A)(\partial_p B) - (\partial_p A)(\partial_x B)  \ \ \equiv \ \ \sum_{i,j} (\partial_i A) J_{i,j} (\partial_j B)
\eeq
where ${J_{i,j}=\text{matrix}\{0,1;-1,0 \}}$.
One easily proves that ${\trc\{[A, B]C\} =\trc\{A [B, C] \} }$.
Note also the chain rule ${[f(H),A]= f'(H)[H,A]}$.     
It is now possible to derive the following identity:
\beq
\trc\left\{ f(H) [A_t, B] \right\}  
\ \ = \ \ \trc\left\{ [f(H), A_t] B \right\}  
\ \ = \ \ \trc\left\{ f'(H) [H, A_t] B \right\}  
\ \ = \ \ -\frac{d}{dt} \trc\left\{ f'(H) A_t B \right\}   
\eeq
where the time evolved function $A_t$ obeys the Hamilton equation of motion ${(d/dt)A_t = -[H,A]_t}$, 
and we use the canonical invariance ${[H,A]_t = [H_t,A_t]}$.   
For a microcanonical distribution ${ f(H) = \gdos(E)^{-1} \delta(H-E)}$ we get
\beq
K_E(t)  \ \ = \ \  -\frac{1}{\gdos(E)} \frac{d}{dE} \Big[ \gdos(E) \ \dot{C}_E(t) \Big]
\hspace{2cm}
\text{[classical microcanonical version]}
\eeq
For a canonical distribution ${ f(H) = Z(\beta)^{-1} \exp(-\beta H) }$ 
the weighted energy derivative is replaced by $1/T$ as anticipated  
from the quantum version. 
Optionally the canonical version can be obtained from the microcanonical 
version by averaging over the energy with the canonical 
weight ${\propto \gdos(E)\exp(-E/T)}$, and integration by parts. 
A physically appealing deduction of the connection
between $K$ and $C$ will be implied by the derivation 
of the classical fluctuation-dissipation relation.

\sheadC{Fluctuations of a many body system}

A single particle dynamics at energy~$\epsilon$ 
can be characterized  by the single particle 
microcanonical fluctuation spectrum $C_{\epsilon}(\omega)$, 
and by the associated response function
\beq
\tilde{K}_{\epsilon}(\omega)  
\ \ = \ \  
i\omega \times
\frac{1}{\gdos(\epsilon)} \frac{d}{d\epsilon} \Big[ \gdos(\epsilon) \ \tilde{C}_{\epsilon}(\omega) \Big]
\eeq
Let us assume that we have a many body occupation
that is described by an occupation function $f(\epsilon_n)$.
If we consider a one-body operator ${\hat{\mathbf{A}} = \sum_{mn}A_{mn} a_m^\dagger a_n}$, 
the expectation function an additive property that relates 
it to the single particle expectation values: 
\beq
\langle \hat{\mathbf{A}} \rangle \ \ = \ \ \sum_n  f(E_n) \ \langle A \rangle_n 
\eeq
If $C$ is the two-body operator such property does 
not in general exist: the total "interaction" is greater 
than the sum of the interactions within subsets.  
But if the two-body operator  $C=[A,B]$ is defined 
as the commutator of two one-body operators, 
it is easy to show that the additive property is re-gained. 
According we deduce that for an $N$ body system  
\beq
\tilde{K}^{[N]}(\omega) \ \ = \ \
\int d\epsilon \ \gdos(\epsilon)f(\epsilon) \ \tilde{K}_{\epsilon}(\omega) 
\eeq
Expression $\tilde{K}_{\epsilon}(\omega)$ using $\tilde{C}_{\epsilon}(\omega)$ we get
\beq
\tilde{K}^{[N]}(\omega) \ \ = \ \
i\omega \times 
\left\{ \amatrix{
\mathsf{g}(\epsilon_F) \ \tilde{C}_{F}(\omega) 
\ \ \ \ \ \ & \text{degenerate Fermi occupation ($T\ll \epsilon_F$)} \cr
(N/T) \ \tilde{C}_T(\omega) 
\ \ \ \ \ \ & \text{dilute Boltzman occupation ($T\gg \epsilon_F$)}
}\right. 
\eeq
where the subscripts $F$ and $T$ implies that $C_{\epsilon}(\omega)$
is evaluated at the Fermi energy, or averaged 
according to the thermal occupation, respectively. 
Note that in practice $ \mathsf{g}(\epsilon_F) \sim N/\epsilon_F$, 
so we have a smooth crossover at ${T \sim \epsilon_F}$. 
The fluctuations of the many-body current 
are deduced from the detailed balance relation:
\beq
\tilde{C}^{[N]}(\omega) 
\ \ = \ \
\frac{\hbar}{2}
\coth\left(\frac{\hbar\omega}{2T}\right)
\ \im\left[\tilde{K}^{[N]}(\omega) \right]
\eeq
Considering a system of Fermions we get 
\beq
\tilde{C}^{[N]}(\omega) 
\ \ = \ \
\left\{ \amatrix{
\frac{\hbar\omega}{2} \coth\left(\frac{\hbar\omega}{2T}\right) \ \mathsf{g}(\epsilon_F) \ \tilde{C}_{F}(\omega) 
\ \ \ \ \ \ & \text{degenerate Fermi occupation} \cr
N \ \tilde{C}_T(\omega) 
\ \ \ \ \ \ & \text{dilute Boltzmann occupation}
}\right. 
\eeq
Contrary to the classical reasoning the zero temperature Fermi sea 
is not noisy ${\nu_N=\tilde{C}^{[N]}(0)=0}$.   
The intensity $\nu_N$ of the current-fluctuations is not simply 
the sum over the one-particle fluctuations. 
The classical result is $\nu_N= N\nu_T$ 
where $\nu_T$ is the thermally averaged one-particle fluctuation-intensity.  
In contrast, the quantum result is $\nu_N = N_T \nu_F$, 
where $\nu_F$ is evaluated at the Fermi energy,  
and ${N_T = T \mathsf{g}(\epsilon_F) \equiv T/\Delta}$ 
is the effective number of particles that contribute to the noise.
Strangely enough (see next lecture) the classical and the 
quantum calculations give the same result within 
the framework of the Drude model: the temperature dependence 
merely shifts from the ``$\nu$" to the ``$N$".

{\bf Many body calculation.-- }
It is interesting to see how the relation between 
many-body fluctuations and single particle fluctuations  
is deduced for a low temperature system of Fermions
using a direct calculation.  
We consider the fluctuations of a general observable~$A$.
If we treat the many body system as a whole
then we have to employ second quantization to 
write ${\hat{\mathbf{A}} = \sum_{mn}A_{mn} a_m^\dagger a_n}$. 
Excluding the irrelevant diagonal $n{=}m$ terms 
we get for a non-interacting system in a thermal state
\beq \label{manybodyPsp}
\tilde{S}^{[N]}(\omega)
\ \ &=& \ \ 
\text{FT} \braket{ \hat{\mathbf{A}}(t) \hat{\mathbf{A}}(0) } 
\\ \nonumber
&=&
\sum_{nm} |A_{mn}|^2 \braket{ a_n^\dagger a_m a_m^\dagger a_n } \, 2\pi\delta(\omega {-}(\epsilon_m{-}\epsilon_n))
\\ \nonumber
&=&
\sum_{nm} (1{-}f(\epsilon_m))f(\epsilon_n) |A_{mn}|^2 \, 2\pi\delta(\omega {-} (\epsilon_m {-} \epsilon_n))
\\ \nonumber
&=&
\int\frac{d\epsilon}{\Delta} (1{-}f(\epsilon{+}\omega))f(\epsilon) \
\tilde{C}_{\epsilon}(\omega)
\eeq
If we had bosons the last expression would be $(1{+}f)f$ instead of $(1{-}f)f$.
In the dilute Boltzmann limit (${f\ll1}$) we recover additivity, 
as expected from classical considerations with regard to uncorrelated motions 
of independent particles. But for non-dilute occupations the results depends 
of the ``many body statistics". For bosons there are enhanced fluctuations 
due to the "bunching" of particles in the same orbital. For fermions it 
is the opposite effect. 
In the latter case let us assume that the single particle power spectrum 
has a well-defined mean level spacing $\Delta$ at the energy range of interest, 
namely at the thermal vicinity of the Fermi energy. At the limit of zero 
temperature the result of the integral is clearly zero for ${\omega<0}$
reflecting that a zero-temperature the system can only absorb energy.
For positive frequencies the zero temperature result ${(\omega/\Delta)\tilde{C}_{F}(\omega)}$ 
is proportional to the number levels $(\omega/\Delta)$ from which transitions 
to empty orbitals can take place. For finite temperature we get: 
\beq
\tilde{S}^{[N]}(\omega)
\ \ &=& 
\ \ \frac{\omega/\Delta}{1-\eexp{-\omega/T}} \
\tilde{C}_{F}(\omega) 
\\
\tilde{C}^{[N]}(\omega) \ \ &=& 
\ \ \frac{\omega}{2\Delta} 
\coth\left(\frac{\omega}{2T}\right) 
\ \tilde{C}_{F}(\omega) 
\\
\tilde{K}^{[N]}(\omega) \ \ &=& \ \ 
i\frac{\omega}{\Delta} 
\ \tilde{C}_{F}(\omega) 
\eeq
Needless to say that these results, that have been deduced here 
from a direct many-body calculation, are consistent with the former 
deduction that has been based on detailed-balance considerations.

\sheadC{Fluctuations of several observables}

Give several observables $\mathcal{F}^j$, 
and assuming that the system is prepared 
in a stationary state, the fluctuations 
can be characterized by the correlation function 
\beq
S^{kj}(t) \ \ = \ \ \langle \mathcal{F}^k(t)\mathcal{F}^j(0) \rangle 
\eeq   
The associated spectral function $\tilde{S}^{kj}(\omega)$ 
is defined by a Fourier transform. For simplicity we use 
the notations ${\mathcal{F}^1=A}$ and ${\mathcal{F}^2=B}$, 
and write the spectral decomposition    
\beq
\tilde{S}^{AB}(\omega) 
= \sum_n  p_n \sum_{m (\ne n)} 
A_{nm} B_{mn}\, 2\pi \delta\left(\omega-\frac{E_m-E_n}{\hbar}\right)
\eeq
It is convenient to write $S^{AB}(t)$ as the sum of 
two {\em real} spectral functions  
that have a good classical limit: 
\beq
S^{AB}(t) \ & = & \ C^{AB}(t) -i\frac{\hbar}{2}K^{AB}(t)
\\
C^{AB}(t)  \ &\equiv& \ \frac{1}{2} \Big\langle A(t)B(0) + B(0)A(t) \Big\rangle
\\
K^{AB}(t) \ &\equiv& \ \frac{i}{\hbar} \Big\langle [A(t),B(0)] \Big\rangle 
\eeq
We use the notations $\tilde{S}^{AB}(\omega)$, 
and $\tilde{C}^{AB}(\omega)$, and $\tilde{K}^{AB}(\omega)$
for their Fourier transforms. 
With regard to the spectral decomposition of $\tilde{C}^{AB}(\omega)$ and $\tilde{K}^{AB}(\omega)$
we note that it is convenient to write $p_n=f(E_n)$. We can simplify 
these expressions by interchanging the dummy indexes $n$,$m$  in the second term. 
Thus we get
\beq
\tilde{C}^{AB}(\omega) \ &=& \
\frac{1}{2} \sum_{n,m} (f(E_n)+f(E_m)) \, A_{nm}B_{mn}
\, 2\pi\delta\left(\omega-\frac{E_m{-}E_n}{\hbar} \right)
\\
\tilde{K}^{AB}(\omega) \ &=& \ 
i\omega \ 
\sum_{n,m} \frac{f(E_n)-f(E_m)}{E_m-E_n} \, A_{nm}B_{mn}
\, 2\pi\delta\left(\omega-\frac{E_m{-}E_n}{\hbar} \right)
\eeq
Note that for a canonical state $f(E_n){-}f(E_m) = \tanh((E_n{-}E_m)/(2T)) \times (f(E_n){+}f(E_m))$.
Note also that in the expression for $\tilde{K}^{AB}(\omega)$, 
the $\omega$ cancels the $E_m{-}E_n$ denominator. The reason for pulling $\omega$ 
out of the sum is to emphasize the low frequency dependence.

\sheadC{Reciprocity relations and detailed balance}

There are some reciprocity relations that should be noticed.
First we note that by definition ${S^{AB}(t) = \big[ S^{BA}(-t) \big]^*}$.
In practice it is more illuminating to write the FTed version 
of this reciprocity relation, which can be directly deduced  
by inspection of the spectral decomposition. Namely,      
\beq
\tilde{S}^{AB}(\omega) \ \ = \ \ \big[\tilde{S}^{BA}(\omega)\big]^*
\eeq 
It follows that  $\tilde{C}^{AB}(\omega) = \big[\tilde{C}^{BA}(\omega)\big]^*$, 
while $\tilde{K}^{AB}(\omega) = -\big[\tilde{K}^{BA}(\omega)\big]^*$.
There is a second reciprocity relation that follows from time reversal invariace. 
Also here it is simpler to look on the spectral decomposition 
and to remember that $[A_{n,m}]^*=A_{n^*,m^*}$, where $n^*$ and $m^*$ 
are the eigenstates of the time reversed Hamiltonian. In practical terms  
it means that one has to reverse the sign of the magnetic field~$h$. 
Consequently  $\big[\tilde{S}^{BA}(\omega;h)\big]^*  = [\pm] \tilde{S}^{BA}(\omega;-h)$, 
where the plus (minus)  applies if the signs of $A$ and $B$ 
transform (not) in the same way under time reversal.
Combining with the trivial reciprocity relation we get the Onsager reciprocity relation
\beq
\tilde{S}^{AB}(\omega;h) \ \ = \ \ [\pm] \ \tilde{S}^{BA}(\omega;-h)
\eeq 
The Kubo formula that we discuss in the next section implies
that the same reciprocity relations hold  
for the response kernel $\alpha^{kj}$, 
to the susceptibility $\chi^{kj}$ 
and to the DC conductance $\bm{G}^{kj}$. 
These are called Onsager reciprocity relations

Finally, we can also generalize what we called the ``detailed balance relation".
In the quantum context this is a relation between ``positive" and ``negative" frequencies.  
Assuming that the system is prepared in a canonical state we have 
\beq
\tilde{S}_T^{AB}(-\omega) 
\ \ = \ \ 
\exp\left(-\frac{\hbar\omega}{T}\right)
\ \tilde{S}_T^{BA}(\omega) 
\eeq
From here it follows that 
\beq
\tilde{K}_T^{AB}(\omega)  \ \ = \ \  
i \, \frac{2}{\hbar}\tanh\left(\frac{\hbar\omega}{2T}\right)  
\  \tilde{C}_T^{AB}(\omega)
\hspace{2cm}
\text{[quantum canonical version]}
\eeq
In the classical limit this relation takes the form
\beq
K_T^{AB}(t)  \ \ = \ \  
- \frac{1}{T} \,  \dot{C}_T^{AB}(t)
\hspace{2cm}
\text{[classical canonical version]}
\eeq
where the dot indicates time derivative.

The Kubo formula that we discuss in the next section 
is expressed using $\tilde{K}^{AB}(t)$. 
But it is more convenient to use~$\tilde{C}^{AB}(t)$. 
The canonical relation between the two is the basis 
for the Fluctuation-Dissipation relation.

\newpage
\makeatletter{}
\sheadB{Linear response theory}

\sheadC{The notion of linear response}

Let us assume that $X(t)$ is an input signal, 
while $F(t)$ is the output signal of some 
black box. Linear response means that 
the two are related by 
\beq
F(t) \ \ = \ \ \int_{-\infty}^{\infty} \alpha(t-t') \ X(t') \ dt'
\eeq
The response kernel $\alpha(t-t_0)$ can be interpreted  
as the output signal that follows a $\delta(t-t_0)$ input signal. 
We assume a causal relationship, meaning 
that $\alpha(\tau)=0$ for $\tau<0$. 
The linear relation above can be written in terms 
of Fourier components as:
\beq
F_{\omega} \ \ = \ \ \chi(\omega) \ X_{\omega}
\eeq
where $\chi(\omega)$ is called the 
generalized susceptibility. 
Because of causality $\chi(\omega)$ is analytic 
in the upper complex plane.  
Consequently its real and imaginary parts 
are inter-related by the Hilbert transform:
\beq
\re [\chi(\omega)] \ \ = \ \ 
\int_{-\infty}^{\infty} \frac{\im[\chi(\omega')]}{\omega'-\omega} \ \frac{d\omega'}{\pi}
\eeq
(the reverse Hilbert transform goes with an opposite sign).
The imaginary part of $\chi(\omega)$
is the sine transforms of $\alpha(\tau)$,
and therefore it is proportional to $\omega$
for small frequencies.
Consequently it is useful to define 
\beq
\chi_0(\omega)  & \ \ \equiv \ \ &  \re [\chi(\omega)] 
\ \ = \ \  
\int_0^{\infty} \alpha(\tau) \cos(\omega\tau) d\tau
\ \ \sim \ \ \int_0^{\infty} \alpha(\tau) d\tau
\\
\eta(\omega) & \ \ \equiv \ \ &  \frac{\im[\chi(\omega)]}{\omega} 
\ \ = \ \ 
\int_0^{\infty} \alpha(\tau) \frac{\sin(\omega\tau)}{\omega}d\tau
\ \ \sim \ \  \int_0^{\infty} \alpha(\tau) \ \tau d\tau
\eeq
The asymptotic expressions apply for small frequencies:
in this "DC~driving" limit one can regard $\chi_0$ and ${\eta=\eta_0}$ 
as constants. Accordingly for small frequencies we write      
\beq
F_{\omega} 
\ \ = \ \ \left[\chi_0(\omega) + i \omega \eta(\omega)\right] \ X_{\omega}
\ \ \approx \ \ \chi_0 \ X_{\omega} \ - \eta_0 \ \dot{X}_{\omega}
\eeq
which implies in time domain ${ F(t) \ \ = \ \ \chi_0 X - \eta_0 \dot{X} }$.

\sheadC{Rate of energy absorption}

Back to Physics, what we called above~$X$ is the deviation of a control parameter 
from a reference value, namely ${X-X_{\text{eq}}}$, and what we called~$F$ 
is the deviation from the corresponding equilibrium value, 
namely, ${F(t) = \langle \mathcal{F} \rangle_t - \langle \mathcal{F} \rangle_{\text{eq}} }$.      
In the ``DC regime" of small frequencies we regard $\chi_0$ and $\eta_0$ 
as constants, and deduce that  
\beq
\langle \mathcal{F} \rangle_t 
\ \ = \ \ 
\langle \mathcal{F} \rangle_{\text{eq}} \ + \chi_0 \ (X-X_{\text{eq}})  \ - \eta_0 \dot{X} 
\ \ = \ \ 
\langle \mathcal{F} \rangle_X \ - \eta_0 \dot{X} 
\eeq
where $\mathcal{F} \rangle_X$ is the canonical $X$-dependent expectation value. 
Thus, the in-phase response gives the conservative effect,  
while the out-of-phase response gives the dissipative term.   
The latter is responsible to the irreversible work as discussed in the "Work" section.
The rate of dissipation is $\dot{\mathcal{W}}=\eta_0 \dot{X}^2$.

The above considerations regarding dissipation 
can be generalized to a source that has a wide 
power spectrum ("AC~driving").
The irreversible work equals the time integral 
over $F(t)\dot{X}$, and hence to the integral 
over $F_{\omega}\dot{X}_{\omega}$. Assuming linear 
response we get an integral over  $\eta(\omega)\dot{X}_{\omega}^2$.
Note that in the context of electrical engineering $X(t)$ might 
represent magnetic flux, hence $\dot{X}_{\omega}$ 
are the Fourier components of the voltage.
For a stationary driving source $\dot{X}_{\omega}^2$ 
is proportional to the measurement time and is 
characterized by a power spectrum $\tilde{S}_{\dot{X}}(\omega)$. 
Consequently the rate of energy absorption is 
\beq
\dot{\mathcal{W}} \ \ = \ \ \int_{-\infty}^{\infty}  \eta(\omega) \ \tilde{S}_{\dot{X}}(\omega) \ \frac{d\omega}{2\pi} 
\ \ \equiv \ \ \bar{\eta}_{AC} \ \text{RMS}[\dot{X}]^2
\eeq
Possibly it is more transparent to consider 
a pure AC source that has a definite frequency~$\Omega$.
In such a case we write
\beq
X(t) \ \ &=& \ \ \re\Big[ A\eexp{-i\Omega t} \Big]  \\
F(t) \ \ &=& \ \ \re\Big[ \chi(\Omega) \ A\eexp{-i\Omega t} \Big] 
\ \ = \ \ \chi_0(\Omega) X - \eta(\Omega) \dot{X}
\\
\dot{\mathcal{W}} \ \ &=& \ \ \langle -\dot{X} \mathcal{F} \rangle_t
\ \ = \ \ \eta(\Omega) \times (1/2)[A\Omega]^2, 
\ \ \ \ \ \text{[averaged over cycle]}
\eeq
Note again that only the out-of-phase response gives dissipation,
and that $A\Omega/\sqrt{2}$ is the RMS value of sinusoidal driving.

\sheadC{LRT with several variables}

Commonly the Hamiltonian  $\mathcal{H}(\bm{r},\bm{p}; \ X_1,X_2,X_3)$  
depends on several control parameters. Then we can define  
generalized forces in the usual way: 
\beq
\mathcal{F}^k \ = \ -\frac{\partial \mathcal{H}}{\partial X_k}
\eeq
Below $X_j$ represent a small deviation from some reference value~${X = X_{\text{eq}} \equiv 0}$.   
The postulated linear-response relation due to small $X(t)$ variation is written as 
\beq
\langle \mathcal{F}^k \rangle_t \ \ = \ \  
\sum_j \int_{-\infty}^{\infty} \alpha^{kj}(t-t') \ X_j(t')dt'
\eeq
The low frequency limit of the linear relation between 
the generalized forces and the {\em rate} of the driving 
can be regarded as a generalized Ohm law that includes 
an additional "geometric" term. 
Disregarding the conservative contribution
and changing notation for the dissipation coefficient 
the one parameter version ${\langle\mathcal{F}\rangle=-G\dot{X}}$ is  
generalized as follows:
\beq
\langle \mathcal{F}^k \rangle 
\ \ = \ \ 
-\sum_{j} \bm{G}^{kj} \ \dot{X}_j 
\ \ = \ \
-\sum_{j} \bm{\eta}^{kj} \ \dot{X}_j 
-\sum_{j} \bm{B}^{kj} \ \dot{X}_j 
\eeq
where $\bm{\eta}^{kj}$ and $\bm{B}^{kj}$ are 
the symmetric and anti-symmetric parts of $\bm{G}^{kj}$.
In an abstract notation this formula can be written as follows:
\beq
\langle \mathcal{F} \rangle 
\ \ = \ \
-\bm{\eta} \dot{\bm{X}} - \bm{B} \wedge \dot{\bm{X}}
\eeq
Note that second term is analogous to a magnetic Lorentz force.
Later we shall see that it can be derived form the  
theory of adiabatic processes, where it can be expressed 
as a "rotor" of the Berry connection $\bm{A}$.
The derivation of the first term requires 
the larger perspective of the Kubo formula
which we discuss in the next section.    
It is the first term that is responsible for dissipation.
Namely, the rate of dissipation is given by 
\beq
\dot{\mathcal{W}} \ \ = \ \ -\sum_k \langle \mathcal{F}^k \rangle \dot{X}_k 
\ \ = \ \ \sum_{k,j}  \bm{\eta}_{kj} \dot{X}_k \dot{X}_j
\eeq

\newpage
\sheadC{The Kubo formula}

The Kubo formula is an expression for 
the response kernel that relates the expectation 
value ${\langle A \rangle_t = \trc(A\rho(t)) }$ of some observable~$A$ 
to driving field~$f(t)$, where the driving term  
in the Hamiltonian ${-f(t)B}$ involves the operator~$B$.    
\beq
\alpha^{AB}(t) \ \ = \ \
\Theta(t) \ \left\langle \frac{i}{\hbar}\Big[A(t),B\Big] \right\rangle 
\ \ \equiv \ \ 
\Theta(t) \ K^{AB}(t)
\hspace{3cm} \text{[Kubo formula]}
\eeq
The formula has a good classical limit, 
and has various derivations. 
See ``Lecture notes in quantum mechanics".
One option is to deduce~${\langle A \rangle_t}$  
from the time evolution of the probability matrix~$\rho(t)$.
Another way is to regard the Kubo formula 
as the interaction picture version 
of the ``rate of change formula".
Namely, the rate of change of the expectation 
value of~$A$ is determined by the expectation value of 
the commutator $[\mathcal{H},A]$, 
hence in the interaction picture it is related to $[B,A]$. 

Yet there is a very simple way to derive the Kubo formula
in ``one line". Assume that the system is prepared in 
a stationary state of $\mathcal{H}_0$, and that we provide 
a pulse ${f(t)=\lambda\delta(t)}$ at ${t=0}$. 
Accordingly the evolution operator after time $t$ is  
is $U(t)=\exp(-i\mathcal{H}_0t) \exp(i\lambda B)$.
Below we use the notation ${ A(t)= \eexp{i\mathcal{H}_0t} A \eexp{-i\mathcal{H}_0t} }$. 
By definition $\alpha(t)$ is the first-order approximation
for the response to this pulse, accordingly we get
\beq
\langle A \rangle_t 
\ \ = \ \ \langle U(t)^{\dag}AU(t) \rangle_{t{=}0}
\ \ = \ \ \langle  \eexp{-i\lambda B} A(t) \eexp{+i\lambda B} \rangle_{t{=}0}
\ \ = \ \ \langle A \rangle_{t{=}0} \ + i \lambda \langle [A(t), B] \rangle_{t{=}0}
\eeq
From here the Kubo formula follows. Note that ${t{=}0}$ refers here to the 
moment that precedes the delta perturbation, at which the state of system is assumed 
to be {\em stationary}, possibly a canonical equilibrium. 
By default the subscript is omitted in the final result.

\sheadC{Memory and Sensitivity}

The so called out-of-time-order correlator (OTOC) of operators $A$ and $B$ is defined as follows:
\beq
K^{ABAB}(t) \ \ = \ \ \Big\langle [A(t),B]^{\dag} [A(t),B] \Big\rangle 
\eeq
In order to understand its significance let us write the this correlator and the Kubo correlator for a classical particle, with the substitutions $A=x$ and $B=p$. 
\beq
K^{AB}(t) \ \ &=& \ \ -\Big\langle [x(t),p]_{\text{PB}}  \Big\rangle   \\
K^{ABAB}(t) \ \ &=& \ \ \Big\langle \Big| [x(t),p]_{\text{PB}} \Big|^2  \Big\rangle 
\eeq
An infinitesimal perturbation at $t=0$ shifts the initial position of the particle a distance $\lambda=\delta x$, and consequently 
${x(t) = x(0) - \lambda [x(t),p]_{\text{PB}} }$.
By definition the Kubo response kernel reflects the memory for the perturbation of the initial conditions, namely  ${K^{AB}(t) = (1/\lambda)\braket{\delta x(t)} }$. 
In contrast to that, the the OTOC reflects the sensitivity for the perturbation of the initial conditions. For chaotic system we have ${ |\delta x(t)| \sim \delta x(0) e^{\gamma t} }$, where $\gamma$ is known as the Lyapunov exponent. 
Accordingly  ${K^{ABAB}(t) = (1/\lambda^2)\braket{|\delta x(t)|^2} }$ reflect the exponential sensitivity of the system to any small perturbation of the initial conditions. The quantum mechanical version of the OTOC in general suppresses this sensitivity.

\sheadC{The Onsager regression formula}

A related deduction of the Kubo formula
is based on the analysis of a ``quench process". We assume that a system 
is described by the Hamiltonian ${\mathcal{H}_{\lambda} = \mathcal{H}_0-\lambda B}$.
For example $B$ might be the volume of the gas, and then the conjugate 
field~$\lambda$ is the applied pressure. At ${t=0}$ the field $\lambda$ 
is instantly turned off, such that the dynamics for ${t>0}$ is described by 
the unperturbed Hamiltonian $\mathcal{H}_0$.  
In the lecture regrading generalized forces it has been 
shown that the compressibility equals $\text{Var}(B)/T$.
This means that the expectation value $\langle B \rangle_t$  
should decay from $\langle B \rangle_{\lambda}$ to $\langle B \rangle_0$, 
where  
\beq
\langle B \rangle_{\lambda} \ \ = \ \  \langle B \rangle_0 \ + \lambda \frac{1}{T}\text{Var}(B) 
\eeq
Here we repeat essentially the same calculation, but in the context 
of a time dependent scenario, considering an arbitrary observable~$A$. 
Not caring about commutation relations ("classical limit") we get
\beq
\braket{A}_t 
\ \ &=& \ \ \trc\left[ A \rho(t) \right] 
\ \ = \ \ \trc\left[   A \ \eexp{-i\mathcal{H}_0t} \rho(0) \eexp{i\mathcal{H}_0t} \right] 
\ \ = \ \ \trc\left[   A(t) \ \rho(t{=}0) \right]  
\\
\ \ &=& \ \ \frac{\trc\left[ A(t) \exp\left(-\beta \mathcal{H}_{\lambda}\right)\right] }{\trc\left[ \exp\left(-\beta \mathcal{H}_{\lambda}\right)\right]} 
\ \ = \ \  \langle A(t) \rangle_0 + \beta \lambda \Big[ \langle A(t)B \rangle_0 -  \langle A(t)\rangle_0\langle B\rangle_0 \Big]
\ + \ \text{higher orders} 
\eeq
Note that here the subscript "0" does not mean ${t{=}0}$, but ${\lambda{=}0}$.  
Initially we have ${ \braket{A}_{t{=}0} = \braket{A}_{\lambda} }$, 
while after a long time the unperturbed equilibrium value is restored, namely, ${ \braket{A}_{t{=}\infty} = \braket{A(t)}_0 = \braket{A}_0 }$.    
Neglecting the higher orders terms the linear-response result is  
\beq
\braket{A}_t   \ \ = \ \  \langle A \rangle_0  \ +  \lambda \frac{1}{T} C^{AB}(t) 
\ \ \ \ \ \ \ \ \ \ \mbox{[$f(t)$ is a step function]}
\eeq
We see that the re-equilibration mimics the decay of the pertinent 
correlation function. This can be regarded as a formal way 
to justify the ``Onsager regression hypothesis" (see later).  
From here we can derive the classical version of the Kubo formula. 
We simply have to notice that the response for a ``delta pulse" is simply 
the derivative of the response for a ``step function". Accordingly
\beq 
\braket{A}_t   \ \ = \ \ \langle A \rangle_0  \ - \lambda \frac{1}{T} \dot{C}^{AB}(t) 
\ \ \ \ \ \ \ \ \ \ \mbox{[$f(t)$ is a delta pulse]}
\eeq
Using the relation $K(\tau)=-(1/T)\dot{C}(\tau)$, 
that has been derived in the lecture regarding fluctuations,  
we deduce the Kubo formula.

\sheadC{The Onsager regression hypothesis}

The Onsager regression hypothesis states that ``the average regression of fluctuations  
should obey the same laws as the corresponding irreversible process". 
The regression scenario is defined as follows. We allow the system to 
equilibrate in the presence of an applied field. Then we turn off the 
field, and watch the time dependence of some observable.  
We already demonstrated that the Onsager regression hypothesis 
can be deduced in the classical limit 
via first order-perturbation theory treatment of the response. 
The quantum generalization of the Onsager regression formula is known as the 
fluctuation-dissipation relation, to be discussed in the next lecture.  

The "regression formula" describes the relaxation of the system back to equilibrium. 
Consider the case of having one fluctuating variable~$A$.  
If we assume that its relaxation obeys an exponential decay law ${ \dot{A} = -\gamma A}$,  
then we can deduce that the rate of relaxation is 
\beq
\gamma \ \ = \ \ -\left. \frac{ \dot{A} } { A } \right|_{t=0} \ \ = \ \ -\frac{\dot{C}(0)}{C(0)} 
\eeq
In the next section we describe the departure from equilibrium with a scaled variable ${X_A= A/C(0)}$, 
and write the relaxation as ${\dot{A} = -\gamma_{AA} X_A}$ with ${\gamma_{AA}=-\dot{C}(0)}$. 
Then, for several variables, it would be possible to get a generalized 
formula  ${\gamma_{ij}=-\dot{C}_{ij}(0)}$ with a reciprocal relation between $\gamma_{ij}$ and $\gamma_{ji}$.

\clearpage
\sheadC{Onsager reciprocity}

Within the Hamiltonian framework the Onsager reciprocity is the 
statement that response coefficients obey relations of the 
type $G^{AB}(h) = G^{BA}(-h)$, where $h$ is the magnetic field.
This follows from the observation that they are related to 
cross-correlation functions that obey reciprocity as discussed 
in previous lecture. Below we provide a more general perspective 
of Onsager reciprocity, which is {\em not} based on Hamiltonian formulation, 
and hence can be applied to a wider range of problems 
in thermodynamics. We no longer assume the canonical Boltzmann ensemble. 
Instead we assume that the probability of a "configuration" is given by 
\beq
p(\varphi_1,\varphi_2,...) \ \ \propto \ \ \eexp{ -\mathcal{A}(\varphi_1,\varphi_2,...) } 
\eeq
For example the $\varphi_j$ might represent a set of chemical reaction coordinates.  
Or it can stand for the amount of energy that is transferred from one body to another body.
For an isolated system that is described by a microcanonical ensemble 
the function $\mathcal{A}(\varphi)$ is the Boltzmann entropy of a given configuration. 
For a system in contact with a heat bath that is described by 
a canonical ensemble ${ \mathcal{A}(\varphi) = \beta F(\varphi)}$, 
where ${F}$ is the Helmholtz function. 

We use the convention that ${\varphi=0}$ is the most probable value if 
there are no constraints nor additional fields. 
We assume that the deviations from equilibrium are small, 
such that $\mathcal{A}(\varphi)$  can approximate by a quadratic expression:
\beq
\mathcal{A}(\varphi) \ \ = \ \ \frac{1}{2}\sum_{ij} A_{ij} \ \varphi_i\varphi_j 
\eeq
As for the temporal aspect we assume that the fluctuations are 
characterized by some correlation function:
\beq
C_{ij}(t) \ \ = \ \ \langle \varphi_i(t) \varphi_j(0) \rangle 
\eeq 
The ${A_{ij}}$ determines the correlations ${ C_{ij}(0) = \langle \varphi_i\varphi_j \rangle = (A^{-1})_{i,j} }$, 
but give no information on the temporal aspect.  
At this point it is convenient to define conjugate variables ${X_k=-\partial_k \mathcal{A} = -\sum_i A_{kj} \varphi_j }$, 
which are like restoring forces, and to realize that ${\langle X_k \varphi_j \rangle = -\delta_{k,j}}$.  
Next we assume that the relaxation of the system is described by a linear relation 
that reflects the tendency of the system to restore equilibrium 
\beq
\dot{\varphi}_i \ \ = \ \ \sum_k \gamma_{ik} \ X_k 
\ \ \ \ \ \ \ \ \ \ \ \ \mbox{[which implies $\dot{\varphi}=\gamma A \varphi$ where both $A$ and $\gamma$ are "nice" matrices]}
\eeq
From here it follows that ${ \langle \dot{\varphi}_i \varphi_j \rangle = -\gamma_{ij}}$.
Thus, as expected, we can derive response coefficients from correlation functions
\beq
\gamma_{ij} \ \ = \ \ -\dot{C}_{ij}(\tau = 0)
\eeq
Note that the absence of the $1/T$ prefactor is because we defined the 
conjugate variables not from the Hamiltonian but from the "Action". 
The Onsager reciprocity relation follows automatically 
from the symmetry of the correlation function.
In the absence of magnetic field, assuming time-reversible dynamics, 
we have ${C_{ij}(\tau)=C_{ij}(-\tau)}$, and hence ${\gamma_{ij}=\gamma_{ji}}$. 
A non-trivial example for the Onsager reciprocity relation is 
discussed with regard to the thermo-electric effect in the kinetic theory lecture.

\sheadC{The Kubo formula for AC/DC driving}
  
The DC value of the dissipation coefficient 
is obtained by integration:
\beq
\eta^{AB} \ \text{[DC limit]}  \ \ = \ \ \int_0^{\infty} K^{AB}(\tau) \ \tau d\tau 
\eeq
More generally, an expression for the generalized susceptibility 
follows from the convolution theorem:  
\beq
\chi^{AB}(\omega) 
\ \ \equiv \ \
\text{FT}\Big[ \alpha^{AB}(\tau) \Big]
\ \ = \ \  
\int_{-\infty}^{\infty} 
\frac{i\tilde{K}^{AB}(\omega')}
{\omega-\omega'+i0} \ \frac{d\omega'}{2\pi}
\eeq
Of particular interest is the case where~${A=B}$ 
is a generalized force that is conjugate 
to the variation of some parameter~$X$ such that ${f(t)=(X(t)-X_0)}$. 
This is  the case of interest in the study of friction 
(where $\dot{f}$ is the displacement velocity) 
and in the study of electrical conductance  
(where $\dot{f}$ is the electromotive field).
From the definition it follows that $K(-\tau)=-K(\tau)$, 
hence $K(\omega)$ is pure imaginary, 
and consequently the friction coefficient is  
\beq
\eta(\omega) 
\ \ \equiv \ \ 
\frac{\im[\chi(\omega)]}{\omega}
\ \ = \ \ 
\frac{1}{i2\omega}\tilde{K}(\omega)  
\hspace{3cm} \text{[Kubo formula for the dissipation coefficient]}
\eeq

\sheadC{The Kubo formula - FGR version}

So far we have used versions of the Kubo formula 
that are "good" both classically and quantum mechanically. 
In the quantum case one can write a version of this formula 
that involves the non-symmetrized correlation function: 
\beq
\eta(\omega) \ \ = \ \ \frac{1}{2\hbar\omega}\left[\tilde{S}(\omega)-\tilde{S}(-\omega)\right]
\hspace{3cm} \text{[Quantum FGR version of the Kubo formula]}
\eeq
This expression can be deduced directly 
from the FGR picture as follows. 
Assume that ${\mathcal{H}_{\tbox{driving}} = -f(t)B}$ 
with ${f(t)=f_0\sin(\Omega t)}$. 
From the FGR it follows that the rate 
of energy absorption due to upward 
transitions is $(f_0/2)^2\tilde{S}(\Omega)\Omega$.
Similarly the rate of energy emission is $(f_0/2)^2\tilde{S}(-\Omega)\Omega$.
The net rate of heating is the difference.
By definition it is written as $\dot{\mathcal{W}} = \eta(\Omega) \overline{[\dot{f}^2]}$, 
where $\overline{[\dot{f}^2]} = (1/2)[f_0\Omega]^2$.
Hence one deduce the above expression for $\eta$. 

Below we discuss the non-trivial generalization of 
the Kubo linear response formalism for the case 
of Hamiltonian that depends on several parameters. 
We start with the dissipation-less quantum adiabatic limit, 
and continue with the full linear response analysis.

\sheadC{Adiabatic response}

For an extended presentation see "Lecture notes in quantum mechanics".
Given an Hamiltonian $\mathcal{H}(\bm{r},\bm{p}; \ X_1,X_2,X_3)$ 
that depends on several control parameters,  
we find the {\em zero order} adiabatic basis $|n(X)\rangle$ with eigenenergies $E_n(X)$.
Then, for a given level, we define in parameter space 
the "Christoffel symbols" that are known in this context as "Berry connection",  
and the associated "curvature field" as follows:
\beq
\bm{A}^{j}_{nm} &=&
i\hbar \left\langle n(X) \left|
\frac{\partial}{\partial X_j} m(X)  \right.  \right\rangle
\\
\bm{B}^{ij}_{n} &=&
\partial_i \bm{A}^{j}_{n} - \partial_j \bm{A}^{i}_{n} 
\eeq
We use the notation $\bm{A}^j_n=\bm{A}^j_{nn}$, 
and note the following identities:  
\beq
\bm{A}^{j}_{nm} 
\ \ &=& \ \ 
\frac{-i\hbar\mathcal{F}^j_{mn}}{E_m{-}E_n}
\ \ \ \ \ \ \ \ \ [n \ne m]
\\
\bm{B}^{ij}_{n} 
\ \ &=& \ \ 
\sum_{m(\ne n)}
\frac{2\hbar \im\left[
\mathcal{F}^i_{nm}\mathcal{F}^j_{mn}\right]}
{(E_m-E_n)^2}
\eeq
If we have 3~control variables it is 
convenient to use notations suggesting  
that we can formally regard ${\bm{A}_n}$ 
as a vector potential whose rotor ${\bm{B}_n}$ 
is formally like a magnetic field: 
\beq
{\bm{X}} & \longmapsto & (X_1,X_2,X_3) \\
{\bm{A}} & \longmapsto & (A^1_{nn},A^2_{nn},A^3_{nn}) \\
{\bm{B}} & \longmapsto & (B^{23},B^{31},B^{12}) 
\eeq
With the above definitions the Schrodinger equation 
can be written as follows:
\beq
\frac{d}{dt}|\psi\rangle \ \ =
\  -\frac{i}{\hbar} \mathcal{H}({X(t)}) \ |\psi\rangle
\eeq
We expand the state in the {\em zero order} adiabatic basis
\beq
|\psi(t)\rangle \ \ = \ \ \sum_n a_n(t) \ |n(X(t))\rangle
\eeq
and get the equation 
\beq
\frac{da_n}{dt} = -\frac{i}{\hbar} (E_n{-}{\dot{X}}\cdot\bm{A}_n) a_n
-\frac{i}{\hbar} \sum_m
{\bm{W}_{nm}} a_m
\eeq
where
\beq
\bm{W}_{nm} \equiv -\sum_j{\dot{X}_j}\bm{A}^j_{nm}
\ \ \ \ \ \ \ \ \text{for} \ n{\ne}m, \ \text{else zero}
\eeq
It follows that the {\em first order} adiabatic state that is associated with the $n$th level is   
\beq
|\psi(t)\rangle \ \ = \ \ |n(X(t))\rangle  + \sum_{m\ne n} \frac{\bm{W}_{mn}}{E_n-E_m} |m(X(t))\rangle
\eeq
Consequently the first order adiabatic response 
of a system that has been prepared  in the $n$th adiabatic state is
\beq
\langle \mathcal{F}^k \rangle 
\ \ = \ \   
-\sum_j \bm{B}_n^{kj} \ \dot{X}_j
\ \ = \ \ -{\bm{B}} \wedge \dot{\bm{X}}
\eeq
We shall explain in the next section that this corresponds 
to the geometric part of the response in the Kubo formula. 
The Kubo formula contains an additional  
non-adiabatic (dissipative) term that 
reflects FGR transitions between levels.

\sheadC{Low frequency response}

Here we go beyond adiabatic response and discuss 
both the adiabatic and dissipative terms that are 
implied by the Kubo formula. Recall that 
the Kubo expression for the response kernel  
is ${\alpha^{kj}(\tau) = \Theta(\tau) \ K^{kj}(\tau)}$, 
whose Fourier transform is the generalized susceptibility:
\beq
\chi^{kj}(\omega) 
\ \ = \ \  
\int_{-\infty}^{\infty} 
\frac{i\tilde{K}^{kj}(\omega')}
{\omega-\omega'+i0} \ \frac{d\omega'}{2\pi}
\eeq
Taking into account that $\re[\chi^{kj}(\omega)]$ 
is symmetric with respect to $\omega$ we have 
\beq
\bm{G}^{kj} \ \ = \ \ 
\lim_{\omega\rightarrow 0}
\frac{\im[\chi^{kj}(\omega)]}{\omega} 
\ \ = \ \ 
\lim_{\omega\rightarrow 0}
\frac{d}{d\omega} \chi^{kj}(\omega)
\ \ = \ \
\int_0^{\infty} K^{kj}(\tau)\tau d\tau
\eeq
The last expression (in time domain) is mentioned for completeness.
In practice it is more convenient to proceed in frequency domain.    
After some straightforward algebra we get 
\beq
\bm{G}^{kj} \ \ = \ \
\frac{1}{2}\lim_{\omega\rightarrow 0}
\frac{\im[\tilde{K}^{kj}(\omega)]}{\omega} 
-
\int_{-\infty}^{\infty}\frac{d\omega}{2\pi}
\frac{\re[\tilde{K}^{kj}(\omega)]}{\omega^2}  
\ \ \equiv \ \ 
\bm{\eta}^{kj} + \bm{B}^{kj}
\eeq
We notice that $\bm{\eta}^{kj}$ is a symmetric matrix
while $\bm{B}^{kj}$ is anti-symmetric. 
Hence in abstract notation the linear-response relation 
can be written as a generalized Ohm law:
\beq
\langle \mathcal{F}^k \rangle 
\ \ = \ \ 
-\sum_{j} \bm{G}^{kj} \ \dot{X}_j 
\ \ = \ \
-\bm{\eta} \dot{\bm{X}} - \bm{B} \wedge \dot{\bm{X}}
\eeq
This is a generalization of the adiabatic response  
formula. The additional term takes into account 
the FGR non-adiabatic transitions between levels.
To see clearly the connection we substitute the 
spectral decomposition of $\tilde{K}^{kj}(\omega)$ 
and get the following expressions:
\beq
\chi^{kj}(\omega) 
\ \ = \ \  
\int_{-\infty}^{\infty} 
\frac{i\tilde{K}^{kj}(\omega')}
{\omega-\omega'+i0} \ \frac{d\omega'}{2\pi}
\ \ = \ \ 
\sum_n f(E_n) \sum_{m} 
\left(\frac{-\mathcal{F}^k_{nm}\mathcal{F}^j_{mn}}{\hbar\omega{-}(E_m{-}E_n){+}i0}
+\frac{\mathcal{F}^j_{nm}\mathcal{F}^k_{mn}}{\hbar\omega{+}(E_m{-}E_n){+}i0}\right)
\eeq
and 
\beq
\bm{\eta}^{kj} \ &=& \
\frac{1}{2}\lim_{\omega\rightarrow 0}
\frac{\im[\tilde{K}^{kj}(\omega)]}{\omega} 
\ \ = \ \ 
-\pi\hbar\sum_{n,m}
\frac{f(E_n)-f(E_m)}{E_n-E_m}
\mathcal{F}^k_{nm}\mathcal{F}^j_{mn}
\delta(E_m-E_n)
\\
\bm{B}^{kj} \ &=& \
{-}\int_{-\infty}^{\infty}\frac{d\omega}{2\pi}
\frac{\re[\tilde{K}^{kj}(\omega)]}{\omega^2}  
\ \ = \ \ 
\sum_{n,m}
(f(E_n)-f(E_m))
\frac{-i\hbar\mathcal{F}^k_{nm}\mathcal{F}^j_{mn}}
{(E_m-E_n)^2}
\eeq
The $\re[]$ and $\im[]$ are not required because 
the doubles summation cares for that.
The expression for $\bm{B}^{kj}$ can be written 
in a way that better emphasize the relation 
to the analysis of the adiabatic response:
\beq
\bm{B}^{kj} \ \ = \ \ 
\sum_n f(E_n)
\sum_{m(\ne n)}
\frac{2\hbar \im\left[\mathcal{F}^k_{nm}\mathcal{F}^j_{mn}\right]}
{(E_m{-}E_n)^2}
\ \ = \ \ \sum_n f(E_n) \bm{B}_n^{kj} 
\eeq

\newpage
\makeatletter{}
\sheadB{The fluctuation dissipation relation}

\sheadC{General formulation}

The essence of the fluctuation dissipation relation (FDR) 
is to relate the response of a system to its fluctuations in equilibrium. 
In order to derive this relation we have to supply 
information on the preparation of the system, 
which is typically assumed to be canonical. 
In the classical context there is a useful microcanonical 
version from which the canonical version can be derived. 
The formal derivation of the FDR is based on the generalized detailed balance relation  
that allows to express $\tilde{K}^{kj}(\omega)$ using $\tilde{C}^{kj}(\omega)$.

We first consider what we call ``the AC version" of the FDR.
For simplicity we consider the one-variable version: 
the driving term in the Hamiltonian is $-X(t)\mathcal{F}$, 
and our interest is in the observable~$\mathcal{F}$
that is conjugated to the driving field~$X$.      
Recall that $\tilde{K}(\omega)$ is imaginary, 
and ${\im[\chi(\omega)]=[1/(2i)]\tilde{K}(\omega)}$.
Assuming canonical preparation 
the detailed balance relation implies that     
\beq
\im[\chi(\omega)]
\ \ = \ \ 
\frac{1}{\hbar} 
\tanh\left( \frac{\hbar\omega}{2T} \right)
\ \tilde{C}^{\scriptscriptstyle\mathcal{F}\mathcal{F}}(\omega)
\hspace{2cm} \text{[FDR, the AC version]} 
\eeq

What we call ``the DC version" of the FDR is obtained 
by taking the small~$\omega$ limit of the AC version, 
which is formally equivalent to the classical limit (small~$\hbar$).   
One deduces that the low frequency dissipation 
coefficient is relate to the equilibrium intensity 
of the fluctuations:  
\beq
\eta \ = \ \frac{\nu_T}{2T},  
\ \ \ \ \ \ \ \ \ \ \ \ \ \ \ 
\nu_T  \ \equiv \ \int_{-\infty}^{\infty} \langle \mathcal{F}(\tau) \mathcal{F}(0) \rangle_T \ d\tau 
\hspace{2cm} \text{[FDR, the DC version]} 
\eeq
Note that in the above writing we assume that the equilibrium value of the 
fluctuating force is ${\langle \mathcal{F} \rangle = 0}$, 
else $\mathcal{F}$ should be re-defined so as to have a zero average. 

For completeness we also point out the multi-variable version of the FDR 
in the DC limit. Here we change notion and use $G^{AB}$ for the 
generalized conductance. In the DC case we know from Kubo 
that~$G$ is an integral over $\tau K(\tau)$.
In the classical treatment $K(\tau)$ is the derivative of $C(\tau)$,  
hence after integration by parts 
\beq
G^{AB} \ \ = \ \ \frac{1}{T}\int_0^{\infty} C^{AB}(\tau) \ d\tau 
\hspace{2cm} \text{[generalized FDR, classical DC version]} 
\eeq
Note that the $(1/2)$ prefactor is absent, and that the integration is over positive~$\tau$, 
and that the cross-correlation function~$C^{AB}(\tau)$ does not have to be symmetric in time. The asymmetry is responsible for the {\em geometric} part of the conductance matrix.

\sheadC{The diffusion-dissipation picture}

We can illuminate the physics of FD for DC driving
using a simple diffusion-dissipation picture. 
We show below that the DC energy absorption rate 
is related to the induced diffusion in energy space. 
To simplify the presentation we use a classical language.
We can deduce that the driving induce 
diffusion in energy space from the relation
\beq
E(t)-E(0) \ = \ - \dot{X} \int_0^t \mathcal{F}(t') dt'
\eeq
leading to 
\beq
\langle (E(t)-E(0))^2 \rangle \ = \ 
\dot{X}^2 \int_0^t\int_0^t \langle \mathcal{F}(t') \mathcal{F}(t'') \rangle dt'dt''
\eeq
where the averaging assumes a microcanonical preparation.  
Thus we get
\beq
\delta E^2(t) \ = \ 2D_E t
\eeq
where the leading order estimate for the diffusion is 
\beq
D_E \ \ = \ \
\frac{1}{2} \dot{X}^2 
\int_{-\infty}^{\infty} \langle \mathcal{F}(\tau) \mathcal{F}(0) \rangle_E \ d\tau  
\ \ = \ \  \frac{1}{2} \nu_E \dot{X}^2
\eeq

On long times we assume that the probability distribution 
in the democratic variable $n=\mathcal{N}(E)$ satisfy 
a standard diffusion equation. Transforming to the 
non-democratic variable~$E$ we get     
\beq
\frac{\partial \rho}{\partial t} \ = \
\frac{\partial}{\partial E}
\left(\gdos(E)D_E \frac{\partial}{\partial E}
\left(\frac{1}{\gdos(E)}\rho\right)\right)
\eeq
where $\gdos(E)$ reflects the ratio between the 
proper phase-space measure~$dn$ and the distorted measure~$dE$.   
For more details see [\href{http://arxiv.org/abs/cond-mat/9902168}{arXiv}]. 
The energy of the system is $\langle \mathcal{H} \rangle=\int E \rho(E)dE$.
Taking the time derivative and integrating by parts,  
it follows that the rate of energy absorption is
\beq
\dot{{\cal W}} \ = \ 
\frac{d}{dt}\langle \mathcal{H} \rangle
= - \int_0^{\infty} dE \ \gdos(E) \ D_E
\ \frac{\partial}{\partial E}
\left(\frac{\rho(E)}{\gdos(E)}\right)
\eeq
For a microcanonical preparation $\rho(E)=\delta(E-\EPS)$.
Substitution and integration by parts leads to  
\beq
\dot{{\cal W}} \ = \ 
\frac{d}{dt}\langle \mathcal{H} \rangle
\ = \
\frac{1}{\gdos(E)}
\frac{d}{dE}\left[\gdos(E) \ D_{E} \right] \Big|_{E=\EPS}
\eeq
By definition ${\dot{{\cal W}}=\eta\dot{X}^2}$
and ${D_E=(1/2)\nu \dot{X}^2}$. 
Consequently the diffusion-dissipation relation reduces
immediately to the microcanonical version of 
the fluctuation-dissipation relation:
\beq
\eta \ = \ \frac{1}{2} \frac{1}{\gdos(E)} \frac{d}{dE}\left[\gdos(E) \nu_E \right]
\eeq
The canonical version $\eta=\nu_T/(2T)$ can be derived 
from the integral expression for $\dot{{\cal W}}$, 
upon the substitution ${\rho(E)=(1/Z)\gdos(E)\eexp{-\beta E}}$.
Optionally it can be obtained from 
the microcanonical version by canonically averaging 
over~$E$ and performing integration by parts.

\sheadC{The wall formula}

The first prototype application of the FD relation 
is to the calculation of the friction in the Brownian 
motion problem. 
Consider a gas of particles in a box.
The system is driven by moving in it a "spoon",
or a "piston" or a "Brownian body". 

\bitem The parameter in $\mathcal{H}(X)$ represents the position of the spoon. \\ 
\bitem The generalized force is the Newtonian force $\langle \mathcal{F} \rangle$ on the spoon. \\
\bitem The DC linear response relation $\langle \mathcal{F} \rangle=-\eta\dot{X}$ describes friction. \\
\bitem The dissipation rate is $\dot{\mathcal{W}} = \eta \dot{X}^2$.

Our purpose below is to find an expression for the friction
coefficient $\eta$ using the FD relation. For this purpose 
we have to calculate the intensity $\nu_T$ of the 
fluctuations of $\mathcal{F}$ at equilibrium,  
and to use the relation ${\eta=\nu_T/(2T)}$. 

Due to random collisions of the gas particles, 
the Brownian body experiences a "random force" that 
can be written as the sum of short impulses: 
\beq
F(t) \ \ = \ \  \sum_j 2\mass v_{j} \ \delta(t-t_j) 
\eeq
Here $t_j$ is the time of the $j$th collision 
with velocity $v_j$ at the $x$ direction.
Note that $|v_j| \sim v_T$, where ${v_T=(T/\mass)^{1/2}}$
is the thermal velocity. 
The rate of collision for $N$ particles is 
\beq
\frac{1}{\tau_0} \ \ = \ \  N \times \left( \frac{\mathsf{A}}{L^2} \right) \times \frac{v_T}{L}
\eeq
where $L^3$ is the volume of the box that holds the gas particles, 
and $\mathsf{A}$ is the effective area of the moving wall.
Accordingly the intensity of fluctuations is 
\beq
\nu_T \ \ = \ \ \tilde{C}(\omega=0) \ \ = \ \  \frac{1}{\tau_0} (\mass v_T)^2 
\ \ = \ \ \mass^2 v_T^3 \ \frac{N}{L^3} \  \mathsf{A} 
\eeq
and for the friction we get
\beq
\eta 
\ \ = \ \  \frac{1}{2T} \tilde{C}(\omega=0)
\ \ = \ \ \rho v_T \times A
\eeq
where $\rho=(N/L^3)\mass$ is the mass density of the gas particles.

We note that if the dynamics of the Brownian body is described by 
a Langevin equation, then ${\nu/\eta=2T}$ implies that a canonical 
equilibrium is reached.  For more details see the Langevin section.
This was in fact the historical deduction of the FD relation 
by Einstein in the context of Brownian motion study.

If the Brownian particle is moving in an incompressible fluid 
the above result does not apply. Instead the friction is given 
by {\em Stokes Law} (see ``Additional topics / The Kinetic picture / Viscosity"), 
and we can use the FD relation ``in reverse" 
in order to deduce the intensity of fluctuations.

\sheadC{The Drude formula}

The second prototype application of the FD relation 
is to the calculation of electrical conductance.
Here we show how to derive the Drude formula for 
a gas of classical particles in an EMF driven ring.
For an extended discussion of electrical conductance 
see the "additional topics" section of the lecture notes.
  
We consider a ring driven by an electro-motive force (EMF).
The interaction term originates from the kinetic term $[p-(e/L)\Phi(t)]^2/(2\mass)$, 
where $\Phi$ is the flux and ${I =(e/L)v}$ is the conjugate current. 
Optionally we can say that the interaction is $-A(t)v$, 
where $A(t)$ is the vector potential and the velocity $v$ 
is the conjugate variable. Summarizing:   

\bitem The parameter in $\mathcal{H}(\Phi)$ represents the magnetic flux. \\ 
\bitem The generalized force is the current $\langle \mathcal{I} \rangle$ in the ring. \\
\bitem The rate in which the flux is varied determines the $\text{EMF}=-\dot{\Phi}$ by Faraday law. \\  
\bitem The DC linear response relation $\langle  \mathcal{I}\rangle = G \times \text{EMF}$ is Ohm law. \\
\bitem The dissipation rate $\dot{\mathcal{W}} = G \dot{\Phi}^2$ describes Joule heating. 

Our purpose below is to find an expression for the conductance~$G$ 
using the FD relation. For this purpose, following Drude, 
we postulate what is the velocity-velocity correlation function;  
calculate the intensity $\nu_T$ of the fluctuations of $\mathcal{I}$ 
at equilibrium, and use the relation ${G=\nu_T/(2T)}$.

Following Drude we assume an exponential velocity-velocity 
correlation function with a time constant $\tau_0$, 
and RMS velocity $v_0$, such that the mean free path is ${\ell=v_0\tau_0}$.
Hence we deduce that $\tilde{C}_{vv}(\omega)$ is a Lorentzian.
The displacement of the particle ${(x(t)-x(0))}$ is the integral 
over the velocity $v(t')$, hence the 
variance is ${\langle (x(t)-x(0))^2 \rangle = 2Dt}$ where
\beq
D \ \ = \ \ \frac{1}{2}\tilde{C}_{vv}(0) \ \ =  \ \ \frac{1}{3}v_0^2 \tau_0 
\hspace{2cm} \mbox{(for a 3D sample)}
\eeq 
For a single particle the current operator is $I=(e/L)v$,  
hence the intensity of the fluctuations of the current 
is $\nu=[(e/L)^2]\, 2D$. 
For $N$ classical particles at thermal 
equilibrium we get $\nu_T = N \, [(e/L)^2] \, 2D_T$, 
where $D_T$ is calculated with the thermal 
velocity that is defined via ${(1/2)\mass v_T^2 = (3/2)T}$. 
For $N$ Fermions at low temperatures we get $\nu_T = N_T \, (e/L)^2 \, 2D_F$,
where $N_T=T/\Delta$ is the effective number of participating  
electron at the Fermi energy. This result has been derived in the 
lecture about fluctuations. Here $\Delta$ is the mean level 
spacing at the Fermi energy. 
Note that ${N=(2/3)(\epsilon_F/\Delta)}$. 
The diffusion coefficient $D_F$ is calculated with 
the Fermi velocity which is determined via ${(1/2)\mass v_F^2 = \epsilon_F}$. 
Either way we get 
\beq
\nu_T \ \ = \ \ N_{\text{eff}} \left(\frac{e}{L}\right)^2  2D_{\text{eff}}
\ \ = \ \  2\left[\frac{N}{L^2} \frac{e^2}{\mass}\tau_0\right] T
\eeq   
and for the conductance we get
\beq
G^{[N]} \ \ = \ \ \frac{1}{2T} \nu_T 
\ \ = \ \ \frac{N}{L^2} \frac{e^2}{\mass}\tau_0 
\ \ \equiv \ \ \frac{\mathsf{A}}{L} \sigma
\eeq
where $\mathsf{A}$ is the cross section of the ring.
As a byproduct of this derivation we see clearly 
why the conductivity $\sigma$ is related to the diffusion
coefficient $D$.

Optionally the Drude expression can be written in 
a way that allows to make an association 
with the Landauer formula of mesoscopic physics.
Considering zero temperature Fermi occupation:   
\beq
G^{[N]} \ \ = \ \  e^2 \left(\frac{N}{\mass v_F L}\right) \frac{\ell}{L}
\ \ \equiv \ \ \frac{e^2}{2\pi} \mathcal{M} \frac{\ell}{L}
\eeq
where $\mathcal{M}$ corresponds to 
the effective number of open modes.
There is a very simple toy model for which the "exponential" 
velocity-velocity correlation can be deduced, 
and hence $\ell/L$ can be evaluated analytically. 
Consider a ring with a single stochastic scatterer
that has a transmission~$g$. The current-current correlation 
function ${C(t)=\langle I(t)I \rangle}$ at given energy~$E$ 
can be calculated as detailed in [\href{http://arxiv.org/abs/cond-mat/0603484}{arXiv}].  
The procedure is to use the identity 
${\langle BA \rangle = \sum_a p_a \langle B \rangle_a a}$, 
where where $A$ and $B$ are any two operators, 
and $\langle B \rangle_a$ is the expectation value 
of $B$ given that $A=a$.  
Applying this rule in our case 
we get ${\langle I(t) I \rangle = \sum_r p_r \langle I(t) \rangle_r I_r}$, 
where~${r=(x,v)}$ labels all the possible states of the particle in the ring,    
and $I=ev\delta(x)$ is the current through the measurement point ${x=0}$. 
The current $\langle I(t) \rangle_{x,v}$,   
given that the particle has been launched at~$x$ with velocity~$v$,   
can be written as a sum of pulses $\sum_j q_j \delta(t-t_j)$. 
If the measurement point ${x=0}$ is situated right across 
the barrier, such that the barrier is at ${x=\pm L/2}$, one obtains 
\beq
C(t) \ \ = \ \ \frac{ev_E}{L} \langle I(t) \rangle_{0,v_E}
\ \ = \ \ e^2\frac{v_E}{L} 
\sum_{n=-\infty}^{\infty} (2g-1)^{|n|} \ \delta\left(t-\left(\frac{L}{v_E}n\right)\right) 
\eeq
which exhibits exponential decay of correlations as in the Drude model.
Assuming low temperature Fermi occupation, 
with $N_T=T/\Delta$ thermal particles, 
that occupy levels whose spacing is ${\Delta = \pi v_F/L}$,   
we use the FD relation ${G = \nu_T /(2T)}$ and get      
\beq
G^{[N]} \ \ = \ \ \frac{1}{\Delta}\int_0^{\infty} C(t) \ \ = \ \ \frac{e^2}{2\pi} \left(\frac{g}{1-g}\right)
\eeq
For small $g$ one can neglect the $1{-}g$ denominator, and this formula becomes 
identical with the Landauer formula.
For larger $g$ the two formulas differ. The reason for this difference 
concerns the geometry: Here we consider the conductance of a barrier 
that is integrated into a closed ring, while Landauer concerns 
the conductance of a barrier that is connected to open reservoirs. 
In the latter case the particle cannot circulate multiple times via the barrier.

\sheadC{Conductor in electric field} 

A straightforward generalization of the driven ring problem  
applies for an extended  piece of metal 
that in placed in a time dependent electric field.
The electric field is described by a vector potential 
such that ${\mathcal{E}=-\dot{A}}$.
The interaction term in the Hamiltonian 
is an extended version of the simplified $-A(t)v$ 
that we have assumed in previous discussion: 
\beq
\mathcal{H}_{\tbox{int}} \ \ = \ \ -\int J(x) \cdot A(x) \ d^3x
\eeq
In linear response theory the current is proportional to the rate
in which the parameters are being changed in time.
Regarding the values of $A$ at different points in space
as independent parameters the postulated linear response relation takes the form
\beq
\langle J(x)  \rangle = \int \bm{\sigma}(x,x') \ {\cal E}(x') \ d^3x
\eeq
where $\bm{\sigma}_{ij}(x,x')$ is called the conductivity matrix.
The FD relation states that the conductivity 
is proportional to the temporal FT of ${\langle J_i(x,t)J_j(x',t')\rangle}$ 
with respect to the time difference ${\tau=t-t'}$. 
The proportionality constant is $1/(2T)$ in the DC limit.

\sheadC{Forced oscillator}

Consider a particle that is (say) bounded to a spring.
Let us assume that the motion of ${x(t)\equiv\langle \hat{x} \rangle_t}$ 
obeys the equation ${\mass \ddot{x}+\eta\dot{x}+\mass\Omega^2x=\mathcal{E}}$, 
where the external driving is due to an interaction term ${-\mathcal{E}(t)\hat{x}}$. 
Accordingly, 

\bitem The parameter in $\mathcal{H}(\mathcal{E})$ represents an electric field. \\ 
\bitem The generalized force is the polarization $\langle x \rangle$ of the particle. \\
\bitem The AC linear response relation is $\langle x \rangle = \chi(\omega) \ \mathcal{E}$.  

The FD relation implies that $x$ has fluctuations at equilibrium, 
that are related to the susceptibility:
\beq
\tilde{C}_{xx}(\omega) \ \ = \ \ 
\hbar \coth\left( \frac{\hbar\omega}{2T} \right) \ \im\Big[\chi(\omega)\Big], 
\hspace{3cm}
\chi(\omega) \ = \ \frac{1}{-\mass\omega^2-i\eta\omega+\mass\Omega^2} 
\eeq
Note that the fluctuations of the velocity 
are $\tilde{C}_{vv}(\omega)=\omega^2\tilde{C}_{xx}(\omega)$.
Integrating over $\omega$ we get $C_{xx}(0)$ and $C_{vv}(0)$, 
from which can deduce the average energy of the oscillator.
The results are consistent with the canonical expectation in the limit of zero damping.

\sheadC{Forced particle}

The limit $\Omega\rightarrow0$ of the forced harmonic oscillator
corresponds formally to a Brownian particle. 
In the classical limit we get for the power spectrum of the velocity:
\beq
\tilde{C}_{vv}(\omega) \ \ = \ \ (T/\mass) \times \frac{2(\eta/\mass)}{\omega^2+(\eta/\mass)^2} 
\eeq
The area of this Lorentzian is ${C_{vv}(0)=T/\mass}$, as expected from the canonical formalism.  
The corresponding velocity-velocity correlation is Drude type (exponential),
with damping constant $\gamma=\eta/\mass$. The integral over the velocity-velocity correlation 
function determines the diffusion coefficient, namely 
\beq
D \ \ = \ \ \frac{1}{2}\tilde{C}_{vv}(0) \ \ = \ \ \frac{T}{\eta} \ \ = \ \ \mu \, T
\eeq  
This is known as the Einstein relation between $D$ and the mobility $\mu=1/\eta$.
There is an optional shortcut in the application of the FD relation, 
that leads directly to the above Einstein relation. 
Let us write the electric field as ${\mathcal{E}=-\dot{A}}$. 
The interaction term is $-A(t)v$. Accordingly, 

\bitem The parameter in $\mathcal{H}(A)$ represents the vector potential. \\ 
\bitem The generalized force is the velocity $\langle v \rangle$ of the particle. \\
\bitem The DC linear response relation $\langle v \rangle = \mu \ \mathcal{E}$ describes drift motion.  \\
\bitem The dissipation rate $\dot{\mathcal{W}} = \mu \mathcal{E}^2$ describes Joule heating (per particle). 

The FD relation in this notations implies that $v$ has fluctuations at equilibrium, 
that are related to the mobility $\mu$. The "intensity" of the velocity 
fluctuations is $2D$. Hence the classical FD relation implies that the ratio 
of the diffusion~($D$) to the mobility~($\mu$) equals the temperature~($T$).

\sheadC{Duality between friction and mobility}

In the ``Forced particle" problem we have considered interaction of the type $-\hat{x} F(t)$, 
and defined the mobility $\mu$, which is the response of $\braket{\hat{x}}_t$ 
to the control field~$F(t)$. In the ``wall formula" problem we have considered interaction of the type $-x(t) \hat{F}$, and defined the friction coefficient $\eta$, which is the response of  $\braket{F}_t$ to the control parameter~$x(t)$. The two points of view on the system are dual, and with the standard definitions we have the identification ${\mu=1/\eta}$.

\sheadC{The fluctuations of an Ohmic system}

{\bf Nyquist Noise.-- } 
The FD relation in the electrical context is known as Nyquist theorem.
It can be used "in reverse" in order to deduce 
the Nyquist noise ${\nu = 2 G T}$, provided $G$ is known from experiment. 
It should be clear that in non-equilibrium conditions we might have 
extra fluctuations, which in this example are known as {\em shot noise}. 

{\bf Ohmic response.-- } 
Sometimes it is convenient to characterize 
the system by its response, and from this to 
deduce the power spectrum of the fluctuations. 
So we regard  $\tilde{K}(\omega)$ as the input. 
Inspired by jargon of electrical engineering,  
so-called Ohmic response is characterized  
by a dissipation coefficient $\eta$ 
that is  independent of $\omega$ 
up to some implicit high frequency cutoff~$\omega_c$.
It follows that the DC intensity of the fluctuations 
is $\nu = 2 \eta T$, and the associated spectral 
functions are:     
\beq
\tilde{K}_{\tbox{ohmic}}(\omega) 
\ &=& \ i2\eta\omega
\\
\tilde{C}_{\tbox{ohmic}}(\omega)
\ &=& \ 
\frac{\hbar}{2} 
\coth\left(\frac{\hbar\omega}{2T}\right)
\im\left[ \tilde{K}(\omega) \right]
\ \ = \ \  \eta \hbar\omega 
\coth\left(\frac{\hbar\omega}{2T}\right)
\\
\tilde{S}_{\tbox{ohmic}}(\omega)
\ &=& \  \tilde{C}_{\tbox{ohmic}}(\omega) - i\frac{\hbar}{2}\tilde{K}_{\tbox{ohmic}}(\omega)  
\ \ = \ \ 2\eta \frac{\hbar\omega }{1-\eexp{-\hbar\omega/T}}  
\eeq

\sheadC{The fluctuations of the potential in metals}

The dielectric constant of a metal is defined via the linear relation
between the total electrostatic potential $U_{\tbox{total}}$ 
and an external test charge density $\rho_{\tbox{ext}}$  
\beq
U_{\tbox{total}} \ \ = \ \ \frac{1}{\varepsilon(q,\omega)} \left(\frac{4\pi e^2}{q^2}\right) \rho_{\tbox{ext}}
\eeq
For simplicity we relate here and below to one component $q$ of the fields. 
The total electrostatic potential is the sum 
of the external potential $U_{\tbox{ext}} = (4\pi e^2/q^2) \rho_{\tbox{ext}}$, 
and the induced potential $U_{\tbox{elct}} = (4\pi e^2/q^2) \rho_{\tbox{elct}}$,  
where $\rho_{\tbox{elct}}$ is the total density of the electrons.  
The dielectric constant can be deduced from the equations of 
motion ${\partial \rho_{\tbox{elct}}/ \partial t = -\nabla J}$  
with ${J=-(\sigma/e^2) \nabla U_{\tbox{total}} - D \nabla \rho_{\tbox{elct}}}$ 
that leads to the relation
\beq
\rho_{\tbox{elct}} \ \ = \ \ \frac{(\sigma/e^2) q^2}{i\omega-Dq^2} \ U_{\tbox{total}}
\eeq
and hence to ${U_{\tbox{total}}=(1/\varepsilon)U_{\tbox{ext}}}$, where 
\beq
\varepsilon(q,\omega) = 1 - \frac{4\pi\sigma}{i\omega-Dq^2}.
\eeq
Note that 
\beq
\im\left[ \frac{-1}{\varepsilon(q,\omega)} \right] 
= 
\frac{4\pi\sigma\omega}{(Dq^2+4\pi\sigma)^2 + \omega^2}
\approx
\frac{\omega}{4\pi\sigma}
\eeq

The interaction between the electrons and an external electrostatic field 
is described by ${\mathcal{H}_{\tbox{ext}} = U_{\tbox{ext}} \rho_{\tbox{elct}}}$ 
which can be also written as ${\mathcal{H}_{\tbox{ext}} = \rho_{\tbox{ext}} U_{\tbox{elct}}}$.
The fluctuation dissipation relation expresses $\tilde{S}^{[N]}(q,\omega)$ 
using the response function  $\alpha(\bm{q},\omega)$  
that relates $U_{\tbox{elct}}$ to $-\rho_{\tbox{ext}}$ which is   
\beq
\alpha(\bm{q},\omega) =
\frac{4\pi e^2}{\bm{q}^2} 
\left[ 1-\frac{1}{\varepsilon(\bm{q},\omega)} \right]
\eeq
Using the fluctuation dissipation relation
\beq
\tilde{S}^{[N]}(\bm{q},\omega) =
\im\Big[\alpha(\bm{q},\omega)\Big]
\, \left(\frac{2}{1-\eexp{-\omega/T}}\right)
\eeq
we deduce 
\beq
\tilde{S}^{[N]}(\bm{q},\omega) \approx
\frac{e^2}{\sigma}
\frac{1}{\bm{q}^2}
\left(\frac{2\omega}{1-\eexp{-\omega/T}}\right)
\eeq
The Ohmic behavior is cut-off 
by $|\omega| \lesssim 1/\tau_c$
and $|\bm{q}|\lesssim {1}/{\ell}$ 
where $\ell=v_{\tbox{F}}\tau_c$ is 
the elastic mean free path, 
and $v_{\tbox{F}}$ is the Fermi velocity.
Recalling the Einstein relation ${\sigma=e^2\nu D}$,
where $\nu=\Delta^{-1}/L^d$ is the density of states per unit volume, 
we can write this result more conveniently 
as follows: 
\beq
\tilde{S}^{[N]}(\bm{q},\omega) \approx
\frac{1}{\nu D\bm{q}^2}
\left(\frac{2\omega}{1-\eexp{-\omega/T}}\right)
\eeq
Note that the electron charge $e$ cancels out from this final result
for the Nyquist noise spectrum. This well-known fact is due to the
effects of screening: A larger value of the charge would be canceled
by a correspondingly stronger suppression of density fluctuations.

\sheadA{System interacting with a bath}

\makeatletter{}
\sheadB{The modeling of the environment}

\sheadC{The Born-Oppenheimer Hamiltonian}

We first discuss system that is coupled to some other degrees 
of freedom that can be eliminated using an adiabatic scheme.
This leads to the Born-Oppenheimer picture. It is strongly related 
to Linear response theory, and the presentation below is 
arranged accordingly. 
Linear response theory is the leading formalism to
deal with driven systems. Such systems are described
by a Hamiltonian
\beq
\mathcal{H} = \mathcal{H}(\bm{Q},\bm{P}; X(t))
\eeq
where $(Q,P)$ is a set of canonical coordinates
(in case that the Hamiltonian is the outcome of "quantization"),
and $X(t)$ is a set of time dependent
classical parameters ("fields").
For example, $X$ can be the position of a piston.
In such case $\dot{X}$ is its velocity.
More interesting is the case where $X$ is the
magnetic flux through a ring. In such a case
$\dot{X}$ is the electro motive force.
The Kubo formula allows the calculation of
the response coefficients. In the mentioned examples
these are the ``friction coefficient"
and the ``conductance of the ring" respectively.

In the limit of a very slow time variation (small $\dot{X}$),
linear response theory coincides with the ``adiabatic picture".
In this limit the response of the system
can be described as a non-dissipative ``geometric magnetism" effect 
(this term was coined by Berry and Robbins). 
If we increase $\dot{X}$ beyond a certain threshold, 
then we get Fermi-golden-rule transitions between levels, 
leading to absorption of energy (``dissipation"). 
Then linear response theory can be regarded 
as a generalization of ``Ohm law".  

The Born-Oppenheimer picture allows 
to deal with Hamiltonians of the type
\beq
\mathcal{H}_{\tbox{total}} \ \ = \ \
\mathcal{H}_0(x,p) \ + \ \mathcal{H}(\bm{Q},\bm{P}; x)
\eeq
Here we replaced the parameter~$X(t)$ by a dynamical variable~$x$.
The standard textbook example is the study
of diatomic molecules. In such case $x$ is
the distance between the nuclei.
It is evident that the theory of driven systems
is a special limit of this problem,
which is obtained if we treat $x$ as a classical variable.
For presentation purpose let us consider the Hamiltonian  
\beq
\mathcal{H}_{\tbox{total}} \ \ = \ \ 
\frac{1}{2M}\sum_j {p_j^2} \ + \ \mathcal{H}(\bm{Q},\bm{P};{x}) 
\eeq
We define the basis $|{x},{n({x})}\rangle = {|x\rangle} \otimes {|n({x})\rangle}$, 
and expand the state as  
\beq
|\Psi\rangle \ \ = \ \  
\sum_{n,x} \Psi_n(x) \ |x,n(x)\rangle
\eeq
Using
\beq
\langle x,n(x)|\mathcal{H}|x_0,m(x_0) \rangle \ &=& \ 
\delta(x{-}x_0)  \times  \delta_{nm}E_n(x)
\\
\langle x,n(x)|p_j|x_0,m(x_0)  \rangle 
\ &=& \
(-i\partial_j \delta(x{-}x_0)) \times  \langle n(x) | m(x_0)  \rangle
\ = \
-i \partial_j\delta(x{-}x_0)\delta_{nm}
- \delta(x{-}x_0)  \bm{A}^j_{nm}(x)
\eeq
we deduce that $p_j  \mapsto -i\partial_j-\bm{A}^j_{nm}(x)$, 
and the Hamiltonian can be written as   
\beq
\mathcal{H}_{\tbox{total}} \ \ = \ \ 
\frac{1}{2M}\sum_j({p_j}-{\bm{A}^j({x})})^2
\ + \ {\bm{E}({x})} 
\eeq
The adiabatic approximation is obtained if one 
neglects the $n\ne m$ terms that couple the 
motion on different energy surfaces.     
These couplings are responsible to the dissipation effect.

\ \\

\putgraph[0.5\hsize]{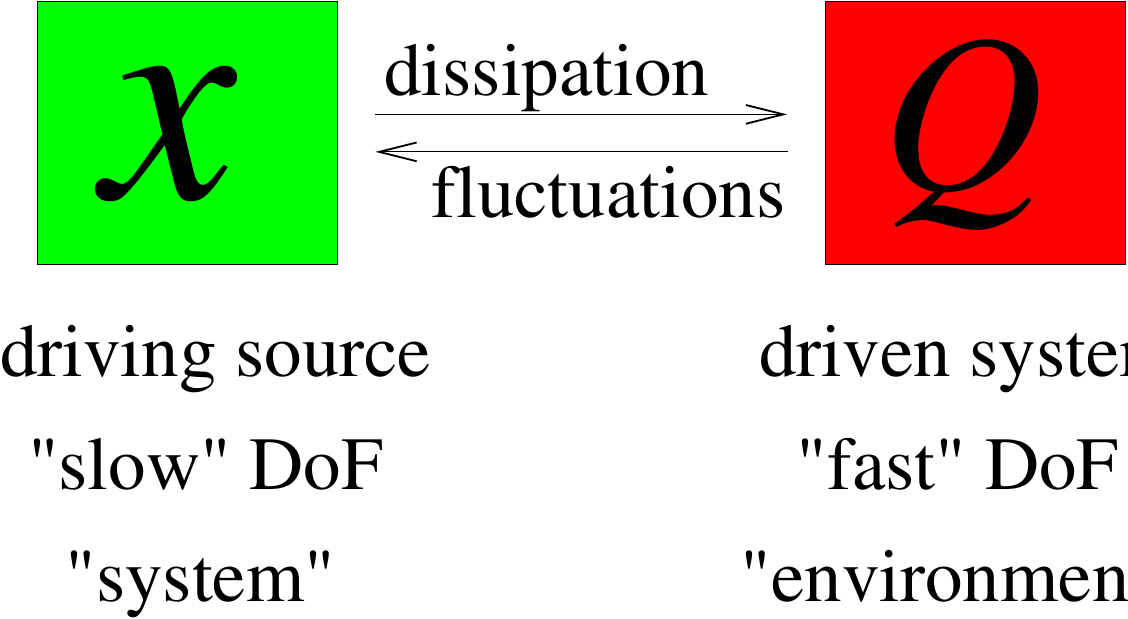}

\ \\

\sheadC{The bath Hamiltonian}

The Hamiltonian of a system that interact with and environment 
is conveniently arranged as 
\beq
\mathcal{H}_{\tbox{total}} 
\ \ = \ \ \mathcal{H}_0(x,p) 
\ + \ \mathcal{H}(\bm{Q},\bm{P};x) 
\eeq
For an interaction with a general (possibly chaotic) environment we write  
\beq
\mathcal{H}_{\tbox{total}}  
\ \ = \ \ 
\mathcal{H}_0(x,p)  \ + \  {x} \bm{B} \ + \ \bm{E} 
\eeq
where $\bm{E}=\{E_n\}$ is the bath Hamiltonian that can be written 
is some diagonal representation, while $\bm{B}=\{B_{nm}\}$ 
represents that interaction term with~$x$.  
Above we assumed that the variation of $x$ is small, 
so we can linearize the interaction term with respect to~$x$.   
More generally we can write  
\beq
\mathcal{H}_0(x,p) 
\ + \ \mathcal{U}(x,Q_{\alpha})
\ + \ \mathcal{H}_{\tbox{bath}}(Q_{\alpha},P_{\alpha}) 
\eeq
It is convenient to model the environment as a huge 
collection of harmonic oscillators. 
For a particle that interacts with such bath we write   
\beq
\mathcal{H}_0(x,p) 
\ \ &=& \ \ \frac{1}{2M} \ p^2 + V(x)
\\
\mathcal{H}_{\tbox{bath}}(Q_{\alpha},P_{\alpha}) 
\ \ &=& \ \ \sum_{\alpha}\left
(\frac{{P_{\alpha}}^2}{2\mathsf{m}_{\alpha}}
+\frac{1}{2} \mathsf{m}_{\alpha}  \omega_{\alpha}^2 {Q_{\alpha}}^2\right)
\eeq
where the interaction is either ZCL-type, or more generally of DLD type: 
\beq
{\cal U}_{\tbox{ZCL}} &=& - {x}\sum_{\alpha} c_{\alpha} {Q_{\alpha}}
\\
{\cal U}_{\tbox{DLD}} &=& -\sum_{\alpha} c_{\alpha} {Q_{\alpha}} u({x}{-}x_{\alpha})
\eeq
The subscripts ZCL and DLD refer to the modeling of the environment as discussed in 
\href{https://journals.aps.org/pre/abstract/10.1103/PhysRevE.55.1422}{PRE 1997}. 
The ZCL (Zwanzig-Cladeira-Leggett) model describes an interaction with a uniform fluctuating field (see figure, upper panels), 
while the DLD (diffusion-localization-dissipation) model allows the possibility of experiencing disordered fluctuations 
that are uncorrelated in space (see figure, lower panels). 
Another possibility is an interaction with chaotic degrees of freedom (see figure, right most panel).

\begin{minipage}{6cm} \includegraphics[width=6cm]{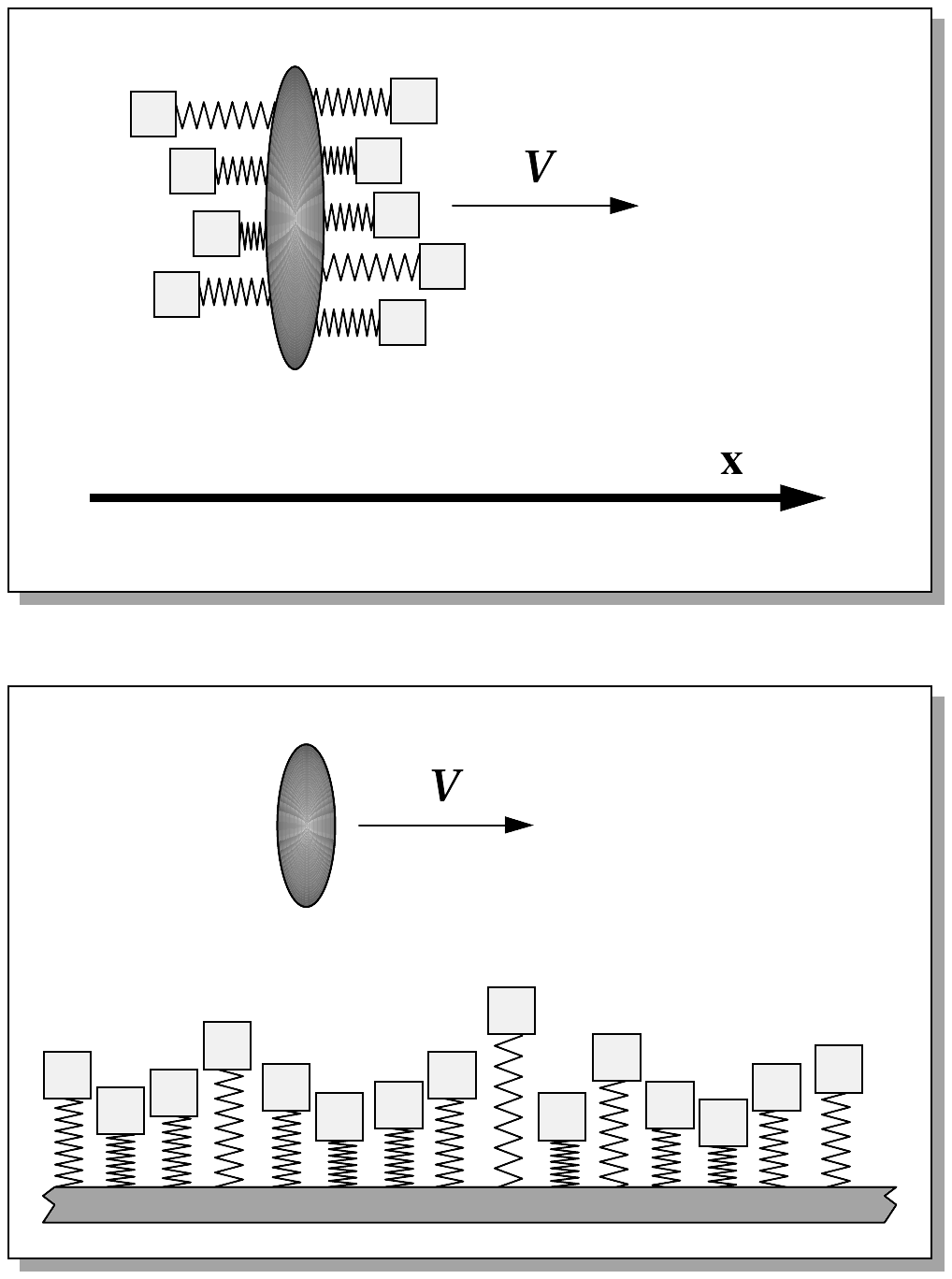} \end{minipage}
\ \ \ \ \ \ \ \ 
\begin{minipage}{3cm} \includegraphics[width=3cm]{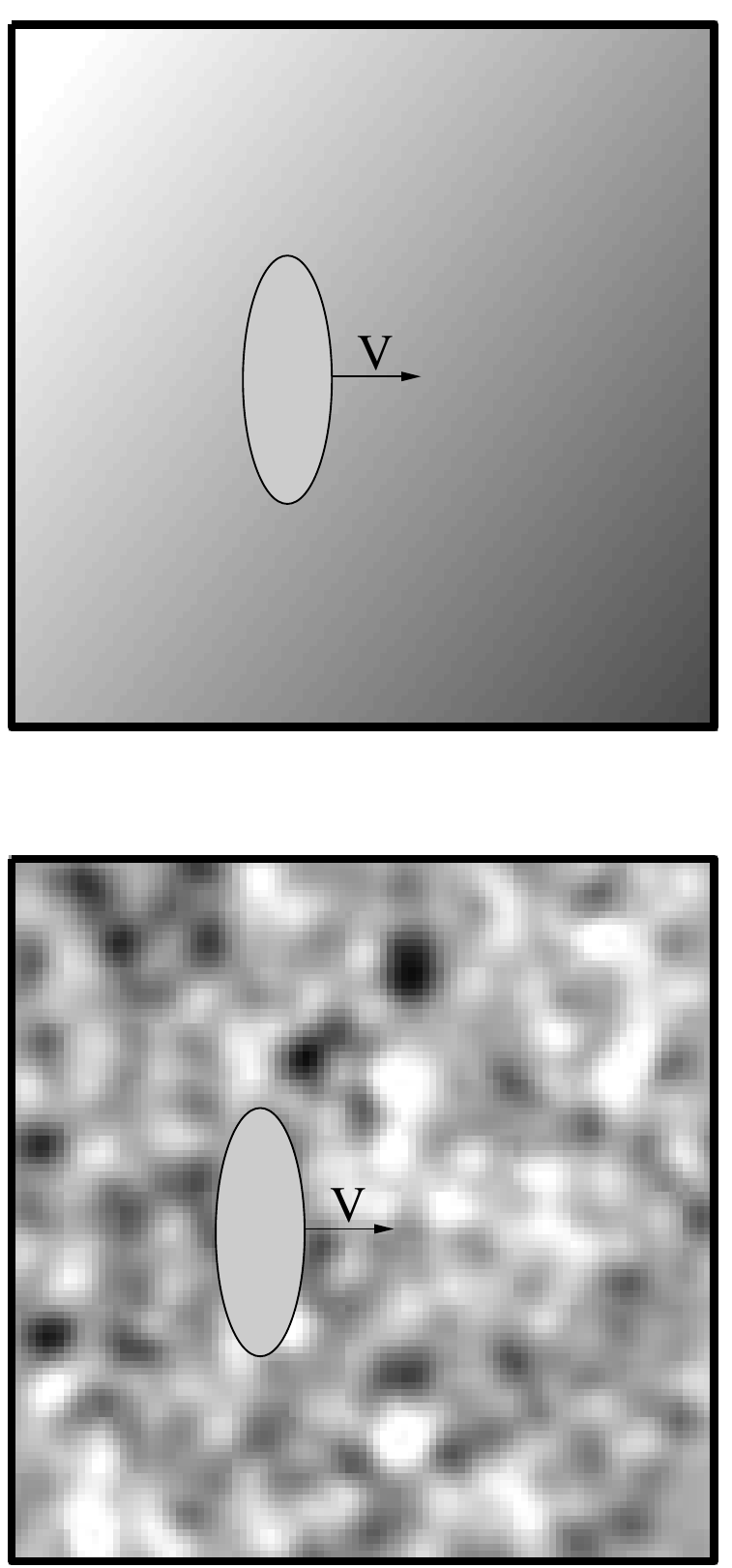} \end{minipage}
\ \ \ \ \ \ \ \ 
\begin{minipage}{5cm} \includegraphics[width=5cm]{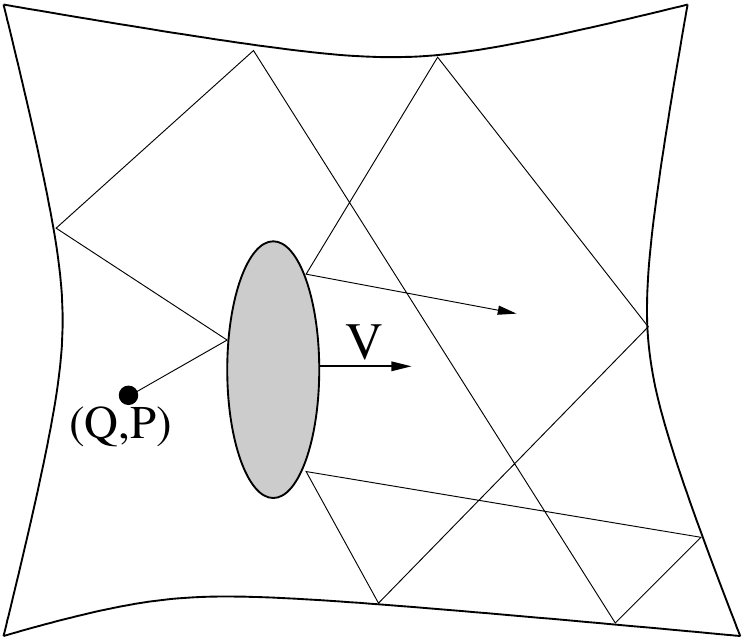} \end{minipage}

\sheadC{The bath fluctuations}

It is common to model the environment as a huge 
collection of harmonic oscillators, and to say that 
the system if subject to the fluctuations 
of a field variable $\mathcal{F}$ which is a linear 
combination of the bath coordinates: 
\beq
\mathcal{F} 
\ \ = \ \ 
\sum_{\alpha} c_{\alpha}Q_{\alpha} 
\ \ = \ \ 
\sum_{\alpha} c_{\alpha} 
\left(\frac{1}{2 \mathsf{m}_{\alpha} \omega_{\alpha}} \right)^{1/2} 
(a_{\alpha} + a_{\alpha}^{\dag})
\eeq
For preparation of the bath 
in state $\bm{n}=\{n_{\alpha}\}$
we get 
\beq
\tilde{S}(\omega)
\ \ = \ \ 
\sum_{\alpha}\sum_{\pm} 
c_{\alpha}^2
\ |\langle n_{\alpha}{\pm}1 | Q_{\alpha} | n_{\alpha} \rangle|^2 
\ 2\pi\delta(\omega\mp\omega_{\alpha})
\eeq
Using
\beq
\langle n_{\alpha}{+}1 | Q_{\alpha} | n_{\alpha} \rangle &=& 
\left(\frac{1}{2 \mathsf{m}_{\alpha} \omega_{\alpha}} \right)^{1/2}
\ \sqrt{1 + n_{\alpha}}
\\
\langle n_{\alpha}{-}1 | Q_{\alpha} | n_{\alpha} \rangle &=& 
\left( \frac{1}{2 \mathsf{m}_{\alpha} \omega_{\alpha}} \right)^{1/2}
\ \sqrt{n_{\alpha}}
\eeq
we get
\beq
\tilde{S}(\omega) 
\ \ = \ \
\sum_{\alpha} 
\frac{1}{2\mathsf{m}_{\alpha} \omega_{\alpha}}
2\pi c_{\alpha}^2  \Big[ 
(1 {+} n_{\alpha}) \delta(\omega-\omega_{\alpha})
+ n_{\alpha} \delta(\omega+\omega_{\alpha})
\Big]
\eeq
For a canonical preparation of the bath 
\beq
\langle n_{\alpha} \rangle \ \ = \ \ f(\omega_{\alpha})  \ \ \equiv \ \ \frac{1}{\eexp{\omega/T}-1} 
\eeq
It follows that  
\beq
\tilde{S}(\omega) 
\ \ = \ \ 
2J(|\omega|) \times 
\left\{\amatrix{   
(1 + f(\omega)) \cr
f(|\omega|)  
}\right\} 
\ \ = \ \ 
2J(\omega)\frac{1}{1-\eexp{-\beta\omega}} 
\eeq
where the lower entry is for ${\omega<0}$.
To get the final compact expression   
we used the identity ${f(-\omega)=-(1+f(\omega))}$, 
and defined the spectral density of the bath as 
\beq
J(\omega) \ = \ \frac{\pi}{2} \sum_{\alpha}
\frac{c_{\alpha}^2}{\mathsf{m}_{\alpha} \omega_{\alpha}}
\delta(\omega-\omega_{\alpha}) 
\ \ \ \ \ \ \ \ \ \mbox{[with anti-symmetric continuation]}
\eeq

{\bf Ohmic response.-- } 
To get an Ohmic bath we set $J(\omega)=\eta\omega$, with some cutoff frequency $\omega_c$. 
Assume that the interaction of the particle with the bath is ${-x F}$, as in the ZCL model. 
Consider a scenario of having a classical particle that is constrained to move with velocity $\dot{x}$.  
Then we get from the fluctuation-dissipation relation (or from a direct calculation) 
that the response of the bath is ${ \braket{\mathcal{F}} = -\eta \dot{x} }$. 
This statement can be generalized for the DLD interaction, 
see \href{https://journals.aps.org/pre/abstract/10.1103/PhysRevE.55.1422}{PRE 1997} for details.

{\bf Einstein coefficient.-- } 
From the formal calculation it comes out that $\tilde{S}(\omega)$ satisfies what we call previously the detailed-balance relation, namely ${[\tilde{S}(-\omega)/\tilde{S}(\omega)] = \exp(-\omega/T)}$. It is the time to illuminate the historical perspective for this terminology. This is related to insights that Einstein had regarding Blackbody radiation. Planck's formula, disregarding $c/4$ factor, is an expression for $u(\omega;T)$, the electromagnetic energy density in space. Consider a two level atom that is immersed in this electromagnetic bath. It experiences fluctuations $\tilde{S}(\omega)$ that induce upward and downward transitions, namely,            
\beq
w_{\downarrow} \ &=& \  A + B u(\omega;T) \\
w_{\uparrow} \ &=& \ \ \ B' u(\omega;T) 
\eeq
Here ${\omega>0}$ corresponds to the energy of the transition, 
and $(A,B,B')$ are the coefficients for spontaneous emission, 
stimulated emission, and stimulated absorption. 
To get equilibrium we expect ${w_{\uparrow}/w_{\downarrow} = \exp(-\omega/T)}$. 
Considering ${T=\infty}$ we realize that we must have ${B'=B}$.
Considering finite temperature we further deduce that the following relation 
between ``Planck" and ``Boltzmann" should hold
\beq
u(\omega;T) \ \ = \ \ \frac{A}{B} \ f(\omega;T)
\eeq
where $A$ and $B$ are temperature independent by definition. 
The theory of electromagnetic field implies that the $A/B$ ratio 
is related to the density of modes:
\beq
\frac{A}{B} \ \ = \ \ \omega \, g(\omega) \ \ = \ \ \frac{1}{\pi^2} \left(\frac{\omega}{c}\right)^3
\eeq  
From a different perspective we can say that the Einstein detailed-balance 
argument implies that we can write
for emission ${ \tilde{S}(\omega) = A [1+f(\omega;T)] }$,  
and for absorption ${ \tilde{S}(-\omega) = A f(\omega;T) }$. 
This is consistent with the direct calculation 
of the power spectrum that we have presented previously.

\sheadC{Spin bath}

We consider the fluctuations of an $\mathcal{F}$ that arise from a bath of spins   
\beq
\mathcal{F} 
= \sum_{\alpha} c_{\alpha}Q_{\alpha} 
= \sum_{\alpha} c_{\alpha} 
(a_{\alpha} + a_{\alpha}^{\dag})
\eeq
Thus $Q_{\alpha}$ is the first Pauli matrix.
Its non-trivial matrix elements are 
\beq
\langle n_{\alpha}{-}1 | Q_{\alpha} | n_{\alpha} \rangle &=& 
\ \sqrt{n_{\alpha}} \\
\langle n_{\alpha}{+}1 | Q_{\alpha} | n_{\alpha} \rangle &=& 
\ \sqrt{1 - n_{\alpha}}
\eeq
In complete analogy we get
\beq
\tilde{S}(\omega)=
\sum_{\alpha} 
2\pi c_{\alpha}^2 \Big[ 
(1 {-} n_{\alpha}) \delta(\omega-\omega_{\alpha})
+ n_{\alpha} \delta(\omega+\omega_{\alpha})
\Big]
\eeq
For canonical preparation 
$\langle n_{\alpha} \rangle = f(\omega_{\alpha})$
where (from here on $\hbar=1$)
\beq
f(\omega) &=& \frac{1}{\eexp{\beta\omega}+1} \\
f(-\omega) &=& \frac{1}{1 + \eexp{-\beta\omega}} =
1 - f(\omega)
\eeq
Thus we get
\beq
\tilde{S}(\omega) \ = \ 
2J(|\omega|) \times 
\left\{\amatrix{   
(1 - f(\omega)) \cr
f(-\omega) 
}\right. 
\ = \
2J(\omega)\frac{1}{1+\eexp{-\beta\omega}} 
\eeq
and
\beq
\tilde{C}(\omega) = J(\omega)
\eeq
where we define
\beq
J(\omega) \ = \ \pi \sum_{\alpha}
c_{\alpha}^2
\delta(\omega-\omega_{\alpha}) 
\ \ \ \ \ \ \ \ \ \mbox{[with symmetric continuation]}
\eeq
For Ohmic bath $J(\omega)=\nu$,  
with some cutoff frequency $\omega_c$.

\sheadC{Spatially extended environment}

In this section we describe fluctuations 
of an extended environment in space and time  
using the form factor $\tilde{S}(q,\omega)$.  
We define 
\beq
\tilde{S}(q,\omega) \ \ = \ \ \text{FT}\Big[ 
\langle \mathcal{U}(x_2,t_2) 
\mathcal{U}(x_1,t_1) \rangle \Big]
\eeq
where the expectation value assumes that the bath 
is in a stationary state of its unperturbed Hamiltonian.   
The force-force correlation function is obtained 
via differentiation. In particular the local power 
spectrum of the fluctuating force is  
\beq
\tilde{S}(\omega) = \int \frac{dq}{2\pi} q^2 S(q,\omega)
\eeq
and the intensity of the fluctuations 
at a given point in space is 
\beq
\nu \ \equiv \ \tilde{S}(\omega{=}0) \ = \ \int \frac{dq}{2\pi} q^2 S(q,\omega{=}0)
\eeq

For the one dimensional DLD bath we get
\beq
{\cal U} \ \  = \ \ -\sum_{\alpha} c_{\alpha} {Q_{\alpha}} u({x}{-}x_{\alpha})
\eeq
Taking into account that the oscillators are independent 
of each other we get
\beq
\langle {\cal U}(x_2,t_2) {\cal U}(x_1,t_1) \rangle 
&=& \sum_{\alpha} c_{\alpha}^2 
\langle {Q_{\alpha}}(t_2) {Q_{\alpha}}(t_1) \rangle 
u({x_2}{-}x_{\alpha}) u({x_1}{-}x_{\alpha}) 
\\
&=& \int dx \left[\sum_{\alpha} c_{\alpha}^2 
\langle {Q_{\alpha}}(t_2) {Q_{\alpha}}(t_1) \rangle 
\delta(x-x_{\alpha}) \right]
u({x_2}{-}x) u({x_1}{-}x) 
\\
&=&
\left[\int u({x_2}{-}x) u({x_1}{-}x) \ dx\right] S(t_2-t_1)
\\
&=&
w(x_2-x_1) \ S(t_2-t_1)
\eeq
Where we have assumed homogeneous distribution of 
the oscillators, and $S(\tau)$ is defined 
implicitly by the above equality. With the convention 
$w''(0)=-1$ it is identified as the local force-force 
correlation function. 
Consequently we get for the form factor
\beq
S(q,\omega) \ \ = \ \ \text{FT}\Big[ 
\langle \mathcal{U}(x_2,t_2) 
\mathcal{U}(x_1,t_1) \rangle \Big]
\ \ = \ \ \tilde{w}(q) \ S(\omega)
\eeq
As an example we may consider the following correlation function:   
\beq
w(r) \ = \ 
\ell^2\exp
\left(-\frac{1}{2}\left(\frac{r}{\ell}\right)^2\right)
\eeq
If the spatial correlation distance 
is very large we get ZCL model:
\beq
w(r) \ = \ \const-\frac{1}{2}r^2 
\eeq
leading to 
\beq
S(q,\omega) = \frac{2\pi}{q^2}\delta(q) \ \tilde{S}(\omega)
\eeq
This means that the force is homogeneous in space, 
and fluctuates only in time, which is effectively the 
case if a particle or an atom interacts with long wavelength modes.

\newpage
\makeatletter{}
\sheadB{Stochastic picture of the dynamics}

There are various "levels" in which the dynamics 
of a non-isolated system can be treated. 
We start with the random walk problem that can 
describe the motion of a Brownian particle 
in the absence of friction. Then we discuss 
the Langevin equation where friction is included.
The dynamics in the above problem is 
described by a diffusion equation and Fokker-Planck  
equation respectively. More generally we can 
talk about Master equations and in particular 
their simplest stochastic version which is known 
as rate equations.

\sheadC{Random walk and diffusion}

Consider a particle that can hope from site to site in a stochastic manner. 
Each step can be represented by a random number $f_t=\pm a$, 
where $a$ is the lattice constant and $t$ is the integer time index.
The total displacement is 
\beq
x(t)-x(0) \ \ = \ \ \sum_{t'=0}^t f(t')
\eeq
Assuming a stationary stochastic process in which
the correlation function is 
\beq
\langle f(t_1) f(t_2) \rangle \ \ = \ \ C(t_1-t_2)
\eeq
we get that the variance is 
\beq
\text{Var}[x] 
\ \ = \ \ \sum_{t_1=0}^t\sum_{t_2=0}^t  \langle f(t_1) f(t_2) \rangle 
\ \ = \ \ \sum_{t'=0}^t \sum_{\tau=-t'}^{+t'}  C(\tau)
\ \ \equiv \ \ \sum_{t'=0}^t 2D(t') \ \ \longrightarrow \ \ 2Dt
\eeq
where the asymptotic value of the diffusion coefficient is 
\beq
D \ \ = \ \ \frac{1}{2}\sum_{\tau=-\infty}^{\infty} C(\tau) 
\eeq
Most significant is to realize that there is a continuum limit of 
the random walk problem where the dynamics is described 
by the following "Langevin" equation of motion 
\beq
\dot{x} \ = \ f(t)
\hspace{2cm} \leadsto \hspace{2cm}  
x(t)-x(0) \ = \ \int f(t') \, dt'
\eeq
and accordingly 
\beq
D \ \ = \ \ \frac{1}{2} \int_{-\infty}^{\infty} C(\tau) d\tau 
\ \ = \ \ \frac{1}{2}\tilde{C}(\omega{=}0)
\eeq

There are various generalizations of the random walk problem, 
where the dwell time or the size of the steps are random variables, 
leading in general to sub-diffusive or super diffusive behavior respectively. 
The latter case is known as Levi-flight.

{\bf Master Equations.-- }   
In the random walk problem the stochastic dynamics can be described 
by an equation for the time evolution of the probabilities~$p_n$ 
to find the particle in site~$n$. This has the form of a {\em rate equation}. 
In the continuum limit it becomes a {\em diffusion equation} 
for the probability density $\rho(x)$.   
More generally this type of master equation is known as 
the {\em Fokker Planck equation}.

\clearpage 
\sheadC{The Langevin equation}

Consider a test particle subject to a homogeneous but fluctuating  
field of force $\mathcal{F}$, leading to stochastic dynamics 
that is described by the Langeving equation $\mass\ddot{x}=\mathcal{F}$.
It is convenient to isolate the average ($=$ ``friction") 
term from $\mathcal{F}$, and accordingly to redefine $\mathcal{F}$ 
as a stochastic variable ($=$ ``noise") that has zero average. 
Consequently the Langevin equation is written as    
\beq
\mathsf{m}\ddot{x} \ = \ -\eta \dot{x} + \mathcal{F}(t)
\eeq
where $\mathcal{F}$ is a stochastic variable that satisfies $\langle \mathcal{F}(t) \rangle = 0$,  and 
\beq
\langle \mathcal{F}(t_2) \mathcal{F}(t_1) \rangle  \ \ = \ \  C(t_2-t_1)
\eeq
It is assumed that $C(\tau)$ has a short correlation time. 
We are interested in the dynamics over larger time scales 
(we have no interest to resolve the dynamics over very short times).    
We also note that if $\mathcal{F}$ were a constant force, 
then the particle would drift with velocity $(1/\eta)\mathcal{F}$. 
The coefficient ${\mu=1/\eta}$ is called {\em mobility}.   
The equation for the velocity $v=\dot{x}$ can be written as 
\beq
\frac{d}{dt} \eexp{(\eta/\mathsf{m})t} v(t) 
\ = \ \frac{1}{\mathsf{m}} \eexp{(\eta/\mathsf{m})t} \mathcal{F}(t)
\eeq
leading to the solution
\beq
v(t) \ = \ \frac{1}{\mathsf{m}} \int_{-\infty}^t dt' 
\eexp{-(\eta/\mathsf{m})(t-t')} \mathcal{F}(t')
\eeq
We see that $\tau_{\eta}=\mathsf{m}/\eta$ is the damping time.  
After time $\gg \tau_{\eta}$ the initial velocity is forgotten, 
hence the lower limit of the integration can be extended to $-\infty$. 
Evidently the average velocity is zero. We turn now to calculate the 
velocity-velocity correlation. "Squaring" and averaging over realizations we get 
\beq
\langle v(t_2) v(t_1) \rangle = 
\frac{1}{\mathsf{m}^2}
\int_{-\infty}^{t_1} \int_{-\infty}^{t_2} dt'dt''  
\eexp{-(\eta/\mathsf{m})(t_1+t_2-t'-t'')} C(t'-t'')
\eeq
We treat $C(t'-t'')$ like a delta function. 
Then it is not difficult to find that 
\beq
\langle v(t_2) v(t_1) \rangle \ = \ 
\frac{1}{2\eta\mathsf{m}}
\eexp{-(\eta/\mathsf{m})|t_2-t_1|}
\int_{-\infty}^{\infty} C(\tau) d\tau 
\ = \  \frac{1}{\mathsf{m}}  \left(\frac{\nu}{2\eta}\right)
\eexp{-|t_2-t_1|/\tau_{\eta}}
\eeq
There is an optional shorter derivation of the latter result:
In Fourier-space the Langevin equation  
is solved easily ${v_{\omega}=[-i\mass\omega+\eta]^{-1}\mathcal{F}_{\omega}}$,  
leading to ${\tilde{C}_{vv}(\omega)=[(\mass\omega)^2+\eta^2]^{-1}\tilde{C}(\omega)}$. 
With ${\tilde{C}(\omega)=\nu}$ we get after FT the same result.

The correlation function $\langle v(t_2) v(t_1) \rangle$ 
for $t_1=t_2=t$ should be consistent with 
$\langle \frac{1}{2}  \mathsf{m} v^2 \rangle = \frac{1}{2}  T$. 
From this one deduces an FD relation $\nu/(2\eta)=T$
with regard to the response characteristics of the bath. 
The displacement $x(t)-x(0)$ of the particle is the 
integral over its velocity $v(t')$. On the average it is zero, 
but the second moment is  
\beq
\langle (x(t)-x(0))^2 \rangle \ = \  
\int_0^t \int_0^t dt'dt''  \langle v(t'') v(t') \rangle 
\ = \ \frac{\nu}{\eta^2} \times t 
\ \equiv \ 2D t
\eeq
Hence we have diffusion in space. 
From the above we deduce the Einstein relation
\beq
\frac{D}{\mu} 
\ \ = \ \ \frac{\nu}{2\eta}   
\ \ = \ \ \text{Temperature}
\eeq
The two results for $D/\mu$, and for $\nu/\eta$, 
can be regarded as special consequences of 
the general FD relation, as demonstrated in a previous lecture.

\clearpage 
\sheadC{The Fokker-Planck Equation}

It is natural to ask what is the "master equation" that describes 
the time evolution of the probability density $\rho_t(x)$ 
in the case of a diffusion process.
We assume that the stochastic equation of motion 
is $\dot{x}=f(t)$ with stochastic~$f(t)$ that has a zero average.
A trivial generalization is to include a drift term 
such that the equation is $\dot{x}=u+f(t)$, 
where $u$ is the so-called drift velocity.
In order to derive the diffusion equation, note that for any 
particular realization of $f(t)$ the probability $\rho_{t+dt}(x_{t+dt})dx_{t+dt}$ 
must equal $\rho_t(x_t)dx_t$. Since the phase space element 
preserves its volume one obtains the Liouville equation ${(d/dt)\rho_t(x_t)=0}$, 
from which one deduces the continuity equation  
\beq
\frac{\partial}{\partial t}\rho_t(x) 
\ \ = \ \ 
- \frac{\partial}{\partial x} \Big[  (u+f(t)) \rho_t  \Big]
\eeq
From this equation it follows that $\rho_{t_0+dt}$ can be expressed 
as an integral that involves $\rho_{t'}$ within ${t_0<t'<t_0+dt}$. 
The equation can be solved iteratively. In order to simplify 
notations we set without loss of generality ${t_0=0}$ and ${t=t_0+dt}$. 
Consequently we get an expansion that involves nested terms with higher 
order $\partial/\partial x$ derivatives of $\rho_0$. 
For sake of clarity we drop the drift term and write      
\beq
\rho_{t} \ \ = \ \ 
\rho_0 
\ - \ \int_0^{t} dt' f(t')  \frac{\partial \rho_0}{\partial x} 
\ + \  {\int_0^{t} dt' f(t') } {\int_0^{t'} dt '' f(t'')} \ \frac{\partial^2 \rho_0}{\partial x^2} 
\ + \ \text{higher order terms}
\eeq
Averaging over realizations of $f()$, and neglecting 
the higher order terms, one obtains a diffusion 
equation, to which we add back the drift term: 
\beq
\frac{\partial}{\partial t}\rho_t
\ \  = \ \  -u\frac{\partial\rho_t}{\partial x} + D\frac{\partial^2\rho_t}{\partial x^2}
\eeq
The neglect of higher order terms, say $\mathcal{O}(dt^3)$ terms, 
is justified in the limit where the correlation time goes to zero.
This is sometimes known as the Markovian approximation.   
It is possible to regard the diffusion equation as a continuity equation 
\beq
\frac{\partial}{\partial t} \rho_t(x)
 = -\frac{\partial}{\partial x} I_t(x), 
\hspace*{2cm}   I_t(x) = u \rho_t(x) -  D \frac{\partial \rho_t(x)}{\partial x}
\eeq
The expression for the current includes a drift term and a diffusion term. 
The diffusion term is known as Fick's law.
Fick's law can be explained heuristically as reflecting 
a non-zero net net flow of particles across a section, 
due to a difference of concentrations between its two sides. 
Ignoring the drift, if we have a sample 
of length $L$ with a steady state current then   
\beq
I \ \ = \ \ -\frac{D}{L} \times \Big[\rho(L)-\rho(0)\Big]
\eeq
This means that there is a strict analogy here to Ohm law, 
implying that $D$ is formally like the conductivity of the chain,
and accordingly can be obtained from a resistor network 
calculation.  This observation is useful in analyzing 
diffusion is non-homogeneous networks.  

The drift velocity is typically related to a the gradient
of an external potential, ${u=-\mu V'(x)}$, 
with a coefficient which is called mobility. Accordingly we write  
\beq
I(x) \ \ = \ \ u\rho(x)  -D \frac{\partial }{\partial x} \rho(x) 
\ \ = \ \ -\mu\rho \frac{\partial V}{\partial x}  -D \frac{\partial \rho}{\partial x}
\eeq
If this expression is applied to a system in canonical equilibrium
with ${\rho(x) \propto \exp(-\beta V(x))}$, it follows from the 
requirement ${I(x)=0}$ that ${\mu=(1/T)D}$. This is called Einstein relation.
It is useful in semiconductor physics. For electrons in metal
it is common to define the conductivity $\sigma=\mu\rho$, 
and postulate that at equilibrium ${\rho(x)=\int dE g(E-V(x))f(E-E_F)}$. 
It follows that the Einstein relation for metals is $\sigma= g(E_F)D$. 
Note that $g(E_F)$ is defined here as the density of one-particle states 
per unit volume, and it is proportional to $\rho/E_F$.

{\bf FPE for Langevin.-- }
As in the case of a "random walk" one can ask what is 
the "master equation" that described the evolution of 
the probability density $\rho(x,p)$.
This leads to the Fokker-Planck equation. 
The derivation is the same as in the case 
of a diffusion process. Here the diffusion 
is in momentum with a coefficient $\nu/2$.  
Including the $v(p)=p/\mass$ drift in the position,
we get the continuity equation 
\beq
\frac{\partial}{\partial t}\rho \ = \  
-\frac{\partial}{\partial x}\Big[ v\rho\Big] 
-\frac{\partial}{\partial p}\Big[
- V'(x)\rho 
- \eta v\rho
- \frac{\nu}{2}\frac{\partial\rho}{\partial p}   
\Big]
\eeq 
There are quantum generalizations of the Fokker-Planck equation
which we discuss in a separate section.

\sheadC{The Ito-Stratonovich interpretation}

Let us try generalize the derivation of the diffusion equation 
for a propcess that is described by the stochastic equation  
\beq
&&\dot{x_j} \ = \ u_j + g_i \, f(t)  
\\ 
&&\braket{f(t)f(t')} \ = \ 2D \delta_{\tau} (t-t')
\eeq
where the $u_j$ and the $g_j$ are some functions of the $x_i$.
The ``noise'' has zero average, namely ${\braket{f(t)} = 0}$, 
and is characterized by a correlation time~$\tau$.  
Accordingly the $\delta_{\tau}(t-t')$ has a short but finite width, 
which is later taken to be zero.
For a particular realization of the noise, 
the continuity equation for the Liouville distribution $\rho(x)$ reads:
\beq
\frac{\partial\rho}{\partial t}  \ \ = \ \ -\sum_j \frac{\partial}{\partial x_j} \left[ (u_j + g_i f(t)) \ \rho \right]
\eeq
We are interested in $\rho(x)$ averaged over many-realizations of the noise.
In its current form the continuity equation cannot be averaged, 
because $\rho$ and $f$ are not independent variables.
To overcome this issue we write for ${\rho(t+dt) - \rho(t)}$  
an integral expression as in the previous section.
Performing the average over realizations of the noise, 
non-vanishing noise-related terms arise from 
the second-order terms and we end up with the equation 
\beq
\pd{\rho}{t} = 
-\frac{\partial}{\partial x_j} 
\Big[   u_j \rho  - g_j D \frac{\partial}{\partial x_i} \left(g_i \rho\right) \Big]
\eeq
Terms that originate from higher orders can be neglected in the $\tau\rightarrow0$ limit.   
It is common to say that the above is the Fokker-Plank equation (FPE) that is associated 
with the stochastic equation according to the Stratonovich interpretation.
Other "interpretations" (as we explain below) provide a similar 
equation with a different order of differentiation. 
The Stratonovich ordering is $[g_j D \partial_i (g_i \rho)]$, 
the Ito ordering is $[D \partial_x (g_j g_i \rho)]$, 
and the Hanggi ordering is $[g_j g_i D \partial_x (\rho)]$.  
All the interpretation are formally equivalent because the difference 
can be absorbed into the definition of the drift velocity $u_j$. 
So one may say that the notion of drift velocity depends 
on the "interpretation".  To make this point physically clear let 
us consider again random walk on discrete lattice with rates of transitions~$w_{x,x'}$. 
One way is to define the drift velocity at ${x=3}$ as ${u = w_{4,3}-w_{2,3}}$. 
Another way would be to define it at ${ x \in [3,4] }$ as ${u = w_{4,3}-w_{3,4}}$. 
The latter definition implies that the steady state for ${u=0}$  
would be uniform in sapce (${\rho=\const}$), and therefore is associated 
with the ordering of Hanggi. Different definitions can lead to different interpretations. 
As a rule of thumb the ${u=0}$ steady state can be used as a guide 
for making  a self-consistent choice.

An observable $X$ is a function of the $x$ variables. 
In order to obtain an equation of motion for $\braket{X}$, 
we multiply both sides of the FPE by~$X$, and integrate over~$x$.
Using integration by parts, and dropping the boundary terms,  
we get the so-called adjoint equation
\beq 
\nonumber
\frac{d}{dt} \braket{X} \ \ &=& \ \ 
\left\langle   
u_j \frac{\partial X}{\partial x_{j}} 
+ g_i \frac{\partial}{\partial x_i} \left[ g_j D  \frac{\partial X}{\partial x_j} \right] \right\rangle  
\eeq
There are cases where instead of handling FPE for the time evolution of $\rho(x)$,  
we can replace it by a reduced set of equations for a complete set of variables 
that characterize the evolving distribution. For example those variables 
might be moments of the distribution, say ${\{ \braket{x},\braket{y},\braket{x^2},\braket{xy}, \braket{y^2},... \} }$. 

\clearpage
\sheadC{Dynamics according to Smoluchowski and Kramers}

The master equation that is associated with the Langeving equation ${\mathsf{m}\ddot{x} = -V'(x) -\eta \dot{x} + \mathcal{F}(t)}$ is an Kramers FPE for $\rho(x,p)$. The stochastic term induces diffusion in momentum with coefficient $\nu/2$, and the friction implies damping with rate ${\gamma =\eta / \mass}$ and mobility ${\mu=1/\eta}$. In the absence of external potential the interplay of noise and friction leads to diffusion in space with coefficient ${D=\nu/(2\eta^2)}$. 
For strong damping the inertial effect can be neglected and the stochastic motion can be described by a simpler equation
${\dot{x} = -\mu V'(x) + \mu \mathcal{F}(t) }$. The master equation that is associated with this simpler version is known as Smoluchowski diffusion equation for the density $\rho(x)$. The drift term is $-\mu V'(x) \rho(x)$, and the diffusion term has coefficient ${D=(1/2) \mu^2 \nu}$ in consistency with Kramers FPE. 
The Smoluchowski diffusion equation can be formally obtained from the Kramers FPE via a leading order expansion in $1/\eta$. The details are described in Section~10 of {\em The Fokker-Planck Equation: Methods of Solution and Applications} by H.Risken.

A major theme in stochastic dynamics is to get the rate of crossing via barrier. This type of activation process is handled within the framework of so-called {\em transition state theory}. In the Smoluchowski approximation (strong damping) the current is given by 
\beq
I(x) \ \ = \ \ - \mu V'(x) \rho(x) - D \frac{\partial \rho(x)}{\partial x}
\ \ = \ \ -\frac{T}{\eta}  e^{-V(x)/T} \frac{\partial}{\partial x}  \left[ e^{V(x)/T} \rho(x) \right]
\eeq  
We assume that ${V(x)=0}$ away from the barrier, and ${V(x)=V_B}$ at the top of the barrier. To be specific we further assume that the curvature at the top of the barrier is $\omega_B$ (inverted harmonic potential).  
Assuming a steay state current ${I(x)=\const}$, we can multiply both sides of the expression above by $e^{V(x)/T}$, and  integrate over interval that contains the barrier. We get that 
\beq
I \ \ = \ \  \frac{T/\eta}{\int_{x_A}^{x_C} e^{V(x)/T} dx} [\rho(x_A) - \rho(x_C)] 
\ \ \approx \ \  \frac{T}{\eta} \sqrt{\frac{\mass \omega_B^2}{2\pi T}} \, e^{-V_B/T} \ [\rho(x_A) - \rho(x_C)] 
\eeq     
Let us assume that the left region is in fact a well that has curvature $\omega_A$ around $x_A$,
and that initially the particle is located there in a state of canonical equilibrium. 
The density $\rho(x_A)$ is determined by normalization, while $\rho(x_C)$ is neglected. 
Changing notation from $I$ to $\Gamma$ we get an expression for the rate of escape:
\beq
\Gamma \ \ = \ \ \frac{\omega_A}{2\pi} \times \left[\frac{\omega_B}{\gamma}\right] \, e^{-V_B/T}  \ \ \equiv \ \  \frac{\omega_A}{2\pi} \times \text{Transmission}
\eeq     
where $\omega_A/(2\pi)$ is known as the {\em attempt frequency}. 
A more refined treatment by Kramers gives the expression
\beq
\Gamma \ \ = \ \ \frac{\omega_A}{2\pi} \times 
\left[ \sqrt{1+\left(\frac{\gamma}{2\omega_B}\right)^2} - \left(\frac{\gamma}{2\omega_B}\right) \right] \, e^{-V_B/T} 
\eeq 
The square brackets in the above formula goes to unity in the formal ${\gamma \rightarrow 0}$ limit. But this is a fallacy. The expression is no longer valid in the weak damping regime. The reason for that is that the escape process is no longer limited by slow diffusion in space (as assumed in the derivation of the Smoluchowski diffusion equation), but rather by the slow diffusion in momentum. So it is natural to write a reduced diffusion equation in momentum, or more precisely in the action variable~$I$. Then one deduces that in the weak damping regime the escape rate is given by 
\beq
\Gamma \ \ = \ \ \frac{\omega_A}{2\pi} \times 
\left[ \frac{\gamma I_B}{T} \right] \, e^{-V_B/T}
\eeq        
where $I_B$ is the action at the escape energy, namely, the enclosed phase-space area, which is given by the $dxdp$ integral over the ${E<V_B}$ region.

\clearpage
\sheadC{Rate equations}

A rate equation is merely a discrete version of the diffusion or Fokker-Planck equation. 
It can be regarded as describing a generalized "random walk" 
problem, where the transition rates $w_{nm}$ are not necessarily 
equal in the ${n\mapsto m}$ and ${m\mapsto n}$ directions. 
The state of the system is described by a column vector $\bm{p}$ 
whose entries are the occupation probabilities $p_n$, such that ${\sum_n p_n =1}$.
The dynamics is determined by the rate equation
\beq
\frac{d\bm{p}}{dt} = \bm{W} \bm{p},  
\hspace{2cm}
\bm{W} = \text{diagonal}\{-\gamma_{n}\} + \text{offdiagonal}\{w_{nm}\}
\eeq
The off-diagonal elements are the rates of transitions, 
namely, $w_{nm}$ is the rate of transition from $m$ to $n$. 
The diagonal elements $-\gamma_i$ of the $\bm{W}$ matrix are determined 
such that each column sums to zero. 
Accordingly ${\sum_n p_n =1}$ is conserved. 
Optionally the rate equation can be regarded as a continuity equation:  
\beq
\frac{dp_n}{dt} 
\ \ = \ \ -\gamma_n p_n + \sum_{m (\ne n)} w_{nm}p_m 
\ \ =  \ \ -\sum_{m (\ne n)} \left[ w_{mn}p_n - w_{nm}p_m \right]
\eeq 
The steady state of the stochastic system is found 
from the equation  ${ \bm{W} \bm{p}^{SS} = 0 }$.
The relaxation modes are the eigenstates, namely ${ \bm{W} \psi = -\lambda \psi }$, 
where ${ \{ -\lambda \} }$ are the eigenvalues. 
Note the sign convention, and note that the ${\lambda = 0}$ mode is the steady-state.   
An arbitrary initial state can be expanded in this basis, 
and consequently the solution of the rate equation is   
\beq
\bm{p}(t) \ \ = \ \ e^{\bm{W}t}\bm{p}(0) \ \ = \ \   \bm{p}^{SS} + \sum_{\lambda \ne 0} C_{\lambda} \, e^{-\lambda t} \ \psi^{(\lambda)} 
\eeq

{\bf Detailed balance.-- } 
In the context of the "system-bath" paradigm it is common to 
model the system as a set of levels~$\{E_{n}\}$ with transition rates
that reflect detailed balance considerations, 
such that ${ p_n^{SS} \propto \exp[-E_{n}/T_B] }$. Namely, 
\beq
\frac{w_{mn}}{w_{nm}} \ \ = \ \ \exp\left[ \frac{E_n-E_m}{T_B} \right] 
\eeq 
A driving noise source or a work agent (see below) can be regarded 
as a bath that has infinite temperature.  
More generally one can regard the average value ${(w_{nm}+w_{mn})/2}$ 
as the "noise" which is introduced into the system by the bath, 
while the difference ${ (w_{nm}-w_{mn}) }$ is the friction.
However this point of view is strictly correct 
only for constant density of states. If the level density grows with 
energy there will be a heating effect even if $T_B{=}\infty$.

{\bf Two level/site system.-- } 
The two level system is the simplest setup for illustration of equilbration process. 
The transition rates  are $w^{\pm}$. The dynamics is generated by the matrix 
\beq
\bm{W} \ \ = \ \ 
\left( \amatrix{ 
-w^{+} & w^{-} \cr
w^{+} &  -w^{-} } \right)
\eeq
Using the notation ${ S = p_2-p_1 }$, recaling that ${ p_1+p_2 = 1 }$, 
and using the noation ${\gamma = w^{+} + w^{-} }$ and ${u = w^{+} - w^{-} }$ 
we get the equation
\beq
\frac{dS}{dt} \ \ = \ \ - \gamma S + u    
\eeq   
which implies exponential relaxation towards the equilibrium value ${ S_{\text{eq}} = u/\gamma }$.
If the transitions are induced by a bath of temperature $T_B$, 
then $S_{\text{eq}}$ corresponds to equilibrium at temperature~$T_B$.
If we expose the system to so-called "work agent", say a sun that has infinite temperature, 
then the new rates are ${ w^{\pm} = w_B^{\pm} + w_A }$,
and the system will reach an equilibrium-like state that corresponds to a higher temperature.   
It is important to realize that the steady state solution features  
energy flow from the work agent via the system to the bath (and not the other way around).
We say the energy is dissipated.

{\bf Three level/site system.-- } 
The three-level system is the simplest setup for illustration of non-equilibrium thermodynamics. 
For example, it can be regarded as a model for a 3-level laser heat engine (see figure) 
or a mathematically equivalent rolling marble machine (see figure). 
The transitions are induced by a hot bath ($T_H$) and by a cold bath ($T_C$).  
In the first example photons can be either emitted or absorbed by a work agent (${T_A = \infty}$).
The second example is further discussed below. 
Either way the dynamics is generated by the matrix
\beq
\bm{W} \ \ = \ \ 
\left( \amatrix{ 
-\gamma_1   &  w_C^{-}  &  w_A  \cr
w_C^{+} &  -\gamma_2  &  w_H^{-} \cr
w_A     &  w_H^{+}  &  -\gamma_3  } \right)
\eeq
where ${\gamma_1 = w_C^{+} + w_A }$, 
and ${\gamma_2 = w_C^{-} + w_H^{+} }$,
and ${\gamma_3 = w_H^{-} + w_A }$.
The affinity of the cycle is defined as 
\beq
\Phi \ \ = \ \ \mathcal{E}_{1 \leadsto 2} + \mathcal{E}_{2 \leadsto 3} + \mathcal{E}_{3 \leadsto 1}
\ \ = \ \   \ln\left[  \frac{ w_{13} w_{32} w_{21} }{ w_{12} w_{23} w_{31} }  \right] 
\ \ = \ \ \frac{\omega_C}{T_C} - \frac{\omega_H}{T_H}
\eeq
In order to have a working engine cycle we require ${\Phi>0}$, 
which implies ${ (\omega_C/\omega_H) > (T_C/T_H) }$.
The efficiency of the engine is  
\beq
\eta \ \ \equiv \ \ \frac{\omega_H - \omega_C }{\omega_H}
\ \ < \ \ 1 - \frac{T_C}{T_H}
\eeq 
The limiting efficiency is the so-called Carnot efficiency.  
We can solve ${\bm{W} \bm{p} = 0 }$ to find the probabilities ${(p_1,p_2,p_3)}$ 
at steady state. Then we can find the probability current ${I(\Phi) = (p_3-p_1) w_A }$ 
at steady state, and the power output of the engine ${ (\omega_H-\omega_C) I(\Phi) }$.

{\bf Work agent.-- }
The mechanical rolling marble machine possibly clarifies better the concept of {\em work agent}. 
Here the task of the engine is to pulls up a weight. The hot bath induce with some probability 
a transition of the marble form position "2" to position "3". 
From there, with some probability, it gets into a car of the roller coaster wheel. 
Then is rolls (trapped in the car) to position "1". The wheel pulls up the weight. 
In order to maximize efficiency it is designed such that the potential energy 
of the whole system (including the weight) is the same at "3" and at "1". 
Consequenltly there is an equal probability to make the ride from "1" to "3". 
However, considering the full cycle, the condition ${\Phi>0}$ ensures that the net work is positive. \\

\includegraphics[width=7.7cm]{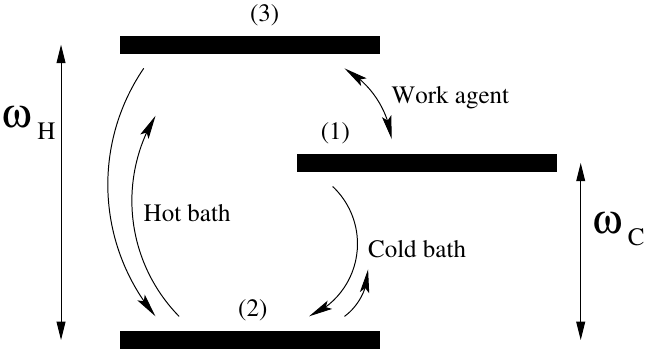}
\hspace*{1cm}
\includegraphics[width=8cm]{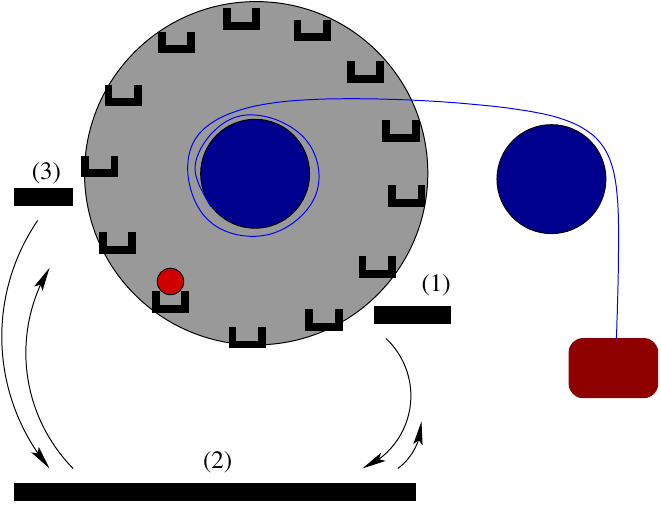}

{\bf The $N$ site ring.-- } 
The dynamics of a particle in an $N$ site ring is generated by the matrix
\beq
\bm{W} \ \ = \ \ \left( \amatrix{ 
-\gamma_1 & w^{-}_{2} & 0 & ...&w^{+}_{1} \cr 
w^{+}_{2} & -\gamma_2 & w^{-}_{3} & 0 & ... \cr 
0 & w^{+}_{3} & -\gamma_3 & w^{-}_{3} & ... \cr ... &...&...&...&w^{-}_{N}\cr 
w^{-}_{1} & 0& ... & w^{+}_{N} & -\gamma _N} \right)
\ \ = \ \ 
- \gamma \bm{1} + w^{+} \bm{D} + w^{-} \bm{D}^{-1}   
\eeq
In the second equality we have assumed that all the anti-clockwise rates equal $w^{+}$, 
and that all the clockwise rates equal $w^{-}$. 
Accordingly ${\gamma = w^{+} + w^{-} }$.
The matrices $\bm{D}$ and $\bm{D}^{-1}$ generate anti-clockwise and clockwise displacements respectively.  
The drift velocity is 
\beq
\frac{d}{dt}\braket{x} \ \ =  \ \ [w^{+} - w ^{-}]a \ \ \equiv \ \ \bar{v}
\eeq
where $a$ is the lattice spacing. 
The proof is as follows:
\beq \nonumber
\frac{d}{dt}\braket{x}  
&=&  \frac{d}{dt} \sum_n p_n x_n  
\ = \ \sum_n x_n w_{nm} p_m 
\ = \ \sum_n x_n \left[w^{+} p_{n{-}1}  + w^{-} p_{n{+}1} -  \gamma p_n \right] \\
&=& \sum_n \left[  w^{+} (x_{n+1} - x_n) p_n +  w^{-} (x_{n{-}1} -x_n)  p_n \right] 
\ = \ ...
\eeq
Irrespective of drift, we have diffusion. Say that we start a distribution at ${x=0}$. 
For simplicity let us assume that ${w^{-}=w^{+}=w }$. We get that the rate of growth 
of the spreading is 
\beq
\frac{d}{dt}\braket{x^2} \ \ =  \ \ 2wa^2 \ \ \equiv \ \ 2D
\eeq
More generally, if the drift velocity is non-zero, 
we can prove that 
\beq
\text{Var}(x) \ \ = \ \ \braket{x^2}-\braket{x}^2 \ \ = \ \ 2Dt
\eeq
where ${D= (1/2)[w^{+}+w^{-}]a^2 }$ is called the diffusion coefficient.

The eigenstates of $\bm{W}$ are the eigenstates of the displacement operator~$\bm{D}$, 
namely {\em momentum states}. It is convenient to write the column representation $\psi_n$ 
as a function, namely ${ \psi_n \equiv \psi(x_n) }$.   
The momentum states are ${ \psi(x) = e^{ik x} }$ with eigenvalues $e^{-ika}$. 
Accordingly 
\beq
\lambda_k \ \ = \ \  \gamma - w^{+} e^{-ika} - w^{-} e^{ika} \ \ = \ \ \gamma [1-\cos(ka)] + i \bar{v} \sin(ka) 
\eeq

{\bf Diffusion.-- } 
Consider a rate equation that describes stochastic motion along chain, 
with transition rates ${w_{nm} = w(r)}$ that depend on the 
hopping distance ${r=(n-m)}$. 
One can deduce the drift velocity and the diffusion coefficient
from the 1st and 2nd moments of the short-time spreading:     
\beq
\bar{v} \ \  &=&  \ \ \frac{1}{2}\sum_{r=-\infty}^{\infty} r \, w(r) \\
D \ \ &=&  \ \ \frac{1}{2}\sum_{r=-\infty}^{\infty} r^2 \, w(r)
\eeq
For unbiased near-neighbor hopping with rate $w$ we get ${ \bar{v} = 0 }$ 
and ${D=w a^2}$,  where $a$ is the lattice constant. 
Accordingly, if we discretize a diffusion equation by slicing 
the $x$~axis into cells of width~$a$, the effective 
hopping rate $w$ should be chosen such that the diffusion coefficient is $D = w a^2$.  
An optionally procedure to determine $\bar{v}$ and $D$ 
is via a Taylor expansion (in $k$) of the eigenvalue $\lambda_k$.

\clearpage
\sheadC{Rate equations - formalism}

In the remaining subsections of this lecture we discuss 
the formal aspects of treating rate equations.  
The state of the system is described by a column vector $\bm{p}$ 
whose entries are the occupation probabilities $p_n$.  
The dynamics is determined by the rate equation
\beq
\frac{d\bm{p}}{dt} = \bm{W} \bm{p},  
\hspace{2cm}
\bm{W} = \text{diagonal}\{-\gamma_{n}\} + \text{offdiagonal}\{w_{nm}\}
\eeq
Probability is conserved hence $\bm{W}$ has the left eigenvector ${q^{0}=\{1,1,...\}}$ 
with eigenvalue ${\lambda_0=0}$.  The associated right eigenvector ${p^{0}}$ is the steady state.
The other eigenvalues $-\lambda_r$ of $\bm{W}$ might be complex, but the 
real part of $\lambda_r$ has to be positive. This follows from the observation that 
for $t\rightarrow\infty$ only the steady state survives, while all the higher has to diminish. 
The proof is based on the Perron-Frobenius theorem with regard to $\bm{U}(t)=\exp(\bm{W}t)$. 
If we have detailed balance (see below) the $\lambda_r$ have to be real and positive. 

In general we can write the transition rates as follows:
\beq
w_{nm} \ \ = \ \ \exp\left[ -\mathcal{B}_{nm} + \frac{\mathcal{E}_{m \leadsto n} }{2} \right] 
\eeq
where $\mathcal{B}$ is a symmetric matrix while  $\mathcal{E}$ is the anti-symmetric part.
The latter can be decomposed into conservative and solenoid components in a unique way:
\beq
\mathcal{E}_{m \leadsto n} \ \ = \ \ \ln\left(\frac{w_{nm}}{w_{mn}}\right) 
\ \ = \ \   (V_m-V_n) \ \ + \ \ \sum_{\alpha} \alpha \ \mathcal{A}^{(\alpha)}_{m \leadsto n}
\eeq 
Note that the ``solenoid gauge" implies that $\mathcal{A}$ is uniform along the $C_{\alpha}$ loop. 
One can use a non-solenoid gauge, e.g. to have it non-zero on one bond only. Anyway we fix its gauge 
and normalized its circulation as follows:
\beq
\sum_{x \in C_{\alpha}} \mathcal{A}^{(\alpha)}_{x} \ \ = \ \ 1
\eeq 
We can define the height of the barrier relative to the potential reference level:
\beq
B_{nm} \ \ = \ \ \mathcal{B}_{nm} \ + \ \frac{V_n+V_m}{2}
\eeq 
Then the expression for the transition rates takes the following form:
\beq
w_{nm} \ \ = \ \ \exp\left[ - \left(B_{nm} - V_m \right) + \frac{1}{2} \sum_{\alpha} \alpha \ \mathcal{A}^{(\alpha)}_{m \leadsto n}  \right] 
\eeq

The detailed balance condition of having no circulations implies that 
there exist a diagonal matrix $\bm{V}$ such that 
\beq
\eexp{\bm{V}} \bm{W} \eexp{-\bm{V}} \ \ = \ \ \bm{W}^{\dag} 
\eeq
Then we can perform a ``gauge" transformation to a symmetric matrix
\beq
\tilde{\bm{W}} \ \ = \ \ \eexp{\bm{V}/2} \bm{W} \eexp{-\bm{V}/2}, \ \ \ \ \ \ \ \ \ \ \tilde{\bm{W}} = \tilde{\bm{W}}^{\dag}
\eeq
It follows that all the $\lambda_r$ have non-negative real values.    
If we spoil the detailed-balance the matrix $\tilde{\bm{W}}$ will becomes 
parameterized by the affinities $\alpha$, and the $\lambda_r$ might become complex.  
Instead of the detailed balance condition we get 
$\eexp{\bm{V}} \bm{W}(\alpha) \eexp{-\bm{V}} = \left[\bm{W}(-\alpha)\right]^{\dag}$
or equivalently $\tilde{\bm{W}}(\alpha) = \left[\tilde{\bm{W}}(-\alpha)\right]^{\dag}$.

\sheadC{Rate equations - counting statistics}

We add a pointer $q$ with conjugate variable $\varphi$ such that ${[q,\varphi]=i}$. 
Now the Hilbert space is spanned by ${|n,q\rangle}$. The dynamics of the joint 
probability distribution ${p_{n}(q)}$ in the presence of affinity~$\alpha$ is generated by the operator  
\beq
\bm{W}(\alpha) \ \ = \ \ \sum_{n,m} |n,q+\mathcal{A}_{m\leadsto n} \rangle \ w_{nm} \ \langle m, q |
\ \ = \ \ \sum_{n,m}  w_{nm} \ \hat{J}^{(m\leadsto n)}  \otimes \eexp{-i\mathcal{A}_{m\leadsto n}\hat{\varphi} }
\eeq
We change basis to  ${|n,\varphi\rangle}$ and use the Laplace transform convention $i \varphi \mapsto \varphi$, such that  
\beq
p_{n}(q) \ \ \equiv \ \ \sum_{\varphi} \tilde{p}_{n}(\varphi) \ \eexp{i\varphi q} 
\ \ \equiv \ \  \sum_{\varphi} p_{n}(\varphi) \ \eexp{\varphi q}       
\eeq
It follows that the moment generating function is  
\beq
Z(\varphi) = \braket{\eexp{-\varphi q}} = \sum_n p_{n}(\varphi),  
\hspace{2cm} \text{with} \ 
p_{n}(\varphi) = \sum_q  p_n(q) \eexp{-\varphi q},   
\eeq
The master equation for $p_n(\varphi)$ is block-diagonal in $\varphi$, with matrix 
\beq
\bm{W}(\alpha; \varphi) \ \ = \ \  w_{mn} \eexp{-\varphi\mathcal{A}_{n\leadsto m}} 
\ \ = \ \  \bm{W}(\alpha-2\varphi)
\eeq 
We no longer have detailed balance but 
${\ \eexp{\bm{V}} \bm{W}(\alpha;\varphi) \eexp{-\bm{V}} = \left[\bm{W}(\alpha; \alpha{-}\varphi)\right]^{\dag}}$.  
The comulants are determined by the lowest eigenvalue $\lambda_0(\varphi)$, leading to the NFT
\beq
g(\varphi)=g(\alpha-\varphi) \ \ \ \ \ \leadsto \ \ \ \ \  P(-q)/P(q)=\exp(-\alpha q)
\eeq

\sheadC{Rate equations - ergodicity}

It is convenient to define a weighted distribution $q_n = p_n / p_n^0$, 
such that $q_n=$uniform once the steady state is reached. 
The rate equation takes the form ${\dot{\bm{p}} = \bm{G}\bm{q} }$, where    
\beq
G_{nm} \ \ = \ \ W_{nm} p_m^0 \ \ \ \ \ \ \ \ \ \ \ \ \sum_n G_{nm} = \sum_m G_{nm} = 0
\eeq
Accordingly 
\beq
\bm{q}^{\dag}\bm{G}\bm{q} \ \ = \ \ -\frac{1}{2}\sum_{nm} G_{nm}(q_n-q_m)^2
\eeq
For an eigen-mode $\bm{G}\bm{q}^r = \lambda_r\bm{p}^r$ and it is implied that ${\lambda_r \sum_n [(p_n^r)^2/p_n^0] > 0}$, 
leading to the conclusion that ${\lambda_r >0}$. 

In order to characterize the approach to steady state we pick a convex function $f(x)$, 
for example ${f(x)= x\ln(x)}$ and define an ergodicity measure 
\beq
H(t) \ \ = \ \ \sum_n p_n^0 \ f(q_n(t))  \ \ = \ \ e.g. \ \ = \ \ \sum_n p_n \ln(p_n/p_n^0) 
\eeq  
Then we get
\beq
\frac{d}{dt}H(t) \ \ = \ \ -\sum_n G_{nm} \ \left[ (f(q_n)-f(q_m)) - (q_n-q_m)f'(q_n) \right] \ \ < \ \ 0
\eeq

\newpage
\makeatletter{}
\sheadB{Quantum master equations}

\sheadC{General perspective}

The description of the reduced dynamics of a system that is coupled to a bath 
using a Master equation is commonly based on the following working hypothesis: 
{\bf (i)}~The bath is fully characterized by a single spectral function. 
{\bf (ii)}~There is a way to justify the neglect of memory effects.
The latter is known as the Markovian approximation.   
In particular it follows that the initial preparation, 
whether it is factorized or not, is not an issue. 
If the master equation is regarded as {\em exact} description 
of the reduced dynamics it should be of the Lindblad form. 
Otherwise is should be regarded merely as an approximation.  

There are two common approximation schemes: 
{\bf (A)}~In the {\em Microscopic regime} of atomic physics (e.g. two level atom)  
it is assumed that the bath induced 
rates are much smaller than the level spacing, 
and a "secular approximation" is employed.
{\bf (B)}~In the {\em Mesoscopic regime} of condense matter physics  (e.g. Brownian motion) 
it is assumed that the bath is Ohmic, and accordingly its effect 
can be treated as a generalization of "white noise".

\sheadC{The general Lindblad form} 

A master equation for the time evolution of the system probability matrix  
is of Lindblad form if it can be written as
\beq 
\frac{d\rho}{dt} \ \ = \ \  -i[\bm{H},\rho] 
\ + \sum_r \nu_r \bm{L}_r \rho \bm{L}_r^{\dagger} 
\ - \frac{1}{2}\left[ \bm{\Gamma}\rho+ \rho\bm{\Gamma} \right], 
\ \ \ \ \ \ \ \ \ \ \bm{\Gamma}= \sum_r \nu_r \bm{L}_r^{\dagger}\bm{L}_r
\eeq
where $\bm{L}_r$ are called Lindblad generators, and $\nu_r$ are positive coefficients. 
An optional style of writing the above master equation is 
\beq
\frac{d\rho}{dt} \ \ = \ \  
-i[\bm{H},\rho] \ + \sum_r \nu_r  
\left[ \bm{L}_r \rho \bm{L}_r^{\dagger} \ - \frac{1}{2} \{ \bm{L}_r^{\dagger} \bm{L}_r, \rho \}  \right], 
\eeq
Lindblad equation is the most general form of a Markovian master equation for the probability matrix.
The time dependence of an expectation values is given by the adjoint equation: 
\beq
\frac{d}{dt} \braket{\bm{Q}} 
\ = \  \trc\left[ \bm{Q} \frac{d}{dt}\rho \right]
\ = \  \trc\left[ \bm{Q} \mathcal{L} \rho \right]
\ = \  \trc\left[ (\mathcal{L}^{\dag}  \bm{Q})  \rho \right]
\ = \  \braket{\mathcal{L}^{\dag} \bm{Q}} 
\eeq
If the master equation is written in the Lindblad form, 
the expression for $\mathcal{L}^{\dag} \bm{Q}$ is the same 
as $\mathcal{L} \rho$ with $\bm{H} \mapsto -\bm{H}$.

\sheadC{Derivation of the Lindblad form}

The most general linear relation between matrices is   
\beq 
\tilde{\rho}_{\alpha\beta} \ \ = \ \ 
\sum_{\alpha'\beta'} \mathcal{K}(\alpha\beta|\alpha'\beta') \, \rho_{\alpha'\beta'} 
\eeq
This linear transformation can be regarded as ``quantum operation" if it preserves  
the hermiticity and the positivity of $\rho$. See [\href{http://arxiv.org/abs/quant-ph/0605180}{arXiv (lecture~53)}] for details. 
Changing notation to $\mathcal{K}(\alpha\beta|\alpha'\beta')=\mathcal{K}_{\alpha\alpha',\beta\beta'}$ 
one observes that $\mathcal{K}_{\alpha\alpha',\beta\beta'}$ should be hermitian, 
with non-negative eigenvalues $\lambda_r$. Accordingly we can find a spectral decomposition with 
a transformation matrix $T(\alpha\alpha'|r)$. 
Changing  notation to ${\bm{K}^r_{\alpha,\alpha'} = T(\alpha\alpha'|r)}$ 
we get the Kraus representation ${ \tilde{\rho} = \sum_r \lambda_r [\bm{K}^r] \rho [\bm{K}^r]^{\dagger}}$. 
Conservation of probability implies ${\sum_r \lambda_r [\bm{K}^r]^{\dagger}[\bm{K}^r]= \bf{1}}$. 
Looking on the incremental change of~$\rho$ during a small time interval~$dt$, 
one obtains the Lindblad form of the Master equation.

Detailed derivation of Lindblad form from the Kraus representation 
can be found in [\href{https://arxiv.org/abs/1204.2016}{arXiv}]. 
For completeness we present here a brief outline.  
We note that for zero time evolution we have as single 
non-zero eigenvalue ${\lambda_{(0)}=N}$ which is associated with the normalized
identity matrix ${ T(\alpha\alpha'|0) = N^{-1/2} \delta_{\alpha,\alpha'} \equiv \bm{L}_0}$, 
where $N$ is the dimension of Hilbert space.  For a small time step we 
substitute ${\lambda_{(r\ne 0)} \equiv \nu_r dt}$ and ${T(\alpha\alpha'|r) \equiv \bm{L}_r}$.
Note that the $\nu_r$ are positive, and that the $\bm{L}_{(r\ne0)}$ are traceless 
due to the orthogonality with $\bm{L}_0$. Accordingly 
\beq
\frac{d\rho}{dt} \ \ = \ \ 
\sum_{r \ne 0} \nu_r \bm{L}_r \rho \bm{L}_r^{\dagger} + ...
\eeq   
The remaining terms in the Lindblad form are the $\bm{H}$-term and the $\bm{\Gamma}$-term, 
that are related to the ${r=0}$ term, and can be straightforwardly deduced. 
However we prefer to point out an indirect approach.
From the same argument as for quantum operations it is clear that any linear expression that preserves hermiticity can be written as $\sum_{r} \nu_r [\bm{L}^r] \rho [\bm{L}^r]^{\dagger}$. 
We transform to some general orthonormal basis where one of the operators is ${\bm{F}^{(0)} = N^{-1/2} \bm{1}}$, while by orthonormality all the other $\bm{F}^r$ are traceless. We get  
\beq
\frac{d\rho}{dt} \ \ = \ \ \sum_{r,s} \tilde{\nu}_{r,s} [\bm{F}^r] \rho [\bm{F}^s]^{\dagger} 
\ \ = \ \ 
-i[\bm{H},\rho] - \frac{1}{2}\left\{ \bm{\Gamma}, \rho \right\} 
+ \sum_{r,s}' \tilde{\nu}_{r,s} [\bm{F}^r] \rho [\bm{F}^s]^{\dagger} 
\eeq
where the last summation excludes the "0" terms.  
The definition of the Hamiltonian ${\bm{H} \propto i\sum_{r}' ([\bm{F}^r] - [\bm{F}^r]^{\dag}) }$
is implied. Also the $\bm{\Gamma}$-term is implied, and optionally can be deduced from 
the requirement of obtaining a trace-preserving map. 
It is now argued that the sum $\sum_{r,s}'$  should coincide, upon diagonalization, 
with the summation over the $\bm{L}^r$ terms. Hence the  $\tilde{\nu}_{r,s}$ must have positive eigenvalues $\nu_r$.

\sheadC{The Ohmic Master Equation} 

Consider the the classical Langevin equation. Using canonical phase-space 
coordinates it reads ${\dot{p} = -V'(x) - \eta v + f(t)}$, 
where ${v=\dot{x}=p/\mass}$ is the velocity, 
and $f(t)$ is white noise that has intensity~$\nu$.
The corresponding master equation for $\rho(x,p)$ 
is the Fokker-Planck equation:
\beq
\frac{d\rho}{dt} = 
-\frac{\partial}{\partial x}\Big[p\rho\Big] 
-\frac{\partial}{\partial p}\Big[
- V'(x)\rho 
- \eta v\rho
- \frac{\nu}{2}\frac{\partial\rho}{\partial p}   
\Big]
\eeq 
This equation can be written with Poisson Brackets, 
which are replaced in the quantum context by commutators:  
\beq
\frac{d\rho}{dt} = 
-i[\mathcal{H},\rho] 
-\frac{\nu}{2}[x,[x,\rho]] 
- i\frac{\eta}{2}[x,\{v,\rho\}]
\eeq 
We shall discuss later the general procedure to derived this master equation 
from an Hamiltonian, where the interaction with the bath is via 
the system operator ${W=x}$. The same procedure can be uses for any~$W$,
leading to 
\beq
\frac{d\rho}{dt} = 
-i[\mathcal{H},\rho] 
-\frac{\nu}{2}[W,[W,\rho]]
-i \frac{\eta}{2}[W,\{V,\rho\}]
-\frac{\nu_{\eta}}{2}[V,[V,\rho]] 
\eeq 
where $v$ has been replaced by $V=i[\mathcal{H},W]$, 
and where ${\nu_{\eta}=0}$.  
This Ohmic master equation does not have the Lindblad form (see below),
and hence in general complete positivity is not guaranteed.
For example: if we consider the relaxation of a wavepacket 
in damped harmonic oscillator, then at low temperatures 
we end up with a sub-minimal wavepacket that violates the uncertainty relation.

The Ohmic master equation involves the bilinear form  $\sum_{r,s} \tilde{\nu}_{r,s} [\bm{F}^r] \rho [\bm{F}^s]^{\dagger}$ 
with ${\bm{F}^{(1)} = W}$, and ${\bm{F}^{(2)} = V}$, and 
\beq
\tilde{\nu}_{r,s} \ \ = \ \ \left( \amatrix{\nu &  -i \frac{\eta}{2} \cr i \frac{\eta}{2} &  \nu_{\eta}  } \right)
\eeq  
In order for this equation to be Lindblad, the matrix $\tilde{\nu}_{r,s}$ should be positive.  
The minimal modification would be to set a non-zero ${\nu_{\eta}=\eta^2/(4\nu)}$. 
With this substitution, after diagonalization, one ends up with a single Lindbald 
term with the generator 
\beq
\bm{L} \ = \ W + i \frac{\eta}{2\nu} V  
\eeq
Note that the pre-factors of the three terms in the modified  Ohmic version  
are $\nu/2$ and $\nu/(2T)$ and $\nu/(32 T^2)$  respectively.
These terms can be regarded as arsing from an expansion 
in powers of $(\Omega/T)$, where $\Omega$ is the frequency of the motion.
Accordingly in the high temperature regime the deviation 
of the standard Fokker-Planck equation from the Lindblad form is negligible.

We can also go in reverse and provide an ``Ohmic interpretation" for each term in the Linblad form.
Namely, consider 
\beq 
\text{Lindblad} \ = \ \bm{L} \rho \bm{L}^{\dagger} -\frac{1}{2}\bm{L}^{\dagger}\bm{L}\rho - \frac{1}{2}\rho \bm{L}^{\dagger}\bm{L}
\eeq
Writing
\beq 
\bm{L} &=& A+iB \\
C &=& i[A,B] \\
D &=& (1/2)\{A,B\} 
\eeq
and using the identity 
\beq 
[A,\{B,\rho\}] = \frac{1}{2}[D,\rho] -\frac{i}{2}\{C,\rho\} + A\rho B -B\rho A
\eeq
we get the following optional expressions for the Lindblad term:
\beq 
\text{Lindblad} &=& -i[D,\rho] -\frac{1}{2}\{A^2+B^2,\rho\} + A\rho A + B\rho B - i[A,\{B,\rho\}]
\\ 
&=& -i[D,\rho] - \frac{1}{2}[A,[A,\rho]] - \frac{1}{2}[B,[B,\rho]] - i[A,\{B,\rho\}]
\eeq
The first term represents ``Lamb shift", the second and the third are "noise" induced diffusion terms, 
and the last is the ``friction" term.

\sheadC{System-bath interaction} 

In the following presentation we assume that the full Hamiltonian is 
\beq
\mathcal{H}_{\tbox{total}} \ \ = \ \ \mathcal{H} - W F + \mathcal{H}_{\tbox{bath}}
\eeq 
where $W$ and $F$ are system and bath operators respectively. 
Neglecting the interaction, the bath is characterized by the spectral function
\beq
\tilde{C}(\omega) \ \ = \ \ \text{FT}\Big[ \langle F(t)F(0) \rangle \Big]
\eeq 
and the convention ${\langle F(t) \rangle=0}$. Whether the bath is composed 
of harmonic oscillators or not is regarded by the working hypothesis as not important.
There is a well known discussion of this point in Feynman-Vernon paper.

The spectral function $\tilde{C}(\omega)$ is characterized 
by temperature~$T$ and by a cutoff frequency $\omega_c$.
The latter is assumed below to be large compared with any other temporal frequency.   
What we call "noise" means $\tilde{C}(-\omega)=\tilde{C}(\omega)$. 
What we call "finite temperature" means 
\beq
\tilde{C}(-\omega)/\tilde{C}(\omega) \ \ = \ \ \exp(-\omega/T)
\eeq 
What we call "white noise" or "infinite temperature Ohmic bath" 
corresponds to $\tilde{C}(\omega)=\nu$, leading to 
\beq
C(t) \ \ = \ \ \langle F(t)F(0) \rangle \ \ = \ \ \nu \delta(t)
\eeq 
What we call ``high temperature Ohmic bath" takes into account 
that $\tilde{C}(\omega)$ possesses an antisymmetric component, 
which is implied by the Boltzmann ratio. Namely, in order to 
have the Boltzmann ratio to leading order in $\omega$ we have 
to add to $\nu$ an antisymmetric term  $\nu \times [\omega/(2T)]$. 
Consequently   
\beq
C(t) \ \ = \ \ \langle F(t)F(0) \rangle \ \ = \ \ \nu \delta(t) + i\eta \delta'(t)
\eeq 
where $\eta=\nu/(2T)$ is the so called friction coefficient. 
If we want to have from first principles an expression 
that holds for arbitrary $\omega$, we can model the bath 
as a collection of harmonic oscillators with spectral density $J(\omega)$. 
Then we get (see ``the modeling of the environment" lecture):
\beq
\tilde{S}(\omega) \ \ = \ \ 2J(\omega)\frac{1}{1-\eexp{-\omega/T}} 
\eeq
This expression is consistent with the above definition of Ohmic bath
provided we set ${J(\omega)= \eta \omega}$. For ${W:=x(t)}$,  
meaning that the bath is driven by a particle that has a given velocity $\dot{x}$, 
we get from the fluctuation-dissipation relation (or from a direct calculation) 
that the response of the bath is ${ \braket{F} = -\eta \dot{x} }$.

For a general bath, non-necessarily Ohmic,  
it is useful to define a bath spectral function 
via a Fourier-Laplace transform
\beq
G(\omega) \ \equiv \ \int_0^{\infty} C(t) e^{-i\omega t} dt  \ \equiv \ \frac{1}{2}\tilde{C}(\omega) - i \Delta(\omega)
\eeq
It is also useful to look on $W$ in the interaction picture:   
\beq
W(t) \ = \  e^{i\mathcal{H} t} W e^{-i\mathcal{H} t}
\ = \ \sum_{n,m} |n\rangle W_{nm} \eexp{i(E_n-E_m)t} \langle m|
\ \equiv \ \sum_{\Omega}\eexp{-i\Omega t} W_{\Omega} 
\eeq
We can say that the unperturbed system Hamiltonian $\mathcal{H}$ 
induces spectral decomposition $W=\sum_{\Omega} W_{\Omega}$ of the interaction. 
For non-degenerated spectrum ${ W_0 = \sum_n \ket{n}\bra{n} }$ 
is the diagonal part of the $W$-matrix in the energy representation, 
and each ${W_{\Omega}^{\dag} = W_{-\Omega}}$ with $\Omega\ne 0$ corresponds 
to a pair of coupled levels.
Additionally it is useful to define
\beq
\tilde{W} \ \equiv \ \int_0^{\infty} C(t) W(-t) dt  \ = \ \sum_{\Omega} G(\Omega) W_{\Omega} 
\eeq  
Coming back to the Ohmic case, 
it is useful to define a ``velocity" operator $V=i[\mathcal{H},W]$. 
Accordingly, in the Ohmic case, we get
\beq
\tilde{W} \ \ \approx \ \ \frac{\nu}{2} \left(W + i \frac{\eta}{\nu}V \right) 
\eeq
The notations above are useful for the purpose of writing down the Master equation 
for the time evolution of the reduced probability matrix.

\sheadC{The Redfield master equation} 

We first demonstrate the derivation of the Master equation in the case of white noise. 
The Hamiltonian is ${\mathcal{H}(t)=\mathcal{H}+f(t)W}$, 
were  $f(t)$ represents white noise: that means that upon 
ensemble average ${\langle f(t)\rangle=0}$, 
while ${\langle f(t)f(t') \rangle = \nu\delta(t-t')}$. 
Given $\rho(t_0) \equiv \rho$, the Liouville von-Neumann equation 
can be solved iteratively to determine $\rho(t_0{+}dt)$, 
where $dt$ is a small time interval. Without loss of generality 
we set ${t_0{=}0}$ and ${t=t_0{+}dt}$ and get:
\beq
\rho(t) \ \ = \ \ \rho -i\int_{0}^{t} dt' \, [\mathcal{H}(t'), \rho] 
-\int_{0}^{t} \int_{0}^{t'}  dt'dt''  \, [\mathcal{H}(t'), [\mathcal{H}(t''), \rho]] + ... 
\eeq
Averaging over realizations of $f(t)$ all the odd orders in this expansion 
vanish, while the leading $dt$ contribution comes only from the 
zero order term that involves $\mathcal{H}$ 
and from the second order term that involves~$W$.
Consequently we get the following Master equation:
\beq 
\frac{d\rho}{dt} 
\ \  = \ \  -i[\mathcal{H},\rho] - \frac{1}{2}\nu[W,[W,\rho]]
\ \  = \ \  -i[\mathcal{H},\rho] - \frac{1}{2}\{\Gamma,\rho\}  + \nu W\rho W 
\eeq
where $\Gamma=\nu WW$. Note that the first two terms in the second expression 
generate so called non-Hermitian dynamics with the 
effective Hamiltonian ${\mathcal{H}_{\tbox{eff}} = \mathcal{H}-(i/2)\Gamma}$, 
while the last term represents ``continuous measurement".

The generalization of the "white noise" derivation for 
a system that is coupled to a high temperature Ohmic bath 
is straightforward. It is based on the assumption that at 
any moment the system-bath state is "factorized", 
which can be justifies if $\omega_c^{-1}$ is sufficiently small.
We define the interactions-representation of the probability matrix 
via ${\rho(t) \equiv U(t) \tilde{\rho}(t) U(-t) }$ where ${U(t)= \eexp{-i\mathcal{H}t}}$.
The iterative procedures provides for $\tilde{\rho}(t)$ the same expansion 
as in the previous subsection with $\mathcal{H}(t)$ 
replaced with $F(t)W(t)$, where ${W(t)=U(-t)WU(t)}$. 
Consequently we get the so-called Redfield equation
\beq 
\frac{d\rho}{dt} 
\ \  = \ \  -i[\mathcal{H},\rho] 
+ \tilde{W} \rho W  + W \rho \tilde{W}^{\dag}  
- W\tilde{W}\rho - \rho \tilde{W}^{\dag}W
\eeq
Note that an optional style of writing this expression is with ${-[W,\, \tilde{W}\rho-\rho\tilde{W}^{\dag}]}$, 
which reduces to  ${-[W,\, [\tilde{W},\rho]]}$ for an hermitian $\tilde{W}$. 
If instead one substitutes the non-hermitian Ohmic expression for $\tilde{W}$, 
one obtains the Ohmic master equation, that contains both noise and friction terms.   
As noted above, one can add to the Ohmic master equation a term $[V,[V,\rho]]$ that represents 
an extra white noise coupled via~$V$.

\sheadC{The secular approximation} 

We come back one step, and consider again general bath, not necessarily Ohmic.
Instead of assuming small correlation time, we shall assume weak interaction. 
Specifically, in atomic physics applications the induced rate of transitions~$w$ 
becomes much smaller compared with the Rabi-Bloch frequency~$\Omega$ of the coherent oscillations. 
Accordingly it is appropriate to write that master equation in the interaction picture:
\beq 
\frac{d\tilde{\rho}}{dt} 
\ \  = \ \ 
\tilde{W}(t)\tilde{\rho} W(t)  + W(t)\tilde{\rho}\tilde{W}(t)^{\dag} 
- W(t) \tilde{W}(t) \tilde{\rho} - \tilde{\rho} \tilde{W}(t)^{\dag}W(t)
\eeq
Substitution of the $\mathcal{H}$-induced spectral decomposition of the $W$-s one observes 
terms that oscillate with frequencies $\Omega+\Omega'$. 
We keep only the terms that oscillate with $\sim 0$ frequency, and hence do not average to zero. 
For example, in ${\tilde{W} \rho W }$ we keep only the ${ G(\Omega) W_{\Omega} \rho W_{-\Omega} }$ terms. 
Consequently we obtain the so called secular approximation
\beq 
\frac{d\rho}{dt} 
\ \  = \ \  -i[\mathcal{H},\rho] 
+\sum_{\Omega} \left[
\tilde{C}(\Omega) \ W_{\Omega} \rho W_{\Omega}^{\dag}
- G(\Omega) \ W_{\Omega}^{\dag}W_{\Omega} \rho 
- G(\Omega)^* \ \rho W_{\Omega}^{\dag}W_{\Omega}
\right]
\eeq
The imaginary part of $G(\Omega)$, aka Lamb shift, 
can be absorbed into the Hamiltonian $\mathcal{H}$, 
so we end up with a simple sum over Lindblad terms 
that are weighted by the spectral intensities $\tilde{C}(\Omega)$, namely, 
\beq 
\frac{d\rho}{dt} 
\ \  = \ \  -i[\mathcal{H}_{\text{eff}},\rho] 
+\sum_{\Omega} 
\tilde{C}(\Omega) 
\left[ W_{\Omega} \rho W_{\Omega}^{\dag}
- \frac{1}{2} \{ W_{\Omega}^{\dag}W_{\Omega}, \rho\} 
\right]
\eeq
In particular one should distinguish the ${\Omega \ne 0}$ terms that induce inter-level transitions 
from the ${\Omega=0}$ term that commutes with the Hamiltonian.

\sheadC{The Pauli master equation} 

For a system that has no degeneracies (for example a few-level atom) it is natural    
to write the secular equation in the $\mathcal{H}$ basis. 
One realizes that the dynamics of the diagonal elements decouples 
from that of the off-diagonal elements.   
Namely, the first term in the secular approximation induces FGR transitions with rates 
\beq
w_{nm} \ \ = \ \  \tilde{C}(-(E_n{-}E_m)) \ |W_{nm}|^2
\eeq
The corresponding decay constants are $\Gamma_n=\sum_m' w_{nm}$. 
The ratio $w_{nm}/w_{mn}$ is not unity unless we consider 
white noise source (infinite temperature). 
For finite temperatures the FGR rates favor downwards transitions.
Consequently we get the so-called Pauli rate equation 
for the probabilities $p_n$  
\beq 
\frac{d\bm{p}}{dt} \ = \ \mathcal{W} \ \bm{p},
\ \ \ \ \ \ \ \ \ \ \ \ 
\mathcal{W} = \left( 
\amatrix{
-\Gamma_1 & w_{12} & ... \cr
w_{21} & -\Gamma_2 & ... \cr
... & ... & ...
} 
\right)
\eeq
For the off-diagonal terms we get 
\beq 
\frac{d\rho_{nm}}{dt} \ \ = \ \ \Big[-i(E_n-E_m) - \gamma_{nm}\Big]\rho_{nm}, 
\hspace*{2cm} \mbox{[for $n\ne m$]}
\eeq  
with dephasing rates  
\beq 
\gamma_{nm} \ \ = \ \ 
\frac{\nu}{2}|W_{nn}-W_{mm}|^2 + \frac{1}{2}(\Gamma_{n}+\Gamma_{m})  
\ \ \equiv \ \  \gamma_{\varphi}  + \Gamma_{\text{rlx}} 
\eeq  
where the first term originates from the ${\Omega=0}$ generator, 
while the second term originates from the  ${\Omega \ne 0}$ transitions.   
We note that the above results can by derived from heuristic consideration,  
without going through the heavy machinery of the master equation formalism.
Taking the white noise master equation as a starting point, it is enough 
to realize that the elements of $\rho_{nm}$ can be classified according 
to their unperturbed frequencies $(E_n-E_m)$.
Elements that are oscillating with different frequencies, have a negligible cross interaction. 
In particular the dynamics of the $p_n$, that have $\sim 0$ frequencies, 
decouple from the dynamics of the off-diagonal elements, leading to FGR picture of transitions. 
For the off diagonal terms the reasoning is similar, and there is an additional 
dephasing $\gamma_{\varphi}$ due to the noisy detuning.
The finite temperature case is merely a variation on the same reasoning.

\sheadC{Damped harmonic oscillator} 

Recall that for a damped particle with coupling $-\bm{x} F$, 
an Ohmic bath has the spectral function ${J(\omega)=\eta\omega}$, 
such that the power spectrum of the fluctuations is 
${S(\omega) = 2J(\omega) [1+f(\omega)]}$ for ${\omega>0}$, 
and  ${S(\omega) = 2J(|\omega|) f(|\omega|) }$ for ${\omega<0}$, 
where ${f(\omega)= 1/(e^{\omega/T}-1)}$. For a particle of mass $\mass$ such bath 
produces friction that leads to damping rate ${\gamma=\eta/\mass}$. 

Consider the the case of damped Harmonic oscillator of frequency~$\Omega$
and damping rate ${\gamma}$. Here it is customary to write 
the interaction as ${ (2\mass\Omega)^{-1/2} [\bm{a}+\bm{a}^{\dag}] F}$. 
Accordingly the $W_{\Omega}$ operators are  ${ (2\mass\Omega)^{-1/2} \bm{a}}$ and its conjugate.  
In the secular approximation we get the master equation 
\beq 
\frac{d\rho}{dt} 
\ \  = \ \  -i [\Omega \bm{a}^{\dag}\bm{a},\rho] 
+ \gamma (1+f(\Omega)) \left[ \bm{a}\rho\bm{a}^{\dag} - \frac{1}{2} \{\bm{a}^{\dag}\bm{a},\rho \}  \right]  
+ \gamma f(\Omega) \left[ \bm{a}^{\dag}\rho\bm{a} - \frac{1}{2} \{\bm{a}\bm{a}^{\dag},\rho \}\right] 
\eeq
The adjoint equation for the expectation value of ${ \bm{n}=\bm{a}^{\dag}\bm{a} }$
implies relaxation towards equilibrium with damping rate~$\gamma$, 
namely, ${(d/dt)\braket{\bm{n}} = -\gamma [\braket{\bm{n}} - f(\Omega)] }$.

\sheadC{The Bloch equation} 

Let us consider a two level system.
The probability matrix is conveniently expressed 
using the Bloch vector ${\vec{S}=(S_x,S_y,S_z)}$, 
were ${S_j = \braket{\bm{\sigma}_j} }$, namely, 
\beq
\rho(t) \ \ = \ \ \frac{1}{2}\Big(1+ S_x \bm{\sigma}_x + S_y \bm{\sigma}_y + S_z \bm{\sigma}_z \Big)
\eeq 
Note that ${S_z=p_{+}-p_{-}}$ is the population probability difference, 
while $S_x$ and $S_y$ are the so called "coherences".
Using the adjoint equation one can easily show that the equation 
of motion for the Bloch vector takes the form 
\beq
\frac{dS}{dt} \ \ = \ \ -\vec{\Omega} \times S  \ - \ \gamma \, (S-S_{\text{eq}}) 
\eeq
where $S$ is regarded as a column vector, 
and ${\gamma=\text{diag}(\gamma_x,\gamma_y,\gamma_z)}$ is a diagonal matrix. 
The first term is generated by the unperturbed Hamiltonian:
We assume ${\mathcal{H}=-(\Omega/2)\sigma_z}$, hence ${\vec{\Omega}=(0,0,\Omega)}$.

We consider the effect of having a coupling term $-WF(t)$, 
where $F(t)$ represents a bath or a noise source. 
Even without going through the master equation formalism 
it is clear that consistency with the canonical formalism 
implies that the equilibrium states is 
\beq
S_{\text{eq}} \ \ = \ \ \left(0,0, \tanh\left(\frac{\Omega}{2T}\right) \right)
\eeq 
We now refer separately to different versions of the Bloch equation.
The different versions are distinguished by the assumptions regarding $W$,  
the intensity $\nu$ of the $F(t)$ fluctuations, and their spectral characteristics.

{\bf Pure dephasing.-- } 
The simplest possibility is to have a so-called pure dephasing effect 
due to a ${W=\sigma_z}$ interaction with a white noise source 
that has an intensity $\nu_{\varphi}$. In the master equation 
it introduces a diffusion term ${(\nu_{\varphi}/2)[W,[W,\rho]]}$.
The implication is to have in the Bloch equation 
\beq
\gamma[\text{Dephasing}] \ \ = \ \ \text{diag}\left(2\nu_{\varphi},2\nu_{\varphi},0\right)
\eeq
The interaction with the noise commutes with ${\mathcal{H}}$ 
therefore there is no equilibration in the $S_z$ direction. 
For this reason if we replace the noise source by a finite 
temperature bath, it will have a similar effect.

{\bf Ohmic version.-- } 
Next in complexity is to consider a high temperature Ohmic bath coupled via $W=\sigma_x$. 
Using the notations of the previous sections we have here a "position" 
coordinate ${W=\sigma_x}$ and a conjugate "velocity" coordinate ${V=\Omega \sigma_y}$.
Consequently, after some straightforward algebra we deduce that
\beq
\gamma[\text{Ohmic}] \ \ = \ \ \text{diag}\left(0,2\nu,2\nu\right)
\eeq
Due to the lack of commutation we have an additional "friction" 
term $2\eta\Omega$ in the master equation for $dS_z/dt$,  
which implies ${S_{eq}=(0,0,\eta\Omega/\nu)}$. 
This is consistent with the canonical expectation,
upon the substitution $\nu/\eta=2T$,   
provided the condition ${(\Omega/T) \ll 1}$ is satisfied. 
This is the regime where the high temperature Ohmic approximation is valid.

The dephasing in the Ohmic version of the Bloch equation is 
non-isotropic in the transverse XY plane.  
Note that the $S_y$ transverse component satisfy the equation 
\beq 
\ddot{S}_y + 2\nu \dot{S}_y + \Omega^2 S_y = 0 
\eeq
which leads to damped frequency $\Omega_{\tbox{eff}}=\sqrt{\Omega^2-\nu^2}$. 
In the secular and NMR versions that we discuss in the next paragrpahs the 
dephasing is isotropic in the XY plane and therefore $\Omega$ is not affected.

{\bf Secular version.-- } 
We now consider what comes out, for the same coupling, within the 
framework of the secular approximation. Note that this 
approximation, unlike the high temperature Ohmic version, 
assumes large $\Omega$. Using the Pauli equation prescription 
we realize that the FGR average transition rate is $\nu$. Hence we get
\beq
\gamma[\text{Secular}] \ \ = \ \ \text{diag}\left(\nu,\nu,2\nu\right)
\eeq
One observes that due to the perturbative nature 
of this approximation the transverse relaxation looks 
isotropic. Disregarding this artifact, 
one should keep in mind that the secular approximation 
allows to consider the case of non-Ohmic bath.
From the general derivation it should be realized 
that $\nu$ in the above equation is determined exclusively  
by the $\tilde{C}(\pm\Omega)$ components of the fluctuations.

{\bf NMR version.-- }
The so called nuclear-magnetic-resonance version 
of the Bloch equation consider a general $W$. 
Formally it is like to add to the secular version 
of the previous paragraph an additional pure dephasing effect.  
Accordingly we write the Bloch equation as 
\beq 
\frac{dS_z}{dt} &=& -\frac{1}{T_1}(S_z-S_{\tbox{eq}}) \\
\frac{dS_{x,y}}{dt}  &=& -[\Omega \times S]_{x,y} -\frac{1}{T_2}S_{x,y} 
\eeq
where the equilibrium value is  
\beq
S_{\tbox{eq}} \ \ = \ \ \frac{w_{+-}-w_{-+}}{w_{+-}+w_{-+}} 
\eeq
as in the secular version.
The rates for the diagonal relaxation and 
for the off-diagonal transverse depahsing are:
\beq 
\frac{1}{T_1} &=& w_{+-}+w_{-+} \ \ \equiv \ \ \gamma_{\text{rlx}} \\
\frac{1}{T_2} &=& \frac{\gamma_{\text{rlx}}}{2} + \gamma_{\varphi} \ \ \equiv \ \ \frac{\gamma}{2}
\eeq  
The pure dephasing rate $\gamma_{\varphi}$ originates 
from the diagonal elements of $W_{nm}$ and hence is formally 
proportional to the intensity $\tilde{C}(0)$, 
while the FGR transition rates originate from the off-diagonal 
elements of $W_{nm}$, and hence are proportional to $\tilde{C}(\pm\Omega)$, 
were $\Omega=|E_+-E_-|$ is the level spacing.

\sheadC{Dicke super-radiance}

Consider $N$ two-level atoms that each of then interact with a local bath, 
namely the interaction term is ${ \sum_j (1/2)\bm{\sigma}_j^x F_j(t) }$. 
The term ``bath" refers here to modes of the electromagnetic field.  
Each of the atoms satisfies a Bloch equation. 
If we sum over over all the Bloch equations we get 
an equation for the expectation value of ${ \bm{S} = \sum_j (1/2)\bm{\sigma}_j }$. 
At zero temperature we have only spontaneous emissions and the equation takes the form 
${ (d/dt) S^z = -\gamma [S^z-S] }$ where ${S = (N/2)}$. 
Note that the zero temperature equilibrium state of all spins "up" 
corresponds in our convention to having all the atoms in the lower level.

If the atoms are packed densely, such that all of them interact with the same bath-modes, 
the interaction term takes the form ${ \sum_j (1/2)\bm{\sigma}_j^x F(t) }$, 
which equals $S^z F(t)$. In the secular approximations we keep 
only the interaction with ${W=S^{+} = (S^z+iS^y)}$. Working out the Lindblad term 
we get the modified Bloch equation 
\beq
\frac{d}{dt} S^z = - \gamma \left[ (1+S^z)S^z - (1+S)S \right] 
\eeq
Note that for ${N=1}$ this is the regular Bloch equation. 
But for ${N\gg 1}$ it can be approximated by
\beq
\frac{d}{dt} S^z =  - \gamma \left[ S_z^2 - \left(\frac{N}{2}\right)^2 \right]
\eeq
Assuming that we start with excited atoms (all spins "down"),  
the rate of decay accelerates, and enhanced by factor $N^2$ during 
the time when ${S^z \sim 0 }$. This is in contrast 
with normal uncorrelated decay where the enhancement factor is $N$.    
The explicit solution of this equation is       
\beq
S^z(t) \ \ = \ \ \frac{N}{2} \tanh\left[ \frac{N}{2} \ \gamma \ (t-t_0)\right] 
\eeq  
where $t_0$ is the time when ${S^z}$ crosses zero, and the emission rate attains 
its maximal super-raddiance value.

\sheadC{The Bloch equations in Laser physics}

The minimal model for a Laser consist of cavity mode that has frequency $\Omega$, 
and $N$ two-level atoms (below for simplicity ${N=1}$) 
that each of them has excitation energy $\mathcal{E}$. 
The cavity mode is like damped harmonic oscillator, 
because it can leak outside with rate $\kappa$, 
and the atoms can decay with rate $\gamma_{\downarrow}$, 
but are also pumped with rate $\gamma_{\uparrow}$. 
We define ${\gamma=\gamma_{\uparrow}+\gamma_{\downarrow}}$ and ${f=\gamma_{\uparrow}-\gamma_{\downarrow}}$.
We also define ${\gamma_{\perp} = \gamma + \gamma_{\varphi} }$ which includes 
an optional pure dephasing effect.  Without the $\kappa$ and the $\gamma$-s 
the system is described by the Hamiltonian 
\beq
\mathcal{H} \ \ = \ \ \Omega \bm{a}^{\dag}\bm{a} +  \frac{\mathcal{E}}{2} \bm{\sigma}^z + g (\bm{a}^{\dag}+\bm{a}) \bm{\sigma}^x  
\eeq
where $g$ is the coupling constant, ans in the so-called Rabi model.  
With the dissipation terms we can derive semi-classical equations 
that couple the Bloch dynamics to the damped oscillator:     
\beq
\frac{d}{dt} \bm{a} &=& -\left( i\Omega + \frac{\kappa}{2} \right)\bm{a}  - ig\bm{\sigma}^x
\\
\frac{d}{dt} \bm{c} &=& -\left( i\mathcal{E} + \frac{\gamma_{\perp}}{2} \right)\bm{c}  + i g (\bm{a}^{\dag}+\bm{a})\bm{\sigma}^z
\\
\frac{d}{dt} \bm{\sigma}^z &=& -\gamma \bm{\sigma}^z + f + 2g (\bm{a}^{\dag}+\bm{a})\bm{\sigma}^y
\eeq
Above we defined the lowering operator ${\bm{c} = (1/2)[\bm{\sigma_x} - i \bm{\sigma_y}]}$, 
such that ${\bm{\sigma}_x = \bm{c}^{\dag}+\bm{c} }$. For $N$ atoms the equations are 
written with ${ \bm{S} = \sum_j (1/2)\bm{\sigma}_j }$. For ${g=0}$ the equations for $\bm{S}$ 
are the standard Bloch equations with steady state at ${S^z=(N/2)[f/\gamma]}$. Below we keep $N{=}1$.   
In the absence of driving (${\gamma_{\uparrow} =0}$) the system relaxes to the normal ground state $(a{=}0, \sigma_z{=}-1)$ provided ${g< \sqrt{\Omega\mathcal{E}}/2 }$. Otherwise it relaxes to a so-called super-radiant ground state with ${ a \ne 0 }$. The term ``super-radiant" is a bit misleading here - there are no oscillations, and therefore no radiation is emitted once the equilibrium is reached. In order to have lasing $f$ should be large enough. Above a threshold value the steady state is a non-equilibrium limit-cycle (NELC), aka the lasing state. In order to find the NELC it is convenient to transform the equations into a ``rotating frame" such that ${\mathcal{E} \sim \Omega \mapsto 0}$. Counter-rotating (non-resonant) terms in the Rabbi interaction term are neglected (so-called Tavis-Jaynes-Cummings approximation). Reduced equations are obtained for the variables ${\bm{n}=\bm{a}^{\dag}\bm{a}}$ and ${\bm{S}^z=(1/2)\bm{\sigma}^z}$, namely 
\beq
\frac{d}{dt} \bm{n} &=& -\left[\kappa  - 2 \frac{g^2}{\gamma_{\perp}}\bm{S}^z \right] \bm{n}
\\
\frac{d}{dt} \bm{S}^z &=& - \left[ \gamma  + 2\frac{g^2}{\gamma_{\perp}} \bm{n} \right] \bm{S}^z + \frac{f}{2}
\eeq
Note that ${\bm{S}^z+\bm{n}}$ is a constant of motion due to the rotating wave approximation.
From ${(d/dt)\bm{n}=0}$ it follows that at steady state either ${\bm{n}=0}$  or ${ \bm{S}^z = (1/2)(\kappa\gamma_{\perp}/g^2)}$. Then from ${(d/dt)\bm{S}^z=0}$ it follows that at steady state 
\beq
\bm{n} [\text{SS}] \ \ = \ \ \frac{f}{2\kappa} - \frac{\gamma\gamma_{\perp}}{2g^2}
\eeq 
The threshold condition $f > \kappa \gamma\gamma_{\perp}/g^2$ to get lasing is implied by positivity of the RHS. Below threshold the attractor is the trivial fixed point at ${\bm{n}=0}$.

\sheadC{Many body rate equations}

In the simplest approximation quantum master equation are approximated by Pauli master equation with Fermi-Golden-Rule rates. In the many body context it is more convenient to consider the adjoint equations, which are  the equations of motion for the expectation values, and possibly for higher moments. Below we consider consider many body rate equations. By this we mean equations of motion for the expectation values of the occupation operators, namely, ${ n_j \equiv \braket{\bm{n}_j} }$. 
For a closed system ${\sum_j n_j = N}$ is a constant of motion.

Recall that the dynamics of a single particle is described by a master equation 
${(d/dt) p_j = \sum_i [I_{i \to j} - I_{j \to i}] }$ 
where the probability current from orbital~$i$ to orbital~$j$
is ${I_{i \to j} = w_{ji} p_i }$. 
If the transitions are induced by a heat bath 
we have 
\beq
\frac{w_{ij}}{w_{ji}} \ \  =  \ \ \exp[-(\varepsilon_i-\varepsilon_j)/T] 
\eeq
This implies that the system relaxes to a canonical equilibrium.     
The simplest many-body variation is to consider a system of classical non-interacting particles.
The adjoint equation for the occupations is 
\beq
\frac{d}{dt} n_j \ \ =  \ \ \sum_i [I_{i \to j} - I_{j \to i}] 
\eeq
where the current of particles that are transported from orbital~$i$ to orbital~$j$ 
is ${I_{i \to j} = w_{ji} n_i }$. 
If the transitions are induced by the same heat bath as in the single particle problem, 
the detailed balance condition ${[I_{i \to j} - I_{j \to i}] = 0}$ 
implies ${n_i/n_j = \exp[-(\varepsilon_i-\varepsilon_j)/T )] }$,  
and therefore we get the Boltzmann distribution ${n_j = f(\varepsilon_j-\mu) }$, 
where $\mu$ is determined by~$N$.  

{\bf Bosons / Fermions.-- }
Consider a system of Bosons or a system of Fermions. The transitions 
are induced by a bath that couples to operators 
that induce hopping, namely, ${ \bm{a}_j^{\dag} \bm{a}_i }$. 
The Fermi-Golden-Rule implies that the  current of particles 
that are transported from orbital~$i$ to orbital~$j$ is   
\beq
I_{i \to j} \ \ = \ \ w_{ji} \ (1\pm n_j) \ n_i   
\eeq
In order to address all possibilities in a compact way we can interpret $\pm$ 
as "0" for classical particles, "+1" for Boson, and "-1" for Fermion.
The detailed balance condition ${[I_{i \to j} - I_{j \to i}] = 0}$ 
is satisfied by the Boltzmann / Bose / Fermi distributions respectively, 
namely  ${n_j = f(\varepsilon_j-\mu) }$, where $\mu$ is determined by~$N$. 
To prove this statement note that the respective distribution functions 
satisfy the identity 
\beq
\frac{f(\omega)}{1 \pm f(\omega)} \ \ = \ \ \exp[-\omega/T] 
\eeq

{\bf Condensation.-- }
Here it is appropriate to recall that condensation of Bosons in the ground orbital 
is implied if $N$ is large. From dynamical point of view the ground orbital, 
labeled by "0", is characterized by ${w_{0,j} > w_{j,0}}$ for any~$j$. 
The Bose function is finite for any ${\varepsilon_j>\mu}$, 
and any excess amount of particles forces ${\mu = \varepsilon_0}$, 
such that ${i=0}$ can accommodate an arbitrary large number of them.

\sheadA{Additional topics}

\makeatletter{}
\sheadB{The kinetic picture}

\sheadC{The Boltzmann distribution function}

The number of one particle states within 
a phase space volume is ${d\mathcal{N}=d^3rd^3p/(2\pi\hbar)^3}$. 
The occupation of this phase space volume is:
\beq 
dN \ \ \equiv \ \  f(\bm{r},\bm{p}) \frac{d\bm{r}d\bm{p}}{(2\pi\hbar)^3}  
\eeq
where $f(\bm{r},\bm{p})$ is called Boltzmann distribution function. 
In equilibrium we have 
\beq
f(\bm{r},\bm{p}) \Big|_{\tbox{eq}} \ \ = \ \ f_{\beta}(\epsilon_{\bm{p}}-\mu) 
\eeq 
where $f_{\beta}(\epsilon-\mu)$ is either the Bose or the Fermi 
occupation function, or possibly their Boltzmann approximation.
If we use ${(\bm{r},\bm{v})}$ with measure $d^3rd^3v$ 
instead of ${(\bm{r},\bm{p})}$ we have 
\beq
f(\bm{r},\bm{v}) \ \  = \ \ \left(\frac{\mass}{2\pi}\right)^3  f(\bm{r},\bm{p}) 
\eeq 
By integrating over $\bm{r}$ and over all directions 
we get the velocity distribution
\beq
F(v) = L^3 \times 4\pi v^2  \left(\frac{\mass}{2\pi}\right)^3 
f_{\beta}\left(\frac{1}{2}\mass v^2 - \mu\right)
\eeq
If we use Boltzmann approximation for the occupation function 
and express $\mu$ using $N$ and $T$ we get 
\beq
F(v) = N \left(\frac{\mass}{2\pi T}\right)^{{3}/{2}}
\ 4\pi v^2 \ \eexp{-\frac{1}{2}\mass v^{2}/T}
\eeq
We note that
\beq
N 
\ = \ \iint \frac{d\bm{r}d\bm{p}}{(2\pi\hbar)^3} f(\bm{r},\bm{p})
\ = \ \int d\epsilon\gdos(\epsilon)f_{\beta}(\epsilon-\mu)
\ = \ \int F(v) dv
\eeq

\sheadC{The Boltzmann equation}

The Liouville equation for $\rho(r,p)$ is merely a continuity equation   
in phase space. It can be written as $(d/dt)\rho=0$ where $d/dt$ 
unlike $\partial/\partial t$ is the total derivative reflecting the 
change in the occupation of a phase space cell.   
The Boltzmann equation for $f(r,p)$ is formally identical 
to the Liouville equation in the absence of collisions,  
and with collisions becomes $(d/dt)\rho = g(r,p)$, 
where $g(r,p)$ is the net rate in which particles are 
generated at $(r,p)$ due to collisions. Accordingly the Boltzmann 
equation is
\beq
\Big[\frac{\partial}{\partial t} + v(p)\cdot\frac{\partial}{\partial x} + \mathcal{F}(r) \cdot \frac{\partial}{\partial p} \Big] 
\ f(r,p) \ \ = \ \ g_{[f]}(r,p)
\eeq
where $v(p)=p/\mass$ is the dispersion relation, and $\mathcal{F}(r)=-V'(r)$ 
is the force due to some external potential. The notation emphasizes that $g(r,p)$  
is a functional of the~$f$ distribution. It can be written as a difference 
of ingoing and outgoing fluxes due to collisions.    
A distribution that gives zero in the left hand side of the Boltzmann equation
is called "ergodic". A distribution that gives ${g=0}$ at any point is "locally equilibrated".
If a locally equilibrated distribution is ergodic it constitutes an equilibrium solution   
of the Boltzmann equation. If there is no such solution, one can look 
for a non-equilibrium steady-state (NESS) solution. For example, assume that $g$ is different 
in two regions of space reflecting the presence baths with different temperatures; 
In such case there exists a non-ergodic NESS solution that features 
a non-zero heat transport through the system.    

The standard Boltzmann expression for $g(r,p)$ is based on 2body collision mechanism and "molecular chaos" assumption.
See chapter~4 of {\bf Huang} or chapter~14 of {\bf Reif}. Assuming that collisions 
from ${(p,p_0)}$ to ${(p_1,p_2)}$ has the same rate as that of the inverse process, 
it takes the following form: 
\beq
g_{[f]}(r,p) \ \ = \ \ 
\int \frac{dp_0}{2\pi} \int \frac{dp_1}{2\pi} \frac{dp_2}{2\pi}
\ w(p,p_0|p_1, p_2) \ 
\left[ f(r,p_2)f(r,p_1) - f(r,p_0)f(r,p) \right]
\eeq
The gas reaches a steady state in accordance with the Boltzmann $H$ theorem.   
The formal solution for the steady state implies the Maxwell-Boltzmann distribution
for the velocities.
A much simpler expression for $g(r,p)$ appears while discussing 
electronic transport. See chapters 16 and 13 of {\bf Ashcroft \& Mermin}. 
Here the scattering mechanics is 1body collisions of the electrons 
with the lattice, leading to 
\beq
g_{[f]}(r,p) \ \ = \ \ \int \frac{dp'}{2\pi} \left[w_{p,p'} f(r,p') - w_{p',p} f(r,p)\right]
\eeq
Note that the first term, that corresponds to electrons that are scattered 
out of the phase-space cell can be written as ${-(1/\tau)f}$ 
where $(1/\tau) \sim \mathcal{N} w$ is the decay rate to the 
other $\mathcal{N}$ cells to which it is connected.  
If we assume that $f$ is close to an equilibrium solution $f_{\beta}$, 
it follows that we can approximate $f \approx f_{\beta}$ 
in the ingoing flux term. Hence we get the so called relaxation time approximation: 
\beq
g_{[f]}(r,p) \ \ \approx \ \ \frac{f_{\beta}(\epsilon(p)-\mu)-f(r,p)}{\tau}
\eeq 
NESS is reached if $\beta(r)$ or $\mu(r)$ 
or the potential $V(r)$ are non-uniform in space. 
Using the relaxation time approximation the solution  
that can be written schematically as follows:
\beq
f(t) \ \ = \ \ \int_{-\infty}^t \left[\frac{1}{\tau}\exp\left(-\frac{t-t'}{\tau}\right)\right]  \ f_{\beta}(t') \ dt'
\eeq
where $f_{\beta}(t')$ stands for $f_{\beta}$ 
that is evaluated at the point ${(r(t'),p(t'))}$, 
which is connected by an unscattered classical 
trajectory that ends at~${(r,p)}$ at time~$t$.
Assuming $\tau$ to be small one can easily  
obtain a first order solution. For example, 
in the presence of a constant field of force 
the NESS becomes 
\beq
f(r,p) \ \ \approx \ \  f_{\beta} - \tau \mathcal{F} v(p) \left(-\frac{ \partial f_{\beta}}{ \partial \epsilon} \right) 
\eeq
where $f_{\beta}(\epsilon(p))$ is the the equilibrium 
occupation function that is calculated in the absence 
of the field. The above NESS carries current whose density
can be written as ${J= \sigma \mathcal{F}}$, where 
\beq
\sigma \ \ = \ \ \int \frac{d^3p}{(2\pi)^3} \tau v^2 \left(-\frac{ \partial f_{\beta}}{ \partial \epsilon} \right)
\ \ = \ \ \int \frac{g(\epsilon)}{\Vol}  \tau v^2 \ \left[-f_{\beta}'(\epsilon-\epsilon_F) \right]
\ \ = \ \ \frac{g(\epsilon_F)}{\Vol} v_F^2 \tau
\eeq
The last equalities assume a zero temperature Fermi occupation.
If $g(\epsilon)$ corresponds to the standard dispersion relation,  
one obtains the Drude formula $\sigma=(N/\Vol)(\tau/\mass)$.

\newpage
\sheadC{The calculation of incident flux}

Given $N$ gas particles that all have velocity ${v}$ 
we can calculate the number of particles that hit 
a wall element per unit time (=flux), and also we can  
calculate the momentum transfer per unit time (=force). 
Using spherical coordinates, such that normal 
incidence is $\theta{=}0$,  
one obtains (per unit area):
\beq
J &=&    
\iint_{|\theta|<\pi/2} 
\left[\frac{d\Omega}{4\pi}\frac{N}{\Vol}\right]
\ v\cos(\theta)  
\ \ = \ \  
\left[\frac{1}{2}\int_0^{1} \cos(\theta) d\cos(\theta)\right] \frac{N}{\Vol} v
\ \ = \ \  
\frac{1}{4} \left(\frac{N}{\Vol}\right) v
\\
P &=&    
\iint_{|\theta|<\pi/2} 
\left[\frac{d\Omega}{4\pi}\frac{N}{\Vol}\right]
\ v\cos(\theta) \ 2\mass v\cos(\theta) 
\ \ = \ \
\left[\int_0^{1} \cos^2(\theta) d\cos(\theta) \right]\frac{N}{\Vol}\mass v^2 
\ \ = \ \ \frac{1}{3} \left(\frac{N}{\Vol}\right) \mass v^2
\eeq
If we have the distribution $F(v)$ of the velocities,
or optionally if  we are provided with the one-particle energy distribution, 
the total flux is given by an integral:
\beq
J_{\tbox{incident}}
\ \ = \ \ \int_{0}^{\infty}\frac{1}{4}\left(\frac{F(v)dv}{\Vol}\right)v
\ \ = \ \ \int_{0}^{\infty} \frac{1}{4}\left(\frac{\gdos(\epsilon)f(\epsilon)d\epsilon}{\Vol}\right)v_{\epsilon}
\eeq
Similar expression holds for the pressure $P$, 
where one can make the identification ${\epsilon = (1/2)\mass v_{\epsilon}^2}$, 
and recover the familiar Grand canonical result.

\sheadC{Blackbody radiation}

The modes of the electromagnetic field 
are labeled by the wavenumber $k$ and the polarization $\alpha$.
For historical reasons we use ${k}$ instead 
of ${p}$ for the momentum and ${\omega}$ instead of ${\epsilon}$ 
for the energy. The dispersion relation is linear $\omega=c|k|$. 
The density of modes is 
\beq
\gdos(\omega) d\omega 
= 2\times\frac{\Vol}{\left(2\pi c\right)^{3}} 4\pi \omega^2 d\omega
\eeq
Recall that the canonical state of oscillators can be 
formally regarded as the grand canonical equilibrium 
of $\mu=0$ Bose particles, with the occupation function   
\beq
\langle  n_{k\alpha} \rangle 
\ \ = \ \ \frac{1}{\eexp{\beta \omega_{k\alpha}}-1}
\ \ \equiv \ \ f\left(\omega_{k\alpha}\right)
\eeq
For the total energy~$E$ we have 
\beq
E \ \ = \ \ \int_0^{\infty} \omega d\omega \ \gdos(\omega) \ f(\omega) 
\ \ = \ \ \Vol \int_0^{\infty} d\omega \frac{1}{\pi^{2}c^3}\left(\frac{\omega^{3}}{\eexp{\beta\omega}-1}\right)
\eeq
For the total number of photons~$N$ we have 
a similar integral but without the $\omega$.
The calculation of the incident flux of photons 
is the same as in the case of non-relativistic particles
with ${v \mapsto c}$. Accordingly we get $J=(1/4)(N/\Vol)c$.
For the flux of energy we just have to replace $N$ by $E$, namely     
\beq
J_{\tbox{incident}}\text{[energy]}
\ \ = \ \ 
\frac{1}{4}\left(\frac{E}{\Vol}\right)c
\ \ = \ \ 
\int_0^{\infty} d\omega 
\left[
\frac{1}{4\pi^{2}c^2}\left(\frac{\omega^{3}}{\eexp{\beta\omega}-1}\right)
\right]
\eeq
The calculation of the pressure (the rate of momentum transfer) 
is again the same as in the case of non-relativistic particles, 
with $\mass v \mapsto (\omega/c)$.  
Accordingly we get for the radiation pressure $P=(1/3)(E/\Vol)$. 
Note that ${P \mapsto (1/2)P}$ for an absorbing surface.

Considering the thermal equilibrium between blackbody 
radiation from the environment, and an object that has 
an absorption coefficient  $a(\omega)$, 
detailed balance consideration implies that 
\beq
J_{\tbox{emitted}}\left(\omega\right)d\omega 
\ \ = \ \ 
a(\omega) J_{\tbox{incident}}\left(\omega\right)d\omega
\eeq
It follows that we can regard $a(\omega)$ as the emissivity of the object. 
From here we get the Planck formula 
\beq
J_{\tbox{emitted}}(\omega) \ \ = \ \
a(\omega) \frac{1}{4\pi^{2}c^2} 
\left(\frac{\omega^{3}}{\eexp{\beta\omega}-1}\right) 
\ \ = \ \ 
\frac{a(\omega)}{4\pi^{2}c^2} \ T^{3} \ 
\left(\frac{\nu^{3}}{\eexp{\nu}-1}\right) 
\eeq
where $\nu=\omega/T$ is the scaled frequency.
See Figure below taken from \href{http://hyperphysics.phy-astr.gsu.edu/hbase/mod6.html}{hyperphysics}.
Note that the peak of a blackbody radiation 
is at $\nu\approx3$ which is known as Wein's law. 
Upon integration the total blackbody radiation is 
\beq
J_{\tbox{emitted}} \ \ = \ \
\int_0^{\infty} J_{\tbox{emitted}}(\omega) d\omega 
\ \ = \ \ \frac{1}{4\pi^{2}c^2}\left(\frac{\pi^{4}}{14}\right) T^4
\eeq
which is know as Stephan-Boltzmann Law. Note that the units 
of the flux are energy per time per unit area.

\includegraphics[height=5cm]{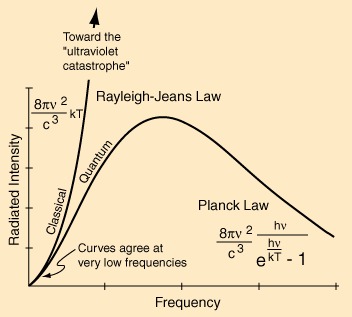}

\ \\

\sheadC{Viscosity}

We have considered above the rate in which momentum is 
transferred to a wall due to {\em ballistic} collisions, 
leading to pressure.
There is a somewhat related effect that is called "viscosity".
It is simplest to explain the concept with regard 
to a gas whose particles have a short mean free path $\ell$, 
such that in equilibrium each gas particle has 
a diffusion coefficient $D=v_T \ell$, 
where $v_T$ is the mean thermal velocity. 

Assume out of equilibrium steady state in which the average 
velocity of the gas particles $\bm{u}(x,y,z)$ is in the $x$~direction 
but its magnitude varies in the $y$~direction.
Due to the transverse diffusion there will be momentum 
transfer across the ${y=0}$ plane, which implies 
that the "upper" flow exerts a force on the "lower" region 
of the fluid (which is possibly the boundary layer of some "wall").    
We shall explain below that if the area of the boundary 
region is $\mathsf{A}$, then the force on it is given 
by the "stress-shear equation" 
\beq
F_{x} \ \ = \ \ \mu \mathsf{A} \frac{du_x}{dy}
\eeq
where $\mu=\rho D$ is the viscosity, 
and $\rho$ is the mass density of the gas.
The argument goes as follows: Divide the $y$ axis 
into layers of width $dy$. Define $w$, the effective 
transition rate of particles between 
layers, such that ${D=w dy^2}$. 
Define the flow of momentum between layers 
as $J(y) = [(\rho dy) u_x(y)] \times w$. 
Hence $J(dy/2)-J(-dy/2)$ is the rate in which 
momentum is transferred across ${y=0}$, 
leading to the desired result.

\clearpage 
\sheadC{The Navier-Stokes equation}

The variables that describe the flow of gas or liquid are the density $\rho(x)$ and 
the velocity ${\mathbf{u}(x)}$. More generally we can add also temperature ${ \theta(x) }$,  
and heat flow ${ \mathbf{q}(x) }$.  
The conservation laws are for the mass, for the momentum, and for the energy.
The conservation of the momentum corresponds to the second law of Newton,   
and formally can be written as [Huang section 5.3]:  
\beq
\frac{d\mathbf{u}_j}{dt} 
\ \ \equiv \ \ 
\left(\frac{\partial}{\partial t} + \sum_i \mathbf{u}_i  \frac{\partial}{\partial x_i} \right)\mathbf{u}_j 
\ \ = \ \ \frac{1}{\rho} \left[ \mathbf{f}_j \ - \ \sum_i \frac{\partial}{\partial x_i} \mathbf{P}_{ij} \right]
\eeq
Following the presentation as in {\bf Huang} one obtains  
the Navier-Stokes equation \href{http://en.wikipedia.org/wiki/Navier-Stokes_equations}{[Wiki]} 
that describes the rate of change of the velocity 
due to momentum transfer in a vicious fluid: 
\beq
\rho \frac{d\mathbf{u}}{dt} \ \ = \ \ \mathbf{f} - \nabla \left(P-\frac{\mu}{3} \nabla\cdot \mathbf{u} \right) + \mu \nabla^2 \mathbf{u} 
\eeq
This equation is valid in the hydrodynamics regime, where the mean free path is small compared 
with to the geometrical length scales. The left hand side contains the non-linear advection term $\rho u \partial u$. The right hand side includes a scalar-pressure term, 
a viscosity term, and an optional external force term (say gravitation).
For incompressible flow ${ \nabla\cdot \mathbf{u} =0 }$,  
and Euler equation is obtained if the viscosity is completely neglected.

{\bf Sound waves.-- }
As we see the viscosity $\mu$ plays a major role in the Navier-Stokes equation.
Usually the equation is supplemented by a continuity 
equations for the mass, and for the energy, as well as by 
state equation that connects the pressure to the density. 
For compressible fluid with state equation ${ P[\rho] }$ 
the continuity and the Euler equations for the time derivatives 
of ${ \rho(x) }$ and  ${ \mathbf{u}(x) }$ lead to sound waves with velocity
\beq
c^2 \ \ = \ \ \frac{\partial P}{\partial \rho} \ \ = \ \ \frac{\kappa}{\rho} 
\eeq 
Once the viscosity is taken into account we get damping 
of the sound waves.

{\bf Stokes law.-- }
A well known result that comes from the Navier-Stokes equation
is Stokes law for the friction force 
that is exerted on a spherical object of radius~$R$ 
\beq
F[\text{stick}] \ = \ -6\pi \mu R \ v_{\tbox{sphere}}, 
\hspace{2cm} 
F[\text{slip}] \ = \  -4\pi \mu R \ v_{\tbox{sphere}} 
\eeq
Roughly the shear is ${1/R}$ while the area is ${ R^2 }$ 
hence the friction is proportional to ${ R }$.
and not to the area of the sphere.
The traditional version (with $6\pi$) assumes no-slip boundary conditions.  
The way to derive it is to find the velocity field 
for the flow, and then to use the ``stress-shear equation" of the previous section. 
For details see Huang p.119 and [\href{http://link.aps.org/doi/10.1103/PhysRevA.2.2005}{PRA 1970}].
The optional derivation via a microscopic theory is quite complicated,  
see [\href{http://link.aip.org/link/doi/10.1063/1.439952}{JCP 1980}].

{\bf Reynolds number.-- }
The dimensionless parameter in the Navier-Stokes equation, that characterizes the effect of the non-linear advection term, is the Reynolds number. Consider for example the Stokes problem where the relevant length scale is $R$, we define 
\beq
\text{Re} \ \ = \ \ \frac{\rho v}{\mu} R \ \ = \ \ \frac{v}{D}R  
\eeq
The original geometry that has been considered by Reynolds refers to flow of fluid via a pipe, where $R$ is the downstream distance from the injection point. A boundary layer of thickness ${\delta \sim \sqrt{Dt}}$ is formed near the walls of the pipe  after a distance that correspond to ${t=R/v}$. The flow is laminar for ${R<\delta}$. At larger downstream distance (larger ``Re") the laminar flow becomes turbulent. This turbulence arises due to loss of stability of the laminar solution. 
Let us consider again the motion of a spherical particle in a fluid. The derivation of Stokes law assumes small ``Re". For large ``Re" a turbulent region is formed downstream after the particle, and the ${ v }$ dependence of the friction acquire a fractional exponent. If the translation velocity $v$ is very large
the friction becomes proportional to  ${ v^2 }$ reflecting transfer of momentum by a moving wall. This should be contrasted by Stokes law that assumes a flow that curves smoothly to the sides of the sphere, and joins at the back of the sphere.

\clearpage 
\sheadC{Heat current in an open geometry}

If we have two boxes, and energy can flow from one to the other, 
then the heat current ${I_Q}$ is simply defined as the rate ${I_E}$ 
of energy transfer. But if we have a flow ${I_N}$ of particles,  
this simple-minded reasoning fails. We would like to argue below 
the the correct expression for the heat current is 
\beq
I_Q \ \ =\ \ I_E-\mu I_N
\eeq
This expression assumes quasi-reversible flow at well-defined 
temperature and energy. 

First of all let us recognize that if $dN$ particles are 
transferred from one box to a second box, then the transferred 
energy $dE$ is ill-defined. Assume for example that the particles 
are transferred from an energy level $\epsilon$ of the first box 
to an energy level with the same energy at the second box.  
We get ${dE=\epsilon dN}$, which depends on the arbitrary energy reference 
of the Hamiltonian, and hence has no physical significance. 
But the quantity ${dQ = dE -\mu dN}$ is well defined. The question is 
how to rationalize that ${dQ}$ is indeed the appropriate definition 
of heat in this context.

Referring to a box with ${N}$ particles and energy ${E}$, recall 
that a quasi-reversible process of taking an energy ${-dE}$ requires 
the supply of energy ${dQ=dE}$, such as  ${dS=0}$. 
In complete analogy, taking ${-dN}$ particles with energy ${-dE}$
requires compensation ${dQ = dE - \mu dN}$, such as  ${dS=0}$.

\sheadC{Thermo-electricity}

Reversible flow through a conductor,
with no entropy production due to Joule heating, 
can be regarded as a sequence of quasi-reversible 
transfer operations.  
In each step heat is taken or given to the phonons 
that dwell in another segment of the conductor.
The net results is the transfer of energy 
from one end of the wire to the other end.    
This reasoning  leads to the Mott analysis of the Peltier effect.
The expression for the electric current of charge ${e}$ carriers 
has the form (Here $J=I/\mathsf{A}$ is the current density):
\beq
J_N \ \ = \ \ e^2 \mathcal{E}  \int c(\varepsilon) \left( -f'(\varepsilon-\mu)  \right) \, d\varepsilon  \ \ \equiv \ \ \sigma \mathcal{E}  
\eeq
while the expression for the quasi-reversible heat current is 
\beq
J_Q = e \mathcal{E}  \int (\varepsilon-\mu) \ c(\varepsilon) \left( -f'(\varepsilon-\mu)  \right) \, d\varepsilon  \ \ \equiv \ \ \sigma S T \mathcal{E}   
\eeq
Using the Sommerfeld expansion we deduce that 
\beq
I_Q  \ \ = \  \ \frac{\pi^2}{3e} T^2 \frac{c'(\mu)}{c(\mu)} \ I_N  \ \ \equiv \ \  ST \, I_N 
\eeq
where ${\Pi=ST}$ is known as the Peltier coefficient, 
and ${S}$ is known as the Seebeck coefficient.
From the Onsager reciprocity it follows that the same coefficient 
appears in the linear relation between ${J_N}$ and ${\nabla T}$, 
as explained below.  

A thermal current can be induced also by a temperature gradient. Namely,
\beq
J_Q \ \ = \ \ - \kappa \nabla T
\eeq
A relation between the thermal conductivity ${\kappa}$ 
and the electrical conductivity ${\sigma}$
can be obtained using a straightforward extension 
of the above derivation (see Ashcroft p.253):
\beq
\kappa \ \ = \ \ \frac{\pi^2}{3e^2} T \sigma  
\eeq
This is known as the Wiedemann-Franz law.
We now turn to discuss more general circumstances  
of having both potential and temperature gradients. 

Considering again two boxes, the natural thermodynamic coordinates  
are  ${\varphi_N=(N_2-N_1)/2}$  and ${\varphi_E = (E_2-E_1)/2}$. 
Note that ${J_N=\dot{\varphi}_N}$ and ${J_E=\dot{\varphi}_E}$.
In the continuum limit $J$ is re-defined as the current density.
The entropy function is ${S(X_N,X_E)}$, and the conjugate  
variables are ${X_N=\nabla(\mu/T)}$ and ${X_E=-\nabla(1/T)}$. 
Here $\nabla$ is the difference, while in the continuum 
limit it becomes the gradient. The linear relation 
between the $J$s and the $X$s involves coefficients ${\gamma_{ij}}$.
The Onsager relation is invariant under the change of reference 
energy, but the coefficients depend on the choice of reference.
It is customary to set ${\mu=0}$ as the reference. Accordingly 
the linear relations take the following form:  
\beq
J_N  \ \ &=& \ \  \gamma_{\sigma}  \frac{\mathcal{E}}{T} +  \gamma_{\perp} \nabla\left( \frac{1}{T} \right) 
\ \ \equiv \ \ \sigma \left(\mathcal{E} - S \nabla T \right)  
\\
J_Q  \ \ &=& \ \ \gamma_{\perp} \frac{\mathcal{E}}{T} +  \gamma_{\kappa} \nabla\left( \frac{1}{T} \right)  
\ \ \equiv \ \ \sigma ST \, \mathcal{E} -  \kappa \nabla T   
\eeq
From the ${J_N}$ equation it follows that in an open circuit 
a temperature gradient would be balanced by an induced 
electric field ${\mathcal{E} = S \nabla T }$. 
This is called Seebeck effect.

\newpage
\makeatletter{}
\sheadB{Scattering approach to mesoscopic transport}

The most popular approach to transport in mesoscopic 
devices takes the scattering formalism rather than the 
Kubo formalism as a starting point, leading to the Landauer   
and the BPT formulas. We first cite these formulas 
and then summarize their common derivation. This should 
be compared with the Kubo-based derivation of the previous section. \\

\sheadC{The Buttiker-Pretre-Thomas-Landauer formula}

We assume without loss of generality that there are three
parameters $(x_1, x_2, x_3)$ over which we have
external control, where $x_3=\Phi$ is the AB flux.
The expression for the current ${\cal I}_{\tbox{A}}$ that goes
{\em out} of lead~$A$, assuming DC linear response, 
can be written as
\beq
I_{\tbox{A}} = -\sum_j \bm{G}^{3j} \dot{x}_j
\eeq
where $-\dot{x}_3=-\dot{\Phi}$ is the EMF,
and therefore $G^{33}$ is the conductance
in the usual sense.
The B\"{u}ttiker-Pr\'{e}tre-Thomas-Landauer 
formula for the generalized conductance matrix is
\beq
\bm{G}^{3j} \ = \ \frac{e}{2\pi i}
\trc\left(P_{\tbox{A}}\frac{\partial S}{\partial x_j}
S^{\dag}\right)
\eeq
In particular for the Ohmic conductance 
we get the Landauer formula:
\beq
\bm{G}^{33} \ = \ \frac{e^2}{2\pi\hbar}\trc(\bm{t}\bm{t}^{\dag})
\eeq

In order to explain the notations in the above formulas 
we consider a two lead system. 
The $S$ matrix in block form is written as follows:
\beq
S = \left( \amatrix{
\bm{r}_{\tbox{B}} & \bm{t}_{\tbox{AB}}\eexp{-i\phi} \cr
\bm{t}_{\tbox{BA}}\eexp{i\phi} & \bm{r}_{\tbox{A}}  }\right)
\eeq
where $\bm{r}$ and $\bm{t}$ are
the so called reflection and transmission
(sub) matrices respectively.
We use the notation $\phi=e\Phi/\hbar$.
In the same representation, we define 
the left lead and the right lead projectors:
\beq
P_{\tbox{A}} = \left( \amatrix{\bm{0} & \bm{0} \cr \bm{0} & \bm{1}} \right),
\ \ \ \ \ \
P_{\tbox{B}} = \left( \amatrix{\bm{1} & \bm{0} \cr \bm{0} & \bm{0}} \right)
\eeq
The following identity is important in order 
to obtain the Landauer formula from the BPT formula:
\beq
\frac{dS}{d\Phi}
\ = \ i\frac{e}{\hbar} ( P_{\tbox{A}} S P_{\tbox{B}} -i P_{\tbox{B}} S P_{\tbox{A}} )
\ = \ i\frac{e}{\hbar} ( P_{\tbox{A}} S - S P_{\tbox{A}} )
\ = \ -i\frac{e}{\hbar} ( P_{\tbox{B}} S - S P_{\tbox{B}} )
\eeq
Another important identity is
\beq
\trc(P_{\tbox{A}} S P_{\tbox{B}} S^{\dag})  
\ = \ \trc(\bm{t}\bm{t}^{\dag})
\ = \ \sum_{a \in A}\sum_{b \in B} |\bm{t}_{ab}|^2 
\eeq
The $\trc()$ operataion is over the channel indexes.

\sheadC{Floque theory for periodically driven systems}

The solution of the Schrodinger equation 
\beq
i\frac{d\psi}{dt}=\mathcal{H}\psi
\eeq
with time independent $\mathcal{H}$ is 
\beq
|\psi(t)\rangle = \sum_{\mathcal{E}} \eexp{-i\mathcal{E}t} |\psi^{(\mathcal{E})}\rangle
\eeq
where the stationary states are found from
\beq
\mathcal{H} |\psi^{(\mathcal{E})}\rangle = \mathcal{E} |\psi^{(\mathcal{E})}\rangle
\eeq
Consider now the more complicated case 
where $\mathcal{H}$ depends periodically on time. 
Given that the basic frequency is $\omega$ 
we can write
\beq
\mathcal{H}(t) = \sum \mathcal{H}^{(n)} \eexp{-in\omega t}
\eeq
The solution of the Schrodinger equation is 
\beq
|\psi(t)\rangle 
= \sum_{\mathcal{E}} \sum_{n=-\infty}^{\infty} 
\eexp{-i(\mathcal{E}+n\omega)t} |\psi^{(\mathcal{E},n)}\rangle
\eeq
where the Flouqe states are found from 
\beq
\sum_{n'} \mathcal{H}^{(n-n')} |\psi^{(\mathcal{E},n')}\rangle 
= (\mathcal{E}+n\omega) |\psi^{(\mathcal{E},n)}\rangle
\eeq
and $\mathcal{E}$ is defined modulo $\omega$.

\sheadC{The Floque scattering matrix}

In scattering theory we can define a Flouqe energy shell $\mathcal{E}$. 
The solution outside of the scattering region is written as 
\beq
|\psi(t)\rangle 
= \sum_{n=n_{\tbox{floor}}}^{\infty}  
\eexp{-i(\mathcal{E}+n\omega)t}
\sum_a \left[
A_{an} \frac{1}{\sqrt{v_{an}}}\eexp{-ik_{an}r} 
- B_{an} \frac{1}{\sqrt{v_{an}}}\eexp{+ik_{an}r} 
\right]
\otimes
|a\rangle
\eeq
where $v_{an}$ and $k_{an}$ are determined 
by the available energy ${\mathcal{E}+n\omega}$.
The current in a given channel is time dependent, 
but its DC component is simply ${\sum_n (|B_{an}|^2-|A_{an}|^2)}$.  
Therefore the continuity of the probability flow implies  
that we can define an on-shell scattering matrix
\beq
B_{bn_b} \ = \ \sum_{an_a} \bm{S}_{bn_b,an_a} \ A_{an_a}
\eeq
We can write this matrix using the following notation 
\beq
\bm{S}_{bn_b,an_a} \ \  \equiv \ \ \bm{S}_{b,a}^{n_b-n_a}(\mathcal{E}+n_a\omega)
\eeq
Unitarity implies that 
\beq
\sum_{bn_b} |\bm{S}_{bn_b,an_a}|^2 
&=& \sum_{bn} |\bm{S}_{b,a}^n(\mathcal{E})|^2 = 1 
\\ 
\sum_{an_a} |\bm{S}_{bn_b,an_a}|^2 
&=& \sum_{an} |\bm{S}_{b,a}^n(\mathcal{E}+n\omega)|^2 = 1 
\eeq
If the driving is very slow we can use the adiabatic relation 
between the incoming and outgoing amplitudes 
\beq
B_b(t) \ = \ \sum_a \bm{S}_{ba}(X(t)) \ A_a(t)
\eeq
where $\bm{S}_{ba}(X)$ is the conventional on-shell scattering 
matrix of the time independent problem.
This relation implies that 
\beq
\bm{S}_{b,a}^{n}(\mathcal{E}) 
= \frac{\omega}{2\pi} 
\int_0^{\omega/2\pi} \bm{S}_{ba}(X(t)) 
\ \eexp{in\omega t}dt 
\eeq
For sake of later use we note the following identity
\beq
\sum_n n |\bm{S}_{b,a}^{n}|^2 
= \frac{i}{2\pi}
\int_0^{2\pi/\omega} dt 
\frac{\bm{S}_{ba}(X(t))}{\partial t}
\bm{S}_{ba}(X(t))
\eeq

\sheadC{Current within a channel}

Consider a one dimensional channel labeled as~$a$. 
Let us take a segment of length~$L$. 
For simplicity assume periodic boundary condition 
(ring geometry). If one state~$n$ is occupied 
the current is   
\beq
I_{a,n} \ \ = \ \ \frac{e}{L} v_{a,n}
\eeq
If several states are occupied we 
should integrate over the energy
\beq
I_a \ \ = \ \ \sum_n f_a(E_n) I_{a,n} \ \ = \ \ 
\int f_a(E) \frac{L dE}{2\pi v_a} \left( \frac{e}{L} v_a\right)
\ \ = \ \ \frac{e}{2\pi } \int f_a(E) dE
\eeq
For fully occupied states withing some energy range we get 
\beq
I_a 
\ \ = \ \ \frac{e}{2\pi } (E_2-E_1)
\ \ = \ \ \frac{e^2}{2\pi } (V_2-V_1)
\eeq
If we have a multi channel lead, then we have 
to multiply by the number of open channels.

\sheadC{The Landauer formula}

Consider a multi channel system which. 
All the channels are connected to a scattering 
region which is described by an $\bm{S}$ matrix. 
We use the notation 
\beq
g_{ba}= |\bm{S}_{ba}|^2
\eeq
Assuming that we occupy a set of scattering states, 
such that $f_a(E)$ is the occupation of those    
scattering states that incident in the $a$th channel,  
we get that the outgoing current at channel $b$ is  
\beq
I_b 
= \frac{e}{2\pi } \int dE  
\left[\left( \sum_a  g_{ba}f_a(E) \right) - f_b(E) \right] 
\eeq
Inserting $1=\sum_a g_{ba}$ in the second term we get    
\beq
I_b 
= \frac{e}{2\pi } \int dE 
\left[\sum_a  g_{ba}(f_a(E)-f_b(E)) \right] 
\eeq
Assuming low temperature occupation with  
\beq
f_a(E) = f(E-eV_a)  \approx f(E) - f'(E)eV_a
\eeq
we get 
\beq
I_b = -\frac{e^2}{2\pi } \sum_a  g_{ba} \ (V_b-V_a)
\eeq
which is the multi channel version of the Landauer formula.
If we have two leads $A$ and $B$ we can write 
\beq
I_{\tbox{B}} = -\frac{e^2}{2\pi } 
\left[\sum_{b\in B} \sum_{a\in A}  g_{ba}\right] \ (V_{\tbox{B}}-V_{\tbox{A}})
\eeq
Form here it follows that the conductance is 
\beq
G = \frac{e^2}{2\pi } 
\sum_{b\in B} \sum_{a\in A}  g_{ba} 
= \frac{e^2}{2\pi } 
\trc(P_{\tbox{B}}\bm{S}P_{\tbox{A}}\bm{S}^{\dag})
\eeq
where $P_{\tbox{A}}$ and $P_{\tbox{B}}$ are the projectors 
that define the two leads.

\sheadC{The BPT formula}

Assuming that the scattering region is periodically 
driven we can use the Floque scattering formalism.   
The derivation of the expression for the DC component $I_b$
of the current in channel~$b$ is very similar 
to the Landauer case, leading to 
\beq
I_b 
&=& \frac{e}{2\pi } \int dE  
\left[\left( 
\sum_{a,n} |\bm{S}^n_{ba}(E-n\omega)|^2 f_a(E+n\omega) 
\right) 
- f_b(E) 
\right] 
\\
&=& \frac{e}{2\pi } \int dE  
\left[\sum_{a,n} |\bm{S}^n_{ba}(E-n\omega)|^2 (f_a(E-n\omega)-f_b(E)) \right] 
\\
&=& \frac{e}{2\pi } \int dE  
\left[\sum_{a,n}  |\bm{S}^n_{ba}(E)|^2 (f_a(E)-f_b(E+n\omega)) \right] 
\\
&\approx& \frac{e}{2\pi } \int dE  
\left[\sum_{a,n}  n\omega |\bm{S}^n_{ba}(E)|^2 (-f_a'(E)) \right] 
=
\frac{e}{2\pi} \left[\sum_{a,n}  n\omega |\bm{S}^n_{ba}(E)|^2 \right] 
\eeq
In the last two steps we have assumed very small $\omega$ 
and zero temperature Fermi occupation.
Next we use an identity that has been mentioned 
previously in order to get an expression that 
involves the time independent scattering matrix: 
\beq
I_b = i\frac{e}{2\pi}  
\sum_a
\frac{\omega}{2\pi}
\int_0^{2\pi/\omega} dt 
\frac{\bm{S}_{ba}(X(t))}{\partial t}
\bm{S}_{ba}(X(t))
\eeq
which implies that the pumped charge per cycle is 
\beq
Q = i\frac{e}{2\pi} 
\oint dX 
\sum_{b\in B} \sum_a
\frac{\bm{S}_{ba}(X)}{\partial X}
\bm{S}_{ba}(X)
\equiv 
-\oint G(X) dX
\eeq
with
\beq
G(X) = -i\frac{e}{2\pi} 
\sum_{b\in B} \sum_a
\frac{\bm{S}_{ba}(X)}{\partial X}
\bm{S}_{ba}(X)
= -i\frac{e}{2\pi} 
\trc\left(P_{\tbox{B}}
\frac{\partial \bm{S}}{\partial X}
\bm{S}^{\dag}\right)
\eeq
Note: since $\bm{S}(X)$ is unitary it follows 
that the following generator is Hermitian 
\beq
\bm{H}(X)=i\frac{\partial \bm{S}}{\partial X} \bm{S}^{\dag}
\eeq 
The trace of a product of two hermitian operators 
is always a real quantity.

\sheadC{BPT and the Friedel sum rule}

If only one lead is involved the BPT formula becomes 
\beq
dN = -i\frac{1}{2\pi} 
\trc\left(
\frac{\partial \bm{S}}{\partial X}
\bm{S}^{\dag}\right) \ dX
\eeq 
where $dN$ is the number of particles that are 
absorbed (rather than emitted) by the scattering region 
due to change $dX$  in some control parameter.
A similar formula known as the Friedel sum rule states that 
\beq
d\mathcal{N} = -i\frac{1}{2\pi} 
\trc\left(
\frac{\partial \bm{S}}{\partial E}
\bm{S}^{\dag}\right) \ dE
\eeq 
where $\mathcal{N}(E)$ counts the number of states 
inside the scattering region up to energy $E$.
Both formulas have a very simple derivation, 
since they merely involve counting of states. 
For the purpose of this derivation we close the lead 
at $r=0$ with Dirichlet boundary conditions.   
The eigen-energies  are found via the equation 
\beq
\det(\bm{S}(E,X)-\bm{1}) \ \ = \ \ 0
\eeq 
Let us denote the eigenphases of $\bm{S}$ as $\theta_r$. 
We have the identity 
\beq  
i\sum_r \delta \theta_r  
= \delta [\ln \det \bm{S}]
= \trc [\delta \ln \bm{S}]
= \trc[\delta\bm{S} \bm{S}^{\dag}]
\eeq 
Observing that a new eigenvalue is found 
each time that one of the eigenphases goes 
via $\theta=0$ we get the desired result.

\newpage
\makeatletter{}
\sheadB{The theory of electrical conductance}

\sheadC{The Hall conductance}

The calculation of the Hall conductance is possibly the simplest 
non-trivial example for adiabatic non-dissipative response. 
The standard geometry is a 2D "hall bar" of dimension ${L_x\times L_y}$. 
In "Lecture notes in quantum mechanics" we have considered what 
happens if the electrons are confined in the transverse direction 
by a potential $V(y)$. Adopting the Landauer approach it is assumed 
that the edges are connected to leads that maintain a chemical 
potential difference. Consequently there is a net current in the~$x$
direction. From the "Landau level" picture it is clear that the 
Hall conductance $G_{xy}$ is quantized in units $e^2/(2\pi\hbar)$.
The problem with this approach is that the more complicated
case of disorder $V(x,y)$ is difficult for handling. We therefore 
turn to describe a formal Kubo approach.  From now on we use   
units such that ${e=\hbar=1}$.

We still consider a Hall bar ${L_x\times L_y}$, but now 
we impose periodic boundary condition such that ${\psi(L_x,y)= \eexp{i\phi_x}\psi(0,y)}$ 
and ${\psi(x,L_y)= \eexp{i\phi_y}\psi(x,0)}$. 
Accordingly the Hamiltonian depends on the parameters ${(\phi_x,\phi_y,\Phi_B)}$, 
where $\Phi_B$ is the uniform magnetic flux through 
the Hall bar in the $z$~direction. The currents ${I_x=(e/L_x)v_x}$ 
and $I_y=(e/L_y)v_y$ are conjugate to $\phi_x$ and $\phi_y$.
We consider the linear response relation ${I_y=-G_{yx}\dot{\phi_x}}$.
This relation can be written as ${dQ_y=-G_{yx}d\phi_x}$. 
The Hall conductance quantization means that a $2\pi$ variation 
of $\phi_x$ leads to one particle transported in the $y$ direction.
The physical picture is very clear in the standard $V(y)$ geometry: 
the net effect is to displace all the filled Landau level 
"one step" in the $y$ direction.   

We now proceed with a formal analysis to show that the Hall conductance
is quantized for general $V(x,y)$ potential. We can define 
a "vector potential" $A_n$ on the $(\phi_x,\phi_y)$ manifold. 
If we performed an adiabatic cycle the Berry phase would be 
a line integral over $A_n$. By Stokes theorem this can be 
converted into a  $d\phi_x d\phi_y$ integral over~$B_n$.  
However there are two complementary domains over which the 
surface integral can be done. Consistency requires that the 
result for the Berry phase would come out the same modulo~$2\pi$.
It follows that 
\beq
\frac{1}{2\pi}\int_0^{2\pi}\int_0^{2\pi} B_n d\phi_x d\phi_y 
\ \ = \ \ \text{integer} \ \ \text{[Chern number]} 
\eeq
This means that the $\phi$ averaged $B_n$ is quantized in units 
of $1/(2\pi)$. If we fill several levels the Hall conductance 
is the sum ${\sum_n B_n}$ over the occupied levels, namely
\beq
G_{yx} \ \ = \ \ \sum_{n \in \text{band}}\sum_m 
\frac{2\im[I^y_{nm}I^x_{mn}]}{(E_m-E_n)^2}
\eeq
If we have a quasi-continuum it is allowed to 
average this expression over $(\phi_x,\phi_y)$.
Then we deduce that the Hall conductance of 
a filled band is quantized. The result is of physical 
relevance if non-adiabatic transitions 
over the gap can be neglected.

\sheadC{The Drude formula}

The traditional derivation of the Drude formula is based 
on the Boltzmann picture. Optionally one can adopt 
a Langevine-like picture. The effect of the scattering 
of an electron in a metal is to randomize its velocity.
This randomization leads to a statistical "damping" of 
the average velocity with rate $1/t_{\ell}$. 
On the other hand the electric field accelerates 
the particle with rate $e\mathcal{E}/\mass$. 
In steady state the drift velocity is ${v_{drift}= (e\mathcal{E}/\mass)t_{\ell}}$, 
and the current density is  ${J= (N/\Vol)ev_{drift}}$ leading to 
the Drude conductivity $\bm{\sigma} = (N/\Vol)(e^2/\mass)t_{\ell}$.
Consequently the conductance of a ring that has a length $L$
and a cross-section $\mathsf{A}$ is 
\beq
G \ \ = \ \ \frac{\mathsf{A}}{L}\bm{\sigma} 
\ \ = \ \ \frac{N}{L^2} \left(\frac{e^2}{\mass}\right)  t_{\ell}
\ \ = \ \ e^2 \left(\frac{N}{\mass v_FL}\right) \frac{\ell}{L}
\ \ \equiv \ \  \frac{e^2}{2\pi\hbar} \mathcal{M} \frac{\ell}{L}
\eeq
where $\ell=v_F t_{\ell}$ is the mean free path at the Fermi energy, 
and $\mathcal{M}$ is the effective number of open modes.
Below we would like to derive this result formally 
from the FD relation. 
 
The canonical version of the FD relation takes the form $G=[1/(2T)]\nu_T$, 
where $\nu$ is the intensity of the current fluctuations 
and $G$ is the conductance. This is known as Nyquist version 
of the FD relation. One way to go is to calculate $\nu_T$ 
for a many body electronic system, see how this is done in 
a previous lecture. But if the electrons are non-interacting 
it is possible to do a shortcut, relating the 
conductance of the non-interacting many body 
electronic system to its single particle fluctuations.
This can be regarded as a generalizations of the canonical Nyquist formula.
The generalization is obtained by re-interpretation of~$f(E)$ 
as the Fermi occupation function (with total occupation~$N$),  
rather than probability distribution.  
Assuming a Boltzmann occupation one obtains ${G^{[N]} = [N/(2T)]\nu_T}$.
A similar generalization holds for a microcanonical occupation, 
from which one can deduce results for other occupations. 
In particular for low temperature Fermi occupation 
of non-interacting particles one obtains: 
\beq
G^{[N]} \ \ = \ \ \frac{1}{2} \gdos(E_F) \ \nu_{E_F}
\ \ = \ \ \frac{1}{2} \gdos(E_F) \left(\frac{e}{L}\right)^2 \int \langle v_{\parallel}(t)v_{\parallel}(0)\rangle dt 
\ \ = \ \ \left(\frac{e}{L}\right)^2 \gdos(E_F) \mathcal{D}_0
\eeq
The crossover from the high temperature "Boltzmann" 
to the low temperature "Fermi"  behavior happens at $T\sim E_F$. 
Assuming exponential velocity-velocity correlation function 
with time constant $\tau_0$, such that the mean free path 
is ${\ell=v_F\tau_0}$, we get ${D_0 = v_F\ell}$.
disregarding numerical prefactors the density of states 
can be written as ${\gdos(E_F)=(L/v_F)\mathcal{M}}$, 
where  $\mathcal{M}$ is the number of open modes. 
From here we get the Drude formula 
\beq
G^{[N]} \ \ = \ \ 
\left(\frac{e}{L}\right)^2 \gdos(E_F) \mathcal{D}_0
\ \ = \ \ 
\frac{e^2}{2\pi\hbar} \ \mathcal{M}  \frac{\ell}{L}   
\eeq
Relating to electrons that are moving 
in a lattice with static disorder,  
the mean free path can be deduced from 
the Fermi Golden Rule (FGR) picture as follows:
Assuming isotropic scattering, the velocity-velocity 
correlation function is proportional 
to the survival probability $P(t)=\eexp{-t/t_{\ell}}$. 
Ignoring a factor that has to do with 
the dimensionality $d=2,3$ of the sample the relation is 
\beq
\langle v(t) v(0) \rangle 
\ \ \approx \ \   v_{\tbox{E}}^2 \ P(t)
\ \ = \ \ v_{\tbox{E}}^2 \eexp{-|t|/t_{\ell}} 
\eeq
where the FGR rate of the scattering is  
\beq
\frac{1}{t_{\ell}} \ \ =  \ \ 2\pi \varrho_{\tbox{E}} |U_{mn}|^2   
\ \ = \ \ \frac{\pi a}{v_{\tbox{E}}}W^2
\eeq
In the last equality we have used $|U_{\bm{n}\bm{m}}|^2 \approx [a/(\mathcal{M} L)]W^2$, 
where $a$ is the lattice spacing, and $W$ is the strength of the disorder. 
Disregarding prefactors of order unity we deduce 
the so-called Born approximation for the mean free path:
\beq
\ell \ \ = \ \ v_{\tbox{E}}t_{\ell} \ \ \approx \ \
\frac{1}{a}\left(\frac{v_{\tbox{E}}}{W}\right)^2 
\eeq

\sheadC{Formal calculation of the conductance}

The DC conductance $G$ of a ring with $N$ non-interacting 
electrons is related by Kubo/FD expression to the density 
of one-particle states $\gdos(E_F)$ at the Fermi energy, 
and to the $\tilde{C}_{vv}(\omega\sim0)$ fluctuations of velocity. 
The latter can be deduced semi-classically from 
the velocity-velocity correlation function,  
or from the matrix elements of the velocity operator 
using the quantum-mechanical spectral decomposition.
Optionally one can use path integral or Green function 
diagrammatic methods for the calculation.   

Let us summarize some optional ways in which 
the Kubo/FD expression for the Ohmic conductance 
can be written. If we use the spectral 
decomposition with $p_n = \gdos(E_F)^{-1} \delta(E_n-E_F)$, 
we get 
\beq
G  
\ \ = \ \ \frac{1}{2} \gdos(E_F) \left(\frac{e}{L}\right)^2\tilde{C}_{vv}(0)
\ \ = \ \ \pi \sum_{nm} \left(\frac{e}{L}\right)^2|v_{nm}|^2 \ \delta(E_n-E_F) \ \delta(E_m-E_n)
\eeq
It is implicit that the delta functions are "broadened" 
due to the assumed non-adiabaticity of the driving,  
else we shall get vanishing dissipation. The calculation
of $G$ is the adiabatic regime requires a more careful treatment, 
and possibly goes beyond LRT.  
As long as the broadening is large compared to the level spacing, 
but small compared with other energy scales, 
the result is not too sensitive to the broadening parameter, 
and small corrections are commonly ignored, unless the ring is very small.          
Schematically the above expression can be written as 
\beq
G \ \ = \ \ \pi \hbar \ \gdos(E_F)^2 \  \left(\frac{e}{L}\right)^2 \overline{|v_{nm}|^2}
\ \ = \ \ \pi \hbar \ \gdos(E_F)^2 \ \overline{|{\cal I}_{nm}|^2}
\eeq
where the bar indicates that an average should be taken
over the near diagonal matrix elements of the 
velocity operator near the Fermi energy. 
A somewhat more fancy way to write the same is 
\beq
G \ \ = \ \ \pi \ \trc\Big[ \mathcal{I} \ \delta(E_F-\mathcal{H}) \ \mathcal{I} \ \delta(E_F-\mathcal{H}) \Big]
\ \ = \ \ \frac{1}{\pi} \ \trc \left[{\cal I} \ \im[{\mathsf G}(E_F)] \  {\cal I} \ \im[{\mathsf G(E_F)}] \right]
\eeq
where $\mathsf{G}=1/(E-\mathcal{H}+i0)$ is the one-particle retarded Green function.
This opens the way to formal calculations 
that are based on path integral or diagrammatic methods.

For a chaotic ring, the dispersion  $\overline{|\mathcal{I}_{nm}|^2}$ 
of the off-diagonal matrix elements is equal, up to a symmetry factor, 
to the dispersion of the diagonal matrix elements. 
Note that $\mathcal{I}_{nn} = -{\partial E_n}/{\partial \Phi}$.
It is common to use the notation $\phi=(e/\hbar)\Phi$.
Hence one obtains the Thouless relation: 
\beq
\bm{G}^{[N]} \ \ = \ \ \text{factor} \times \frac{e^2}{\hbar} \times
\frac{1}{\Delta^2} \ \overline{\left| \frac{\partial E_n}{\partial \phi}  \right|^2}
\eeq
where the numerical factor depends on symmetry considerations,
and $\Delta$ is the mean level spacing at the Fermi energy.
There is a more refined relation by Kohn.
The Thouless relation is a useful staring point for 
the formulation of the scaling theory for localization.

\sheadC{Conductivity and Conductance}

Consider a ring geometry, and assume that the current is driven
by the flux $\Phi$. In order to have a better defined model
we should specify what is the vector potential ${\cal A}(\bm{r})$ along the ring.
We can regard the values of ${\cal A}$ at different points in space
as independent parameters (think of tight binding model).
Their sum (meaning $\oint {\cal A}(\bm{r}) {\cdot} d\bm{r}$) should be $\Phi$.
So we have to know how $\Phi$ is "distributed" along the ring.
This is not just a matter of "gauge choice" because
the electric field ${{\cal E}(\bm{r}) = - \dot{{\cal A}}(\bm{r})}$
is a gauge invariant quantity. So we have to know how
the voltage is distributed along the ring. However,
as we explain below, in linear response theory this information
is not really required. Any voltage distribution that results in
the same electro-motive force, will create the same current.

In linear response theory the current is proportional to the rate
in which the parameters are being changed in time.
Regarding the values of ${\cal A}$ at different points in space
as independent parameters, linear response theory postulates 
a linear relation between $\langle J(\bm{r}) \rangle $ and ${\cal E}(\bm{r}')$ 
that involves the conductivity matrix $\bm{\sigma}(r,r')$ as a kernel. 
The current density has to satisfy the continuity
equation $\nabla\cdot\langle J(r)  \rangle=0$.
From here it follows that if we replace ${\cal A}$  by ${\cal A}+\nabla\Lambda(r)$,
then after integration by parts we shall get the same current.
This proves that within linear response theory the current
should depend only on the electromotive force~$-\dot{\Phi}$,
and not on the way in which the voltage is distributed.
Note that ${\cal A} \mapsto {\cal A}+\nabla\Lambda(\bm{r})$
is not merely a gauge change: A gauge transformation
of time dependent field requires a  compensating replacement
of the scalar potential (which is not the case here).

In the following it is convenient to think of a device
which is composed of "quantum dot" with two long leads,
and to assume that the two leads are connected
together as to form a ring. We shall use
the notation ${\bm{r}=(\ar,s)}$, where $\ar$ is the coordinate
along the ring, and $s$ is a transverse coordinate.
In particular we shall distinguish a left "section" $\ar=\ar_{\tbox{B}}$
and a right section $\ar=\ar_{\tbox{A}}$ of the two leads,
and we shall assume that the dot region is described by
a scattering matrix $S_{ab}$.

We further assume that all the voltage drop is
concentrated across the section $\ar=\ar_{\tbox{B}}$,
and we measure the current ${\cal I}_{\tbox{A}}$ through the
section $\ar=\ar_{\tbox{A}}$. With these assumptions we
have two pairs of conjugate variables,
which are $(\Phi_{\tbox{A}}, {\cal I}_{\tbox{A}})$ and  $(\Phi_{\tbox{B}}, {\cal I}_{\tbox{B}})$.
Note that the explicit expression for the
current operator is simply
\beq
{\cal I}_{\tbox{A}} = e\frac{1}{2}  (v \ \delta(\ar-\ar_{\tbox{A}}) + \delta(\ar-\ar_{\tbox{A}}) v)
\eeq
where $v$ is the $\ar$ component of the velocity operator.
We are interested in calculating the conductance,
as define through the linear response relation
$\langle {\cal I}_{\tbox{A}} \rangle = - G^{AB} \dot{\Phi}_{\tbox{B}}$.
The Kubo expression takes the form
\beq
G^{AB} = \frac{\hbar}{\pi}\ \trc \left[{\cal I}_{\tbox{A}} \ \im[{\mathsf G}] \  {\cal I}_{\tbox{B}} \ \im[{\mathsf G}] \right]
\eeq
This is yet another version of the Kubo formula.
Its advantage is that the calculation of the trace involves
integration that is effectively restricted to two planes,
whereas the standard version (previous section)
requires a double integration over the whole "bulk".

\sheadC{From the Kubo formula to the Landauer formula}

Before we go on we recall that it is implicit that
for finite system $\im[{\mathsf G}]$ should be "smeared".
In the dot-leads setup which is described
above, this smearing can be achieved by assuming
very long leads, and then simply "cutting" them apart.
The outcome of this procedure is that
${\mathsf G}^{\pm}$ is the Green function of an
open system with outgoing wave (ingoing wave)
boundary conditions. As customary we use
a radial coordinate in order to specify locations
along the lead,  namely  $\ar=\ar_a(r)$, with ${0<r<\infty}$.
We also define the channel basis as
\beq
\langle \ar, s|a, r \rangle \ = \ \chi_a(s) \ \delta(\ar-\ar_a(r))
\eeq
The wavefunction in the lead regions can be expanded as follows:
\beq
|\Psi\rangle = \sum_{a,r} \left(
C_{a,+} \eexp{ik_a r} + C_{a,-} \eexp{-ik_a r} \right)
\  |a, r\rangle
\eeq
We define projectors $P^{+}$ and $P^{-}$ that project
out of the lead wavefunction the outgoing and the ingoing
parts respectfully. It follows that
$P^{+}{\mathsf G}^{+} = {\mathsf G}^{+} $,
and that $P^{-}{\mathsf G}^{+} = 0 $,
and that ${\mathsf G}^{-}P^{-} = 0 $ etc.
We define the operator
\beq
\Gamma_{\tbox{A}} &=& \sum_{a \in A} |a, r_{\tbox{A}}\rangle \hbar v_a  \langle a, r_{\tbox{A}}|
\\ \label{e68}
&=& \delta(r-r_{\tbox{A}}) \otimes \sum_{a \in A} |a\rangle \hbar v_a  \langle a|
\eeq
where $v_a=(\hbar k_a/\text{mass})$ is the velocity in channel ~$a$.
The matrix elements of the second term in Eq.(\ref{e68}) are
\beq
\Gamma_{\tbox{A}}(s,s') =   \sum_{a \in A}  \chi_a(s) \ \hbar v_a \ \chi_a^{*}(s')
\eeq
The operator $\Gamma_{\tbox{B}}$ is similarly defined for the other lead.
Note that these operators commute with the projectors $P^{\pm}$.
It is not difficult to realize that the current operators can be written as
\beq
I_{\tbox{A}} &=& (e/\hbar)[ -P^{+}\Gamma_{\tbox{A}} P^{+} + P^{-}\Gamma_{\tbox{A}} P^{-} ] \\
I_{\tbox{B}} &=& (e/\hbar)[ +P^{+}\Gamma_{\tbox{B}} P^{+} - P^{-}\Gamma_{\tbox{B}} P^{-} ]
\eeq
Upon substitution only two (equal) terms survive
leading to the following version of Kubo formula:
\beq
G^{BA} = \frac{e^2}{2\pi\hbar}\ \trc \left[\Gamma_{\tbox{B}} \ {\mathsf G}^{+} \  \Gamma_{\tbox{A}} \ {\mathsf G}^{-} \right]
\eeq
There is a well known expression (Fisher-Lee) that relates
the Green function between plane $A$ and plane $B$ to the $S$ matrix. Namely:
\beq
{\mathsf G}^{+}(s_{\tbox{B}},s_{\tbox{A}}) = -i\sum_{a,b}
\chi_b(s_{\tbox{B}}) \frac{1}{\sqrt{\hbar v_b}} S_{ba}
 \frac{1}{\sqrt{\hbar v_a}} \chi_a^{*}(s_{\tbox{A}})
\eeq
Upon substitution we get
\beq
G^{BA} = \frac{e^2}{2\pi\hbar}\ \sum_{a \in A} \sum_{b \in B}  |S_{ba}|^2
\eeq
This is the Landauer formula. Note that the sum gives the
total transmission of all the open channels.

\sheadC{From the Kubo formula to the BPT formula}

It should be emphasized that the original
derivations of the Landauer and the BPT formulas
are based on a scattering formalism
which strictly applies only in case of
an open system ($=$ system with leads
which are connected to reservoirs).
In contrast to that Kubo formula is
derived for a closed system.
However, it can be shown that by taking  
an appropriate limit it is possible 
to get the BPT formula from the Kubo formula. 
Namely, 
\beq
\eta^{kj} &=&
\frac{\hbar}{\pi}\ \trc \left[F^k \ \im[{\mathsf G}^{+}] \  F^j \ \im[{\mathsf G}^{+}] \right]
\\ 
&=& \frac{\hbar}{4\pi}
\trc \left[ \frac{\partial S^{\dag}}{\partial x_i} \frac{\partial S}{\partial x_j}\right]
\eeq
\beq
\bm{B}^{3j} &=&
-\frac{i\hbar}{2\pi}\ \trc \left[F^3 \ ({\mathsf G}^{+}{+}{\mathsf G}^{-}) \  F^j \ \im[{\mathsf G}^{+}] \right]
\\
&=& \frac{e}{4\pi i}
\trc \left[ P_{\tbox{A}}\left(
\frac{\partial S}{\partial x_j} S^{\dag}
-\frac{\partial S^{\dag}}{\partial x_j} S
\right)\right]
+ \text{intrf}
\eeq
So the sum is 
\beq
\bm{G}^{3j} \ = \ \frac{e}{2\pi i}
\trc\left(P_{\tbox{A}}\frac{\partial S}{\partial x_j}
S^{\dag}\right)
\eeq
For more details see Phys. Rev. B {\bf 68}, 201303(R) (2003).

\newpage
\makeatletter{}
\sheadB{Irreversibility and Nonequilibrium processes}

\sheadC{The origin of irreversibility}

Assume an isolated system with Hamiltonian  $\mathcal{H}(X)$, 
where $X$ is a set of control parameters that determine  
the ``fields". For simplicity assume that at $t=0$ 
the system is in a stationary state. A driving process 
means that $X=X(t)$ is changed in time. In particular 
a cycle means that ${X(t_{\tbox{final}})=X(t{=}0)}$. 
A driving process is called {\em reversible} is we can 
undo it. In the latter case the combined process 
(including the "undo") is a closed cycle, such that at the 
end of the cycle the system is back in its initial state. 
Generally speaking a driving cycle becomes reversible 
only in the adiabatic limit. Otherwise it is irreversible.

{\bf Micro-reversibility.-- }
One should not confuse reversibility with micro-reversibility. 
The latter term implies that the mechanical evolution has 
time reversal symmetry (TRS). This TRS implies that if we could 
reverse that state of the system at some moment (and also the 
magnetic field if exists)  then ideally the system would come 
back to its initial state. This is called Lodschmit Echo. 
In general it is impossible to reverse the state 
of the system, and therefore in general micro-reversibility 
does not imply reversibility!         

{\bf Sudden process.-- }
The irreversibility of typical systems is related to {\em chaos}. 
The simplest example is free expansion. In this example 
$X$ is the location of a piston. At $t=0$ the system is prepared 
in an ergodic state, say a microcanonical state on the 
energy surface $\mathcal{H}(X_A)=E$.    
The piston is moved outwards abruptly form $X_A$ to $X_B$.   
After some time of ergodization the system will 
become ergodic on  $\mathcal{H}(X_B)=E$. 
There is no way to reverse this process.

{\bf Slow driving.-- } 
The more interesting scenario is a slow cycle. 
Using the assumption of chaos it can be argued that at the end of the cycle the state 
will occupy a shell around  $\mathcal{H}(X_A)=E$.
If the system is driven periodically (many cycles),  
the thickness of this shell grows like $\sqrt{D_Et}$ 
with ${D_E \propto \dot{X}^2}$.  
This diffusion in energy space implies
(with some further argumentation) 
monotonic increase of the average energy.  
Thus irreversibility implies dissipation of energy: 
The system is heated up on the expense of the work 
which is being done by the driving source. 

{\bf Non equilibrium steady state.-- } 
Another reason for irreversibility is having a "frustrated" system 
that is connected to several baths, each in different temperature, 
as in the prototype problem of heat conduction.
Typically, after a transient, a steady state is reached.
But this steady state is not a canonical thermal equilibrium state.
With such configuration one can associate a rate of "entropy production".

\sheadC{The notion of Entropy}

The term "entropy" is used in a diverse way. In order to avoid confusion 
we distinguish between the Shanon entropy, the Von-Neumann entropy, 
the Boltzmann entropy, and the Thermodynamic entropy. 
All are calculated by the same look-alike formula ${S = -\sum_r p_r\log(p_r)}$, 
but the context and the meaning of the $p_r$ is in general not the same.      

{\bf Information entropy:-- }
If $\{p_r\}$ are the probabilities to get an output $r$ of a measurement, 
then $S$ provides a measure for the {\em uncertainty} which is involved in our knowledge 
of the statistical state. This point of view that regards $S$ as an information 
measure has been promoted by Shanon. In the quantum mechanical context 
we define ``classical state" as implying 100\% certainty for {\em any} measurement.
Such states do not exist in Nature. Rather the states of minimum uncertainty 
in $N$ dimensional Hilbert space are the pure states, and they have 
finite information entropy. See \href{http://arxiv.org/abs/quant-ph/0401021}{quant-ph/0401021}.  
They should be contrasted with the worst mixed state whose entropy is ${S=\log(N)}$.

{\bf Von-Neumann entropy:-- }
Von-Neumann has used a Shanon look-alike formula in order 
to characterize the {\em purity} of a quantum state. 
In the Von-Neumann definition, the $p_r$ are the weights 
of the pure states in the mixture, namely $S=-\trc[\rho \ln \rho]$, 
where $\rho$ is the probability matrix.     
It is important to realize that the Von-Neumann entropy has nothing 
to do with the theory of irreversibility. If we consider for example 
the free expansion of a wavepacket in a big chaotic box, 
then we have ${S=0}$ at any moment.
Still it is nice that the Von-Newman entropy of a canonical state 
coincides with the thermodynamic definition.

{\bf Boltzmann entropy:-- }
Boltzmann has defined $S$ is a way that allows to discuss irreversibility. 
The idea was to divide the phase space of a system into small cells,  
and to characterize the statistical state in this representation using~$S$. 
Then Boltzmann has proposed  that $S$~has the tendency to increase with time 
if the motion is chaotic (a variation of this idea is the ``H~theorem"  
that refers to the reduced one-particle description of a gas particle). 
The same reasoning can be carried out in the quantum mechanical context 
where the division of Hilbert space into ``cells" is defined by a complete 
set of projectors. Obviously in the latter context recurrences imply 
that the dynamics of a quantized closed chaotic system  
looks irreversible only for a limited time.  

{\bf Thermodynamic entropy:-- }
Using ideal gas thermometer we have identified the empirical temperature ${\theta=1/\beta}$, 
a notion that is postulated by the 0th law of thermodynamics.   
Later we have shown that $\dbar Q = \sum_r dp_r E_r$ has an 
integration factor $T=1/\beta$ as postulated in thermodynamics (see below).   
Thus we can write $\dbar Q = TdS$, where the definition of $S$ is implied.
It turns out that $S(T)=-\trc[\rho_{eq} \ln \rho_{eq}]$, 
where $\rho_{eq}$ is a canonical state of temperature~$T$.

\sheadC{Digression - traditional thermodynamics}

Let us discuss how "entropy" is defined in "traditional thermodynamics" 
without relaying on Statistical Mechanics. The first step is to characterize 
any thermal state by empirical temperature $\theta$. 
This is well defined by the "zeroth law" of thermodynamics. 
The second step is to represent all the thermal states as points 
in a ${\bm{X} = (X,\theta)}$ space. 
Now we can define "adiabatic surface" as the set of
states that can be reached via a reversible adiabatic 
process that does not involve exchange of energy with the 
environment. We can label each surface by an number ${S(\bm{X})}$, 
that we call "entropy" (with quotations marks).
The convention would be that $S[A]<S[B]$ 
if we can get from $A$ to $B$ via an irreversible process.

If we have a reversible process that starts at point $\bm{X}$, 
and ends at point ${\bm{X}+d\bm{X}}$ the change in "entropy" 
is ${dS=\bm{\nabla} S \cdot d\bm{X}}$. 
At the same time we can write for the 
heat ${\dbar Q = \bm{F} \cdot d\bm{X}}$.      
By definition both ${dS=0}$ and ${\dbar Q =0}$ define
the same adiabatic surfaces. It follows that there is 
an "integration factor" such that ${\bm{F}=T(X,\theta) \bm{\nabla} S }$, 
and hence one can write $\dbar Q=TdS$.
We now postulate that there is a possibility 
to define $S$ such that $T$ is a function of $\theta$ alone.
This leads to the definitions of the ``absolute temperature" 
and of the ``thermodynamic entropy".

Let us rephrase the thermodynamic postulate in a more 
illuminating way. Consider a reversible isothermal process 
at temperature $\theta_H$ that connects two adiabatic surfaces.
Consider a second reversible isothermal process 
at temperature $\theta_C$ that connects the same surfaces.
To say that $\dbar Q$ has an integration factor 
that depends on $\theta$ alone means that the ratio $\dbar Q_H/\dbar Q_C$
depends only on the temperatures $\theta_H$ and $\theta_C$. 
Hence we can define ``absolute temperature" using the definition 
of Carnot, and the definition of $S$ is implied. 

In Carnot's picture the ratio $\dbar Q_H/\dbar Q_C$ 
has to do with the efficiency of the heat transfer process.  
According to Carnot the maximal ratio $\dbar Q_H/\dbar Q_C$
depends only on the temperatures $\theta_H$ and $\theta_C$. 
In "traditional thermodynamics" Carnot's statement 
is regarded as the consequence of either Clausius or Kelvin's statements 
that we derive later. If Carnot's statement were false, 
one would be able to combine two reversible processes that 
do not have the same "efficiency" in order to produce a device 
that can pump heat from cold to hot bath without investing work.     

During a reversible quasi-static process the change of the entropy 
of a system $a$ is ${\int \dbar Q_a / T_a}$, 
while that of a second system is ${\int \dbar Q_b / T_b}$.
If we have $T_a=T_b=\theta$, it follows that 
the total entropy change has an additive property, 
hence entropy is an extensive quantity.

\sheadC{The space of all possible states}

{\bf Canonical states:-- } 
The following visualization is useful. Consider a systems that has 
energy levels $\epsilon_n(X)$. Any canonical state~$\rho$ 
of the system can be represented as a point in a ${(X,T)}$ plane, 
and has some entropy $S(T;X)$. Note that ${T=0}$ states 
have zero entropy and energy $E=\epsilon_0(X)$.  
We can use $S$ as an optional coordinate instead of~$T$, 
and define $E_{\text{eq}}(X,S)$ as the energy of the canonical state that 
has entropy~$S$.   

{\bf Excited states.-- }
We now add a 3rd vertical axis for the energy. 
In this extended ${(X,S,E)}$ representation the canonical states 
form a surface $E_{\text{eq}}(X,S)$. We refer to this surface as the floor. 
Non-canonical states with the {\em same} entropy 
as the canonical state have a higher energy 
and accordingly are represented by points {\em above} the floor.
\beq
E^*[\text{energy of an excited state that has entropy $S$}] \ \ > \ \ E_{\text{eq}}(X,S) 
\eeq
These excited states are represented in the extended ${(X,S,E)}$ 
space as points that reside "above" the canonical state $E_{\text{eq}}(X,S)$. 
Accordingly, all states along a vertical line have the same entropy, 
but only the lowest state "on the floor" is canonical. 
The trivial example is of course the excited pure states, 
that by definition have zero entropy, 
while their energy ${E^*=\epsilon_n(X)}$ is larger than ${E_{\text{eq}}=\epsilon_0(X)}$.

{\bf General processes:-- } 
We visualize a thermodynamic process as a trajectory in the ${(X,S,E)}$ space, 
or optinaly we can project is on the ${(X,T)}$ plane. 
A reversible quasi-static process 
that connects points $A$ and $B$ on the floor is represented 
by a {\em solid} line in the ${(X,T)}$ plane.  
An actual non-reversible process, 
that resides "above" the floor,
is represented by a {\em dashed} line in the ${(X,T)}$ plane.    
In a closed system Boltzmann told us 
that the entropy during a process always increases. 
Loosely speaking this means that the probability at the end of 
the process is scattered on more "energy levels".

{\bf Thermodynamic processes:-- } 
If the system can be attached to baths we can consider 
a more restricted set of processes that we call "Thermodynamic processes".  
Such processes start and end at the "floor".
In other words, we assume that before and after the process 
the system is found in equilibrium with a heat bath. 
The process is irreversible if during the intermediate stages 
it is represented by a dashed line that resides "above" the floor.

\sheadC{The Statistical-Mechanics version of the second law}

The Boltzmann entropy is defined as ${S = -\sum_r p_r \ln p_r}$, 
where $p_r$ is the probability to be in the $r$th cell in phase space.
The Boltzmann version of the second law states that 
for any process from state~"$A$" to state~"$B$" 
\beq
S^{universe}[B] - S^{universe}[A] \ \ > \ \ 0
\eeq 
The Boltzmann entropy is a theoretical construct and hence 
the statistical version of the second law has no practical value.
We have to "translate" both the definition of entropy 
and the "second law" into a thermodynamic language.
For this purpose it is essential to assume that both $A$ and $B$ 
are equilibrium states (while during the process the system may be out of equilibrium). 
Then we can identify the Boltzmann entropies $S(A)$ and $S(B)$ with the 
thermodynamic entropies of states~$A$ and~$B$. We shall see in the  
next section how it helps to formulate a thermodynamic version of the second law 
in terms of "Heat" and "Work".

\sheadC{The thermodynamic version of the second law}

In order to translate this microscopic formulation of Boltzmann into the practical 
language of thermodynamics one assumes: {\bf \ (1)}~In the initial and final states 
the system is in equilibrium with bodies that have well defined 
temperatures $T_A$ and $T_B$ respectively; {\bf \ (2)}~During the process the system 
absorbs heat from equilibrated bodies that have well defined temperatures,  
while the system itself might be out of equilibrium; 
{\bf \ (3)} The change in the entropy of an  equilibrated body that has 
a well defined temperature $T$ is $-\dbar Q/T$, where $\dbar Q$ is the heat transfer 
with the usual sign convention. With these assumptions we get 
the thermodynamic version of the second law:
\beq
\Big[S^{sys}[B] - S^{sys}[A]\Big] - \int_A^B \frac{\dbar Q}{T_{baths}} \ \ > \ \ 0
\eeq 
In particular for a closed cycle we get the Clausius inequality 
\beq
\text{Entropy production} \ \ \equiv \ \ - \oint \frac{\dbar Q}{T_{baths}} \ \ > \ \ 0
\eeq

{\bf Clausius statement:-- }
The simplest application of the Clausius inequality concerns the direction of heat flow.
Consider a cycle $(AB)^{\#}$ in which the system is in contact with~$T_A$, and later in contact with~$T_B$ 
(work is not involved). The result of such cycle is the transfer of 
an amount~$q$ of energy from~$T_A$ to~$T_B$. Assuming $T_B<T_A$ it follows from the Clausius inequality 
that $q$ must be positive, which loosely speaking means that heat can flow only from the high to the low temperature.
(work-free heat pumps do not exist).

{\bf Kelvin statement:-- }
Another immediate implication of the Clausius inequality is that there exist no 
process whose sole result is to transfer heat into work. If such process 
existed one would have at the end of each cycle a single bath with ${Q > 0}$, 
and hence the total entropy of the universe would decrease. 
Also the inverse statement is true: if it were possible to device a work-free pump 
that violates Clausius statement,  then it would be possible to violate Kelvin's statement. 
The proof is based on the possibility to combine such pump device with a Carnot engine.

{\bf Maximum work principle:-- }
Consider an isothermal process. We use the standard assumptions: 
the temperature of the bath is~$T_0$, the initial state is equilibrium, 
and also at the end of the process we wait until an equilibrium is reached. 
Using the first law of thermodynamics (energy conservation) we can 
substitute ${Q =(E(B)-E(A))-\mathcal{W}}$, where $\mathcal{W}$ is the work that  
has been done on the system. 
Using $F(A)=E(A)-T_0S(A)$ and $F(B)=E(B)-T_0S(A)$ we deduce 
from the second law 
\beq
\mathcal{W} \ \ > \ \ [F(B)-F(A)] \ \ = \ \ \text{minimal work required to carry out the process}
\eeq
The work that can be extracted from an engine is $W=-\mathcal{W}$. 
Obviously in order to extract positive work~$W$ we need ${F(A)>F(B)}$.
The {\em maximum} work that can be extracted is equal 
to the free energy difference ${[F(A)-F(B)]}$.
In particular it follows that if the universe included only one bath, with one temperature, 
it would not be possible to extract work from a closed cycle.

{\bf Irreversible work:-- }
Assuming that the state of the system 
is canonical-like at {\em any} instant of time, 
with a well define temperature $T_{\text{sys}}$ 
at any moment along the $A\mapsto B$ process.
We have established that the change of energy 
can be written as $dE=-ydX+T_{\text{sys}}dS^{\text{sys}}$.
The second term originates from transitions between levels. 
These transitions are induced 
by the coupling to the environment and/or 
by the non-adiabaticity of the driving.
On the the other hand by definition ${dE=\dbar\mathcal{W}+\dbar Q}$. 
We have identified $dW=ydX$ as the reversible 
work that could be done by the system.
The irreversible work is the difference ${\dbar{\cal W}_{\text{irvrs}} = \dbar\mathcal{W} - (-dW)}$. 
Accordingly $T_{\text{sys}}dS^{\text{sys}}$ is identified 
as the sum of heat $\dbar Q$ and irreversible work $\dbar{\cal W}_{\text{irvrs}}$.
Namely, 
\beq
T_{\text{sys}}dS^{\text{sys}}  \ \ = \ \ \dbar\mathcal{Q} \ + \ \dbar{\cal W}_{\text{irvrs}}
\eeq 
In an actual experiment the irreversible work ${\cal W}_{\tbox{irvrs}}$   
can be determined by subtracting $-ydX$ from the total work 
that has been done on the system, or it can be deduced 
from the above relation by integrating over ${T_{sys}dS^{\text{sys}}-\dbar\mathcal{Q}}$. 
Obviously the result would be the same, which reflects 
the first law of thermodynamics (conservation of energy).

Optionally the above relation can help us to 
express $dS^{\text{sys}}$ using $\dbar\mathcal{Q}$ 
and $\dbar{\cal W}_{\text{irvrs}}$. Then it is 
possible to rewrite the Clausius statement as follows:
\beq
\int_A^B \frac{\dbar{\cal W}_{\text{irvrs}}}{T_{sys}} 
\ \ + \ \ \int_A^B   \left(\frac{1}{T_{sys}} - \frac{1}{T_{baths}} \right) \dbar Q \ \ > \ \ 0
\eeq 
We see that the origin of reversibility is 
(i)~irreversible work, e.g. frictional effects; 
(ii)~temperature difference between the system 
and the bath during heat conduction.

\newpage
\sheadC{The Carnot Cycle paradigm}

A {\em strict adiabatic process} is a quasi-static process during 
which the system is totally isolated from the environment. 
For such process we have the adiabatic theorem. 
Namely, assuming that the motion is chaotic the system that has 
been prepared with definite energy~$E$ will remain on the  
the same adiabatic surface (classical version) 
or in the same energy level (quantum version) 
if a parameter~$X$ is being changed very slowly.
If the system is prepared with probability $p_n$ in some energy 
shell (classical) or energy level (quantum) then this 
probability will not change during the process, and hence also 
the entropy will remain constant.  In the classical 
version $n$ is the phase space volume of the evolving energy surface, 
while in the quantum mechanical formulation it is the index 
that labels the energy levels. In the classical limit ${n\in[0,\infty]}$, 
and the associated energy is denoted as $E=\epsilon_n(X)$.  
  
We can represent all the possible states of a system as points 
in ${(X,S,E)}$ space as described in a previous section.  
The thermo-adiabatic lines connects canonical points that 
have the same entropy. Such lines are going along the "floor" 
of the ${(X,S,E)}$ space. 
A {\em thermo-adiabatic process} is defined as a quasi-static 
process along a thermo-adiabatic line. We can think of such 
process as composed of many infinitesimal steps, where each 
step consists of a strict adiabatic process followed by a contact 
interaction with a bath that has the appropriate temperature.   

To see that the quasi-static limit exists, note the following: 
If a system is prepared in a canonical state $E_{\text{eq}}(X_0,S)$, 
Then its energy after a strict adiabatic process 
is ${E^* > E_{\text{eq}}(X,S)}$ for any $X$ away from~$X_0$. 
For a small variation $dX$ the energy difference can be expanded 
as ${dE^* \propto dX^2}$. If after such a variation 
the system is connected to a bath that has the {\em appropriate} temperature,
such that $S(T;X)=S$, it would relax to a canonical state 
with the same entropy, but with the lower energy $E_{\text{eq}}(X,S)$.
This relaxation involves an entropy production ${dS^{\text{env}} = dE^*/T}$
due to the release of energy to the bath. 
Integrating $dS^{\text{env}}$ over the whole process we see that 
in the quasi-static limit the entropy production goes to zero.  

A strict Carnot cycle involves only two heat baths.
The cycle $(ABB^*CDD^*)$ is illustrated in the Figure.  
The initial preparation is canonical at~$A(T_1,X_A)$.
The process from $A(T_1,X_A)$ to $B^*(X_B)$ is {\em strictly} adiabatic.
At the end of this stage the obtained state is not canonical. 
The process from $B^*(X_B)$ to $B(X_B,T_2)$ is the equilibration 
due to contact with a bath that has the temperature~$T_2$. 
It is an irreversible relaxation process in which the system 
goes to a lower energy with the same entropy. 
At the end of this process the obtained state is canonical. 
The process form $B(X_B,T_2)$ to $C(X_C,T_2)$ is quasi-static 
in {\em contact} with the same heat bath. 
The process from $C(X_C,T_2)$ to $D^*(X_D)$ is {\em strictly} adiabatic. 
The process from $D^*(X_D)$ to $D(X_D,T_1)$ and later 
back to $A(T_1,X_A)$ is in contact with the heat bath~$T_1$. 

\ \\ 

\includegraphics[width=0.9\hsize]{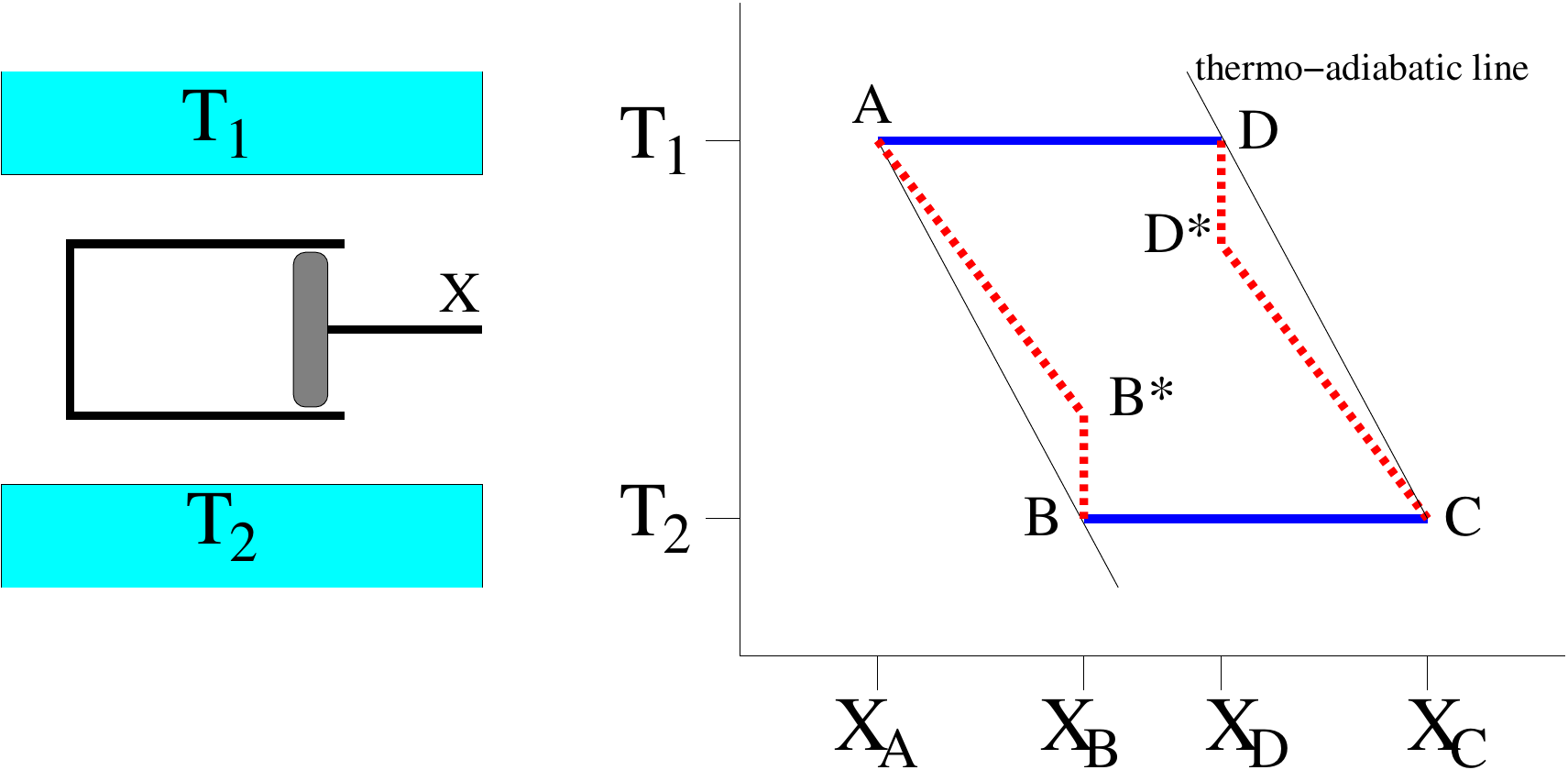}

\newpage
\sheadC{Fluctuations away from equilibrium}

It is customary to say that at equilibrium the expectation  
value of an observable reflects the typical value  
of this observable, while the fluctuations are 
relatively small. If the central limit theorem applies 
the RMS/mean should scale as~$1/\sqrt{N}$. 
However, it turns out that the {\em full} statistics 
might reveal interesting information about the underlying dynamics. 
In the following we shall discuss processes  
where the distribution function of work 
or entropy production does not satisfy 
the symmetry relation $P(-s)=P(s)$. 
Rather it satisfies a detailed-balance look-alike relation:  
\beq
P(-s) \ \ = \ \ \eexp{-\beta s}  \ P(s), \ \ \ \ \ \ \ \ \ \
\text{[beta-symmetric distribution]}
\eeq
It follows that $P(s)$ can be written as a product of 
a symmetric function and an exponential factor $\eexp{\beta s /2}$. 
Another consequence of the $\beta$-symmetry is 
\beq
\langle \eexp{-\beta s} \rangle \ \ = \ \ 1,   \ \ \ \ \ \ \ \ \ \
\text{[convex average]}
\eeq  
The latter equality can be re-phrased as follows:  
In analogy with the definition of {\em harmonic} average and {\em geometric} average 
that are defined as the inverse of $\langle (1/s) \rangle$ 
and as the exp of $\langle \log(s) \rangle$ respectively, 
here we can define a {\em convex} average that is defined 
as the log of the above expression. The convex average is zero 
for a $\beta$-symmetric distribution, while the standard algebraic 
average is positive
\beq
\langle s \rangle \ \ > \ \ 0,   \ \ \ \ \ \ \ \ \ \
\text{[convex inequality]}
\eeq  
While for a symmetric distribution the average 
value ${\langle s \rangle}$ has to be zero, 
this is no longer true for a $\beta$-symmetric distribution. 
Rather the average should be related to the variance.
To be specific let us assume that~$s$ has Gaussian distribution.
It can be easily verified that such distribution 
has $\beta$-symmetry with $\beta=2\mu/\sigma^2$,  
where ${\mu =\langle s \rangle}$  is the average value
and $\sigma^2 =\text{Var}(s)$ is the variance.   
This relation between the first and second moment 
can be regarded as a fluctuation dissipation relation:
\beq
\langle s \rangle  \ \ = \ \ \frac{1}{2} \beta \ \text{Var}(s),   \ \ \ \ \ \ \ \ \ \
\text{["fluctuation dissipation" relation]}
\eeq 
We can formalize this relation for non-Gaussian distribution  
in terms of comulant generating function $g(\lambda)$ which is defined through 
\beq
\langle \eexp{-\lambda s} \rangle  \ \ \equiv \ \ \eexp{g(\lambda)} 
\eeq
Note that due to normalization $g(0)=0$, 
while $g'(0)=-\mu$ and $g''(0)=\sigma^2$. 
In particular for a Gaussian ${g(\lambda)=-\mu\lambda+(1/2)\sigma^2\lambda^2}$. 
For a symmetric distribution $g(-\lambda)=g(\lambda)$. 
But for $\beta$-symmetry we must have  
\beq
g(\beta-\lambda) \ \ = \ \ g(\lambda), \ \ \ \ \ \ \ \ \ \
\text{[characterization of beta-symmetric distribution]}
\eeq
Again we see that for a Gaussian $\beta$-symmetry implies 
a relation between the mean and the variance.

In the following we shall consider two versions of the non-equilibrium 
fluctuation theorem. In one version we consider the statistics $P(\mathcal{W})$ 
of the work $\mathcal{W}$ that is done by an agent during a cycle 
that involves a thermally isolates system.  
In the second version we consider the statistics $P(\mathcal{S})$ 
of the entropy production $\mathcal{S}$ during a cycle that involves 
exchange of energy with several heat baths.

\newpage
\sheadC{The distribution function of the work}

The Crooks relation and Jarzynski equality
concern the probability distribution of the {\em work} 
that is done during a non-equilibrium process. 
For presentation purpose let us consider a gas 
in cylinder with a movable piston.
Initially the piston is in position~$A$, and the gas 
in equilibrium with temperature~$T_0$. The canonical 
probabilities are
\beq
p_r^{(A)} \ \ = \ \ \frac{1}{Z(A)} \eexp{- (1/T_0) E_r^{(A)}},
\hspace{2cm} \text{where} 
\   
Z(A) \ = \ \exp\left[-\frac{F(A)}{T_0}\right] 
\eeq
Now we displace the piston to position~$B$ doing work~$\mathcal{W}$. 
After that we can optionally allow the system to relax to 
the bath temperature~$T_0$, but this no longer affects~$\mathcal{W}$.  
The distribution of work is defines as 
\beq
P_{A\leadsto B}(\mathcal{W}) 
\ \ = \ \ \sum_r p_r^{(A)} \ \delta\Big(\mathcal{W}-(E_r^{(B)}-E_r^{(A)})\Big)
\eeq
It is implicit here that we assume a conservative deterministic classical system 
with a well-defined invariant measure that allows division of phase space 
into "cells". The phase-space states $|r^{(B)}\rangle$ are associated with $|r^{(A)}\rangle$ 
through the dynamics in a one-to-one manner. In other words, the index~$r$ in the above 
definition labels a trajectory that starts at~$r$. If the dynamics is non-adiabatic 
the order of the cells in energy space is likely to be scrambled: if the $E_r^{(A)}$ are indexed 
in order of of increasing energy; it is likely that $E_r^{(B)}$ will become disordered.

If the dynamics is not deterministic the above definition can be modified 
in an obvious way. To be specific let us consider the quantum case, 
where the probability to make a transition form an eigenstate $|n^{(A)}\rangle$ 
of the initial Hamiltonian, to an eigenstate $|m^{(A)}\rangle$ of the final Hamiltonian,  
is given by 
\beq
\mathrm{P}_{A\leadsto B}(m|n) \ \ = \ \  
\Big|\langle m^{(B)}|U_{A\leadsto B}| n^{(A)} \rangle\Big|^2 
\eeq
Then we define the spectral kernel: 
\beq 
P_{A\leadsto B}(\omega) 
\ \ = \ \ 
\sum_{n,m} p_n^{(A)}
\ \mathrm{P}_{A\leadsto B}(m|n) 
\ \delta\Big(\omega-(E_m^{(B)}-E_n^{(A)})\Big) 
\eeq
Since we consider here a closed system, we can identify 
the work as the energy difference $\mathcal{W}=\omega$.
For further discussion of how work can be defined in the 
quantum context see \href{http://arxiv.org/abs/1202.4529}{arXiv:1202.4529}

\sheadC{The Crooks relation}

We have defined the probability distribution $P_{A\leadsto B}(\mathcal{W})$ 
for a process that starts at equilibrium with the piston at position $A$. 
The probability distribution ${P_{B\leadsto A}(\mathcal{W})}$  
is defined in the same way for a reversed process: 
initially the piston is in position~$B$, and the gas 
in equilibrium with temperature~$T_0$, then the piston 
is displaced to position~$A$. The Crooks relation states that 
\beq
\frac{P_{B\leadsto A}(-\mathcal{W})}{P_{A\leadsto B}(\mathcal{W})} 
\ \ = \ \ \exp\left[-\frac{\mathcal{W} - (F(B)-F(A))}{T_0}\right]
\eeq
The derivation of this relation using the "quantum" language is trivial
and follows exactly the same steps as in the derivation of the 
detailed balance relation for any spectral function $\tilde{S}(\omega)$.
The only difference is that here we have an extra factor $\exp[F(B)-F(A)]$, 
on top of the Boltzmann factor, that arises because the $p_n^{(A)}$ 
in the forward process involves a normalization factor $1/Z(A)$, 
while the $p_m^{(B)}$ is the reversed process involves a normalization factor $1/Z(B)$.

\sheadC{The Jarzynski equality}

Multiplying both sides of the Crooks relation by $P_{B\leadsto A}(\mathcal{W})$, 
integrating over $\mathcal{W}$, and taking into account the normalization 
of $P(-\mathcal{W})$, one obtains the Jarzynski equality
\beq
\left\langle \exp\left[-\frac{\mathcal{W}}{T_0}\right] \right\rangle 
\ \ = \ \ \exp\left[ - \frac{F(B)-F(A)}{T_0} \right]
\eeq
It follows from the Jarzynski equality that ${\langle \mathcal{W} \rangle > [F(B)-F(A)]}$, 
which is equivalent to the maximum work principle.
It reduces to  ${\mathcal{W} = (F(B)-F(A))}$ in the 
the case of a quasi-static adiabatic process.  

An optional one line derivation of the Jarzynski equality
in the context of deterministic classical dynamics is as follows:
\beq
\left\langle \exp\left[-\frac{\mathcal{W}}{T_0}\right] \right\rangle 
\ \ = \ \ \frac{1}{Z(A)} \sum_r \eexp{-(1/T_0)E_r(A)} \ \exp\left[-\frac{E_r(B)-E_r(A)}{T_0} \right] 
\ \ = \ \ \exp\left[ - \frac{F(B)-F(A)}{T_0} \right]
\eeq
The Crooks relation could have been derived in a similar way, 
but we had preferred to get it using the "quantum" language, 
and to regard the Jarzynski equality as its implication.

\sheadC{The fluctuation dissipation relation}

Let us see what is the implication on the Crooks relation with regard 
to a simple closed cycle for which ${F(B)=F(A)}$.  
In such case $P(\mathcal{W})$ is a $\beta$-symmetric distribution. 
It follows that there is a "fluctuation dissipation relation"  
\beq
\langle \mathcal{W} \rangle  \ \ = \ \ \frac{1}{2T} \ \text{Var}(\mathcal{W})
\eeq
Considering a multi-cycle process ${\text{Var}(\mathcal{W})=2D_Et}$
and $\langle \mathcal{W} \rangle = \dot{\mathcal{W}}t$, 
leading to the dissipation-diffusion relation that we have 
derived in past lecture ${\dot{\mathcal{W}} = (1/T)D_E}$, 
from which follows the dissipation-fluctuation relation ${\eta=\nu/(2T)}$.

\sheadC{The non-equilibrium fluctuation theorem}

The non-equilibrium fluctuation theorem 
(Bochkov, Kuzovlev, Evans, Cohen, Morris, Searles, Gallavotti) 
regards the probability distribution of the {\em entropy~production} 
during a general non-equilibrium process. 
The clearest formulation of this idea assumes that the 
dynamics is described by a rate equation. The transition rates 
between state~$n$ and state~$m$ satisfies 
\beq
\frac{w(m|n)}{w(n|m)} \ \ = \ \ \exp\left[-\frac{E_m-E_n}{T_{nm}} \right]
\eeq
Where $T_{nm}$ is the temperature that controls the~$nm$ transition.
We can regard the rate equation as describing a random walk process.
Consider a trajectory $x(t)$. If the particle makes a transition 
from~$m$ to~$n$ the entropy production is $(E_m-E_n)/T_{nm}$.
Hence we get for example
\beq
\frac{w(1|2)w(2|3)w(3|4)}{w(4|3)w(3|2)w(2|1)}
\ \ = \ \ \exp\left[-\frac{E_1-E_2}{T_{1,2}}  -\frac{E_2-E_3}{T_{2,3}} -\frac{E_3-E_4}{T_{3,4}} \right]
\ \ \equiv \ \  \eexp{-\mathcal{S}[1 \leadsto 2 \leadsto 3 \leadsto 4]}
\eeq
In general we write  
\beq
\frac{P[x(-t)]}{P[x(t)]} \ \ = \ \ \exp\Big[ -\mathcal{S}[x] \Big]
\eeq
From this "microscopic" relation we deuce that the probability 
distribution of the energy production $\mathcal{S}$ is a $\beta$ 
symmetric function. A simple example for the practicality 
of this relation concerns the fluctuations of the current~$I$ 
that emerge due to the motion of a particle in a ring. 
Given a trajectory $q \equiv It$ is the winding number 
and $\mathcal{S} \equiv qS_{\circlearrowleft}$ is the entropy production.
The non-equilibrium fluctuation theorem implies 
that the ratio $P(-q)/P(q)$ should equal ${\exp(-qS_{\circlearrowleft})}$. 
Note that in the case of an electric current $S_{\circlearrowleft}=eV/T$,
where $V$ is the electro-motive force.

\newpage
\sheadC{Analysis of heat conduction}

A prototype application of the non-equilibrium fluctuation theorem 
concerns the analysis of heat flow form hot bath~$T_H$ to cold bath~$T_C$. 
The temperature difference is ${\epsilon=T_H-T_C}$. We assume that the conductor 
that connects the two baths can be modeled using a master equation. 
The transition between states of the conductor are induced by the bath 
and are like a random walk. With any trajectory we can associate 
quantities $Q_H$ and $Q_C$ that represent that heat flow from the 
baths into the conductor. From the fluctuation theorem it follows that 
\beq
\frac{P(-Q_H,-Q_C)}{P(Q_H,Q_C)} \ \ = \ \  \exp\left[ \frac{Q_C}{T_C}+\frac{Q_H}{T_H}   \right] 
\eeq
Next we define the absorbed energy ${\bar{Q}=Q_H+Q_C}$ 
and the heat flow ${Q=(Q_H-Q_C)/2}$. We realize that 
in the long time limit $Q\sim t$ while the fluctuations 
of $\bar{Q}$ are bounded. Accordingly we get 
\beq
\frac{P(-Q)}{P(Q)} \ \ = \ \  \exp\left[ -\left(\frac{1}{T_C}-\frac{1}{T_H}\right) Q   \right] 
\eeq
If we use a Gaussian approximation, we get a "fluctuation-dissipation" relation  
\beq
\langle Q \rangle \ \ = \ \ \frac{1}{2} \left(\frac{1}{T_C}-\frac{1}{T_H}\right) \text{Var}(Q)
\eeq
The relation can be linearized with respect to ${\epsilon=T_H-T_C}$.
The thermal conductance is defined through ${\langle Q \rangle = K \epsilon \times t}$, 
and the intensity of fluctuations through ${\text{Var}(Q)=\nu \times t}$.
Thus we deduce that 
\beq
\langle \dot{Q} \rangle \ = \ K \times (T_H-T_C), 
\ \ \ \ \ \ \ \ \ \ \ \ \ \ \ 
\text{with} \ K = \frac{1}{2T^2} \nu 
\eeq

\clearpage

\Cn{\LARGE Detailed Table of Contents \\ \ \\ \ \\ } 

\immediate\closeout\tempfile

\makeatletter{} 
{\ \newline ======= \Large Thermal Equilibrium. (page 2)}
 
{\ \newline \bf \large The statistical picture of Mechanics: }
{Random variables; }
{Several random variables; }
{The statistical description of a classical particle; }
{Dynamics in phase space; }
{The route to ergodicity; }
{Stationary states; }
{The microcanonical and canonical states; }
{Mathematical digression; }
 
{\ \newline \bf \large Spectral functions: }
{The definition of counting and partition functions; }
{Two level system or spin; }
{Two spins system in interaction; }
{Harmonic oscillator; }
{Particle in a 1D box; }
{A particle in 3D box, or higher dimensions; }
{Classical particle in magnetic field; }
{Gas of classical particles in a box; }
{Two quantum identical particles; }
{Two quantum particles in a box with interaction; }
 
{\ \newline \bf \large The canonical formalism: }
{The energy equation of state; }
{The Equipartition theorem; }
{Heat capacity; }
{Generalized forces; }
{Susceptibility and fluctuations; }
{Empirical temperature; }
{The Virial theorem; }
{Pressure on walls; }
{Tension of a polymer; }
{Polarization; }
{Magnetization; }
 
{\ \newline \bf \large Thermodynamics: }
{Absolute temperature and entropy; }
{The Thermodynamic potentials; }
{The Gibbs Hamiltonian approach; }
{The chemical potential; }
{The extensive property; }
{Work; }
{Heat; }
{Quasi static process; }
{Cycles; }
 
{\ \newline \bf \large Chemical equilibrium and the Grand Canonical state: }
{The Gibbs prescription; }
{Chemical equilibrium; }
{The law of mass action; }
{Equilibrium in pair creation reaction; }
{Equilibrium in liquid-gas system; }
{Site system; }
{The grand canonical formalism; }
{Fermi occupation; }
{Bose occupation; }
{Bosonic mode occupation; }
 
{\ \newline \bf \large Quantum ideal gases: }
{Equations of state; }
{Explicit expressions for the state equations; }
{Ideal gases in the Boltzmann approximation; }
{Bose Einstein condensation; }
{Fermi gas at low temperatures; }
 
{\ \newline ======= \Large Systems with interactions. (page 46)}
 
{\ \newline \bf \large Interactions and phase transitions: }
{Gas of weakly interacting particles; }
{The grand canonical perspective; }
{The cluster expansion; }
{The Virial coefficients; }
{The Van-der-Waals equation of state; }
{From gas with interaction to Ising problem; }
{Yang and Lee theorems; }
 
{\ \newline \bf \large The Ising model: }
{Model definition; }
{The spatial correlation function; }
{Critical behavior and the scaling hypothesis; }
{Digression regarding scaling; }
{Solution of the 1D Ising Model; }
{Solution of the 2D Ising model; }
 
{\ \newline \bf \large Phase transitions - heuristic approach: }
{The ferromagnetic phase transition; }
{The anti-ferromagnetic phase transition; }
{Beyond the Ising model; }
{The mean-field Hamiltonian; }
{Coupled rotors; }
{The variational approach; }
{The Bragg Williams formulation; }
{The Gaussian approximation; }
{The importance of fluctuations; }
 
{\ \newline \bf \large Phase transitions - field theory: }
{The Landau model; }
{Related models; }
{The Gaussian approximation; }
{Digression - Gaussian integrals; }
{The mean field equation; }
{Symmetry breaking ; }
{The one dimensional model; }
{Coarse graining and scaling; }
{Renormalization Group (RG) analysis; }
{Implications of the RG results; }
{The Heisenberg model; }
{The XY model; }
 
{\ \newline ======= \Large Fluctuations and Response. (page 82)}
 
{\ \newline \bf \large Fluctuations: }
{The classical power spectrum; }
{The quantum power spectrum; }
{The detailed balance relation; }
{The classical version of "detailed balance"; }
{Fluctuations of a many body system; }
{Fluctuations of several observables; }
{Reciprocity relations and detailed balance; }
 
{\ \newline \bf \large Linear response theory: }
{The notion of linear response; }
{Rate of energy absorption; }
{LRT with several variables; }
{The Kubo formula; }
{Memory and Sensitivity; }
{The Onsager regression formula; }
{The Onsager regression hypothesis; }
{Onsager reciprocity; }
{The Kubo formula for AC/DC driving; }
{The Kubo formula - FGR version; }
{Adiabatic response; }
{Low frequency response; }
 
{\ \newline \bf \large The fluctuation dissipation relation: }
{General formulation; }
{The diffusion-dissipation picture; }
{The wall formula; }
{The Drude formula; }
{Conductor in electric field; }
{Forced oscillator; }
{Forced particle; }
{Duality between friction and mobility; }
{The fluctuations of an Ohmic system; }
{The fluctuations of the potential in metals; }
 
{\ \newline ======= \Large System interacting with a bath. (page 103)}
 
{\ \newline \bf \large The modeling of the environment: }
{The Born-Oppenheimer Hamiltonian; }
{The bath Hamiltonian; }
{The bath fluctuations; }
{Spin bath; }
{Spatially extended environment; }
 
{\ \newline \bf \large Stochastic picture of the dynamics: }
{Random walk and diffusion; }
{The Langevin equation; }
{The Fokker-Planck Equation; }
{The Ito-Stratonovich interpretation; }
{Dynamics according to Smoluchowski and Kramers; }
{Rate equations; }
{Rate equations - formalism; }
{Rate equations - counting statistics; }
{Rate equations - ergodicity; }
 
{\ \newline \bf \large Quantum master equations: }
{General perspective; }
{The general Lindblad form; }
{Derivation of the Lindblad form; }
{The Ohmic Master Equation; }
{System-bath interaction; }
{The Redfield master equation; }
{The secular approximation; }
{The Pauli master equation; }
{Damped harmonic oscillator; }
{The Bloch equation; }
{Dicke super-radiance; }
{The Bloch equations in Laser physics; }
{Many body rate equations; }
 
{\ \newline ======= \Large Additional topics. (page 129)}
 
{\ \newline \bf \large The kinetic picture: }
{The Boltzmann distribution function; }
{The Boltzmann equation; }
{The calculation of incident flux; }
{Blackbody radiation; }
{Viscosity; }
{The Navier-Stokes equation; }
{Heat current in an open geometry; }
{Thermo-electricity; }
 
{\ \newline \bf \large Scattering approach to mesoscopic transport: }
{The Buttiker-Pretre-Thomas-Landauer formula; }
{Floque theory for periodically driven systems; }
{The Floque scattering matrix; }
{Current within a channel; }
{The Landauer formula; }
{The BPT formula; }
{BPT and the Friedel sum rule; }
 
{\ \newline \bf \large The theory of electrical conductance: }
{The Hall conductance; }
{The Drude formula; }
{Formal calculation of the conductance; }
{Conductivity and Conductance; }
{From the Kubo formula to the Landauer formula; }
{From the Kubo formula to the BPT formula; }
 
{\ \newline \bf \large Irreversibility and Nonequilibrium processes: }
{The origin of irreversibility; }
{The notion of Entropy; }
{Digression - traditional thermodynamics; }
{The space of all possible states; }
{The Statistical-Mechanics version of the second law; }
{The thermodynamic version of the second law; }
{The Carnot Cycle paradigm; }
{Fluctuations away from equilibrium; }
{The distribution function of the work; }
{The Crooks relation; }
{The Jarzynski equality; }
{The fluctuation dissipation relation; }
{The non-equilibrium fluctuation theorem; }
{Analysis of heat conduction; }

\end{document}